\documentclass[11pt, oneside]{mnthesis}
\usepackage{epsfig,epic,eepic,units}
\usepackage{hyperref}
\usepackage{url}
\usepackage{longtable}
\usepackage{mathrsfs}
\usepackage{bigstrut}

\usepackage{graphicx}
\usepackage{multirow}
\usepackage{colortbl}
\usepackage{pdflscape}
\usepackage{graphicx}
\usepackage{capt-of}
\usepackage{xcolor}
\usepackage{soul}
\usepackage{multicol}
\usepackage{multirow}
\usepackage{booktabs}
\usepackage{caption}
\usepackage{subcaption}
\usepackage{hhline}
\usepackage{mathtools}
\usepackage{algorithm}
\usepackage{algpseudocode}
\usepackage{amssymb}
\usepackage[numbers]{natbib}
\usepackage{pdfpages}
\usepackage{rotating}

\usepackage{tabularray}

\newcommand{\NS}{\mathrm{ns}}

\begin{document}
\bibliographystyle{plainnat}

\phd 

\title{\bf Towards an Effective Organization-Wide Bulk Email System}
\author{Ruoyan Kong}
\campus{University of Minnesota} 
\program{Computer Science} 
\degree{DOCTOR OF PHILOSOPHY}
\director{Advised by Prof. Joseph A. Konstan} 


\abstract{

Bulk email (emails sent to a large list of recipients) is widely used in organizations to communicate messages to employees. It is an important tool in making employees aware of policies, events, leadership updates, etc. However, in large organizations, the problem of overwhelming communication is widespread. Ineffective organizational bulk emails waste employees’ time and organizations’ money, and cause a lack of awareness or compliance with organizations’ missions and priorities.

Prior research mainly studied commercial bulk emails from a single stakeholder’s perspective, such as helping senders improve open rates or helping recipients filter unsolicited bulk emails. However, within organizations, bulk email communication involves multiple stakeholders (employees, communicators, managers, leaders, the organization itself, etc.) with different priorities. The goal of organizational bulk email system is to both reach organization-wide communication effectiveness and provide positive experiences for all the stakeholders. 

This thesis focuses on understanding and improving organizational bulk email systems by 1) conducting qualitative research to understand different stakeholders’ perceptions of the system and its current effectiveness; 2) proposing economic models to describe stakeholders’ actions/cost/value; 3) conducting field studies to evaluate personalization methods’ effects on getting employees to read bulk messages; 4) designing tools to support communicators in evaluating, designing, and targeting bulk emails.

We performed these studies at the University of Minnesota, interviewing 25 employees (both senders and recipients), and including 317 participants in our studies in total. We found that the university's current organizational bulk email system is ineffective as only 22\% of the information communicated was retained by employees. The failure of this system was systemic — it had many stakeholders, but none of them necessarily had a global view of the system or the impacts of their own actions. Then to encourage employees to read high-level information, we implemented a multi-stakeholder personalization framework that mixed important-to-organization messages with employee-preferred messages and improved the studied bulk email's recognition rate by 20\%.
On the sender side, we iteratively designed and deployed a prototype of an organizational bulk email evaluation platform (CommTool). In field evaluation, we found several features (such as bulk emails' message-level performance) of CommTool helped communicators in designing bulk emails. At the same time, to enable these message-level metrics, we collected ground-truth eye-tracking data and developed a novel neural network technique to estimate how much time each message is being read using recipients' interactions with browsers only, which improved the estimation accuracy from 54\% (heuristics) to 73\%. In summary, this work sheds light on how to design organizational bulk email systems that communicate effectively and respect different stakeholders' value.

}
\words{331}    
\copyrightpage 
\acknowledgements{
Thanks to my awesome advisor, Prof. Joseph Konstan. Joe always gave me insightful guidance, opportunities to work on interesting topics, and much encouragement. Sometimes after I completed the experiment, I realized why Joe suggested I revise my experiment design at first (my original experiment design would not work)! 

I am also grateful for working with Prof. Loren Terveen, Prof. Haiyi Zhu, Prof. Lana Yarosh, and Prof. Stevie Chancellor, as well as my committee member Prof. Gedas Adomavicius and Prof. Evan Suma Rosenberg. They all spent much time discussing research projects with me and gave me helpful feedback. 

I am also very lucky to share my Ph.D. journey with all my labmates, including Irene Ye Yuan, Haofei Cheng, Chuankai Zhang, Ruixuan Sun, Qiao Jin, Haiwei Ma, and many many other my friends at Grouplens Lab, who always listened to my practice talks, reviewed my drafts, and conducted studies together with me.

I’d also like to acknowledge funding
from National Science Foundation (IIS-2001851,
IIS-2000782, CNS-1952085, and CNS-2016397), and my study participants, communicators, and employees of the University of Minnesota.

}
\dedication{To my parents Hongxia Zeng and Zhangming Kong. To my grandmother Hewen Hui and grandfather Tengfang Zeng. To my partner Zitao Shen. 
}


\beforepreface 

\figurespage
\tablespage

\afterpreface            


\chapter{Introduction}
\label{intro_chapter}
\section{Background and Problem Statement}
\begin{figure}[!htbp]
\centering
  \includegraphics[width=0.7\columnwidth]{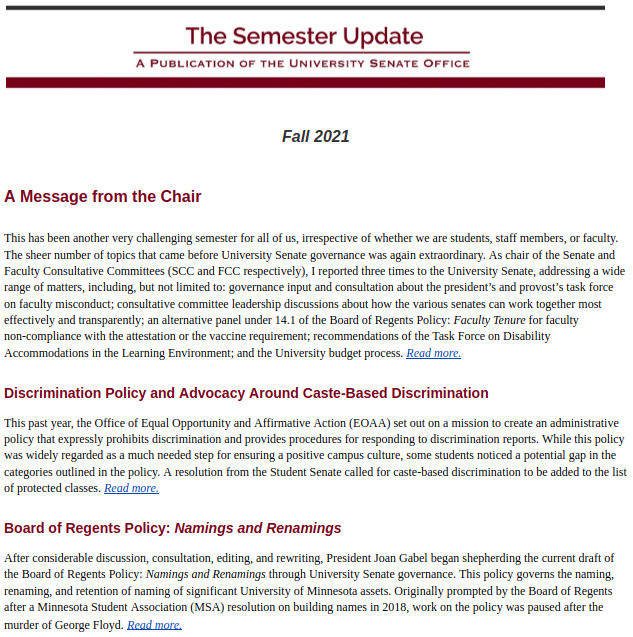}
  \caption[An example organizational bulk email of the studied site]{An example organizational bulk email of the studied site. The whole email includes 19 messages and 1809 words. }~\label{fig:example_email}
\end{figure}

\textbf{Bulk email (or mass email) is email that is sent to a large group of recipients \cite{berghel1997email}}.\footnote{See the appendix \ref{jargonapp} for a full list of terms we define in this thesis.}Bulk email is an essential part of organizational communication. Organizations, such as universities or companies, frequently use internal bulk emails (e.g., newsletters, single bulk emails) in delivering information to their employees. Example messages include changes of leadership, summaries of organizational meetings, announcements of organizational events, updates of organization's policies, etc. We refer to bulk emails within organizations as \textbf{organizational bulk emails} and the systems that generate, design, distribute, process, and evaluate organizational bulk emails as \textbf{organizational bulk email systems}. Figure \ref{fig:example_email} is an example organizational bulk email of our studied site – the University of Minnesota. It is from the university senate office, which wants to inform employees of the policies and plans the university senates are discussing. This office's communicator organizes, designs, and distributes the information. In this thesis, I study such organizational bulk email systems. Specifically, I am motivated by the three insights below.

\textbf{First, organizational bulk emails too often are ineffective.} For example, the email above is long, unpersonalized, and untargeted --- it has 19 pieces of information (\textbf{organizational bulk messages}) within the email; the content is the same for all recipients; it is sent to all faculty and staff of the university. In this university, I found that in a typical week, employees on average receives over 30 such organizational bulk emails with over 250 pieces of information in total per week. Not just in universities, but also in many large organizations, the problem of overwhelming and ineffective communication is widespread. The size of inboxes, the number of unread messages, and the employees’ feeling of email overload keep increasing \cite{grevet2014overload}. Employees are receiving emails irrelevant to them while missing important ones \cite{doi:10.1080/02680513.2018.1556090,Jackson:2003:UEI:859670.859673}. 

\textbf{Second, organizational bulk emails have high costs in time and money to their organizations}. An estimate of the cost of the email above would be
24,000 employees * 2 min (reading and interruption time) * \$0.75 salary+fringe+overhead/min = \$36,000. However, this cost is not calculated, not visible to the sender, and not considered or managed anywhere in the current organizational bulk email system.

\textbf{Third, organizational bulk email systems are interesting theoretically as a multi-stakeholder system \cite{andriof2002introduction}}, within which stakeholders have different priorities. The stakeholders in an organizational bulk email system include:
\begin{itemize}
   \item Communicators (information gatekeeper): the staff in charge of designing and distributing organizational bulk emails.
  \item Communicators' clients (information producer): the original sources of organizational bulk emails, who request the communicators to distribute the information.
   \item Employees (information recipient): the organization's staff who receive organizational bulk emails from the communicators.
   \item Management: the direct managers of the employees.
\end{itemize}

There is also a key stakeholder --- the organization itself --- that has priorities not always recognized by the stakeholders above. Communicators and their clients naturally focus on their own needs --- getting the word out, establishing evidence of notice or compliance, or preserving a record of communication. But the organization, on another side, pays for employees' time in handling those emails and wants to protect the reputation of their communication channels. Employees, faced with more communication than they can handle, start to ignore these communications. In turn, organizational goals around compliance, informed employees, and employee productivity may suffer. 

For example, an announcement of a new child care center would reflect the complexity of this multi-stakeholder problem. The stakeholders might have different perceptions of and actions on this message. The communicators need to complete this task quickly and might achieve that by sending this message as a single email to all the employees. However, those employees who do not have children at that age would have a time cost of reading this message and might question the relevancy of that communication channel. Then the organization's communication channels' reputation might suffer. But on another side, the communicators' clients, such as organization leaders, might perceive this message's value of showing that the organization is family-friendly and are willing to incur that cost. In summary, this communication system needs to carefully balance different stakeholders' value and cost to maximize its productivity.
 
Previous studies on bulk emails are mainly about external bulk emails, which focus on either the senders or the recipients'
 perspectives on bulk emails' value. Work on the sender-side \cite{sahni2018personalization, trespalacios2016effects, wattal2012s} focused on exploring personalization or targeting to improve open rates, click-through rates, conversion rates, etc. Work on the recipient-side focused on helping recipients filter bulk emails \cite{christina2010study, mohammed2013classifying, gangavarapu2020applicability}. However, within organizations, employees may be responsible for reading and knowing about some bulk messages even if they do not like them --- we should not simply filter out organizational bulk emails that employees view as irrelevant. On the other hand, optimizing open rate or other metrics for the organizational bulk email senders could bring time/money cost to the organization when some bulk emails are actually irrelevant to their recipients. 
 
 Given the motivations above, I designed and conducted a series of studies on improving organizational bulk email system with the consideration of multiple stakeholders' perspectives. 

\section{Summary of Chapters}
In this thesis, I will first introduce the related work around organizational communication, bulk email personalization, bulk email evaluation (performance tracking) in Chapter 2. Then in Chapter 3, I introduce the first step of my dissertation work to understand this system and identify research opportunities. I investigated this system's effectiveness, failings, and stakeholders' experiences through surveys and interviews. I found that organizational bulk emails were ineffective and expensive because of two main reasons --- the employees do not like to read some information that is viewed as important by the organization; the senders do not have enough information on employees' interests and cost. In Chapter 4, based on the findings of the empirical study, I further proposed an economic model to describe the value, cost, and actions of this system's stakeholders and how their diverse perspectives cause ineffectiveness. The proposed model enables me to clarify the relationships between this system's stakeholders and identify a bunch of potential interventions, enable to do the next two studies.

With these 2 observations, this thesis explores the opportunity of a personalization tool (chapter 5) and a communication tool (chapters 6 \& 7) for communicators. I started with communicators because they are important gatekeepers between the original sender and the vast majority of employees and the control point of routing information in this system. 

Specifically, the personalization tool I built in chapter 5 employed the mechanism to encourage employees to read the important-to-organization messages. I also built a rule-based model to estimate employees' and communicators' preferences on bulk messages. The communication tool prototype in chapter 7 (chapter 6 is a pre-requisite for chapter 7 in enabling the tool to estimate employees' reading time) evaluated different features' usefulness in supporting communicators to estimate bulk emails/messages' performance, cost, and reputation. I designed that prototype to encourage communicators and original senders to make editing/targeting decisions based on the whole system's effectiveness.

In summary, my dissertation work would provide 1) a better understanding of organizational bulk email system's effectiveness and its stakeholders' perceptions through a survey study, a case study, and an economic model; 2) a personalization tool that encourages employees to read important bulk messages; 3) a communication prototype with a field test that aims to find features that can enable communicators to consider bulk emails' performance, cost, and impact on communication channels' reputations. There are other topics left to be explored on this system, e.g., allocating communication budget, studying who should decide whether a message should be sent out or not, etc. Many of these studies will involve not only HCI community but also the disciplines of management, communication, psychology, etc. I will discuss these opportunities in the discussion section.

In the following, I introduce the studies of my dissertation work, including their background, related work, study methods, and findings. I concluded this thesis by discussing this series of studies' implications and future opportunities.

\chapter{Related Work}
\label{related_work_chapter}

In this chapter, I summarize related work around organizational communication, organizational email systems, and bulk emails that inspire our study.

\section{Organizational Communication}
\citeauthor{morgan1997images} in the book \textit{Images of organization} defined an organization (e.g., a company, a university) as a system with “a loose network of people with divergent interests who gather together to pursue common goals” \cite{morgan1997images}. From the perspective of information theory, \citeauthor{path1968communication} proposed that an organization could be viewed as a communication system that processes and distributes information to coordinate internal tasks and adapt to external changes \cite{path1968communication}. With this definition, \textbf{communication is a key factor in an organization’s coordination and production process}. In fact, organizational communication has been studied for more than a century by various disciplines including sociology, psychology, anthropology \cite{conrad2017history}, etc. Communication has been called \textit{``the life blood of an organization''} \cite{goldhaber1990organizational}, and \textit{``the glue that binds it all together''} \cite{katz2008communication}.

There are various theories on the process of organizational communication, including how to model the process of generating, distributing, and interpreting messages. Many studies aim to motivate organizational communication's various stakeholders to reach organizational goals. For example, \citeauthor{lewis2007organizational} proposed an organizational communication model that connects the selection of communication strategies with stakeholders’ concerns, assessments, and interactions with the system \cite{lewis2007organizational}. \citeauthor{rajhans2012effective} interviewed employees of Vanaz Engineers Ltd., and found that employees' commitment to and trust in the organization increased if they felt that the organization's communication is effective \cite{rajhans2012effective}. \citeauthor{baker2007organizational}  identified diverse stakeholders of organizational communication, including the people who “generate” information (information producers), people who “distribute” information (information gatekeepers), and people who “interpret” information (information recipients) \cite{baker2007organizational}. \citeauthor{welch2007rethinking} also summarized the stakeholder groups of organizational communication and modeled how the achievement of organizational communication's goals is determined by all those stakeholders' actions \cite{welch2007rethinking}.

Given the various stakeholder groups, \citeauthor{jones2004organizational} proposed that handling the relationships between various stakeholders within organization is a major task for the discipline of organizational communication in the 21st century \cite{jones2004organizational}. Several studies have shown that organizational communication affects employees' job performance, job satisfaction, motivation, and the feeling of job security. For example, \citeauthor{giri2010assessing} surveyed 380 employees working at various organizations in India and found that job satisfaction and performance are very much dependent on the communication behavior of the organization \cite{giri2010assessing}. \citeauthor{jiang2014organizational} surveyed 639 employees in six different companies and found that employees who perceived higher levels of positive
organizational communication practices reported fewer negative consequences of job insecurity
compared with employees who reported lower levels of organizational communication \cite{jiang2014organizational}.
\citeauthor{ruck2012valuing} proposed that the goals of organizational communication should include motivating employee commitment, promoting a positive sense of belonging, and developing the awareness of organization needs \cite{ruck2012valuing}. For information producers, Randall \cite{SCHULER1979268} did a case study in a large manufacturing firm, finding that if information producers failed to provide clear information, it was unlikely that the information recipients would continue to seek out information. For management, Stohl and Redding \cite{stohl1987messages} studied a steel construction corporation, finding that the employees' cognitive failures were significantly related to their managers' communication styles. 

Motivated by these studies, for this thesis, we sought to design organizational bulk email systems that respect different stakeholders' preferences.

\section{Organizational Emails}
 Organizational internal emails (we refer to them as organization emails in this thesis) are emails whose senders and recipients are within the same organization. Organizational emails create what Sproull and Kiesler called a “networked organization” \cite{sproull1991computers}, in which people can be available even when they are physically absent. 
 
 There are many studies on organizational emails focusing on employees (recipients) and employee-perceived effectiveness, including:
 
 \begin{itemize}
     \item How organizational email burdened employees: Dabbish and Kraut proposed the concept \textbf{email overload} to describe employees’ perceptions that their own use of
email has gotten out of control because they receive more emails than they can handle effectively, in a study on white-collar workers \cite{10.1145/1180875.1180941}. \citeauthor{whittaker1996email} interviewed 20 office workers in 1996 and found that the problem of email overload is overwhelming \cite{whittaker1996email}. \citeauthor{fisher2006revisiting} revisited this study in 2006 and found that employees' email archives have grown tenfold compared to 1996 \cite{fisher2006revisiting}. Merten and Gloor \cite{merten2010too} found that employee job satisfaction went down as internal email volume increased in a case study in a 50 people company. Huang and Lin \cite{huang2009factors} surveyed three universities and found that knowledge workers were “ruled by email.” 
    \item  Datasets on recipient's data: employees' log data \cite{yi2020natural}, inbox data, and behavior data with organizational emails were collected, including Avocado dataset \cite{yang2017characterizing}, Enron Corpora \cite{bekkerman2004automatic}, and Outlook Dataset \cite{alrashed2019evaluating}, etc.
    \item How the information producers and communicators should change their email actions according to recipients’ needs to reduce email overload: Jackson et al. \cite{Jackson:2003:UEI:859670.859673} studied 16 employees at the Danwood Group and proposed that email frequency should be controlled. Lu et al. \cite{lu2012epic} described EPIC, an email prioritization tool that combined global priority with individual priorities. Reeves et al. \cite{reeves2008marketplace} experimented with attaching virtual currency to emails to signal importance in a large company.
 \end{itemize}
 
These studies show that information overload is a widespread challenge for organizations in email communications. In this thesis, we aim to reduce the information overload caused by a specific type of email --- organizational bulk emails.

\section{Bulk Email}
The development of email is accompanied by the usage of bulk emails (or mass email) \cite{berghel1997email}. There are many studies about \textbf{external bulk emails} --- the bulk emails sent to the recipients outside organizations. External bulk emails usually have goals like advertising products to customers, promoting services to clients, or creating the awareness of the brand to the general public \cite{hartemo2016email}. According to the Data \& Marketing Association’s annual email marketing report, the average return on investment (ROI) of email marketing is 3,800\% in 2020 (\$38 for every \$1 invested) \cite{marketing}. 

\subsection{Filtering External Bulk Email}
Many external bulk emails are unsolicited bulk emails --- their recipients do not opt-in to receive them, which we often called as spam \cite{cranor1998spam}. According to Rao and Reiley’s estimation, American companies and individuals lose \$20 billion annually due to spam for their time filtering these messages \cite{rao2012economics}. Many studies focused on helping recipients filter unsolicited bulk emails by 1) building white lists and black lists \cite{albrecht2005spamato, christina2010study} based on email addresses, IP addresses, domain names, etc.; 2) content-based filtering based on extracted features and words of emails and classification models \cite{mohammed2013classifying, 10.1145/3411763.3451524, kong2021nimblelearn}; 3) agents that check whether the email is from a human, a robot, or malicious senders by peer-to-peer networks \cite{metzger2003multiagent}, CAPTCHA \cite{he2008filtering}, etc. It is worth noting that many organizational bulk emails are also unsolicited --- employees would often automatically be subscribed to some internal mailing lists and receive bulk emails from them. 

\subsection{Personalizing / Targeting External Bulk Email }
The studies on the sender-side focused on developing personalization and targeting methods \cite{marketing} to get higher open rates, click-through rates (CTR), conversion rates, return on investment (ROI), etc. The personalization content of bulk emails could be demographic information like names \cite{sahni2018personalization, wattal2012s}, majors, departments \cite{trespalacios2016effects}; or preference information like browsing history \cite{wattal2012s}, deals or tools recommendation \cite{singh2015email, hawkins2008understanding}. Though Sahni et al.'s experiments found adding recipients' names to subject lines useful \cite{sahni2018personalization}, many other studies supported that the personalization based on preferences performed better than the personalization based on demographics. Wattal et al. \cite{wattal2012s} found that customers responded negatively to emails with identifiable information. Trespalacios and Perkins also found the effect of adding identifiable demographic information insignificant in an experiment with a university email \cite{trespalacios2016effects}. Hawkins et al. pointed out that personalized messages need to provide the recipients with new information about themselves instead of simply adding names or addresses \cite{hawkins2008understanding}. Wattal et al. \cite{wattal2012s} personalized email content based on customers' purchasing preferences and received positive responses. Many personalization designs have been tested on external bulk emails. We summarized them into the following five categories:
\begin{enumerate}
\item Subject line: An informative subject line was believed to be a key factor in successful email marketing \cite{waldow2012rebel}. Sahni et al. added recipients' names to subject lines \cite{sahni2018personalization}, and that method increased a marketing email's open rate by 20\%. However, there are also studies showing that an uninformative subject line could create an information gap that attracts recipients to open emails \cite{doi:10.1080/13645579.2015.1078596,callegaro2009effect}. The ``Long vs. Short Email Subject Line Test'' of WhichTestWon.com in 2011 \cite{waldow2012rebel} found that a longer subject line led to a higher open rate. But Alchemy Worx's test on a discount promotion email found that a longer subject line led to lower open rates \cite{worx2016subject}. An explanation for the contradictory results is that the longer subject lines only influence by providing more information. Under this theory, the factor that actually matters here is whether the topic of the subject line matches the recipients' preferences. The longer subject lines might get lower open rates but higher action rates, because only those who are interested in it will open it \cite{worx2016subject,jaidka2018predicting}. 

\item Top section: The traditional theory is that users would pay more attention to the top positions during browsing \cite{shrestha2007eye}. Wattal et al. \cite{wattal2012s}, and Trespalacios and Perkins \cite{trespalacios2016effects} tried adding recipients' names, majors, or departments to the greetings or the first paragraphs of emails in their studies and Wattal et al. found that greetings influenced their customers' response rates significantly. 

\item Selection of contents: Many studies personalized commercial bulk emails by selecting the most interesting content for recipients. Wattal et al. \cite{wattal2012s} put products that the customer might like most in the email. By analyzing 30 email-marketing campaigns, Rettie \cite{rettie2002email} found that response rate is negatively correlated with email length. Carvalho \cite{carvalho2006personalization} proposed a personalization algorithm that put news liked by similar users in e-newsletters.

\item Order of contents: Besides the theories supporting putting important information on top \cite{shrestha2007eye}, there were also many studies supporting different arrangements. Wojdynski and Nathanie \cite{wojdynski2016going} examined 12 web page designs and found that the advertisements in the middle or bottom positions got better recognition. Heinz and Mekler \cite{heinz2012influence} found that banner placement did not influence recognition and recall.

\item Visual designs: Several studies focused on how to highlight the important content. Rettie \cite{rettie2002email} found that response rate is positively correlated with the number of images. Wojdynski and Nathanie \cite{wojdynski2016going} found that \textit{``users tend to gaze at a target object that is surrounded by objects with weaker “demand for attention” values.''}  

\end{enumerate}

Informed by these studies, we also explore personalization in the context of organizational bulk emails.

\subsection{Bulk Email Platforms}
As a part of CRM (customer relationship management), many large business platforms are supporting bulk email senders in designing and targeting bulk emails, such as Salesforce, Mailchimp, Revue, Constant Contact, etc \cite{bernstein2015research, monitor2012campaign}. 

In terms of designing, targeting, testing, and evaluating bulk emails, the features supported by the two most popular CRM platforms, Salesforce and MailChimp, include:
\begin{itemize}
    \item Targeting: 1) target by user fields that already existed in the database or predicted demographics based on their behavior logs on the websites, 
    
    https://www.salesforce.com/products/marketing-cloud/best-practices/
    
    email-audience-segmentation/, 
    
    https:// help.salesforce.com/s/articleView?
    
    id=sf.networks\_audiences\_rbc.htm\&type=5; 
    
    2) target by behavioral segmentation, which usually needs to connect to the users' behavior logs on websites, 
    
    {https://mailchimp.com/features/predicted-demographics}; 
    
    3) target by users' active status, {https://www.salesforce.com/news/stories/
    
    salesforcemakes-email-marketing-smarter-with-new-einstein-ai-innovations/}.
    \item Personalization: 1) personalize strings based on user fields; 2) trig emails by users' actions on the websites, 
    
    https://help.salesforce.com/s/articleView?language=en\_US 
    
    \&type=5\&id=sf.mc\_es\_available\_perso nalization\_strings.htm, 
    
    https://mailchimp.com /features/personalization/.
    \item Costs: 1) unsubscribe rates; 2) the trend of open rates/click rates/number of subscriptions 
    
    {https://help.salesforce.com/s/articleView?id=sf.pardot\_emails\_open\_ tracking
    
    .htm\&type=5}.
    \item A/B Tests: test the emails on several subscribers and compare their open rates 
 /click rates/unsubscribe rates 
    
    {https://help.salesforce.com/s/articleView?language=en\_US \&type=5\&
    
    id=sf.mc\_es\_ab\_testing.htm}.
\end{itemize}

There are still challenges for these bulk email platforms in data integration, personalization, and evaluation \cite{hartemo2016email}. Several CRM platform designs have been proposed to resolve these challenges:

\begin{itemize}
    \item Data Integration, Targeting, and Personalization: Sukarsa et al. integrated the bulk email platform with a hotel's customer reservation system \cite{sukarsa2020software}. They also reported their whole Software Development Life Cycle for this integration. Kokkalis et al. developed MyriadHub, a crowd-sourced mail client where the crowd workers extracted conversational patterns and got matching templates and suggested responses from MyraidHub \cite{10.1145/3025453.3025954}. Du et al. proposed EventAction, a marketing tool that could identify similar records based on the customers' previous product usage and recommend marketing actions like sending promotional emails \cite{10.1145/3170427.3188531}.
    \item A/B Testing: Xu et al. introduced XLNT, a large-scale A/B test platform on Linkedin \cite{xu2015infrastructure}. XLNT focused more on testing the experiments on Linkedin.com instead of emails. 
    \item Customer Feedback: Garbett et al. \cite{garbett2016app} implemented a promotion platform for the campaigns of mobile applications. They sent promotional emails to ``the most influential'' community members and reached 11 successes in 27 campaigns.
\end{itemize}

These studies on external bulk emails show that personalization and evaluation tools helped bulk email senders complete their tasks and improved bulk email recipients' experience. In this thesis, we will also develop these tools within the context of organizational bulk emails to improve this system's effectiveness.




\chapter{Mixed-methods Study on Understanding An Organizational Bulk Email System}

\newcommand{\hlc}[2][yellow]{{%
    \colorlet{foo}{#1}%
    \sethlcolor{foo}\hl{#2}}%
}
\label{study1}




\section{Introduction}
First and foremost, we want to \textbf{understand an example organizational bulk email system}. In this chapter, we introduce a mixed-methods study (interviews and surveys) we conducted to understand our study site's (University of Minnesota) organizational bulk email system's effectiveness and its various stakeholders. \footnote{Ruoyan Kong, Haiyi Zhu, and Joseph A Konstan. Learning to ignore: A case
study of organization-wide bulk email effectiveness. Proceedings of the ACM on
Human-Computer Interaction, 5(CSCW1):1–23, 2021.
\cite{kong2021learning}; Ruoyan Kong, Haiyi Zhu, and Joseph Konstan. Organizational bulk email systems: Their role and performance in remote work. New Future of Work, 2020 \cite{kong2020organizational}.}

As we discussed in the introduction, a feature of organizational bulk email system is its multiple stakeholders. Previous research on organizational emails mainly studied general emails from recipients' perspectives, such as which emails employees would delay reading \mbox{\cite{Sarrafzadeh2019CharacterizingAP}}, the probability of an email being retained/deleted \mbox{\cite{10.1145/765891.766073}}, and \mbox{recipients’} productivity \mbox{\cite{mark2016email}}. However, bulk email communication within organizations is not only a problem involving email recipients but is also an example of a multi-stakeholder problem \cite{andriof2002introduction}. The stakeholders in organizations include communicators, their clients, recipients (employees), and their direct managers.
 Not only are there different goals for communicators, their clients, and recipients, but also there is a key fourth stakeholder --- the organization itself --- which has its priorities not always recognized by communicators, their clients, and recipients.  Communicators and their clients naturally focus on their own needs --- getting the word out, establishing evidence of notice or compliance, or preserving a record of communication. But recipients, faced with more communication than they can handle, have to scan, filter, or simply ignore organizational bulk emails. In turn, organizational goals around compliance, informed employees, and employee productivity may suffer.
 
  


 Thus to \textbf{understand the multiple stakeholders' perspectives} within organizational bulk email systems, we conducted a mixed-methods case study. Case studies are the predominant method of conducting organizational studies \mbox{\cite{swanborn2010case, buchanan2012case}} and organizational communication studies \mbox{\cite{Sarrafzadeh2019CharacterizingAP, 10.1145/765891.766073, mark2016email, SCHULER1979268, stohl1987messages}}. A case study enables us 1) learn domain experts' (communicators) information management techniques and how the bulk email system is used within the organization; and 2) understand the links between different stakeholders within the same organization \mbox{\cite{lazar2017research}}. We selected a representative organization (University of Minnesota) with hierarchical structures, and centralized and decentralized communication offices. We aim to 1) inform the design of the organizational bulk email systems \mbox{\cite{yin2003case}}; 2) provide interview protocols for future studies on such systems; and 3) be generalizable to similar organizations.

\textbf{We conducted a survey study and an interview study} at the University of Minnesota, to learn the effectiveness of its organizational bulk email system, the communicators' practice of designing and distributing organizational bulk emails, and the communicators/recipients/managers' experience and assessments. We analyzed self-reported data, logged inbox data, and inbox-review data collected from surveys and interviews (artifact walkthroughs \cite{10.1145/227614.227615}). This study inspired us to develop personalization tools and evaluation tools to support communicators in designing / targeting in Chapters 5 to 7.


\section {Gaps and Research Questions}
From the related work on organizational communication and bulk email (see Chapter 2), we identified 2 major gaps:
\begin{itemize}
    \item  The studies on organizational emails focused on recipients: though studies on organizational communication pointed out that when measuring organizational communication’s effectiveness, we should consider not only recipients, but also information providers (communicators and their clients) and the organization itself \cite{stohl1995organizational,jones2004organizational}. Previous studies on organizational emails have not investigated whether recipients and information providers within an organization have different preferences about emails, and how these mismatches affect an organization.
    \item  The studies on organizational bulk emails have focused on external bulk emails \cite{hartemo2016email}: few studies of bulk emails have examined instances where the bulk email sender and recipients are part of the same organization, where the ultimate goal is maximizing the whole organization's interests rather than the information pro- viders' or the recipients' interests \cite{10.1145/1180875.1180941, merten2010too}. Within organizational context, whether a recipient (employee) should receive an organizational bulk email may not be determined by whether the recipient likes the email. Sometimes employees may be responsible for knowing about this email though they might not be interested in it. We see a need to understand bulk emails from an organizational context. 
\end{itemize}

Based on these gaps we identified a need to undertake a systemic investigation of bulk email communications from a whole organization's view. Notice that we \hlc[white]{referred} to organizational bulk email as \textbf{bulk email} and organizational bulk email system as \textbf{bulk email system} in the rest of this chapter. We \hlc[white]{posed} the following research questions:

\noindent\textbf{Q1:}  How effective is the organizational bulk email system at communicating
information to employees?

\noindent\textbf{Q2:}  What are communicators' current practices for designing and distributing bulk emails? 

\noindent\textbf{Q3:}  What are the experiences with and assessments of bulk emails of different stakeholders?

We conducted a survey study and an in-depth case study of a representative organization. We interviewed both communicators and recipients within an organization using an artifact walkthrough approach. We delved into email cases and got their assessments of email values. This chapter will discuss communicators, recipients, communicators' clients, and recipients' managers.

\section{Study Site}
We carried out our studies at the University of Minnesota, a large university with approximately 24,000 employees and five campuses. We met with a group of 9 communicators within this university at the beginning of this project. In the meeting, we found agreement that the current bulk email system was not what the university wanted it to be, since the communicators felt that bulk emails were ignored too often. They encouraged us to move forward. 

\begin{figure}[!htbp]
\centering
  \includegraphics[width=1\columnwidth]{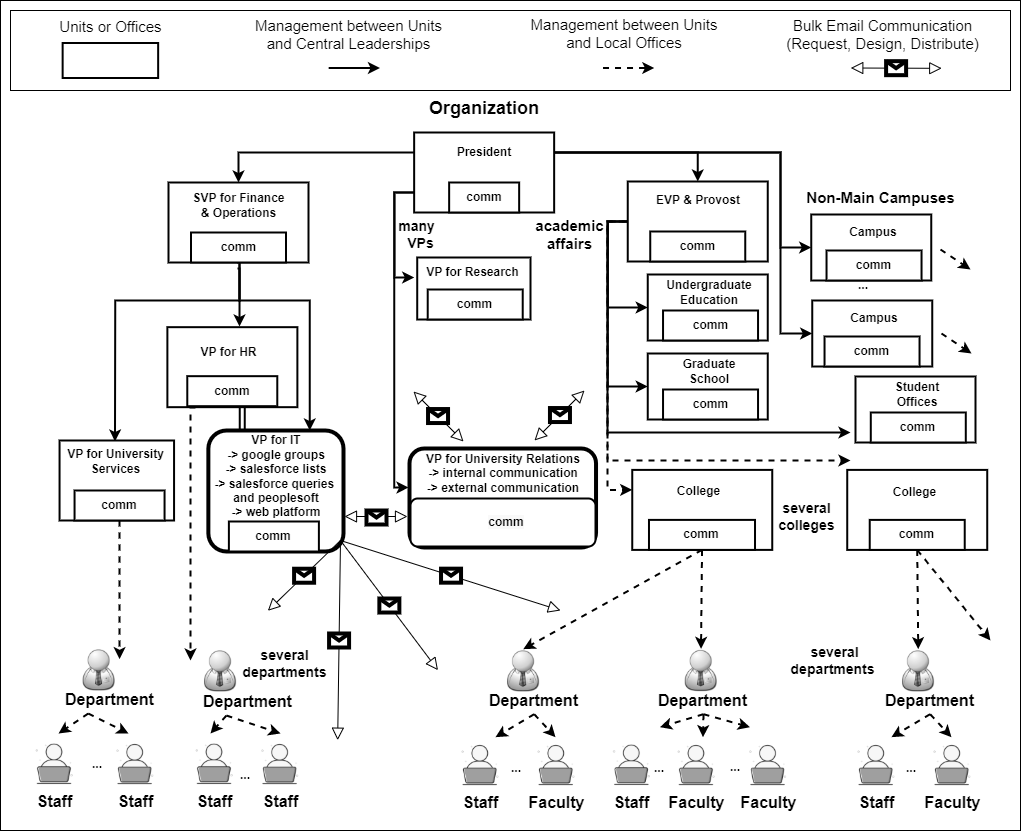}
  \caption[Organization structure of the University of Minnesota related to bulk communication]{Organization structure of the University of Minnesota related to bulk communication; ``SVP'', ``EVP'', ``VP'': (Senior/Executive) Vice President’s offices; “comm”: communication units or communicators. Details of other campuses and many central units are not shown. Some units also conduct external bulk communication (e.g., emails to alumni and the general public), which is not shown here.}~\label{fig:bulk_structure}
\end{figure}

Figure \ref{fig:bulk_structure} is the structure of the organizational bulk email system. Like many large universities, it has both centralized and decentralized governing structures. Each major unit has a communication office or communicators (communication director/staff). There are central units led by university leaders responsible for the core business functions. The University Relations office (under the Vice President for University Relations) is in charge of internal communications for the university. Each Vice President’s office has a communication unit or several communication staff, who draft and send emails for university leaders and conduct university-wide bulk email communications through newsletters. Many central units run their own newsletters. Employees are subscribed to many of these newsletters by default. 

The non-central units like colleges' Deans’ offices also perform bulk communications locally. Their communicators conduct local communications and also work with central communication offices, such as collecting and submitting content to the central newsletter editors, or redistributing the messages from the central units.  

The IT infrastructure provides technical support for the bulk email system, including an email system (Gmail), an email marketing platform (Salesforce), and a human resource system (Peoplesoft). The email marketing platform controls the access to sending bulk emails to campus-wide groups via mailing lists or query-based databases. Only trained communicators have access to the corresponding authorized mailing lists. It is worth noting that querying the exact group of employees can often be difficult and time-consuming. For example, there is no clear definition of ``researchers'' in the PeopleSoft database. To send a message to all the ``researchers'' of the university, the communicators need to read the definitions of hundreds of job codes to judge whether each job code should be included in the mailing list.

\section{Survey Study Design }
To test the effectiveness of channels without the influence of message content, we designed survey questions in pairs: a real message and a corresponding fake message. We built 11 such message pairs and asked the participants to indicate whether they could recognize each message. 162 responses were collected.\footnote{Ruoyan Kong, Haiyi Zhu, and Joseph Konstan. Organizational bulk email systems: Their role and performance in remote work. New Future of Work, 2020.
 \cite{kong2020organizational}} 

The participant pool was defined as employees who had received the real messages we selected in the past 2 weeks before the survey was distributed, were not senior leaders of the university, and were part of the university's volunteer pool for conducting usability studies on university systems. We randomly selected 3000 potential participants from the participant pool.

We distributed surveys to all of the potential participants through invitation emails with survey links. Participation was voluntary and uncompensated. We collected 162 completed responses, and there were 30 additional responses that were incomplete. The data we show below are calculated based on the completed responses. It is worth noting that the participants who completed the survey were a population likely to be more engaged with organization email than average. The effectiveness calculated based on their data would likely be higher than the effectiveness based on the whole recipient population within this university.

To test the effectiveness of channels without the influence of message content, we designed survey questions in pairs: a real message and a corresponding fake message. We wanted to test whether the recipients could recognize the real messages from the fake messages. The real and fake messages had similar content features. We built 11 such message pairs with different channel and content features.

For the real message, the standards of selection were:
\begin{itemize}
    \item It was received by all the potential participants (employees).
    \item It had general importance to the university and the participants.
\end{itemize}

For the fake message, the standards of selection were:
\begin{itemize}
    \item It had a similar importance level as the corresponding real message.
    \item It had the same actionable, relevancy features as the corresponding real message.
\end{itemize}

After we designed the message pairs, we worked with the communicator group at the university to verify the messages we selected matched the standards above. 

In the survey, we asked the participants to indicate whether they could recall the message on a 5 point likert scale where 1 is ``I Have Not Seen it'' and 5 is ``I Have Seen it''. The purpose of this investigation was to test how well participants recalled the message they received, with the corresponding fake message with similar content features serving as a control. When analyzing survey responses, a score of 4 or 5 was considered as having seen the message, while 1 to 3 indicated having not seen it.

\section{Survey Study Findings}
\textbf{Employees did not retain most of the bulk messages.} On average, the real messages were claimed seen by $38\%$ participants, but the fake messages were also claimed seen by $16\%$ participants. That suggested a $22\% = 38\% - 16\%$ average effectiveness. $22\%$ of the real bulk messages could be recalled by participants. $62\%$ of real messages were not claimed as seen. $16\%$ of real messages claimed as seen could not be discerned from the fake messages. See Table \ref{tab: effect} for effectiveness and percentage of recognition of all messages.

\begin{table}[!htbp]
\centering
\caption[Effectiveness of each test message]{Effectiveness of each test message. For each message group, effectiveness of the real message in it was defined as $\% real \ message \ claimed \ seen \ by \ participants - \% fake \ message \ claimed \ seen \ by \ participants$.}
~\label{tab: effect}
\scalebox{0.7}{
\begin{tabular}{|c|c|c|c|c|c|c|c|c|c|c|c|c|} 
\hline
\textbf{message no.}               & 3    & 7    & 1    & 2    & 4    & average  & 11   & 10   & 5    & 9     & 8     & 6     \\ 
\hline
\textbf{effectiveness}           & 0.94 & 0.39 & 0.34 & 0.3  & 0.25 & 0.22 & 0.22 & 0.13 & 0.12 & -0.01 & -0.02 & -0.1  \\ 
\hline
\textbf{\%real claimed seen} & 0.98 & 0.42 & 0.51 & 0.38 & 0.41 & 0.38 & 0.36 & 0.17 & 0.18 & 0.08  & 0.31  & 0.38  \\ 
\hline
\textbf{\%fake claimed seen} & 0.04 & 0.03 & 0.16 & 0.08 & 0.16 & 0.16 & 0.14 & 0.03 & 0.06 & 0.09  & 0.34  & 0.49  \\
\hline
\end{tabular}
}
\end{table}

The message that was most effective was the single message that was relevant and actionable to most of the recipients (``Your 2019 W-2 tax-reporting form is now available online. (message 3)'') --- $98\%$ of the recipients claimed that they had seen it, and it was the only one that had effectiveness over $50\%$. The real message that the least recipients claimed seen is the message about another campus in the newsletter (message 9) --- only $8\%$ of participants claimed that they had seen this message. The message that was most ineffective was from the president (message 6) --- though $38\%$ participants claimed they had seen the real message, $49\%$ claimed they had seen the fake one, which might mean that the participants remember the types of bulk messages they received, but did not retain these messages' content.

\section{Artifact Walkthrough Design}

 What are the practices, experiences, and assessments of different stakeholders in the bulk email system at this university? We conducted artifact walkthroughs with 17 stakeholders --- 6 communicators, 9 recipients, and 2 managers within the university.\footnote{Ruoyan Kong, Haiyi Zhu, and Joseph A Konstan. Learning to ignore: A case
study of organization-wide bulk email effectiveness. Proceedings of the ACM on
Human-Computer Interaction, 5(CSCW1):1–23, 2021. \cite{kong2021learning}} An artifact walkthrough is used to walk the interviewer through the process of completing a task. It helps the interviewer infer possible design intent by
asking probing questions regarding the decision between various options \cite{beyer1999contextual}. In our study, for example, recipients used certain queries in their inbox to retrieve specific types of emails; answered questions about their actions (trashed/opened/scanned/read in detail) and assessments (sometimes they were asked to reread the email) on each email case retrieved.

\subsection{Recruitment and Participants}
We interviewed 17 participants in the same university using the following process:
\begin{enumerate}
    \item   We worked with the same group of communicators in section 3.3 to generate a list of potential participants by a stratified sampling approach,  which included 8 communicators and 20 recipients, based on length of work experience, nature of the office, and job responsibility.
    \item   We began inviting participants for a 30 to 60 minute interview in our lab or their own offices.
    \item   We stopped inviting after we interviewed 6 communicators and 9 recipients as we found the addition of new interviewees did not add any new observations (i.e., we had reached data saturation).
    \item   To avoid recipients' answers being influenced by their awareness of manager involvement, we waited until after the interview was complete to ask the recipient if they would allow us to invite their direct manager to discuss the non-personal emails we had collected from them. Two recipients agreed and we invited their managers for an extra 30-minute interview.
\end{enumerate}


This study was reviewed and approved by the IRB of the University of Minnesota \footnote{IRB ID: STUDY00006417}. 17 individuals  participated ( Table \ref{tab:demographic}). These include: 

\begin{enumerate}
\item 6 communicators --- 2 communication directors who had over 10 years of experience and 4 communication staff, from 5 different central offices and 1 college office; their responsibilities included editing newsletter, writing drafts for university leaders, leading technical support, etc.

\item 9 recipients --- 5 staff (1 from a central office, 2 from college offices, 2 from departmental offices) and 4 faculty; 5 recipients had been at the university for over 10 years.

\item 2 managers --- R5's manager (a college office's director) and R6's manager (a program director) were invited.
\end{enumerate}

\subsection{Interview Protocol}
\noindent\textbf{A. Communicators.} The artifact-walkthroughs with communicators were composed of 3 parts: 
\begin{itemize}
\item General practice questions on their duties, goals, and mechanisms of distributing bulk emails.
\item Email case questions on how an email was designed, sent and measured. We asked communicators to select important/unimportant cases from their points of view.
\item General assessment questions, see Table ~\ref{tab:gatekeeper_protocol} in the appendix.
\end{itemize}

\begin{table}[!htbp]
\centering
\footnotesize
\begin{tabular}{|p{5cm}|p{8cm}|} 
\hline
 \textbf{Questions}                                                                                  & \textbf{Command}                                                                                                                                                                                                          \\ 
\hline
How many emails did the participant receive within 1 week?                                      & newer\_than:7d,in:anywhere AND NOT from:me                                                                                                                                                                                 \\ 
\hline
How many emails did the participant receive and not read within 1 week?                              & newer\_than:7d,in:anywhere AND label:unread AND NOT from:me                                                                                                                                                                \\ 
\hline
How many emails did the participant receive from their organizations within 1 week?                  & newer\_than:7d from:umn.edu AND NOT from:me ,in:anywhere                                                                                                                                                                   \\ 
\hline
How many emails did the participant receive from their organizations and unread within 1 week?       & newer\_than:7d,in:anywhere AND NOT from:me AND label:unread AND from:umn.edu                                                                                                                                               \\ 
\hline
How many mass emails did the participant receive from the organization within 1 week?                & newer\_than:7d,from:umn.edu AND NOT from:me AND (list:(local) OR list:(list) OR to:(lists) OR (category:(Forums\textbar{}Promotions) )) ,in:anywhere                                                                       \\ 
\hline
How many mass emails did the participant receive from the organization and unread within 1 week?     & newer\_than:7d,from:umn.edu AND NOT from:me AND label:unread AND (list:(local) OR list:(list) OR to:(lists) OR (category:(Forums\textbar{}Promotions))) ,in:anywhere                                                       \\ 
\hline
How many newsletters did the participant receive from the organization within 1 week?                & newer\_than:7d,from:umn.edu AND NOT from:me AND (list:(local) OR list:(list) OR to:(lists) OR (category:(Forums\textbar{}Promotions) )) AND (subject:(news\textbar{}update\textbar{}brief)) ,in:anywhere                   \\ 
\hline
How many newsletters did the participant receive from the organization and unread within 1 week?     & newer\_than:7d,from:umn.edu AND NOT from:me AND label:unread AND (list:(local) OR list:(list) OR to:(lists) OR (category:(Forums\textbar{}Promotions))) AND (subject:(news\textbar{}update\textbar{}brief)) ,in:anywhere  \\ 
\hline
How many personal emails did the participant receive from the organization within 1 week?            & newer\_than:7d,from:umn.edu AND NOT from:me AND to:umn.edu AND NOT (list:(local) OR list:(list) OR to:lists OR (category:(Forums\textbar{}Promotions))) ,in:anywhere                                                      \\ 
\hline
How many personal emails did the participant receive from the organization and unread within 1 week? & newer\_than:7d,from:umn.edu AND NOT from:me AND to:umn.edu AND NOT (list:(local) OR list:(list) OR to:lists OR (category:(Forums\textbar{}Promotions))) ,in:anywhere ,label:unread                                        \\
\hline
\end{tabular}
\caption[Email searching commands for Gmail]{Email searching commands for Gmail. The University of Minnesota used a Gmail system based on G Suite for Education.}
\label{tab:command}
\end{table}

\noindent\textbf{B. Recipients.} The artifact-walkthroughs with recipients were composed of 3 steps: 
\begin{itemize}
    \item {Collect inbox logged data: They were asked to copy and paste 10 email queries in their Gmail's search box (see Table \mbox{\ref{tab:command}}) to retrieve subsets of organizational emails, non-mass emails, mass emails, and newsletters they had received in the previous week. We recorded the number of emails left unread/opened of each category (see the actions' definitions in Table \ref{tab:action_definition} --- logged left unread/opened).
    
      Notice that we retrieved emails in all categories, including those archived and in trash/spam, by command ``in:anywhere'', except those permanently deleted from the trash. By default, only emails that have been in the trash for more than 30 days would be automatically deleted by Gmail. We did not observe the case that the trash was manually cleaned by the recipients. We assumed that we could retrieve all emails received in the past 7 days.
    
      These queries were pretested in one of the authors' inbox to ensure they retrieve the right subset of emails. We recorded the size of each subset, and how many were opened.}
    \item Collect self-report data: For the mass/newsletters we retrieved above, we investigated up to 12 emails for each category. For each email, we recorded the recipient's actions on: 1) whether the email was left unread/opened/scanned/read in detail/trashed (see the actions' definitions in Table \ref{tab:action_definition} --- self-report left unread/opened/ scanned/read in detail/trashed); 2) whether the email asked for actions/whether the recipient took actions; 3) importance/urgency/relevance on scale 1 to 5; 4) reasons for the answers/actions, and 5) whether they would change their mind about answers/actions if they reread it now (see appendix \ref{app4}). We investigated 163 cases in total.
    \item General questions: We asked recipients some general questions such as how frequently they checked their email accounts, how many emails accounts they had, did they feel that the number of emails they received was too many to read all of them, and how often they did not read an email. 
\end{itemize}


\noindent\textbf{C. Managers.} The answers from the employee-manager pair were confidential from each other. We showed the managers the emails we collected from their employees, and asked them to give their preferred actions that their employees ought to have done with those emails --- trashed/left unread/opened/scanned/read in detail, and estimations on importance/urgency from 1 to 5. We investigated 31 emails in total with the two managers. \hlc[white]{Note that the term ``manager'' here means the direct supervisor of the recipients, and is not a reflection of a particular position in the organization. Furthermore,
the recipients listed as "director" in Table \mbox{\ref{tab:demographic}} are heads of their offices or programs but are not listed as "manager" because we interviewed them about their own received messages, not about the messages of one of their employees.} 

\hlc[white]{To summarize, for participants invited as communicators/recipients/managers, we discussed the bulk emails they sent/received/their employees received correspondingly.}

\begin{table}[!htbp]
\centering
\resizebox{\textwidth}{!}{%
\begin{tabular}{|c|c|} 
\hline
 \textbf{The Actions of Recipients}  & \textbf{Definition}                                                                                                                                                                                                 \\ 
\hline
Logged Left Unread                   & The email was labeled unread by Gmail.                                                                                                                                                                              \\ 
\hline
Self-Report Left Unread              & The recipient claimed that they did not open the email.                                                                                                                                                             \\ 
\hline
Logged Opened                        & The email was labeled read by Gmail.                                                                                                                                                                                \\ 
\hline
Self-Report Opened                   & The recipient claimed that they opened the email.                                                                                                                                                                   \\ 
\hline
Self-Report Scanned                  & The recipient claimed that they read the email quickly to get its general idea only.                                                                                                                                \\ 
\hline
Self-Report Read in Detail           & The recipient claimed that they read the whole email thoroughly and carefully.                                                                                                                                      \\ 
\hline
Self-Report Trashed                  & \begin{tabular}[c]{@{}c@{}}The recipient claimed that they archived the email or moved it to the trash folder; \\untrashed email was left in the inbox or moved to a non-trash folder.\\ \end{tabular}  \\ 
\hline
Self-Report Opened and Trashed       & The recipient claimed that they opened the email first then trashed it.                                                                                                                                             \\
\hline
\end{tabular}}
\caption[The definitions of recipients' actions with emails]{The definitions of recipients' actions with emails. Logged data was collected by using queries to retrieve subsets. Self-report data was collected by asking recipients their actions directly. Each participant provided both of the logged data and self-report data.}\vspace{-0.3in}
\label{tab:action_definition}
\end{table}

\begin{table}[!htbp]
\centering
\resizebox{\textwidth}{!}{%
\begin{tabular}{|c|c|c|c|c|c|c|c|c|c|} 
\hline
\textbf{\#} & \begin{tabular}[c]{@{}c@{}}\textbf{Stakeholder}\\\textbf{Type}\end{tabular} & \begin{tabular}[c]{@{}c@{}}\textbf{Years at }\\\textbf{University}\end{tabular} & \textbf{Position} & \begin{tabular}[c]{@{}c@{}}\textbf{Level of}\\\textbf{Office}\end{tabular} & \textbf{\#} & \begin{tabular}[c]{@{}c@{}}\textbf{Stakeholder}\\\textbf{Type}\end{tabular} & \begin{tabular}[c]{@{}c@{}}\textbf{Years at}\\\textbf{University}\end{tabular} & \textbf{Position} & \begin{tabular}[c]{@{}c@{}}\textbf{Level of}\\\textbf{Office}\end{tabular}  \\ 
\hline
C1          & Communicator                                                                & 6 - 10                                                                          & Staff             & Central                                                                    & R1          & Recipient                                                                   & 1 - 5                                                                          & Staff             & Central                                                                     \\ 
\hline
C2          & Communicator                                                                & 11- 20                                                                          & Director          & Central                                                                    & R2          & Recipient                                                                   & 1 - 5                                                                          & Staff             & Departmental                                                                \\ 
\hline
C3          & Communicator                                                                & 1 - 5                                                                           & Staff             & College                                                                    & R3          & Recipient                                                                   & 11 - 20                                                                        & Director          & Departmental                                                                \\ 
\hline
C4          & Communicator                                                                & 11 - 20                                                                         & Director          & Central                                                                    & R4          & Recipient                                                                   &  20                                                                            & Staff             & College                                                                     \\ 
\hline
C5          & Communicator                                                                & 1 - 5                                                                           & Staff             & Central                                                                    & R5          & Recipient                                                                   & 1 - 5                                                                          & Staff             & College                                                                     \\ 
\hline
C6          & Communicator                                                                & 6 - 10                                                                          & Staff             & Central                                                                    & R6          & Recipient                                                                   &  20                                                                            & Professor         & \textbackslash{}                                                            \\ 
\hline
M1          & R5's Manager                                                                & 6 - 10                                                                          & Director          & College                                                                    & R7          & Recipient                                                                   & 11 - 20                                                                        & Professor, Director         & \textbackslash{}                                                            \\ 
\hline
M2          & R6's Manager                                                                &  20                                                                             & Professor         & \textbackslash{}                                                           & R8          & Recipient                                                                   & 11 - 20                                                                        & Professor         & \textbackslash{}                                                            \\ 
\hline
\multicolumn{5}{|c|}{}                                                                                                                                                                                                                                                       & R9          & Recipient                                                                   & 1 - 5                                                                          & Professor, Director         & \textbackslash{}                                                            \\
\hline
\end{tabular}}
\caption[Demographics of participants]{Demographics of participants. C1 --- C6, R1 --- R5, M1 are staff and the rest are faculty. Central-level offices are in charge of university-wide affairs, like university services, information technology; college-level or departmental offices are located within a college or department, such as the dean's office.}
\label{tab:demographic}\vspace{-0.3in}
\end{table}

\subsection{Data Analysis}
To compare the experience and assessments of the effectiveness of the bulk email system from different stakeholders, we took a grounded theory approach \cite{charmaz2014constructing}, specifically with the following iterative procedure:
\begin{enumerate}
    \item Identifying the themes from the communicators' transcripts.
    \item Searching for relevant text and themes in the recipients' transcripts.
    \item Comparing the actions taken by the recipient and preferred by the manager (if applicable).
    \item Inviting new participants and repeating the steps above until we reached data saturation. We stopped when we interviewed 17 stakeholders because we found strong repetition in the themes identified.
\end{enumerate}

\section{Artifact Walkthrough Findings}

\subsection{Number of Emails Received and Actions Taken}
Table~\ref{tab:email_number} reported the statistics of the 9 recipients for emails received, opened, read, and trashed by category. The recipients, on average, received 376 emails in the week prior: 153 of them were organizational emails and 30 of them were bulk emails (25 mass emails and 5 newsletters). The number of messages in one of the 55 newsletters investigated in the interview could be as many as 35, with an average of 8.5. Participants (recipients) reported that they, in general, received too many bulk emails. R6 said \textit{``Sometimes I felt overwhelmed.''}. R4 and R8 said that a large number of bulk emails became a burden to them.

Faculty received 175 organizational emails a week on average (148 non-mass emails, 23 mass emails, 4 newsletters), and for staff, this average number was 136 (103 non-mass emails, 27 mass emails, 6 newsletters). The average logged open rate of mass/newsletters of faculty recipients ($>90\%$) was higher than their self-report open rate (<70\%). Sometimes they clicked an email's title and removed its unread tag, but they did not think that they ``opened'' the email.

Though many bulk emails were opened, few of them were read in detail. In fact, $58\%$ of mass emails were reported being opened by staff while only $28\%$ of them were read in detail; $67\%$ of mass emails were reported as being opened by faculty although only $13\%$ of them were read in detail.

Many bulk emails were trashed, $52\%$ of mass emails and $22\%$ of newsletters were reported as being trashed by faculty, while for staff the percentages were $27\%$ and $15\%$.

\begin{table*}[!htbp]
\centering
\resizebox{\textwidth}{!}{%
\begin{tabular}{|l|c|c|c|c|c|c|} 
\hline
\multicolumn{1}{|c|}{\textbf{Faculty}} & \textbf{Email Type}                         & \textbf{Overall} & \textbf{Organizational} & \begin{tabular}[c]{@{}c@{}}\textbf{Organizational}\\\textbf{Non-Bulk}\end{tabular}  & \begin{tabular}[c]{@{}c@{}}\textbf{Organizational}\\\textbf{Mass}\end{tabular}  & \begin{tabular}[c]{@{}c@{}}\textbf{Organizational}\\\textbf{~Newsletter}\end{tabular}  \\ 
\hline
\multicolumn{1}{|c|}{no.= 4}           & \textbf{\# Received Per Week}               & 560              & 175                     & 148                                                                                 & 23                                                                              & 4                                                                                \\ 
\hline
                                       & \textbf{\% Logged Open Rate}                       & 55\%             & 96\%                    & 97\%                                                                                & 92\%                                                                            & 94\%                                                                             \\ 
\hline
                                       & \textbf{\% Self-report Open Rate}           & \textbackslash{} & \textbackslash{}        & \textbackslash{}                                                                    & 67\%                                                                            & 59\%                                                                             \\ 
\hline
                                       & \textbf{\% Self-report Read in Detail Rate} & \textbackslash{} & \textbackslash{}        & \textbackslash{}                                                                    & 13\%                                                                            & 9\%                                                                              \\ 
\hline
                                       & \textbf{\% Self-report Trash/Archive Rate}  & \textbackslash{} & \textbackslash{}        & \textbackslash{}                                                                    & 52\%                                                                            & 22\%                                                                             \\ 
\hline
\multicolumn{1}{|c|}{\textbf{Staff}}   & \textbf{Email Type}                         & \textbf{Overall} & \textbf{Organizational} & \begin{tabular}[c]{@{}c@{}}\textbf{Organizational}\\\textbf{~Non-Bulk}\end{tabular} & \begin{tabular}[c]{@{}c@{}}\textbf{Organizational}\\\textbf{~Mass}\end{tabular} & \begin{tabular}[c]{@{}c@{}}\textbf{Organizational}\\\textbf{~Newsletter}\end{tabular}  \\ 
\hline
\multicolumn{1}{|c|}{no. = 5}          & \textbf{\# Received Per Week}               & 229              & 136                     & 103                                                                                 & 27                                                                              & 6                                                                                \\ 
\hline
                                       & \textbf{\% Logged Open Rate}                       & 54\%             & 73\%                    & 78\%                                                                                & 52\%                                                                            & 50\%                                                                             \\ 
\hline
                                       & \textbf{\% Self-report Open Rate}           & \textbackslash{} & \textbackslash{}        & \textbackslash{}                                                                    & 58\%                                                                            & 58\%                                                                             \\ 
\hline
                                       & \textbf{\% Self-report Read in Detail Rate} & \textbackslash{} & \textbackslash{}        & \textbackslash{}                                                                    & 28\%                                                                            & 30\%                                                                             \\ 
\hline
                                       & \textbf{\% Self-report Trash/Archive Rate}  & \textbackslash{} & \textbackslash{}        & \textbackslash{}                                                                    & 27\%                                                                            & 15\%                                                                             \\
\hline
\end{tabular}}
\caption[Average per-capita email volume by category]{Average per-capita email volume by category; \% logged open rate is calculated by the inbox logged data; \% self-report open/read-in-detail/trash/archive rate is calculated by the self-report data, based on all the bulk emails investigated in the corresponding category, for example, self-report read in detail rate of mass email = \# mass email read in detail/\# mass email investigated.}
\label{tab:email_number}
\end{table*}

\subsection{Nature of Bulk Email System as an Organizational System}
The bulk email system was part of the university's administrative system, with communicators trying to achieve the communication goals of this university. The communicators in this system had clients, performance metrics, and tools.~\\

\noindent\textbf{ Communicators as a gateway for organizational leaders to reach employees.}

Communicators had clients from their own offices or other offices who were not professionals in communication and who wanted to promote events, notify staff of changes of policies, etc. Many clients were the communicators' leaders:
\begin{quote}
\textit{We work with our leadership whoever sends the message, so for example, we send a few messages for the associate vice president for research, who's also the head of responsive administration. (C1)}
\end{quote}

Naturally, clients wanted to send more content and reach more people. A communicator talked about their clients:

\begin{quote}
    \textit{
We're working with other departments coming up with certain final messages that may be close to the amount of information that people can digest. But some of them think that people have a more of an appetite for reading. (C1)
}
\end{quote}

Thus communicators usually needed to narrow mailing lists and shorten the length of emails. One mechanism they used was letting clients fill in  templates:

\begin{quote} 
 \textit{   The U Relations developed a partnerships with emergency management. When there is a safety alert, there are certain blanks in the template they fill in. That's why you can see the safety messages [emails] are often similar. Because we give them the authority to send them, so we want to limit the scope of the content they can choose to put in there. (C4)
}\end{quote}

Sometimes communicators could not modify the emails. Some bulk emails were required by law:

\begin{quote}

\textit{
     We have things called safety advisories. Maybe somebody is grabbed from Ford Hall last night and we just found out about it. That person is gone but we are required by federal law to tell you about that. (C4)}\\
\end{quote}

\noindent\textbf{Communicators are anchored to the ``open rate'' metric}

Optimizing a system depends on using appropriate metrics. The metrics used by communicators to measure bulk email's effectiveness did not match recipients' assessments. We asked communicators about the metrics they used in the bulk email system, and measured recipients' assessments for each bulk email from various metrics like whether trashed/opened/scanned/read in detail, and ratings on urgency/importance /relevance. For communicators, the priority order of the metrics used to report a bulk email's performance was: open rate $>$ click rate $>$ replies and others. They only needed to report a good open rate to their clients. Communicators indicated that $50\%$ was a good open rate to them:

\begin{quote}    
\textit{For all campus messages [emails], open rates are usually at least 50\%, which is good for us, so pretty high.  We look at click rate but not closely. Most times we look at open rates. (C5)
}\end{quote}

\begin{figure}[!htbp]
\centering
  \includegraphics[width=1\columnwidth]{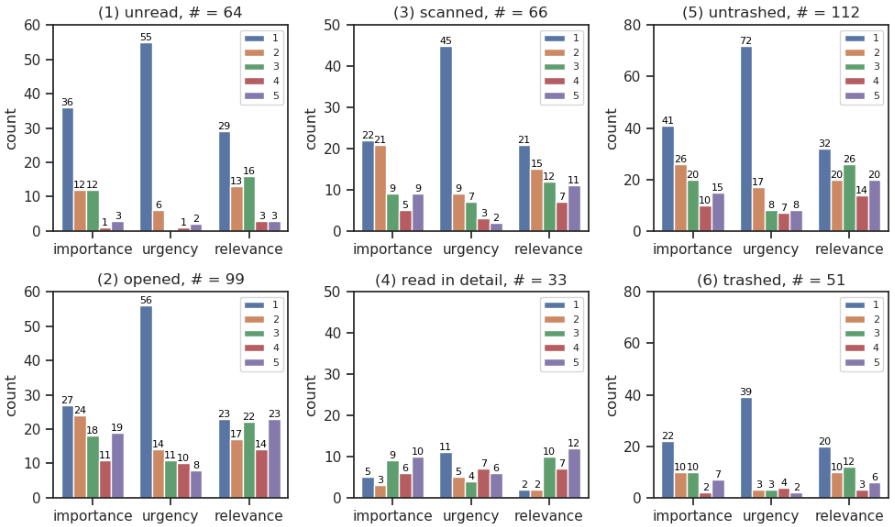}
  \caption[Counts of ratings on importance/urgency/relevance of the investigated bulk emails]{Counts of ratings on importance/urgency/relevance of the investigated bulk emails which were reported being left unread/opened/scanned/read in detail/left untrashed/trashed by the recipients.}~\label{fig:rating_2}
\end{figure}

However, ``open rate'' can be a flawed metric for measuring recipients' feedback, because:
\begin{enumerate}
    \item ``Being opened'' did not always mean ``being read''. As we discussed above, the logged open rate, which was used by communicators, might be higher than the self-report open rate --- as sometimes recipients clicked an email's title and removed its unread tag, without even reading any of the email.  Some faculty recipients, like R7, opened 17 of the 22 bulk emails we investigated, but only read 2 of them in detail, see Table \ref{tab:user_rate}.
    
    Recipients were not opening bulk emails because they considered them relevant but because they were unsure of their relevance and wanted to verify:
    \begin{quote} \textit{
I read the first line then deleted it because I've seen similar information in other news. (R7)
}\end{quote}

    \item Most ``opened'' bulk emails were only scanned and got low ratings in importance/urgency/ relevance. As shown in Figure \mbox{\ref{fig:rating_2}} (3), 43/54/36 bulk emails \hlc[white]{got} 1 or 2 on importance/urgency/ relevance ratings out of 66 scanned emails.

    \item Some bulk emails were opened, found irrelevant, and then trashed. As shown in Table \mbox{\ref{tab:user_rate}}, R1, R3, R5, R6, R7, R9 all trashed some bulk emails after they opened them. 
    
    Among participants, trashing email as an inbox tool was a matter of personal preference. For example, R1 trashed almost every bulk email after they opened and read it (see Table \mbox{\ref{tab:user_rate})}, \textit{``If it doesn't require actions, or not related to my work, I just trash it.''}. In contrast, R2 did not trash any email. 
    
    Trashing a bulk email \hlc[white]{did} not necessarily mean that it was useless. For example, R5 trashed a relevant bulk email after reading it: \textit{``I skimmed it because I was going to see whether we have all clear on the water issue ... I delete it because the water is good, no action needed.''}, and trashed a relevant bulk email without opening it: \textit{``I didn't need to open it because all the useful information are in the title --- come and get [person name] ... I trashed it.''}.
    
\end{enumerate}



Though ``open rate'' was a flawed metric, communicators had few tools in the bulk email system to measure performance beyond an email's open/click rate: \textit{
    ``I don't know, how long people spend, whether they share with other people.'' (C1)
}

\noindent\textbf{Implication 1: Provide End-to-End Metrics.} Focusing on ``open rate'' distracted communicators from real measurements of both channel and bulk email's effectiveness. Communicators usually were not involved in the transactional process of the bulk emails. They did not know about the recipients' level of awareness of the content --- the recipients read the emails and took corresponding actions to complete the organization's goals. Thus showing metrics on the transactional process of bulk emails to communicators could be helpful. For example, a measure showing how many people missed the due dates of events that were sent to them via bulk emails. When connecting heterogeneous systems is difficult, we could build systems similar to some human resources analytic systems \mbox{\cite{rasmussen2015learning}}, such as collecting assessments from clients/leaders for each bulk email.

\begin{table}[!htbp]
\centering
\resizebox{0.85\textwidth}{!}{%
\begin{tabular}{|c|c|c|c|c|c|c|} 
\hline
\textbf{\#} & \textbf{type} & \textbf{\# investigated} & \textbf{\# opened} & \textbf{\# read in detail} & \textbf{\# trashed} & \textbf{\# opened and trashed}  \\ 
\hline
R1          & staff         & 20                           & 5                         & 3                                 & 14                         & 5                              \\ 
\hline
R2          & staff         & 23                           & 19                        & 12                                & 0                          & 0                              \\ 
\hline
R3          & staff         & 12                           & 4                         & 1                                 & 1                          & 1                              \\ 
\hline
R4          & staff         & 8                            & 5                         & 2                                 & 1                          & 0                              \\ 
\hline
R5          & staff         & 20                           & 15                        & 6                                 & 3                          & 2                              \\ 
\hline
R6          & faculty       & 17                           & 12                        & 1                                 & 7                          & 4                              \\ 
\hline
R7          & faculty       & 22                           & 17                        & 2                                 & 8                          & 5                              \\ 
\hline
R8          & faculty       & 24                           & 10                        & 2                                 & 3                          & 0                              \\ 
\hline
R9          & faculty       & 17                           & 12                        & 4                                 & 14                         & 9                              \\
\hline
\end{tabular}
}
\caption[The number of bulk emails investigated/opened/read in detail/trashed/opened and trashed]{The number of bulk emails investigated/opened/read in detail/trashed/opened and trashed. The data was self-reported by participants on the specific bulk emails we investigated during the interviews.}
~\label{tab:user_rate}
\end{table}

\noindent\textbf{ Diverse types of communication with different content and delivery channels, and different matchings between content and channels.}

\noindent\textbf{A. Bulk email's contents.} Communicators mentioned three types of bulk email's content:

\begin{itemize}
    \item Transactional content: This was content aimed at facilitating the work of the university, and which asked recipients to take action. Communicator C4 talked about the goals of transactional bulk emails: \textit{
       ``Generally, we want you to change your behavior or be aware of something happening, or think differently about behavior or things like that.''
    }
    
   Some transactional content was urgent and required immediate actions:
    
    \begin{quote} \textit{
         Around something we want people to immediately know about, we usually send those transactional versus commercial, we have that option within the salesforce.  \footnote{See Saleforce's transactional message API: \url{https://developer.salesforce.com/docs/marketing/marketing-cloud/guide/transactional-messaging-api.html}.}(C1)
    }\end{quote}
    

    \item Highlighted News: These were the most important updates that offices \hlc[white]{wanted} employees to be aware of (e.g., announcements by the university president of new officials), though for most employees these would not be actionable. 
    
    \item Good-to-know News: These were other updates that offices felt employees would benefit from knowing, or that the institution would benefit from having more people aware of. Communicators \hlc[white]{recognized} that individual employees could miss this component without consequence. A communicator introduced a e-newsletter with good-to-know and highlighted news:
    
    \begin{quote} \textit{
        The broadest thing we do is that we have a blog, that we produce content and stories every month, and every month we sent highlights of the blog, that we want people to know about, to be aware of. (C1)
    }\end{quote}
    
    

\end{itemize}

A bulk email's content might be good-to-know for some recipients, while being transactional for others. We asked a communicator whether all the recipients (all students, faculty, and staff) of a bulk email about campus-safety tips should read it: \textit{
   ``Students for sure. For faculty and staff, it's a reminder, but it's not something you have to know.'' (C5)}

\noindent\textbf{B. Bulk email's distribution mechanisms.}  There were 3 types of distribution mechanisms of bulk emails: single bulk emails (mass emails), newsletters (newsletter emails), and redistribution (sent to the communicator's contacts in decentralized units). Rather than being universal, the matching of type of content to type of distribution mechanism varied across communicators. Some communicators sent good-to-know news through newsletters and some sent it through single bulk emails. Some communicators sent highlighted news and transactional content through single bulk emails and some communicators sent it through newsletters:
    \begin{quote} 
        \textit{We send single messages [individual bulk emails] when something is really timely we need to get it out, and when it’s critical that people receive and see it. If something is a story, less critical then that might find its way into our newsletter instead. (C1)}
        
        \textit{If there is an actionable thing we want people to do, that typically will go into a newsletter bulletin. The single message [individual bulk email] would be like a topic that needs to have a lot more detail but not necessarily an action needed to come out of it. So it’s an awareness thing. (C5)
    }\end{quote}
    
    
    

As communicators use bulk email distribution mechanisms differently, good-to-know news, highlighted news, or transactional content might be put into a newsletter at the same time. We asked C5 whether the recipients of an email should read all of the 10 messages in it: \textit{``
Not really. So we really want the message ``have the ID card with you'' get out, that was important, the most actionable thing, but the rest of it is to communicate that we care.'' (C5)
}



Recipients needed to go through all the messages in each email to screen messages relevant to them. A recipient talked about a newsletter with over 30 messages:
\begin{quote} \textit{
    What I do is I go through it, ok, [program name] ... I'm not interested in it, upcoming programs ... if I see something interesting then I'll read and click on that. (R6)
}\end{quote}

This lack of consistency between bulk email's content and distribution mechanisms across different communicators caused confusion about what type of content newsletters should be used for and whether recipients should view newsletters as important.\\

\noindent\textbf{Communicators' various roles --- Communicator Directors vs. Communicator Staff.}

\noindent Communicators' concerns regarding bulk emails varied based on their roles. When asked about improving the bulk email system, the communicator directors prioritized the engagement of the community, \textit{``We struggle on the engagement, make people know the service we have.'' (C2)}, \textit{``Our hardest things are how to let people feel that we all have a part of responsibility in safety, each individual have to make effort, it's much easier for us to keep the campus safe.'' (C4)}

Communicator staff, however, would like more tools within the bulk email system to support their operational work, \textit{``It's great to have people to say `I don't want this type of thing' or `Please send more content like this.''' (C5), ``From a user perspective, I think that maybe adapting some common templates could be helpful.'' (C6)}

\subsection{Different Perspectives on Bulk Email's Values}
\noindent\textbf{Communicators' assessments --- high-level bulk email is important.} 

High-level bulk emails were source of general university information, such as changes in benefits, leadership, public safety, administration, or working tools, and were usually from the university-wide offices (see Table \ref{tab:sender_level} for the definitions of email levels). They were classified as bulk emails with general importance by communicators. Communicators usually sent these types of emails to all faculty and staff: \textit{
         ``Public safety is everybody who works at the U should know about, and any big change in benefits or leadership will affect everybody.'' (C1)
    }
    
Communicators viewed bulk emails from university leaders as important, as we asked C5 what kind of emails were considered more important: \textit{
        ``Message from the president, from the vice president of university services about campus safety, those types of things.'' (C5)\\
    }

\noindent\textbf{Managers' assessments --- have awareness about high-level information.}

Managers sometimes agreed with their employees (recipients) that some bulk emails were not important. Within the 31 emails we asked managers to assess, there were 5 emails that the recipient and the manager both agreed should be scanned; and there were 13 emails that both the recipient and their manager agreed should be left unread. There were times when managers disagreed with their employees. There were 8 cases where employees opened these emails while their managers thought they should skip them. These 8 cases involved emails of interest to employees, but that were not particularly job-related, such as alumni marketing, a company's on-campus recruitment feedback questionnaire, etc.

Managers thought that employees should have some sense about what was going on at the high level of this university, specifically leadership and administration changes, as in the 5 cases shown in Table \mbox{\ref{tab:disagree}}. That table represents the bulk emails that managers 1) thought their employees should read while the employees did not, or 2) thought the employees should read in detail while the employee scanned. For example, M2 thought R6 should read the bulk email with the title \textit{``Your Help Matters Now''} and content \textit{``Legislative session ends Monday. Please take 60 seconds to make sure that when the final negotiations are done.''} in detail, but R6 left it unread.\\

\noindent\textbf{Employees' assessments --- high-level bulk email is irrelevant.} 

 Employees often skipped bulk emails about legislation of the university and leadership changes (see Table \ref{tab:disagree}) because 1) they had time pressure; 2) they felt these emails were too high level to be related to their role as non-leader staff: \textit{
    ``I skimmed and deleted it. It's a program that's important to my boss (the dean), but not directly related to my office.'' (R7)
    }
    
Surprisingly, the higher the bulk email's level, the lower the probability that the bulk email would be opened/read in detail; Figure \ref{fig:open_level} is a graph showing the levels of the investigated bulk emails (see Table \ref{tab:sender_level}) with their open/read in detail rates. A bulk email from a university-wide office had 48\% chance of being opened and 14\% chance of being read in detail, while those statistics for a bulk email from a department's office were 95\% and 57\%. 

Ignoring high-level bulk emails made employees miss some useful emails. When asked to read previously unread bulk emails their managers had deemed important, employees found some of these emails helpful. R6 found that bulk email \#1, on Table \mbox{\ref{tab:disagree}}, should be read and was actually easily actionable: \textit{``When I read it now, I may click on here and take 30 seconds to make sure the final negotiation. So I didn't realize that it was so easy to click this button. I was thinking it might involve more like writing a short letter.''}

\begin{table}[!htbp]
\centering
\resizebox{\textwidth}{!}{%
\begin{tabular}{|p{0.2cm}|p{1.8cm}|p{1.6cm}|p{1.8cm}|p{4.5cm}|p{2cm}|p{5cm}|} 
\hline
\textbf{\#} & \textbf{Manager \& Recipient} & \textbf{Manager Expectation} & \textbf{Employee Reaction} & \textbf{Title}                                                                                            & \textbf{Sender}                  & \textbf{Employee's Reason}                                                                                       \\ 
\hline
1           & M2 - R6                       & Read in Detail                 & Unread                     & Your Help Matters Now                                                                                     & University Government Relations  & I just deleted it because I think it's important to advocate for the university but I don't do that for myself.  \\ 
\hline
2           & M2 - R6                       & Scan                           & Unread                     & [College Name]'s Update • Summer 2019                                                                        & Newsletter Center of [College Name] & I was too busy, this email is not like so important.                                                             \\ 
\hline
3           & M1 - R5                       & Scan                           & Unread                     & Reminder - Incorporating GoldPASS into your Career Courses session tomorrow                                & University Employer Relations    & I will not open it, because it's for undergrad department.                                                       \\ 
\hline
4           & M1 - R5                       & Scan                           & Unread                     & PeopleSoft Navigation Update: MyU  PeopleSoft Outage This Weekend                                         & PeopleSoft Update Team           & I already learn it from other sources.                                                                           \\ 
\hline
5           & M2 - R6                       & Read in Detail                 & Scan                         & {[}[College Name] - Staff] [College Name] Highlights, Leadership Announcements and Resources to Share, 05/20/19 & [College Name] Leader's Office      & I am busy and I will spend less time on this.                                                                    \\
\hline
\end{tabular}}
\caption[Title/sender of the bulk emails which managers thought their employees should read]{Title/sender of the bulk emails which managers thought their employees should read while they did not, or thought the employees should read in detail while they just scanned,  and the employees' reasons.}
~\label{tab:disagree}
\end{table}

\begin{table}[!htbp]
\begin{minipage}[b]{0.48\linewidth}
\centering
    \small{
    \begin{tabular}{|p{0.2cm}|p{2cm}|p{3cm}|} 
    \hline
     \textbf{\#}  & \textbf{Level}  & \textbf{Meaning}                                                               \\ 
    \hline
    1             & University      & Sent from university wide offices.  \\ 
    \hline
    2             & College         & Sent from college wide offices.     \\ 
    \hline
    3             & Department      & Sent from department wide offices.  \\
    \hline
    \end{tabular}}
    \caption[The email levels' definitions]{The email levels' definitions. For example, ``university-wide'' offices indicates that the office was in charge of sending bulk emails to recipients across the university.}
    \label{tab:sender_level}
\end{minipage}\hfill
\begin{minipage}[b]{0.5\linewidth}
\centering
    \centering
      \includegraphics[width=0.9\columnwidth]{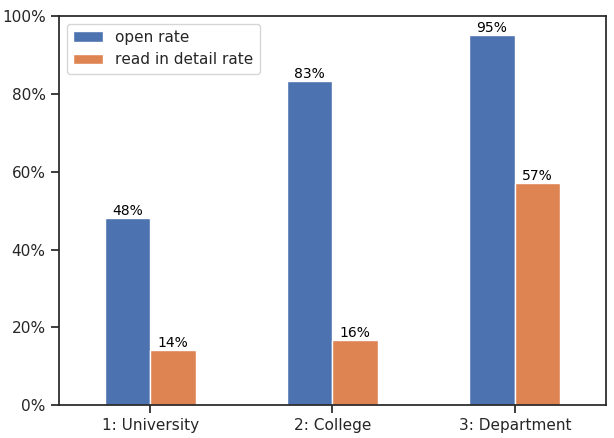}
      \captionof{figure}[The self-report open/read in detail rate ]{The self-report open/read in detail rate of bulk emails of different levels collected in the interviews with recipients.}~\label{fig:open_level}
\end{minipage}
\end{table}

\subsection{Costs of Communication}

\noindent\textbf{Recipients’ feeling of burden and their reading strategies.}
~\\
Controlling the burden on recipients was not one of the goals of the current bulk email system, as stated by a communicator when asked whether a recipient's time cost was considered: \textit{``No, the email we sent are most not nice-to-know emails. They are required information for people.'' (C2)}

Receiving many emails did not necessarily make recipients feel burdened. R9 received over 774 emails per week but did not get a sense of email overload. R9 talked about their strategy: \textit{``I can handle it (bulk email) ...  I scan it right away ... I delete the things I know I don't have to read ... I look at the subject ... Mentally you do it in hierarchy --- you determine what's the next, what's the first.''}


The reading strategy "checking the content whenever a new email came in" might bring on a sense of burden, like R6 said: \textit{`` I checked my email pretty quickly. Even when I am working on something, I may go back to my email to see is there anything important that's coming, and I respond pretty quickly ... To be honest, sometimes I felt overwhelmed.''}. This result mirrored Mark's findings on work emails and productivity \mbox{\cite{mark2016email}}.
 
Retrieving the correct bulk email from inboxes can be a burden: \textit{``I might miss something. Sometimes my supervisor ask me whether you received it --- I trashed it then I try to find it. That's frustrating because in Gmail sometimes I can't find it. You try to think what's in the subject line, date ...\mbox{''} (R4)}.


\noindent\textbf{Implication 2: Make Recipients' Burden Visible to Communicators.} The time cost caused by bulk email and the collective burden put on recipients was inapproachable for the communicators. They did not know whether they sent a bulk email to too many recipients and let them spend unnecessary time, or collectively let some recipients feel overloaded. For example, employees' time spent on bulk emails and the corresponding financial costs could be estimated \cite{jackson2006simple}.\\

\noindent\textbf{Communicators' Costs --- Loss of channel credibility (organizational communication capital).}

Communicators have, as requested by their clients, had to send some bulk emails at the cost of the sender's credibility. We asked a communicator whether the credibility of bulk email channel was considered:
\begin{quote} \textit{
 Yes, we definitely think about it (the credibility of the sender's name). But we work with different clients, they request the audience they want. (C1)
}\end{quote}

%
Recipients learned from experience and stopped opening bulk emails from low-credibility senders. For example, R9 trashed an email from the university fitness center without opening it: \textit{
``I recognized the sender and trashed it because I am not interested in it.''
}

\noindent\textbf{Implication 3: Make the Credibility of Bulk Email Channels Visible to Communicators.} Without a tool to tell communicators how credible a channel is to its recipients, communicators discarded ineffective channels and replaced them with new channels, which was a time consuming strategy. Quantifying channels' credibility and setting standards of usage for channels could be a helpful tool. For example, calculating the credibility and success probability of communication \mbox{\cite{dewatripont2005modes}}, or avoiding sending optional bulk emails via high-credibility channels.


\subsection{Understanding Current Practice and Its Failings}
\noindent\textbf{ Lack of personalization tools}

 Communicators tried to help recipients find important messages in a newsletter by putting them in the top position or putting their keywords in the subject of the email: \textit{``Sometimes we put important messages in the subject line. It's not the same importance (to all recipients), but they can recognize that it's important or not. We do try to keep it relevant and make it clear at the top.'' (C1)}
 
 However, as C1 said, importance of the messages in a bulk email may vary for different recipients. Recipients tended to close a newsletter whose first message was not relevant; in the process they often missed the later messages which were useful to them. R6 told us why a bulk email was unread and R6 found it useful when we asked them read the email:
 \begin{quote}\textit{
    They started with sports, then I thought it's not relevant to me. But when I go back with you, I found out that it has research that I am interested in. (R6)
    }\end{quote}
    
Given this situation, communicators thought that personalization of email designs would be helpful, but they did not have the technology to do so. Currently, all recipients receive the same subject lines/subtitles/content for each bulk email. We asked a communicator whether they had felt the need for personalization of emails):
\begin{quote} \textit{
     We haven't, although it's something we would like to explore doing. I don't think our team has technical expertise. I don't know where and how to get that type of thing. (C6)
}\end{quote}

\noindent\textbf{Implication 4: Explore Personalization.} Develop personalization tools for communicators to design bulk emails. For instance, have users estimate the relevance of messages and rank messages from high to low relevance \cite{trespalacios2016effects}.\\

\noindent\textbf{Flawed Email Targeting Tools}

\noindent\textbf{A. Target through mailing lists --- recipients cannot opt-out.} There were some mailing lists built for certain groups, e.g. all faculty of the university, all staff of a department. Communicators assumed that recipients would unsubscribe from mailing lists they were not interested in. Communicators also designated certain lists without opt-out choice to avoid core recipient opt-out. For example, a communicator talked about a monthly newsletter:
\begin{quote} \textit{
    For the people who work as administrators, they cannot opt-out of that. I check this. Everyone else is optional, they can unsubscribe. (C3)
}\end{quote}


However, there were cases where a mailing list was used to send both important and unimportant bulk emails at different times. Recipients could not opt-out from this mailing list, even though unimportant bulk emails were sent through it, because they would not receive important emails sent through that mailing list in the future. For example, a communicator talked about an optional bulk email sent to communication directors:

\begin{quote} \textit{
It's optional for them to take action to do what I ask them to do. This is meant to be kind of helpful service I'm giving ... They could (opt-out of this mailing list) but this is a mailing list for the communication directors so they probably won't. (C5)
}\end{quote}


\noindent\textbf{B. Target through querying position-database --- limited to the human resource system.} If there were no mailing lists existing for the target population, communicators would query the database of employees when sending emails. However, the current bulk email system only has access to the human resource system. The HR system solely supports querying based on positions, not on recipients' networks or interests. A communicator introduced this process:

\begin{quote} \textit{
 We will go to HR, and ask them to pull a list from PeopleSoft. Sometimes, for example, we were trying to reach everyone with family, so we go to HR, and they were not able to pull out. They can only do if the data is in PeopleSoft and we can get it by job code. (C5)
}\end{quote}

A communicator talked about a bulk email whose target population was all employees who had not downloaded documents from Webex. But the communicator did not have such information in the bulk email system, it was sent to all Webex users instead:

\begin{quote} \textit{
    This is the final reminder to Webex users: action required to download reports. Receivers are people who have been using Webex in the last year. That went to 9000 receivers.} 
    
    \textit{We cannot tell who has downloaded their file or not. The audience is who is using Webex and have files. Anybody who has already moved to Zoom, downloaded their files don't need to take action, but we don't know who are them. (C2)
}\end{quote}

\noindent\textbf{C. Redistribution mechanisms --- no tracking of bulk emails.} Communicators without enough information to target recipients often assumed that lower-level units would have more employee information asking unit-level contacts to forward emails to the related recipients:

\begin{quote} \textit{
     We have in each unit one person in charge of the facility things. And they may have people in their department, in their buildings. Sometimes we will send information to the unit facility leads, which is a small set of people (author's note: this is a set of 400 to 500 people). That's when we want other people to distribute something more locally versus from the University Relations. We want people in [college] to get some messages [emails] from their unit. We may ask them to distribute that message [email]. (C4)
}\end{quote}

However, communicators could not track whether and how the contacts sent these bulk emails, nor the performance of these bulk emails. When we asked a communicator if they could track the performance of the bulk emails sent to the contacts, the response was: \textit{``Unfortunately we can not track that.'' (C6)}



\noindent\textbf{D. When it is hard to target --- overwhelm everyone.} As both mailing lists and querying were limited to certain scenarios, precise targeting was difficult for communicators. C6 introduced a group difficult to narrow: \textit{``Open enrollment is tough because we don't know how many people need to take action. That's kind of a dilemma. Because we try to make that system easier for people so you don't have to do anything. But that makes it harder for us to do communications.''}

Motivated by ``getting the proof of delivery'', communicators thought that it was better to overwhelm everyone than to miss a single person. The bulk email mentioned above finally went to all employees: \textit{``We moved everyone to do it to ensure that no one is left behind.'' (C6)}





``Sent to all'' was one of the reasons that recipients received irrelevant bulk emails. R4 talked about a bulk email they classified as not useful: \textit{``This is what I call blanket message [email] --- it goes to everyone, it's common knowledge, it's not that related to me.''}

\noindent\textbf{Implication 5:} Precise targeting is difficult. Previous research has shown that an email's label can change recipients' behaviors \mbox{\cite{reeves2008marketplace}}. Given those findings, a mechanism could be designed to hint at an email's value to its recipients before they open it, for example, labeling whether a bulk email is optional or mandatory, action-needed or not \cite{10.1145/1978942.1979456}. \\








\noindent\textbf{Communicator transferred the responsibility of being aware}

Communicators transferred the responsibility of knowing bulk emails' content to their recipients. A communicator talked about a newsletter from the college's dean that they sent to all employees weekly, with 17 messages and 5 messages from the dean:

\begin{quote} \textit{
They are all important emails, sent once a week. So our expectation is --- ``This is the important email you get from your college, regarding your job. This email is the business of the college, you should read it.'' (C3)}\end{quote}

However, recipients found the bulk emails were too many or too long for them to filter out unimportant ones, find important ones, and be aware of all of the content. They felt that it was not the recipient's responsibility to be aware of all the content. A recipient talked about a weekly newsletter sent to all employees about government activities:

\begin{quote} \textit{
It took so much time to read, 10 paragraphs, nobody gonna read that, it's gambling. People use emails like a safety cover, but there is no understanding from the recipients' side. The responsibility shifts from the sender to the recipient, that's unfair. (R3)}\\\end{quote}

\noindent\textbf{Vicious cycle --- recipients did not read then communicators sent more.}

Similar to Randall's findings in organizational communication \mbox{\cite{SCHULER1979268}}, the ineffectiveness of the bulk email system fed into a vicious cycle --- 1) recipients did not read bulk emails on time; 2) to get recipients to read bulk emails/take actions on time, communicators sent bulk emails multiple times/more widely; 3) recipients received more irrelevant bulk emails, lost trust in bulk email channels, and read fewer bulk emails in the future. A communicator talked about the collection process of the employee engagement survey: \textit{``Those messages [emails] start from September, October, in terms of emails, go through January ... we do have to do emails more than once because our email open rate is about 61 or 62 percent.'' (C6)
}

Communicators did not realize the existence of the cycle and thought repeated bulk emails would be appreciated: \textit{
``My assumption is if somebody is emailing me about a service that is going away, and I have taken actions, I will never need to open it, but I will be grateful if I get reminders.'' (C2)}

However, recipients found they usually opened a bulk email and found that they already read it/still not interested/already took action. For example, a recipient trashed a bulk email about reimbursement:

\begin{quote} \textit{
It is important and relevant but this is not the first time she sent. I already picked up what I need ... I only read the first email a couple of months ago ... She even attached the same document again ... so I just deleted it. (R9)}\end{quote}

\noindent \textbf{Implication 6: Provide Feedback to Communicators.} To interrupt the vicious cycle, recipients' feedback could be collected. This could include whether the recipient wanted more like this/less like this (similar to the applications in social medias \cite{khan2017social}), how much time the recipient had spent on it, whether they had already seen/replied to this bulk email. Communicators could use that feedback to decide whether to keep sending these bulk emails to the recipient, or only distribute reminders to people that have not read or confirmed them.

\noindent \textbf{Limitations:} The observations might not be generalizable to all kinds of organizations with different cultures/structures.

\section{Discussion}


\subsection{Align Different Stakeholders' Priorities}
We found a large gap between different stakeholders' priorities. Communicators thought they were sending important high-level information about the organization, and communicators' clients wanted their messages to be sent broadly. Regardless of those intentions, recipients did not view high-level bulk emails as important. This gap in perception was caused by different stakeholder priorities.

On one side, the major task for communicators was to get the ``proof of delivery'' of bulk emails to their clients, whose first priority was to make the target population aware of certain information and take appropriate actions. On the other side, recipients, who were collectively burdened by the communicators, wanted to receive bulk emails that were interesting and relevant to them, and fewer messages overall.

Outside of the organizational environment, recipients can unsubscribe or not open email --- forcing senders to modify their practices to reach audiences. However, within the organization, where employees have responsibilities to complete the organization's tasks and minimize the organization's costs, this misalignment of priorities brings costs to the organization. The communication channels' credibility end up harmed, and the organization's tasks do not get done.

Thus it is important to align different stakeholders' priorities, while also including organizational priorities in the bulk email system. Recipients should be reminded that besides information they find interesting, they have the responsibility of being aware of what is going on in the organization. Communicators and their clients should be reminded that, beyond distributing their information, they have the responsibility of maintaining the effectiveness of organizational communication channels and should avoid wasting employees' time on unnecessary emails. At an extreme, a Key Performance Indicator System \mbox{\cite{kueng2000process}}, which takes the organization's priorities into consideration, could be integrated into the bulk email system. For communicators, there could be indicators of the time cost, the credibility cost, as well as clients' feedback for each bulk email. For recipients, there could be indicators of the speed of taking actions required by the senders.

\subsection{Limitations --- Organizational Culture/Structure's Influence on Bulk Email System}
This is a qualitative case study of one study site, as such the observations may not be generalizable across organizations. It is possible that stakeholders in other organizations have different practices/perspectives on bulk email systems with respect to their organizations' cultures/structures.

The organization we studied had a hierarchical management structure --- from university leadership (Board of Regents, Office of the President, Office of the Provost), to the college offices (deans, academic affairs), down to the department offices (heads, program directors). Thus the information flow of the bulk email system was mostly one directional --- low-level employees received information from the high-level leaders via bulk emails, while high-level leaders rarely needed to get information from low-level employees via bulk emails.

As for stakeholders within an organization, bidirectional interaction proved to be an important factor in the engagement levels of organizational affairs \mbox{\cite{norris2017stakeholders}}. Due to the hierarchical management structure within the university, the bulk email recipients may have less willingness to participate, causing the bulk email system to be ineffective. It is possible that recipients in organizations with a flat structure have a higher willingness to participate in bulk email communication.

Though hierarchical, the organization's communication system was also decentralized. Each communication office worked for its own units, had its own goals, and served its own clients. There were no central alignments of these communication offices and information would often be sent multiple times from different offices. These repeated bulk emails became a burden that multiple communication offices collectively placed on recipients. As the previous research found, \textit{``decentralized organization is not a very efficient organization
model for work that requires a lot of intensive interaction
between different employees.''} \mbox{\cite{maki2004communication}}. It is possible that a bulk email system with a central communication office to arrange different communication offices' tasks and collaboration would work more efficiently.

\section{Conclusion and Future Work}
\noindent\textbf{Conclusion:} This study indicated a systemic failure of the organizational bulk email system. The organizational bulk email system had many stakeholders, but none of them had a global view of the system or the impacts of their own actions. First, the communicators' high-level clients wanted their messages sent broadly. Second, the communicators felt they were sending important high-level information through bulk emails; they got high ``open rates'' and only reported that to their clients. Third, the recipients viewed most of the bulk emails not relevant; they often opened and then rapidly discarded bulk emails without reading the details. Last, though the communicators across the organization worked together in a network that attempted to improve the quality of communication practice, there was little consistency in how bulk email channels were used and there was a limited set of targeting and feedback tools to support their task.

In theory, the answer is simple; there is a long history of information filtering technology to help recipients avoid spam based on their preferences, and email marketing technology to help communicators advertise products to external recipients. But in an organizational setting, recipients' or communicators' preferences are not enough; the organization's priorities matter. Recipients may not have an interest in certain information, but their employer expects an awareness of that information regardless. At the same time, communicators and their clients may want to send bulk emails broadly, but they need to consider the cost to the organization.

\noindent\textbf{Impact:} We hope this case study provides detailed information on how an organizational bulk email system works and fails, with respect to this organization's structure and nature. To the best of our knowledge, this is the first work focusing on multi-stakeholders' perspectives on bulk emails within an organization. We hope our protocols on inbox artifact-walkthroughs are useful for future research.

Based on these findings, we plan to 1) frame an economic model for organizational bulk email systems, including the value, cost, and action of its stakeholders (chapter 4); 2) design an organizational bulk email system to support multi-stakeholder prioritization. This system would include personalization tied to both the interests of the recipient and the needs of the organization (chapter 5). Finally, we plan to 3) support communicators in evaluating their bulk communications' value and cost (chapters 6, 7).

\chapter{Economic Model}
\label{economic_model}

\section{Introduction}
In this chapter, we propose an economic model to describe the organizational bulk email system to identify this system's control points, potential interventions, and research opportunities. \footnote{Joseph A Konstan and Ruoyan Kong. The challenge of organizational bulk email
systems: Models and empirical studies. In The Elgar Companion to Information
Economics. Edward Elgar Publishing, 2023. \cite{economic_model}}

We consider this system's economics from this simple scenario. Imagine that the CIO of a university wants to inform everyone that voicemail will be down Sunday from 5-7 am, and does so by asking a communicator to send an email to 20,000 employees. The CIO may find this email free and convenient. But it is not free to the organization. An estimate of the cost of employee time would be 20,000 * 2 min (reading and interruption time) * \$0.75 salary+fringe+overhead/min = \$30,000. Plus, there is a future cost in the sender's reputation. If only 3\% of the employees find this message useful, then the other 97\% who read this email might view future emails from CIO as possibly irrelevant and fail to open them. In this case, the stakeholders (CIO, communicator, employees) interact with the system differently, and each perceives various costs to and value from the operation of the system. Moreover, the organization as a whole is a stakeholder, seeking to maximize both its employees' productivity and their organizational knowledge. 

 In this chapter, we frame the ongoing economics (such as the value, cost, and actions of the stakeholders above) within organizational bulk email systems based on our empirical studies (Chapter 3).
 
 We will first look at related work on organizational communication models. We then propose an economic model based on our empirical studies and explain the model with specific bulk email cases. Finally, we discuss design implications and research opportunities identified from the proposed model.

\section{Related Work and Gaps}
Several economic models around organizational communication have been proposed. Davenport and Beck proposed an economic model of employees' attention \cite{davenport2001attention}. It described an ``attention budget'' for each person and applied the law of supply and demand to it --- the larger the amount of information employees received, the greater the demand for their attention, the higher the likelihood more information would be ignored, and ultimately the greater the likelihood business goals will not be attained. Blazenaite treated organizational communication as a series of treatments of measurable communication variables and relationships. They framed the value organizations receive from communication at each stage and pointed out that communicators are the most powerful components of this system \cite{blazenaite2012effective}. Ruck and Welch suggested that organizational communication should consider employees' information needs on designing the contents instead of focusing on the communication process only. They proposed a model to reflect employees' individual (job) and social (organizational identification) communication needs \cite{ruck2012valuing}. It is also worth noting that the stakeholders in these economic models might not be fully rational. For example, Glynn surveyed a symphony orchestra and found the stakeholders' perceived value (``capability'') of communication is nonlinear, nonrational, and socially constructed \cite{glynn2000cymbals}.

Though there are several economic models for organizational communication, the economics within organizational bulk email systems have not been specifically explored. What makes organizational bulk email interesting is the variety of choices that exist for distribution and targeting (including newsletters, individual messages, and routing messages through trusted individuals), and the degree to which some of these choices become known "channels" with \textbf{reputations.} The reputation of a channel is defined as the recipients' perception of the credibility, quality, and relevance of the emails sent through this channel. It is associated with the identifiable characteristics of the channel---the sender's name, the sender's address, or the newsletter brand in the subject line. The recipients can identify this channel from these characteristics before opening its messages.  Then the recipients make decisions on which messages to open and read. Reputation can be analyzed either accumulatively (the overall reactions of all the audience of a channel) or individually (each employee's individual reaction to different channels). Bulk email also comes with a set of tools that are better capable of tracking the opening and reading of messages (compared with hardcopy distribution, for instance) allowing more detailed study of the practices of recipients and more direct feedback to senders.  

To bridge this gap, in this chapter, we propose an economic model for an organizational bulk email system based on our study in chapter 3 which frames 1) the value, costs, and preferences of the stakeholders in the organizational bulk email system and how they lead to different actions; 2) how these actions affect the organization's cost, value, and communication channels' reputations. For simplicity, we refer to organizational bulk email as bulk email, and organizational bulk email system as bulk email system below.

\section{Channels and Distribution Mechanisms}
What makes organizational bulk email interesting is the variety of choices that exist for distribution and targeting. In this section, we introduce the available channels and distribution mechanisms for bulk emails. The~\textbf{channel}~of a bulk email refers to its sender and brand. They could be newsletters' names or university leaders' (offices') names. They are characteristics visible to recipients before opening.~

 Let's use the email ``\textit{U of M Brief (January 26, 2022)}'' and~\textit{``Senior Leader Search Update''}~in Figure \ref{fig:brief_sample} and \ref{fig:president} as examples. They are sent to all the employees across all the university campuses. ``\textit{Brief}'' is a weekly newsletter from the University Relations Office. It shows its brand ``\textit{U of M Brief}'' in its subject line and template. Employees will observe that this email is sent from ``U of M Brief $\mathrm{<}$brief@umn.edu$\mathrm{>}$'' in their inboxes.~\textit{``Senior Leader Search Update''}~has its brand ``\textit{Office of the President}'' in its template. This email's sender is shown as ``Office of the President $\mathrm{<}$noreply.gabel@umn.edu$\mathrm{>}$''. The senders, brands, and templates of these bulk emails mean they usually do not look like personal emails from the same ``people'' (probably intentionally). Different channels have different levels of ``\textbf{grouping}'' --- the number of messages in that channel's email. ``\textit{Brief}'' is high-grouping (around 30 messages embedded in each email). ``\textit{Office of the President}'' is low/medium-grouping (1 to 5 messages in each email).

\begin{table}
\begin{minipage}{.48\textwidth}

\centering
  \includegraphics[width=1\columnwidth]{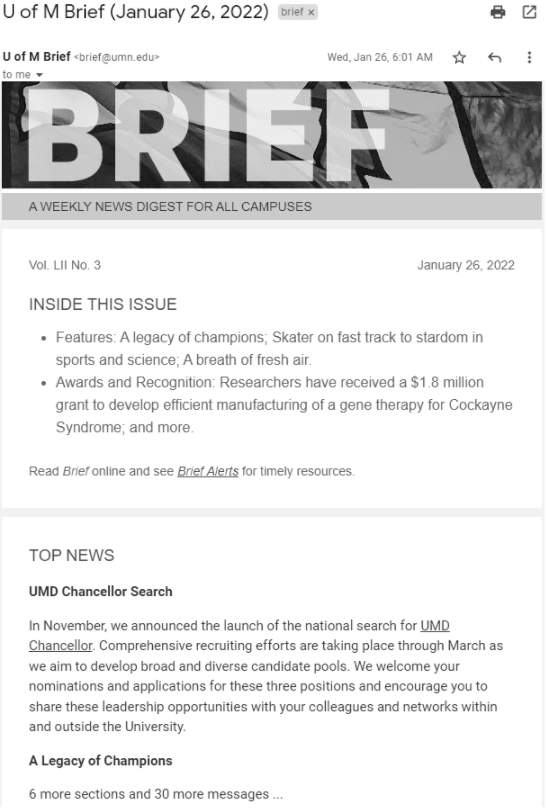}
  \captionof{figure}[An example message ``\textit{UMD Chancellor Search}'']{An example message ``\textit{UMD Chancellor Search}'' in an example email ``\textit{U of M Brief''.} Several paragraphs are omitted in this figure.}~\label{fig:brief_sample}

\end{minipage}%
\hspace{0.05in}
\begin{minipage}{.46\textwidth}
\centering
  \includegraphics[width=1\columnwidth]{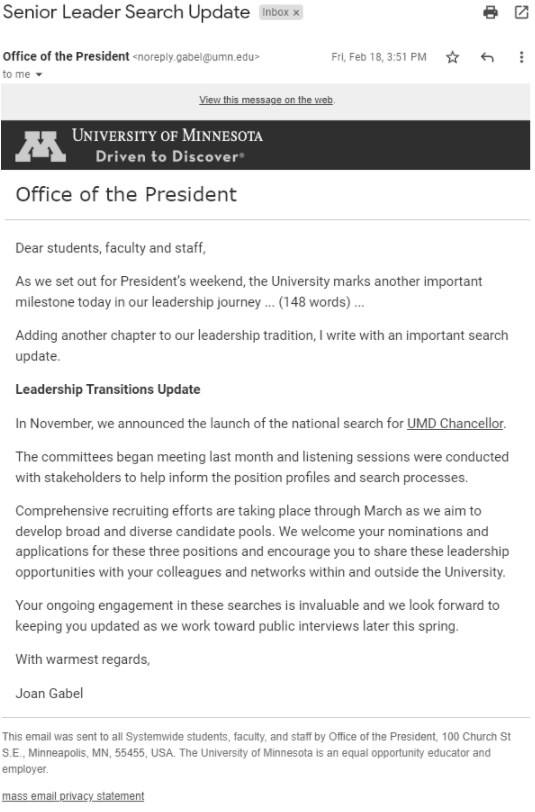}
  \captionof{figure}[An example email \textit{``Message from the President''}]{An example email \textit{``Message from the President''. }It contains two messages. }~\label{fig:president}
\end{minipage}%
\end{table}

The~\textbf{distribution mechanism}~of a bulk email refers to how communicators send it to recipients. Communicators use different distribution mechanisms for different channels. With different distribution mechanisms, channels have different levels of ``\textbf{targeting}'' --- the granularity of matching between the messages in that channel and its audiences. There are three major distribution mechanisms: newsletter, mailing to a list, email tree.

\subsection{ Newsletter }

\begin{figure}[!htbp]
\centering
  \includegraphics[width=0.8\columnwidth]{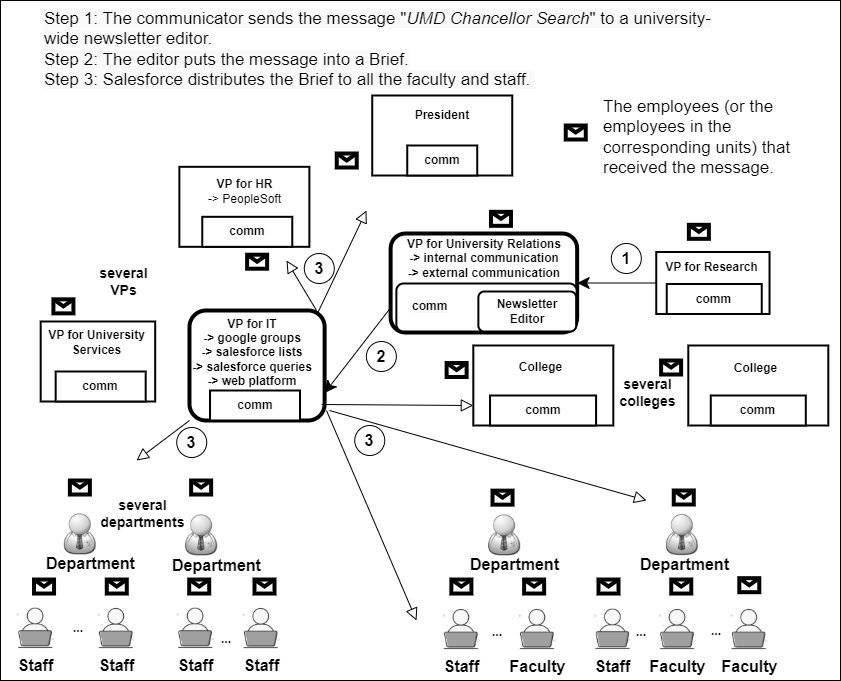}
  \caption{Distribute bulk messages through newsletters.  }~\label{fig:newsletter}
\end{figure}

 Suppose that a communicator at the President's office wants to announce the search for the Chancellor of the UMD campus (University of Minnesota -- Duluth [UMD]) and feels that this message has a broad audience (see Figure \ref{fig:brief_sample}). They could submit the content to a central newsletter editor (Figure \ref{fig:newsletter}, step 1). The editor then puts the message into a newsletter, such as the U of M Brief above (step 2). The communicator distributes the Brief to its subscribers (--- all the staff and faculty) through Salesforce (step 3). From a low level to a high level of targeting, the communicator has choices within all-employee newsletters (e.g., \textit{Brief }or \textit{Message from the President}), more narrowly targeted university-wide newsletters (E.g., Synthesist, a biweekly newsletter of the Graduate School sent to all graduate faculty \& staff with around 5 messages embedded), and affinity group newsletters (E.g., the research newsletter (a google group) of the college of science and engineering, see Figure \ref{fig:mask}, 2c)

\subsection{Mailing to a List}

\begin{figure}[!htbp]
\centering
  \includegraphics[width=0.8\columnwidth]{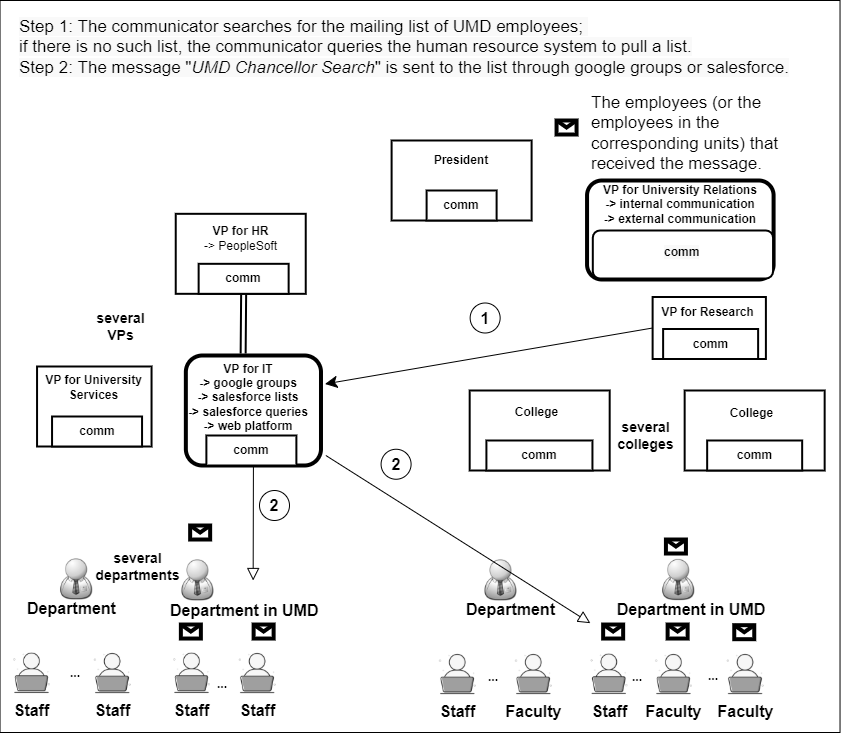}
  \caption{Distribute bulk messages through mailings lists or querying databases. }~\label{fig:query}
\end{figure}

 Suppose the communicator above wants to be more targeted by only sending the message to the employees of that specific campus UMD. The communicator first searches for the mailing list of UMD employees in Google Groups or Salesforce. If there is no such list, the communicator queries the human resource system to pull a list (Figure \ref{fig:query}, step 1). The email is sent to the mailing list by Gmail or Salesforce (step 2). From a low level to a high level of targeting, the communicator has choices within all-employee lists, departments or units' employee lists, and filtered lists based on more fields (such as job codes). For example, the communicator could choose send the message only to directors, department heads, and deans on the UMD campus.

\subsection{Email Tree (Send to Local Units to Redistribute)}

 We use the term~\textbf{email tree}\textit{~}to refer to the process of distributing a message through the management structure from higher-level managers down to lower-level management and eventually to individual employees. The goals of using such a tree are to gain employees' attention (because they listen to their local managers) and sometimes also to filter the recipients to those who need to receive the message (since managers know which units or employees are likely to need it). For example, to reach all faculty in the university, a communicator could send the message to Deans and Chancellors, asking them to forward it to their faculty; in turn they would send it on to Department Heads who would send it to the faculty (who would in turn pay more attention because they know their department head), or send to Vice Presidents to forward within their staffs to keep parallel.

\begin{figure}[!htbp]
\centering
  \includegraphics[width=1\columnwidth]{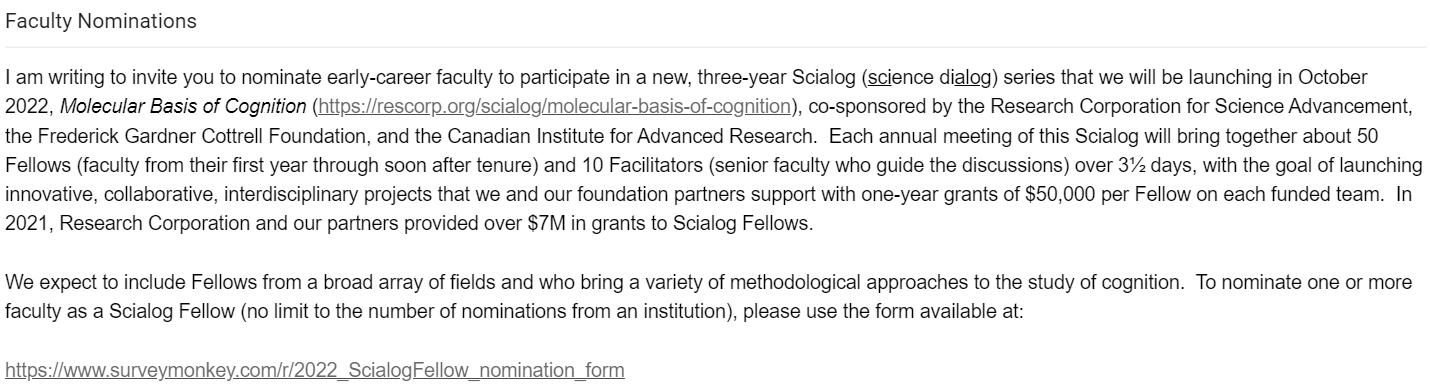}
  \caption{An example message ``\textit{Nominations for Cognition Faculty}''. }~\label{fig:nomination}
\end{figure}

\begin{figure}[!htbp]
\centering
  \includegraphics[width=0.6\columnwidth]{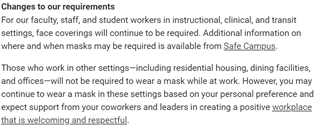}
  \caption{An example message ``\textit{Changes to Face Mask Requirements}''. }~\label{fig:mask}
\end{figure}

 For filtering, suppose that a communicator from the Office of Vice President for Research has a message on a grant opportunity for the early-career faculty in the field of cognitive science (see Figure \ref{fig:nomination}) and could not get this list from their databases. They could distribute the message through an email tree. The communicator sends the message to several relevant Deans' offices, such as the Associate Deans for research of the college of biological sciences, the college of science and engineering (Figure \ref{fig:mask}, step 1a), etc. If the college is large, the Associate Dean might ask several department heads to redistribute the message to the relevant faculty (step 2a). If the college is small enough that the Associate Dean knows each faculty member's interests, they might send individual messages to all the early-career cognition faculty they know (step 2b). The Associate Dean could also put the message in a college research newsletter (step 2c). The faculty and staff who subscribe to the newsletter's google group will receive the newsletter. The degree of grouping and targeting depends on the local units' actions.

\begin{figure}[!htbp]
\centering
  \includegraphics[width=1\columnwidth]{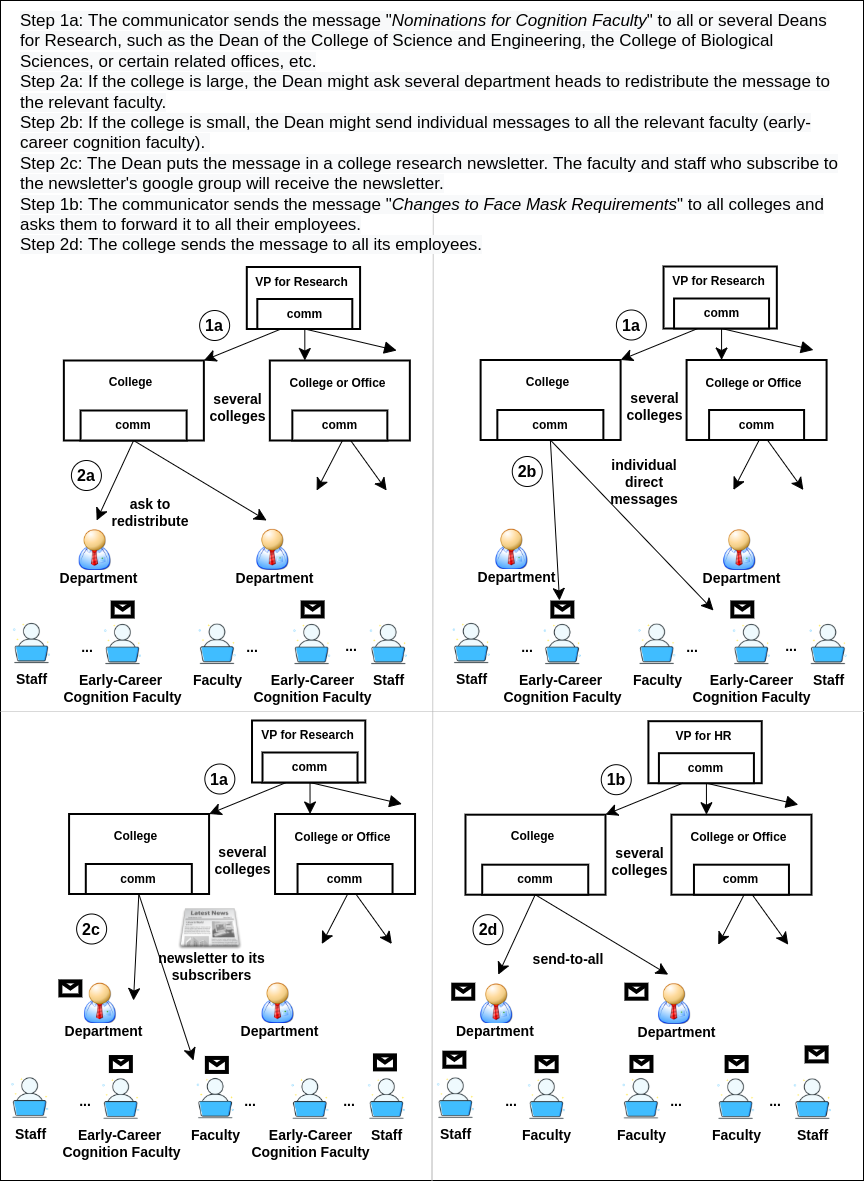}
  \caption{Distribute bulk messages through an email tree. }~\label{fig:tree}
\end{figure}

To consider how an email tree can borrow channels of higher reputation and hence increase the likelihood that a message is read, let us examine the case of a communicator from the Office of Vice President for Human Resources who has a message about face mask requirements (see Figure \ref{fig:mask}), and they would like everyone to know about this message. The communicator could use an email tree to utilize local attention since employees are more likely to open emails from their own units, as discussed in the mixed-methods study section. The communicator sends the message to all the colleges and administrative units (Figure \ref{fig:tree}, 1b), and asks them to forward this message to all of their employees (step 2d).

 We summarize various channels' distribution mechanisms and levels of grouping and targeting in Table \ref{tab:distribute_mechanism}. When a communicator receives a message request, they will have to decide which channels and distribution mechanisms (the \textbf{control point} that influences the corresponding stakeholders' cost and value retrieved from this message) to employ. We propose an economic model below to discuss how this control point influences the actions, costs, and values of various stakeholders and the reputations of communication channels.
\begin{table}
\centering
\caption[Summary of distribution mechanisms and their levels of targeting]{Summary of distribution mechanisms and their levels of targeting, and the example channels with the corresponding distribution mechanisms and their levels of grouping.}~\label{tab:distribute_mechanism}
\scalebox{0.7}{
\begin{tabular}{|p{1.2in}|p{1.2in}|p{0.5in}|p{0.7in}|p{1.3in}|} \hline 
Distribution Mechanism & \multicolumn{4}{|p{3.7in}|}{Low Targeting                               \`{a}                                 High Targeting} \\ \hline 
Newsletter & All-employee newsletters. E.g., \textit{Brief} (high grouping), \textit{Message from the President }(low/medium grouping) & \multicolumn{2}{|p{1.3in}|}{University-wide newsletters to all employees. E.g., \textit{Synthesist} (medium grouping)} & Affinity-group newsletters. E.g., the research newsletter (a google group) of the college of science and engineering (Figure \ref{fig:mask}, 2c) (medium grouping) \\ \hline 
Mailing to a List & All-employee lists E.g., the voicemail down message from the CIO to all the employees (low grouping)  & \multicolumn{2}{|p{1.3in}|}{College / Department / unit lists, E.g., sending the message ``\textit{Nominations for Cognition Faculty}'' to the list of faculty in the college of science and engineering (low grouping)} & Filtered/selected listsE.g., sending the message ``\textit{Nominations for Cognition Faculty}'' to the early-career faculty with research interests in cognitive science (low grouping) \\ \hline 
Email Tree & \multicolumn{2}{|p{1.7in}|}{Use local status / attention to reach all employees, E.g., \textit{Changes to Face Mask Requirements }(Figure \ref{fig:mask}, 2d -- low grouping)} & \multicolumn{2}{|p{2.0in}|}{Use hierarchy to identify all relevant employees, E.g., \textit{Nominations for Cognition Faculty }(Figure \ref{fig:mask}, 2a /2b -- low grouping)} \\ \hline 
\end{tabular}
}
\end{table}

\section{Economic Model}
\begin{figure}[!htbp]
\centering
  \includegraphics[width=1\columnwidth]{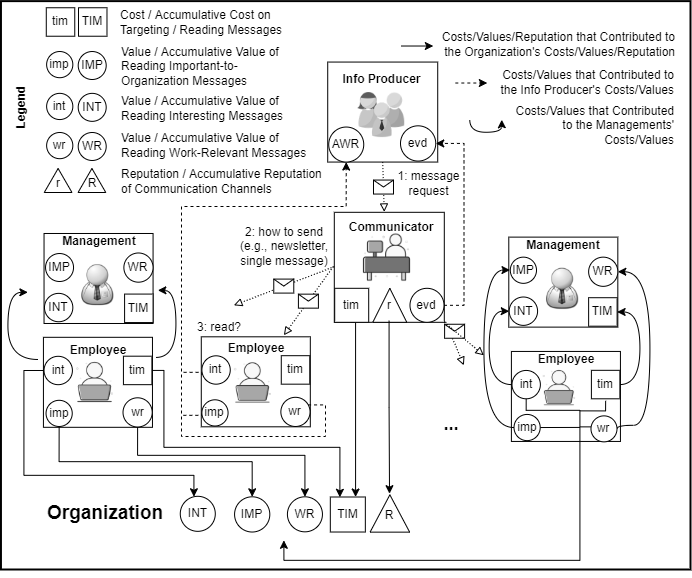}
  \caption[The correlation between the stakeholders' costs and value]{The correlation between the stakeholders' costs and value in the organizational bulk email system.  }~\label{fig:all_stakeholder}
\end{figure}

 In this section, we propose an economic model for organizational bulk email systems based on our study site's organization structure and bulk email structure above. Figure \ref{fig:all_stakeholder} shows the correlation between the stakeholders' costs and value in the organizational bulk email system. One information producer sends a request to the communicator to distribute a message (step 1). The communicator then makes decisions on their actions, including whether, how, and whom to send it (step 2). The employees who receive the messages then decide whether to read or not (step 3) and then receive related costs and value. These cumulative value and costs contribute to the organization's total value and cost. The time cost of the communicator also contributes to the organization's total cost.

 Table \ref{tab:cost_summary} summarizes the stakeholders' costs and value that they could receive from their actions with bulk emails. These value and cost are inspired by our interview study above --- communicators want to get their tasks done within reasonable efforts; employees want to get relevant information; information producers want their information to reach all the relevant employees). ${\mathrm{t}}_{\mathrm{e,m}}$ refers to whether employee $\mathrm{e}$ is targeted as message $\mathrm{m}$'s recipient (${\mathrm{t}}_{\mathrm{e,m}}$ = 1 if employee $\mathrm{e}$ receives message $\mathrm{m}$ otherwise 0). ${\mathrm{p}}_{\mathrm{e,m}}$ refers to employee $\mathrm{e}$'s probability of reading message $\mathrm{m}$ (we'll discuss its relationship with channel reputation below). In general, the information producers value that employees receive their messages; the communicators want to meet their clients' requirements with minimal time and effort; the employees would like to get useful information efficiently. The stakeholders try to maximize their returns ($\mathrm{g}\mathrm{=}$ value -- cost). The model in Table \ref{tab:cost_summary} has the following assumptions:

\begin{enumerate}
\item  Different stakeholders have different understandings of the value of a message being work-relevant / important / interesting to an employee. For example, we use ${\mathrm{v}}^{\mathrm{wr,p}}_{\mathrm{e,m}}$ to represent information producer $\mathrm{p}$'s assessment on how work-relevant message $\mathrm{m}$ is for employee $\mathrm{e}$ to know about it. This value could be different from ${\mathrm{v}}^{\mathrm{wr,b}}_{\mathrm{e,m}}$ --- the manager $\mathrm{b}$'s assessment on message $\mathrm{m}$'s work-relevance to employee $\mathrm{e}$. And these two values might all be different from ${\mathrm{v}}^{\mathrm{wr}}_{\mathrm{e,m}}$ --- the assessment on message $\mathrm{m}$'s work-relevance from employee $\mathrm{e}$ themselves.

\item  The time cost ${\mathrm{c}}^{\mathrm{tim}}$ contains multiple elements. For the information producers, it contains the time needed for composing and communicating with the communicators. For the communicators, it contains the time cost for designing and targeting the messages. 

 For employees, it contains the time cost for interruption (same for the messages in each email), deciding (same for the messages in each email), and reading (proportional to the corresponding message's length and position in the email). An employee's time cost with a message will be zero if they do not receive it. If they receive it, they might be interrupted and take several seconds to decide to read it or not based on a message's sender and subject line, even if they decide not to read it later. Time costs for employees also depend on their reading strategy. For example, the employees who process their email in batches might spend less total time due to less interruption.

\item  Communication channel $\mathrm{ch}$'s reputation ${\mathrm{r}}_{\mathrm{ch}}$ is the weighted sum of its reputation to each employee $\mathrm{e}$. ${\mathrm{r}}_{\mathrm{ch}}$ also contributes to the reputation $\mathrm{R}$ of the organization's communication channels as a whole. ${\mathrm{r}}_{\mathrm{ch}}$ is not only influenced by the messages sent through channel $\mathrm{ch}$ but by also other communication channels.

\item  Each information producer, communicator, manager is also an employee. For the simplicity of this model, we do not consider the interactions between employees and do not consider the nonrationality of stakeholders' actions. It would be helpful to validate these simplifications or explore these factors to build a more complete model in future work.
\end{enumerate}

 We will discuss how we come up with the return formulas in this table below in this section.

\begin{table}
\centering
\caption[Summary of the costs, value, return, and reputations]{Summary of the costs ($\mathrm{c}$), value ($\mathrm{v}$), return ($\mathrm{g}$), and reputations ($\mathrm{r}$) of the organizational bulk email system's stakeholders/communication channels. }~\label{tab:cost_summary}
\scalebox{0.65}{
\begin{tabular}{|p{4in}|p{4in}|} \hline 
\textbf{Information Producer }$\boldsymbol{\mathrm{p}}$\textbf{'s Value and Cost on Message m}\newline ${\mathrm{v}}^{\mathrm{awr}}_{\mathrm{p,m}}$: the relevant employees know what they are trying to disseminate ( $\mathrm{=}\sum_{\mathrm{e}}{{\mathrm{t}}_{\mathrm{e,m}}{\mathrm{p}}_{\mathrm{e,m}}\mathrm{(}{\mathrm{v}}^{\mathrm{wr,p}}_{\mathrm{e,m}}\mathrm{+}{\mathrm{v}}^{\mathrm{imp,p}}_{\mathrm{e,m}}\mathrm{+}{\mathrm{v}}^{\mathrm{int,p}}_{\mathrm{e,m}}\mathrm{)}}$)\newline ${\mathrm{v}}^{\mathrm{evd}}_{\mathrm{p,m}}$: provide the evidence that the required information was disseminated to the relevant employees ( $\mathrm{=}\sum_{\mathrm{e}\mathrm{\in }\mathrm{relevant\ employees}}{{\mathrm{t}}_{\mathrm{e,m}}}$)\newline ${\mathrm{c}}^{\mathrm{tim}}_{\mathrm{p,m}}$: time needed for composing and communicating with the communicator\newline ${\mathrm{v}}_{\mathrm{p,m}}$: information producer $\mathrm{p}$'s total value from message $\mathrm{m}$ ($\mathrm{=}{\mathrm{v}}^{\mathrm{awr}}_{\mathrm{p,m}}\mathrm{+}{\mathrm{v}}^{\mathrm{evd}}_{\mathrm{p,m}}$)\newline ${\mathrm{g}}_{\mathrm{p,m}}\mathrm{=}{\mathrm{v}}_{\mathrm{p,m}}\mathrm{-}{\mathrm{c}}^{\mathrm{tim}}_{\mathrm{p,m}}$\textbf{} & \textbf{Communicator }$\boldsymbol{\mathrm{c}}$\textbf{'s Value and Cost on Message }$\boldsymbol{\mathrm{m}}$\newline ${\mathrm{v}}^{\mathrm{evd}}_{\mathrm{c,m}}$: meet their clients' requirements (send the message to all the relevant employees,  $\mathrm{=}\sum_{\mathrm{e}\mathrm{\in }\mathrm{relevant\ employees}}{{\mathrm{t}}_{\mathrm{e,m}}}$)\newline ${\mathrm{c}}^{\mathrm{tim}}_{\mathrm{c,m}}$: the time cost for sending\newline ${\mathrm{\Delta }\mathrm{r}}_{\mathrm{ch}}$: communication channel $\mathrm{ch}$'s reputation's change\newline ${\mathrm{g}}_{\mathrm{c,m}}\mathrm{=}{\mathrm{v}}^{\mathrm{evd}}_{\mathrm{c,m}}\mathrm{-}{\mathrm{c}}^{\mathrm{tim}}_{\mathrm{c,m}}\mathrm{+}\mathrm{\Delta }{\mathrm{r}}_{\mathrm{ch}}$\textbf{} \\ \hline 
\textbf{Employee }$\boldsymbol{\mathrm{e}}$\textbf{'s Value and Cost on Message }$\boldsymbol{\mathrm{m}}$\newline ${\mathrm{v}}^{\mathrm{wr}}_{\mathrm{e,m}}$: get work-relevant information (if they read it)\newline ${\mathrm{v}}^{\mathrm{int}}_{\mathrm{e,m}}$: get enjoyment/interesting information (if they read it)\newline ${\mathrm{v}}^{\mathrm{imp}}_{\mathrm{e,m}}$: get important-to-organization information (if they read it)\newline ${\mathrm{c}}^{\mathrm{tim}}_{\mathrm{e,m}}$: the time cost of interruption, deciding, and reading (would be 0 if they do not receive this message)\newline ${\mathrm{v}}_{\mathrm{e,m}}$: employee $\mathrm{e}$'s total value from message $\mathrm{m}$ ($\mathrm{=}{\mathrm{t}}_{\mathrm{e,m}}{\mathrm{p}}_{\mathrm{e,m}}\mathrm{(}{\mathrm{v}}^{\mathrm{wr}}_{\mathrm{e,m}}\mathrm{+}{\mathrm{v}}^{\mathrm{imp}}_{\mathrm{e,m}}\mathrm{+}{\mathrm{v}}^{\mathrm{int}}_{\mathrm{e,m}}\mathrm{)}$)\newline ${\mathrm{g}}_{\mathrm{e,m}}\mathrm{=}{\mathrm{v}}_{\mathrm{e,m}}\mathrm{-}{\mathrm{c}}^{\mathrm{tim}}_{\mathrm{e,m}}$ & \textbf{Manager }$\boldsymbol{\mathrm{b}}$\textbf{'s Value and Cost}\newline ${\mathrm{v}}^{\mathrm{wr}}_{\mathrm{b,m}}$: the employees maintain awareness of the work-relevant messages sent through communication channels ($\mathrm{=}\sum_{\mathrm{e}\mathrm{\in }\mathrm{E}}{{\mathrm{t}}_{\mathrm{e,m}}{\mathrm{p}}_{\mathrm{e,m}}{\mathrm{v}}^{\mathrm{wr,b}}_{\mathrm{e,m}}}\mathrm{\ )}$, $\mathrm{E}\mathrm{=}$ the employees who report to $\mathrm{b}$ \newline ${\mathrm{v}}^{\mathrm{imp}}_{\mathrm{b,m}}$: the employees have good feelings about the organization through getting important-to-organization information ($\mathrm{=}\sum_{\mathrm{e}\mathrm{\in }\mathrm{E}}{{\mathrm{t}}_{\mathrm{e,m}}{\mathrm{p}}_{\mathrm{e,m}}{\mathrm{v}}^{\mathrm{imp,b}}_{\mathrm{e,m}}}\mathrm{)}$\newline ${\mathrm{v}}^{\mathrm{int}}_{\mathrm{b,m}}$: the employees have good feelings about the organization through getting interesting information ($\mathrm{=}\sum_{\mathrm{e}\mathrm{\in }\mathrm{E}}{{\mathrm{t}}_{\mathrm{e,m}}{\mathrm{p}}_{\mathrm{e,m}}{\mathrm{v}}^{\mathrm{int,}\mathrm{b}}_{\mathrm{e,m}}}$)\newline ${\mathrm{c}}^{\mathrm{tim}}_{\mathrm{b,m}}$: the employees' time cost ($\mathrm{=}\sum_{\mathrm{e}\mathrm{\in }\mathrm{E}}{{\mathrm{c}}^{\mathrm{tim}}_{\mathrm{e,m}}}\mathrm{)}$\newline ${\mathrm{v}}_{\mathrm{b,m}}$: manager $\mathrm{b}$'s value from message $\mathrm{m}$ ($\mathrm{=}{\mathrm{v}}^{\mathrm{wr}}_{\mathrm{b,m}}\mathrm{+}{\mathrm{v}}^{\mathrm{imp}}_{\mathrm{b,m}}\mathrm{+}{\mathrm{v}}^{\mathrm{int}}_{\mathrm{b,m}}$)\newline ${\mathrm{g}}_{\mathrm{b,m}}\mathrm{=}{\mathrm{v}}_{\mathrm{b,m}}\mathrm{-}{\mathrm{c}}^{\mathrm{tim}}_{\mathrm{b,m}}$ \\ \hline 
\multicolumn{2}{|p{8in}|}{\textbf{Organization }$\boldsymbol{\mathrm{o}}$\textbf{'s Value and Cost on All Messages\newline }${\mathrm{v}}_{\mathrm{o,m}}$: the weighted sum of all stakeholders' value ($\mathrm{=}\sum_{\mathrm{p,m}}{{\mathrm{w}}_{\mathrm{p}}{\mathrm{v}}_{\mathrm{p,m}}}\mathrm{+}\sum_{\mathrm{e,m}}{{\mathrm{w}}_{\mathrm{e}}{\mathrm{v}}_{\mathrm{e,m}}}\mathrm{+}\sum_{\mathrm{c,m}}{{\mathrm{w}}_{\mathrm{c}}{\mathrm{v}}_{\mathrm{c,m}}}\mathrm{+}\sum_{\mathrm{b,m}}{{\mathrm{w}}_{\mathrm{b}}{\mathrm{v}}_{\mathrm{b,m}}}$)\newline ${\mathrm{c}}^{\mathrm{tim}}_{\mathrm{o,m}}$: the accumulative cost of all stakeholders ($\mathrm{=}\sum_{\mathrm{e,m}}{{\mathrm{c}}^{\mathrm{tim}}_{\mathrm{e,m}}}\mathrm{+}\sum_{\mathrm{c,m}}{{\mathrm{c}}^{\mathrm{tim}}_{\mathrm{c,m}}}\mathrm{+}\sum_{\mathrm{p,m}}{{\mathrm{c}}^{\mathrm{tim}}_{\mathrm{p,m}}}$)\newline $\mathrm{\Delta }\mathrm{R}$: the accumulative change of communication capital ($=\sum_{\mathrm{ch}}{\mathrm{\Delta }{\mathrm{r}}_{\mathrm{ch}}}$)\newline ${\mathrm{g}}_{\mathrm{o}}\mathrm{=}\sum_{\mathrm{p,m}}{{\mathrm{w}}_{\mathrm{p}}{\mathrm{v}}_{\mathrm{p,m}}}\mathrm{+}\sum_{\mathrm{e,m}}{{\mathrm{w}}_{\mathrm{e}}{\mathrm{v}}_{\mathrm{e,m}}}\mathrm{+}\sum_{\mathrm{c,m}}{{\mathrm{w}}_{\mathrm{c}}{\mathrm{v}}_{\mathrm{c,m}}}\mathrm{+}\sum_{\mathrm{b,m}}{{\mathrm{w}}_{\mathrm{b}}{\mathrm{v}}_{\mathrm{b,m}}}\mathrm{-}\mathrm{(}\sum_{\mathrm{e,m}}{{\mathrm{c}}^{\mathrm{tim}}_{\mathrm{e,m}}}\mathrm{+}\sum_{\mathrm{c,m}}{{\mathrm{c}}^{\mathrm{tim}}_{\mathrm{c,m}}}\mathrm{+}\sum_{\mathrm{p,m}}{{\mathrm{c}}^{\mathrm{tim}}_{\mathrm{p,m}}}\mathrm{)+}\sum_{\mathrm{ch}}{\mathrm{\Delta }{\mathrm{r}}_{\mathrm{ch}}}$\textbf{}} \\ \hline 
\end{tabular}
}
\end{table}

\subsection{ Employee's Value and Cost}

 Employee $\mathrm{e}$ will only receive value from a message $\mathrm{m}$ if 1) they receive that message (${\mathrm{t}}_{\mathrm{e,m}}$ = 1), and 2) they read that message, or they can get value from reading the title of that message (${\mathrm{p}}_{\mathrm{e,m}}$ = 1), otherwise they receive zero value. Whether they receive message $\mathrm{m}$ or not depends on communicator $\mathrm{c}$'s actions. Their probability of reading a message $\mathrm{m}$ they receive (${\mathrm{p}}_{\mathrm{e,m}}$) is positively correlated with the message's communication channel's reputation ${\mathrm{r}}_{\mathrm{ch}}$ and the reputation $\mathrm{R}$ of the organization's communication channels as a whole. For example, an employee might trust and tend to read an email sent within their own unit compared to an email sent from central units; or an employee might trust \textit{President's Message} more compared to \textit{Brief.}  

 Message $\mathrm{m}$'s chance of being read might also depend on its position in the email. For example, a message in the Top News (the top section of \textit{Brief}) might have a higher probability of being read compared to the messages in the later sections of \textit{Brief}.

 Employee $\mathrm{e}$'s time cost with message $\mathrm{m}$ would be 0 if they do not receive it (if ${\mathrm{t}}_{\mathrm{e,m}}=0,{\mathrm{c}}^{\mathrm{tim}}_{\mathrm{e,m}}=0\ $). If they receive it, they will pay the corresponding time cost ${\mathrm{c}}^{\mathrm{tim}}_{\mathrm{e,m}}$ given their actions with this message, including the time of interruption, deciding, and reading. The factors that influence the size of ${\mathrm{c}}^{\mathrm{tim}}_{\mathrm{e,m}}$ are still left to be studied. For example, will employees spend more time on the channels they trust? Will they spend more time on the messages they feel work-relevant or interesting to them, or important to the organization? Will they spend more time reading a single-message email compared to a message in a newsletter? Will they spend more time reading the Top News of a newsletter compared to the later sections? These questions need to be answered for quantifying employee's costs.

 For the value employee $\mathrm{e}$ receives from reading message $\mathrm{m}$, there are three types of potential value. The first type is the value of message $\mathrm{m}$ being work-relevant to employee $\mathrm{e}$ (${\mathrm{v}}^{\mathrm{wr}}_{\mathrm{e,m}}$). The second type is the value of message $\mathrm{m}$ being interesting to employee $\mathrm{e}$ (${\mathrm{v}}^{\mathrm{int}}_{\mathrm{e,m}}$). The third type is the value of the message $\mathrm{m}$ being important-to-organization ${\mathrm{v}}^{\mathrm{imp}}_{\mathrm{e,m}}$. This value generally seems to be smaller than the value from the organization's perspective (${\mathrm{v}}^{\mathrm{i}\mathrm{mp}}_{\mathrm{e,m}}\mathrm{\le }{\mathrm{v}}^{\mathrm{imp,o}}_{\mathrm{e,m}}\mathrm{)}$. As according to our prior interview study in chapter 3, some employees feel that this type of information, such as the messages from the University Senate above, is too high-level to be actionable or relevant to them. 

 In summary, the final return ${\mathrm{g}}_{\mathrm{e,m}}$ the employee receives from that message would be 
\begin{equation} \label{GrindEQ__1_} 
{\mathrm{g}}_{\mathrm{e,m}}\mathrm{=}{\mathrm{t}}_{\mathrm{e,m}}{\mathrm{p}}_{\mathrm{e,m}}\mathrm{(}{\mathrm{v}}^{\mathrm{wr}}_{\mathrm{e,m}}\mathrm{+}{\mathrm{v}}^{\mathrm{imp}}_{\mathrm{e,m}}\mathrm{+}{\mathrm{v}}^{\mathrm{int}}_{\mathrm{e,m}}\mathrm{)}\mathrm{-}{\mathrm{c}}^{\mathrm{tim}}_{\mathrm{e,m}} 
\end{equation} 

\subsection{ Information Producer's Value }

 The information producer would like the relevant employees to know about their messages (${\mathrm{v}}^{\mathrm{awr}}_{\mathrm{p,m}}\mathrm{\ =}\sum_{\mathrm{e}}{{\mathrm{t}}_{\mathrm{e,m}}{\mathrm{p}}_{\mathrm{e,m}}\mathrm{(}{\mathrm{v}}^{\mathrm{wr,p}}_{\mathrm{e,m}}\mathrm{+}{\mathrm{v}}^{\mathrm{imp,p}}_{\mathrm{e,m}}\mathrm{+}{\mathrm{v}}^{\mathrm{int,p}}_{\mathrm{e,m}}\mathrm{)}}$). For example, if they are announcing a museum exhibit, they would like all the employees who might be interested in it to know about the message.

 Driven by compliance, the information producer sometimes needs to get the evidence that the required information was disseminated to all the potentially relevant employees (${\mathrm{v}}^{\mathrm{evd}}_{\mathrm{p,m}}\mathrm{=}\sum_{\mathrm{e}\mathrm{\in }\mathrm{relevant\ employees}}{{\mathrm{t}}_{\mathrm{e,m}}}$).

 The information producer spends time composing the message and communicating with communicators on their message request. In summary, the information producers' return of a message would be 
\begin{equation} \label{GrindEQ__2_} 
{\mathrm{g}}_{\mathrm{p,m}}\mathrm{=}{\mathrm{v}}^{\mathrm{evd}}_{\mathrm{p,m}}+\sum_{\mathrm{e}}{{\mathrm{t}}_{\mathrm{e,m}}{\mathrm{p}}_{\mathrm{e,m}}\mathrm{(}{\mathrm{v}}^{\mathrm{wr,p}}_{\mathrm{e,m}}\mathrm{+}{\mathrm{v}}^{\mathrm{imp,p}}_{\mathrm{e,m}}\mathrm{+}{\mathrm{v}}^{\mathrm{int,p}}_{\mathrm{e,m}}\mathrm{)}}\mathrm{-}{\mathrm{c}}^{\mathrm{tim}}_{\mathrm{p,m}} 
\end{equation} 

\subsection{Communicator's Value and Cost}

 When communicator $\mathrm{c}$ receives a message request, they first decide between sending it versus rejecting it/asking their clients to narrow the recipients or shorten the contents (control point 1). However, they are very likely to receive pushback from their clients if they reject the requests. 

 If they select to send this message, they have several channels across the dimensions of targeting and grouping (control point 2, see Table \ref{tab:distribute_mechanism}). Generally speaking, with a higher level of targeting, the communicator will spend more time sending message $\mathrm{m}$ (${\mathrm{c}}^{\mathrm{tim}}_{\mathrm{c,m}}\mathrm{\uparrow }$), and it reduces the problem of \textbf{oversend} (employees receive less irrelevant messages) and saves employee's time (${\mathrm{c}}^{\mathrm{tim}}_{\mathrm{e,m}}\mathrm{\downarrow }$). The communicator's time cost may be higher if they want to put the message in a channel under other communicators' controls. For example, the communicator from the President's office needs to spend time communicating with the editor of \textit{Brief} to put the message ``\textit{UMD Chancellor Search}'' in it. 

 The communicator is an important control point in the organizational bulk email system. Through communicator $\mathrm{c}$ cares about the influences of their actions on channel $\mathrm{ch}$'s reputation ${\mathrm{r}}_{\mathrm{ch}}$, they do not have access to this influence, and they need to meet the information producer's requirements (${\mathrm{v}}^{\mathrm{evd}}_{\mathrm{c,m}}$).  Especially when they have limited information in the databases and targeting tools, the high-targeting choices might bring a problem of \textbf{undersend} (miss relevant employees), causing them to fail to get ${\mathrm{v}}^{\mathrm{evd}}_{\mathrm{c,m}}$. Therefore, from the communicator's return formula with message $\mathrm{m}$ (equation 3), they might tend to get ${\mathrm{v}}^{\mathrm{evd}}_{\mathrm{c,m}}$ with a minimal time cost by some low-targeting actions.
\begin{equation} \label{GrindEQ__3_} 
{\mathrm{g}}_{\mathrm{c,m}}\mathrm{=}{\mathrm{v}}^{\mathrm{evd}}_{\mathrm{c,m}}\mathrm{-}{\mathrm{c}}^{\mathrm{tim}}_{\mathrm{c,m}}\mathrm{+}\mathrm{\Delta }{\mathrm{r}}_{\mathrm{ch}} 
\end{equation} 

\subsection{ Management's Value and Cost}

 Management is the direct manager/supervisor of the information recipients. They basically share the same value/costs with their employees. They care about whether their employees read and get the information they need to complete organization tasks (${\mathrm{v}}^{\mathrm{wr,b}}_{\mathrm{e,m}}$) while controlling the time cost of processing bulk messages (${\mathrm{c}}^{\mathrm{tim}}_{\mathrm{e,m}}$). For example, for an employee who often uses voicemails during the weekend, their manager would like them to read the voicemail down notification email to prepare earlier. But managers and employees might disagree on ``what is relevant or important to know'', which means that ${\mathrm{v}}^{\mathrm{wr,b}}_{\mathrm{e,m}}$, ${\mathrm{v}}^{\mathrm{int,b}}_{\mathrm{e,m}}$, ${\mathrm{v}}^{\mathrm{imp,b}}_{\mathrm{e,m}}$ might not equal to the employee's own assessments ${\mathrm{v}}^{\mathrm{wr}}_{\mathrm{e,m}}$, ${\mathrm{v}}^{\mathrm{int}}_{\mathrm{e,m}}$, ${\mathrm{v}}^{\mathrm{imp}}_{\mathrm{e,m}}$. In our interviews with managers, they indicated that high-level information, like legislation news, what is going on generally in the organization, and leadership updates, could be helpful, and the employees have responsibilities in knowing about these messages. Managers might also disagree with the organization / information producers on these values since they know better about their employees' job responsibilities.  All these factors make manager $\mathrm{b}$'s return with message $\mathrm{m}$ as
\begin{equation} \label{GrindEQ__4_} 
{\mathrm{g}}_{\mathrm{b,m}}\mathrm{=}\sum_{\mathrm{e}\mathrm{\in }\mathrm{E}}{{\mathrm{t}}_{\mathrm{e,m}}{\mathrm{p}}_{\mathrm{e,m}}\mathrm{(}{\mathrm{v}}^{\mathrm{wr,b}}_{\mathrm{e,m}}\mathrm{+}{\mathrm{v}}^{\mathrm{imp,b}}_{\mathrm{e,m}}\mathrm{+}{\mathrm{v}}^{\mathrm{int,b}}_{\mathrm{e,m}}\mathrm{)}}\mathrm{-}\sum_{\mathrm{e}\mathrm{\in }\mathrm{E}}{{\mathrm{c}}^{\mathrm{tim}}_{\mathrm{e,m}}} 
\end{equation} 

\subsection{Organization's Value and Cost}

 The organization's value is a weighted sum of all stakeholders' value. We represent the organization's perspective by the weights ${\mathrm{w}}_{\mathrm{p}},\ {\mathrm{w}}_{\mathrm{e}},\ {\mathrm{w}}_{\mathrm{c}},\ {\mathrm{w}}_{\mathrm{m}}$. The organization may not weigh each ${\mathrm{w}}_{\mathrm{p}},\ {\mathrm{w}}_{\mathrm{e}},\ {\mathrm{w}}_{\mathrm{c}},\ {\mathrm{w}}_{\mathrm{m}}$ the same. For example, they may decide that the President's weight is higher than Vice Presidents, or it cares more about the value of higher-paid employees. An important question left to be studied here is who and how to decide these weights. 

 The organization's cost is an accumulative sum of all stakeholders' costs. At the same time, the organization would also like to maintain or improve the reputation of its communication channels ($\sum_{\mathrm{ch}}{{\mathrm{r}}_{\mathrm{ch}}}$). With these factors, the organization's total return with all messages is defined as 
\begin{footnotesize}
\begin{equation} \label{GrindEQ__5_} 
{\mathrm{g}}_{\mathrm{o}}\mathrm{=(}\sum_{\mathrm{p,m}}{{\mathrm{w}}_{\mathrm{p}}{\mathrm{v}}_{\mathrm{p,m}}}\mathrm{+}\sum_{\mathrm{e,m}}{{\mathrm{w}}_{\mathrm{e}}{\mathrm{v}}_{\mathrm{e,m}}}\mathrm{+}\sum_{\mathrm{c,m}}{{\mathrm{w}}_{\mathrm{c}}{\mathrm{v}}_{\mathrm{c,m}}}\mathrm{+}\sum_{\mathrm{b,m}}{{\mathrm{w}}_{\mathrm{b}}{\mathrm{v}}_{\mathrm{b,m}}}\mathrm{)-(}\sum_{\mathrm{e,m}}{{\mathrm{c}}^{\mathrm{tim}}_{\mathrm{e,m}}}\mathrm{+}\sum_{\mathrm{c,m}}{{\mathrm{c}}^{\mathrm{tim}}_{\mathrm{c,m}}}\mathrm{+}\sum_{\mathrm{p,m}}{{\mathrm{c}}^{\mathrm{tim}}_{\mathrm{p,m}}}\mathrm{)+}\sum_{\mathrm{ch}}{\mathrm{\Delta }{\mathrm{r}}_{\mathrm{ch}}} 
\end{equation} 
\end{footnotesize}
Equation \eqref{GrindEQ__5_} indicates that, though designing and targeting bulk emails is time-consuming for communicators, it might make sense for the organization to encourage a communicator to spend 1000 minutes targeting and designing a message, compared to spending 1 minute direct sending that message to a list of 20, 000 employees, each of whom might spend over 1 minute reading it on average. However, the organization's costs and value are not measured, reflected, or incorporated anywhere in current organizational bulk email systems.

\section{Case Study}
 Let's use the example message ``\textit{UMD Chancellor Search}'' (Figure \ref{fig:brief_sample}) and ``\textit{Nominations for Cognition Faculty''} (Figure \ref{fig:nomination}) to discuss the scales of the stakeholders' value and costs given different channels. 

\subsection{``\textit{UMD Chancellor Search}''}

 This message is sent from the President's office to advocate the search for the Chancellor of the Duluth campus. The employees who are interested in the nomination or application (mainly from the Duluth campus) would view this message as important or interesting. For most employees from other campuses, it might be considered irrelevant high-level information. For the information producer (the President), they want 1) all the employees who might be interested in it to receive and read this message (get ${\mathrm{v}}^{\mathrm{awr}}_{\mathrm{p,m}}$); 2) to show that they value this search. Therefore, they tend to send this message from the central offices. It is worth noting that \textbf{``show the importance of a message'' might be different from ``a message is really important}''. Because if a message is really important, communicators might use an email tree to utilize the local attention to reach more employees. And for the communicator $c$ who receives this request, they will need to select a channel for this message which could show that they send the message to all the potentially relevant employees (get ${\mathrm{v}}^{\mathrm{evd}}_{\mathrm{c,m}}$).

\noindent \textbf{ \textit{``Message from the President''}}\textit{:} If the communicator sends it as a single-message email (\textit{``Message from the President''}), they will definitely get the proof-of-delivery ${\mathrm{v}}^{\mathrm{evd}}_{\mathrm{c,m}}$ because all the employees across all the campuses will receive this message (${\mathrm{t}}_{\mathrm{e,m}}\mathrm{=1}$). 

 For the information producer, a single-message email from this channel could also show the importance of this message. They will get a higher value of awareness because whoever opens this single-message email will be exposed to this message (${\mathrm{p}}_{\mathrm{e,m}}$ and ${\mathrm{c}}^{\mathrm{tim}}_{\mathrm{e,m}}$ might be large). From equations \eqref{GrindEQ__2_} and \eqref{GrindEQ__3_}, the information producer and communicator would get a positive return.

 However, the majority of the recipients will not perceive the value of this message (${\mathrm{v}}^{\mathrm{wr}}_{\mathrm{e,m}}\mathrm{+}{\mathrm{v}}^{\mathrm{imp}}_{\mathrm{e,m}}\mathrm{+}{\mathrm{v}}^{\mathrm{int}}_{\mathrm{e,m}}$ is small). From equation \eqref{GrindEQ__1_}, most employees would get a negative return from this message, with a high time cost and low value in general. And the small part of employees interested in this message will get a positive return, as they have a large chance to open and read this email.

 For the organization as a whole, from equation \eqref{GrindEQ__5_}, the return might be negative as most employees receive a negative return from this message. This choice also hurts the reputation of its communication channel --- employees might think that \textit{``Message from the President'' }sends irrelevant information and stop reading them in the future. The reputation of the organization's communication channels as a whole might also be hurt if the employees then also start to untrust the messages from other university leaders (to be studied). 

 It is worth noting that the damage to the channel's reputation might affect every stakeholder in the end. Because it would gradually make employees stop reading messages, then the information producer's message will be neglected in the future (${\mathrm{v}}^{\mathrm{awr}}_{\mathrm{p,m}}$ will be small), and the employees might also miss the messages that could be interesting or work-relevant to them.

\noindent \textbf{ \textit{``U of M Brief''}}\textit{:} The communicator could also put the message in \textit{Brief}, they will also get the proof-of-delivery ${\mathrm{v}}^{\mathrm{evd}}_{\mathrm{c,m}}$ as \textit{Brief} will be sent to all the employees (${\mathrm{t}}_{\mathrm{e,m}}\mathrm{=1}$). With a minimal cost on designing and targeting, the communicator would get a positive return from equation \eqref{GrindEQ__3_}.

 To show that this message is viewed as important by the organization, the communicator might put the message in Top News. Then the effects of this message would be similar to the effects of putting it in a single-message email.

 If the communicator puts the message in the later sections of \textit{Brief}, the employees who are interested in it might miss this message if they skip the later sections\textit{. }Then the information producer might get less value on employee's awareness (${\mathrm{v}}^{\mathrm{awr}}_{\mathrm{p,m}}$ is small). From equation \eqref{GrindEQ__1_}, with low value and low time cost, most employees who are not interested in this message would get a zero or small negative return. For the organization as a whole, its return would be a small (negative) value --- less time cost and less attention.

\noindent \textbf{\textit{Precisely targeted mailing list}}: To save employee's time, the communicator could query the database to pull a list of employees from the Duluth campus and directly send a single-message email to them. In this case, the employees outside that campus would not spend time on the message. Then the total time cost of all the employees would be smaller. 

 However, there is also a risk for the communicator and the information producer. There might be employees from other campuses who are also interested in nominating the UMD's chancellor. They might fail to get the proof-of-delivery if they miss these relevant recipients. Moreover, even with a powerful database, this approach is time-consuming, and communicators may not want to spend days generating a mailing list.

 The best approach here might be combining Brief and targeted mailing lists. We could put the message in a Brief to collect the proof-of-delivery, while sending it as a single-email message to the UMD employees to catch their attention. By this approach, the organization might receive a large value with a small cost because the employees who spend much time processing and reading this message would mainly be the potentially relevant employees.

\subsection{\textit{``Nominations for Cognition Faculty''}}

 This message is sent from the Office of the Vice President for Research to notify a fellowship for the early-career faculty in the field of cognitive science --- only those faculty would receive a value on work-relevance if they read this message. For the information producer, they want all these faculty to be aware of this message. But they won't view this message as of general importance to the organization. Therefore this message should not be sent as a single-message email to all the employees from the central units. For the communicator who receives this message request, their task is to select a channel to guarantee that they deliver the message to all the relevant employees.

\noindent \textbf{\textit{Precisely targeted mailing list}}: The communicator might not be able to get a precisely targeted list for this message if the database does not have a complete record of faculty's research interests. To guarantee that the recipient lists will cover all the relevant employees (get ${\mathrm{v}}^{\mathrm{evd}}_{\mathrm{c,m}}$), the communicator's best option is to send the message to the list of faculty in several colleges, such as all the faculty in the college of science and engineering, or all the faculty in the college of biological sciences (or the college of liberal arts, medical school, etc.), which would still be a large list. Sending the message as a single-message email might increase its chance of being read by the several early-career cognition faculty. However, this approach would bring time costs to a large list of irrelevant employees, and the organization as a whole. Besides, it will hurt the reputation of this channel. The employees who receive this message might feel the messages from the Office of the Vice President for Research as possibly irrelevant and stop reading them. In summary, the organization's return would be negative given the limited capability of targeting by querying the database.

\noindent \textbf{``\textit{UMN Research}''}: The communicator could contact their office's newsletter editor of UMN research, which is sent to all faculty and research-related staff monthly. For the communicator, they could get the proof of delivery within a minimal time cost. However, the information producer and relevant faculty might not get the value of awareness if the relevant faculty are too busy to read this newsletter. In short, this method would get a small value (less attention) with a small cost for the organization. The relevant faculty are affected most if they miss this information.

\noindent \textbf{\textit{Email tree}}\textit{: }Another way is to distribute the message to several relevant College Deans, such as the Dean of the College of Biological Sciences, the College of Science and Engineering, etc. Then the employees' value and costs depend on how their college distributes this message. If the Dean's office just send-to-all, the employees' time cost would still be high. Suppose the college is small and the Dean knows about the faculty's research interests and years of experience. In that case, they could directly send individual messages to the relevant employees, which are then very likely to be opened and read. Then the information producer and the relevant employees will get the value of awareness, and the organization suffers a minimal cost.

 However, the risk for the communicator and information producer is that they can't guarantee that all Deans will notice this request and forward this message --- Deans are busy. There might be relevant faculty being missed. This is why there are relevant people in each college --- for example, the communicator could send the message to all the Research Associate Deans to show openness / inclusion.

 In summary, we could not find an optimal method to deal with this type of message in the current organizational bulk email system. Our databases do not have enough information to support precisely targeting the recipients. All the current channels might cause a negative return or risk to some stakeholders.

\section{Opportunity and Future Research Directions}
 From an economic perspective, we should only send a bulk email when the organization's return with that email is positive from equation \eqref{GrindEQ__5_}. We should improve the system's mechanism towards maximizing this return. Proposing a model is not the same as instantiating/calibrating it. Therefore we propose several research directions and potential interventions below, which might make organizational bulk email system's economic mechanism more effective.

\subsection{ Preliminary step: operationalize and measure value and cost}

 Many of the value and costs proposed in the model are not operationalized and measured in the current CRM platforms. The CRM platforms mainly track open rates by inserting an invisible pixel in the email and track click rates by rewriting the links uniquely.\footnote{ \url{https://help.salesforce.com/s/articleView?id=sf.pardot\_emails\_open\_tracking.htm} } However, awareness, engagement, time, money, and reputation are not recorded and measured anywhere. Operationalizing these effects is necessary for building mathematical models assessing bulk email system performance.

 Measuring these effects is also challenging. For example, due to privacy concerns, many browsers do not allow access to the precise time of loading an image and the access to the time gap between loading two images, which makes it difficult to measure the reading time of emails. Testing the bulk messages' awareness with minimal effort from the recipient's side also needs to be solved. After that, calculating money cost would need a model to link reading time and salary.

\subsection{ Interventions on the sender-side}

 The motivation of getting the proof-of-delivery ($v^{evd}_{c,m}$) within a minimal time cost ($c^{tim}_{c,m}$) is an important part of the return formulas of communicators and clients. This is why they often select to send-to-all and then employees receive many irrelevant organizational bulk emails. To incentivize communicators and clients to spend more time targeting and designing bulk emails, there are three kinds of interventions from the equations \eqref{GrindEQ__2_} and \eqref{GrindEQ__3_}: reducing the targeting cost (decrease $c^{tim}_{c,m}$), making the change of reputation observable to communicators (measure $\mathrm{\Delta }r$), and incorporating employees' costs ($\sum_{\mathrm{m,e}}{I_{e,m}p_{e,m}c^{tim}_{e,m}}$) into the return formula.

\noindent\textbf{Reducing the targeting/filtering cost.} The databases of the current bulk email systems only provide minimal data like units and job titles. However, this information is not enough to represent employees' preferences. If we can measure open rates, click rates, and reading time, we could predict employees' preferences through collaborative filtering techniques \cite{chen2020item}, deep learning models \cite{liu2021sg, cui2021tf}, weakly supervised learning \cite{liu2021densernet, cui2023faq}, etc. More importantly, we need to understand the organization's preferences. A possible direction is to learn from the employees' job descriptions. When the organization recruits an employee for a position, it posts a job description to introduce the responsibility of this job. We could build a knowledge management system \cite{tiwana2000knowledge} along with natural language processing models like Bert, XLM \cite{10.1145/3543873.3584633, 10.1145/3543873.3584622, tian2023fashion, tian2020cross} to learn which messages would be work-relevant to the corresponding employees with these job descriptions. After we build such models, we could either choose to help communicators better target messages and narrow the range of recipients or implement better filtering on the recipient-side to help them select those relevant messages. For example, we could send a message in several rounds. And in the later rounds, we only send the message to the employees who are similar to those who have clicked the message in the previous rounds.

\noindent\textbf{Making the change of reputation observable to communicators.} As we discussed in the model, the outcomes of organizational bulk emails influence themselves, their channels' future emails, other bulk emails, and other channels. Each time we send a message, we could measure its influence on the channels' reputations (e.g., by the difference between the open rates/reading time of this email and the next email from this channel). And when we detect a decrease in reputation, we remind the communicators and clients that they should put more effort in targeting and designing the email.

\noindent\textbf{Incorporating employees' reading cost into communicators and clients' return.} As we discussed above, for the communicators, it is convenient to select ``send-to-all'' and expensive to select ``filter and target''. Besides the technical reasons in targeting, selecting ``send-to-all'' is also caused by the invisibility of the organization's cost. For example, though the communicators need to spend 300 minutes to filter the recipients while ``send-to-all'' only costs them 3 minutes, ``send-to-all'' would cost over 10,000 minutes for the whole organization. We need to measure and present the overall cost to the communicators given their actions on ``send via newsletter'' versus ``send via single-message email'', ``send-to-all'' versus ``precisely targeting''. The cost could be utilized in 3 ways: first, showing the cost of different approaches in the training sessions with communicators on incentivizing them to pay more effort in targeting; second, showing the estimated communication cost to support communicators in selecting communication channels for each message; third, building a budget system to allocate time budgets to departments and their communicators centrally. For example, an employee could at most spend 20 minutes in bulk emails per week, or a department could at most use 5 minutes of each employee per week.

\noindent\textbf{ Employing a human-centered approach.} The discussion in this chapter indicates the necessity of employing a human-centered approach \cite{guo2022human} to the problem of organizational bulk email communication. For example, in the case study of ``\textit{UMD Chancellor Search}'', we pointed out that for the communicators and information producers, ``showing a message is important'' is different from ``a message is really important''. This inconsistency indicates senders' disagreements and misperceptions on the optimal communication channels for the corresponding messages --- there is no common criterion for the communicators to understand the value of a message, and which channel they should use. Therefore, besides technical solutions, human-centered approaches are necessary for such social-technical systems, like educating the senders on the bulk email practices they should apply.

\subsection{ Intervention on the recipient-side}

 From equation \eqref{GrindEQ__1_}, we could see that employees care more about the messages that are interesting or work-relevant to them but less about the messages that are viewed as important-to-organization. Then many high-level messages, such as the messages from the University Senate (Figure 1), would not be read. There is a need to coordinate different stakeholders' preferences. At first, we need to study the size of the preference conflicts between the organization and the employees, i.e., the value of $\mu $ and $\lambda $ in equations \eqref{GrindEQ__1_} and \eqref{GrindEQ__5_} --- to how much extent the employees perceive the value of a message that is important to the organization and to how much extent the organization perceives the value of a message that is interesting to an employee. After that, there are three possible mechanisms for incentivizing employees to read the ``important'' messages. First, we could incentivize employees to read those messages by putting them next to the ``interesting messages''. Second, we could explain why a bulk message is important to employees by explainable recommendation \cite{zhang2018explainable, chen2022relax, chen2022explain}. Third, we could design performance evaluation systems \cite{ma2022traffic} to encourage employees to read those important messages \cite{ahmed2013employee}, such as giving badges or rewards to employees if they paid enough attention and sending reminders if they forgot to read.

\section{Conclusion} 

 We proposed an economic model to describe different stakeholders' actions, value, and costs in organizational bulk email systems. The model shows that, with different goals, none of the stakeholders have the ability to see the global effects of their actions on the system's effectiveness and the organization's costs and value. Based on the model, we identified two main opportunities: 1) design mechanisms for encouraging employees to read the “important” messages (Chapter 5); 2) make the change of reputation and employees' cost visible to senders (Chapters 6 \& 7).

\chapter{Multi-Objective Personalization of Organizational Bulk Emails}
\label{study2}


\section{Introduction}
As we found in the interviews (Chapter 3), many bulk messages are viewed as ``important-to-organization'' information by organization leaders and communicators. They build the awareness of the community and promote the organization's missions. However, normal employees may prefer messages more relevant to their jobs or interests \cite{kong2021learning}. Organizations face the challenge of balancing prioritizing the messages they prefer employees to know (tactical goals) while maintaining employees' positive experiences with these bulk emails, then they continue to read these emails in the future (strategic goals). In this study, we evaluate the use of personalization to help organizations reach these goals. \footnote{Ruoyan Kong, Charles Chuankai Zhang, Ruixuan Sun, Vishnu Chhabra, Tanush-
srisai Nadimpalli, and Joseph A. Konstan. Multi-objective personalization in
multi-stakeholder organizational bulk e-mail: A field experiment. Proc. ACM
Hum.-Comput. Interact., 6(CSCW2), nov 2022. doi: 10.1145/3555641 \cite{10.1145/3555641}}




Figure \ref{fig:brief}, the newsletter U of M Brief of our study site, is an example of the preference mismatch above. It is a weekly newsletter sent to all the employees across all 5 campuses with around 30 messages and 7 sections (top news, u-wide news, and each campus' news) in each newsletter. For example, the university wants employees to know what its governing board is doing, so it puts Boards of Regents updates as the first message of the Brief with the hope that all the employees who opened this Brief will notice this message (see Figure \mbox{\ref{fig:brief}}). However, the employees might perceive it as unactionable \textbf{high-level information} and view this Brief as irrelevant to their job. They might stop reading Brief after seeing this message.



Therefore, designing organizational bulk emails is a \textbf{multi-objective problem} for organizations. Their \textbf{tactical} objective is to use bulk emails to make employees aware of ``important'' messages, such as the board of regents meeting updates. At the same time, they have the \textbf{strategic} objective of maintaining the effectiveness of their communication channels by ensuring employees see messages they perceive as relevant and continue to read in the future (the value of protecting channel reputation in chapter 4). Organizations need to balance their short-term tactical goals and long-term strategic goals in designing these emails.

\begin{table}[!htbp]
\begin{minipage}{.48\textwidth}

\centering
  \includegraphics[width=0.9\columnwidth]{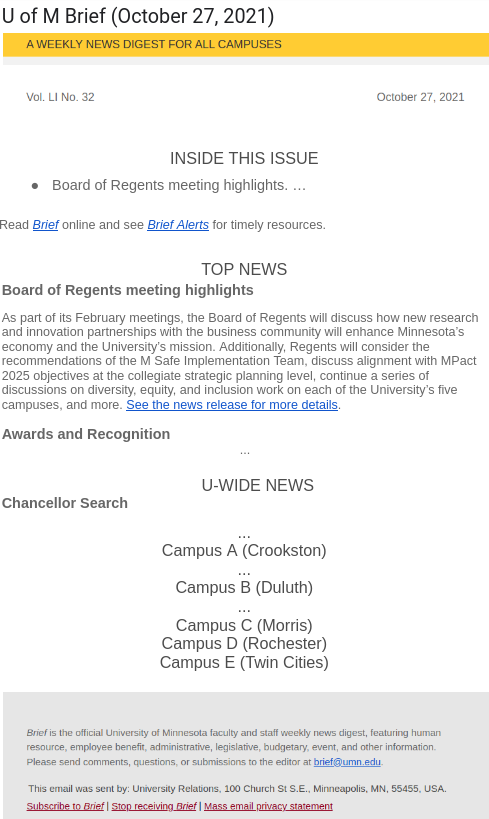}
  \captionof{figure}[A sample Brief on Oct 27, 2021]{A sample Brief on Oct 27, 2021. The messages within each section are hidden in this picture. For a full email of a personalized Brief, please see Appendix \ref{app6}.}~\label{fig:brief}

\end{minipage}%
\hspace{0.05in}
\begin{minipage}{.46\textwidth}
\centering
  \includegraphics[width=0.9\columnwidth]{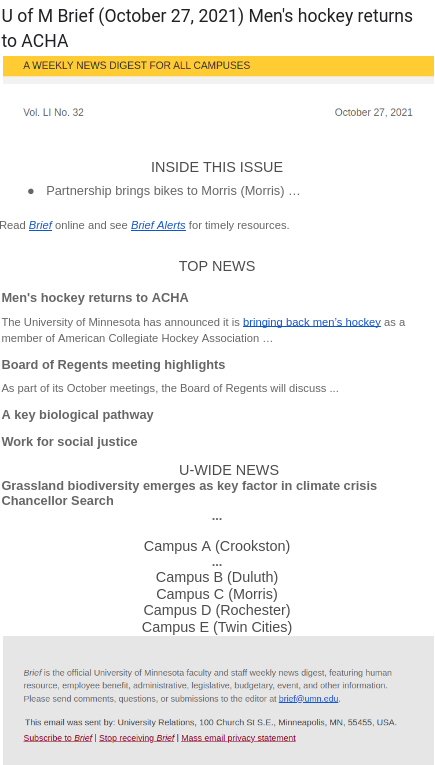}
  \captionof{figure}[A sample personalized Brief]{A sample personalized Brief. The messages within each section are hidden in this picture.}~\label{fig:per_brief}
\end{minipage}%
\end{table}

\hlc[white]{Here we see the opportunity of exploring personalization based on both \textbf{organization's preference} (organization's view on message's work-relevance and importance to employees as assessed by the newsletter editors) and \textbf{employee's preference} (employee's view on message's work-relevance and interest level to themselves) to help organizations reach these communication goals. For example, a design we tried with the newsletter above (see Figure \mbox{\ref{fig:per_brief}}) is to put \textbf{organization-preferred messages} (e.g., board of regents meetings) adjacent to \textbf{employee-preferred messages} (e.g., Men's hockey, in a case where the specific employee like sports) to balance these two interests. Specifically, we conducted an 8-week field experiment with a university newsletter and 141 employees. We employed a 4x5x5 factorial design in personalizing subject line/top news/message order. We measured these designs' influences on the open/interest/recognition rate of the whole newsletter (strategic goals) and the recognition/read-in-detail rate of the messages within it (tactical goals).}

Our contributions include two parts. Regarding the fundamental theoretical advances, we studied the unique personalization problem with organizational bulk emails where the short-term tactical goals might not align with the longer-term strategic goals, while the previous studies are most about commercial bulk emails, whose tactical goals and strategic goals are often aligned --- these studies usually prioritize the messages that recipients would like, such as the products the recipients might be interested in \cite{sahni2018personalization, wattal2012s}. Also, we conducted an 8-week field experiment to enable employees to get used to the experimental newsletters, while the research on organizational emails is mainly lab experiments (for example, \citeauthor{10.1145/3290605.3300604} invited participants to their lab to try an email visualization tool \cite{10.1145/3290605.3300604}), dataset analysis (the datasets often used include Enron \cite{klimt2004enron} and Avocado dataset \mbox{\cite{yang2017characterizing}}), and observational field studies (for example, \citeauthor{mark2016email} observed how email use influences employee's stress \cite{mark2016email}). Regarding the practical advances, we provided tradeoffs and suggestions for organizations in designing bulk emails. We designed a personalization framework for organization's bulk emails, including the process to collect stakeholders' preferences, the algorithms to generate personalized bulk emails, and the mechanisms to evaluate their performance. The rest of this chapter includes related work (2), background (3), methods (4), results (5), and discussion (6). This study was approved by the IRB of the University of Minnesota (STUDY00012816).

\section{Related Work \& Gaps}
\subsection{Bulk Email Personalization's Data Sources}
Personalization has been pointed out as a solution for email overload \cite{cecchinato2014personalised} and several studies used personalization to improve commercial bulk emails' performance. The personalization data sources could be demographic information like names \cite{sahni2018personalization, wattal2012s}, majors, departments \cite{trespalacios2016effects}; or preference information like browsing history \cite{wattal2012s}. Though Sahni et al.'s experiments found adding recipients' names to subject lines useful for improving open rate \cite{sahni2018personalization}, Wattal et al. \cite{wattal2012s} found that customers responded negatively to emails with identifiable information. Trespalacios and Perkins also found the effect of adding identifiable demographic information insignificant in an experiment with a university email \cite{trespalacios2016effects}. Hawkins et al. pointed out that personalized messages need to provide the recipients with new information about themselves instead of simply adding names or addresses \cite{hawkins2008understanding}. Wattal et al. \cite{wattal2012s} personalized email content based on customers' purchasing preferences and received positive responses. We personalize based on preferences instead of demographics in this study. We do not want to simply add the recipients' names to every internal newsletter of the organization. The employees could quickly learn that seeing their names in the organizational bulk emails means nothing special.

\subsection{Bulk Email Personalization's Design Choices}

We reviewed different ways to personalize commercial bulk emails in Chapter \ref{related_work_chapter}, including subject line, top section, selection of contents, order of contents, and visual designs. \hlc[white]{In this chapter, we focused on leveraging personalization to improve bulk email's performance in this paper. But it is worth noting that there are other potential factors influencing recipients' engagement with bulk emails. For example, the email marketing platform Mailchimp found that sending frequency, writing (like the use of emojis), and sender's industry influence bulk email's open rate (ranging from 15\% to 28\% according to their report) \mbox{\cite{chimp2018average}}. Bulk email's from lines, signature lines \mbox{\cite{jenkins2008truth}}, and sending time and day \mbox{\cite{abrahams2010multi, bilovs2016open}} are also found to influence open rate by around 10\% in the previous experiments from marketing, communication, and management science.}

\subsection{Gaps in Personalizing Organizational Bulk Emails}


Personalizing organizational bulk email is different from personal, commercial bulk email in several ways. Most important,
employees may have an obligation to read, know, and act upon information from their employer --- even information
they may personally not find interesting --- in a manner that does not apply to typical commercial bulk email. While
commercial bulk email may use one-off branding (to focus on one-time response rates) or recurring branding (to 
build a reputation and encourage repeat reading), organization bulk email nearly always uses recurring branding
tied to the structure and leadership of the organization. Accordingly, organizations always have to balance the
tactical goal of having employees read the messages they choose to send (and view as important) while also
maintaining the effectiveness of the communication channel by having employees perceive the messages as 
relevant. There is no study on how different ways of personalization could help organizations achieve these two types of goals. We introduce a field experiment to bridge this gap.

\section{Study Overview and Background}
\begin{figure}[!htbp]
\centering
  \includegraphics[width=0.85\columnwidth]{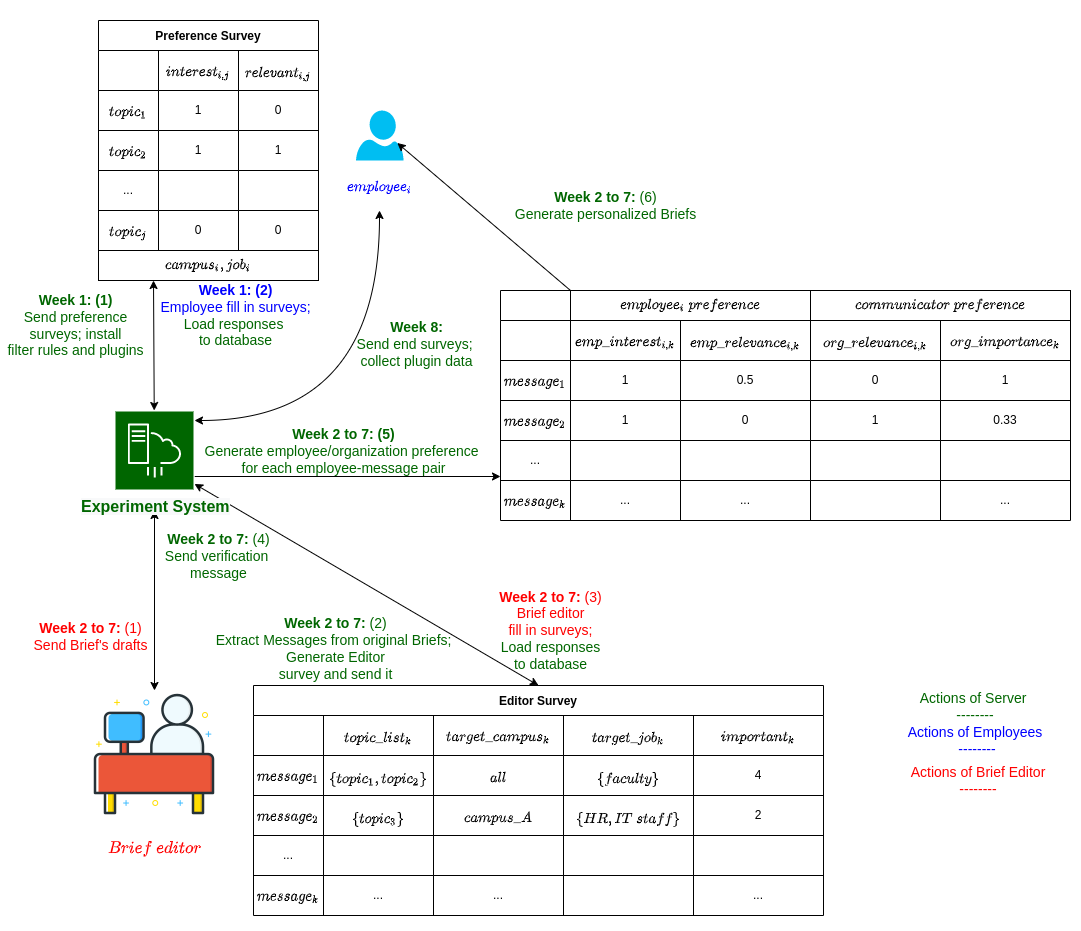}
  \caption[The outline of the experiment procedure]{The outline of the experiment procedure. Weeks 2 to 7's steps were repeated weekly. We sent the original Brief in week 2.}~\label{fig:procedure}
\end{figure}

This study is aimed to use personalization to help organizations achieve their tactical goals (getting employees to know important messages) and strategic goals (keeping employees interested in the newsletters) of bulk communication. Next, we first introduce the newsletter we personalize. After that, we introduce the different ways of personalizing the newsletter we tried, and propose our hypothesis on these designs' influence on the communication goals. Then in the method section, we introduce the 8-week field experiment we carried out to evaluate these hypotheses, including the participant recruiting process, how we calculate the employees'  and organizations' preferences with the newsletter, and how we implement the different personalization mechanisms and measure employees' feedback. Figure \ref{fig:procedure} is the outline of this experiment.

\subsection{Studied Newsletter}

Before the study, we met with three communicators from the central offices of the study site and decided to experiment with the newsletter U of M Brief (see figure \ref{fig:brief}), which is sent to all the employees weekly (with subject line \textit{``U of M Brief [Date]''}). Each brief contains around 30 messages and 7 sections. Regarding the organization's preference for messages, Brief encourages people to submit information about:
\begin{itemize}
    \item ``need-to-know'' administrative news
    \item messages that aimed to make the university more accessible and that create connections among faculty and staff
    \item promoting healthy lives
    \item the university’s mission of outreach, research, teaching, and education
\end{itemize}


\subsection{Designs}
\hlc[white]{We considered several personalization designs, including original/random/employee-preferred/organization-preferred/mixed designs. It is worth noting that we focus on personalizing the order of messages in a newsletter in this chapter, because we are asked to send out all the messages. There are other personalization designs we have not tried, such as filtering out irrelevant messages, highlighting important messages, personalizing the time of sending messages, etc.

Let us use Figure \mbox{\ref{fig:brief}}'s Brief as an example. Suppose a faculty member is interested in biology stories and sports while the organization wants them to know about administrative and social justice updates.}

\noindent\textbf{A Subject line}: which message to be added to the Brief's subject line.

\textbf{A1} Original subject line: \textit{``U of M Brief (October 27, 2021)''}.

 \textbf{A2} Random subject line: the original subject line with a random message from the newsletter, e.g., \hlc[white]{\textit{``U of M Brief (October 27, 2021) - Fall 2021 Capstone presentations''}}.

 \textbf{A3} Organization-preferred subject line: the original subject line with the message that the organization mostly preferred the faculty member to read (see 5.4.2 for the definition of organization/employee's preference), e.g., \hlc[white]{\textit{``U of M Brief (October 27, 2021) - Board of Regents meeting highlights''}}.

 \textbf{A4} Employee-preferred subject line: the original subject line with the message that the faculty member mostly preferred, e.g., \hlc[white]{\textit{``U of M Brief (October 27, 2021) - Men's hockey return to ACHA''}}.

\noindent\textbf{B Top news}: which 4 messages are to be selected as the Brief's top news.

\textbf{B1} Original top news: the same top news as Figure \ref{fig:brief}.

\textbf{B2} Random top news: use 4 random messages.

\textbf{B3} Organization-preferred top news: use the 4 messages the organization most preferred the faculty member to know. \hlc[white]{E.g., Board of Regents meeting highlights, Work for social justice, etc}. B3's messages might be different from B1's messages as B3 is personalized --- the organization might set different priorities for different employees.

\textbf{B4} Employee-preferred top news: use the 4 messages the faculty member mostly preferred as top news. \hlc[white]{E.g., Men's hockey return to ACHA, A key biological pathway, etc}.

\textbf{B5} Mixed top news: mix 2 employee-preferred messages and 2 organization-preferred messages in top news. \hlc[white]{E.g., Men's hockey return to ACHA, Board of Regents meeting highlights, A key biological pathway, Work for social justice}.

\noindent\textbf{C Message Order}: how to sort the messages in the non-top sections of this Brief.

\textbf{C1} Original order: use the original Brief's messages' order.

\textbf{C2} Random order: sort the messages randomly.

\textbf{C3} Organization-preferred order: sort the messages by the organization's preference (see 5.4.2 for the definition of organization / employee's preference).

\textbf{C4} Employee-preferred order: sort the messages by the faculty member's preference.

\textbf{C5}\hlc[white]{ Zipper order: repeat this process --- select the message with the highest employee-preference score (see 6.4.2), the message with the highest organization-preference score, the message with the 2nd highest employee-preference, etc}.

If a message is added to the subject line but not selected to top news, we add it to the end of top news to avoid the employees feeling deceived if they click into the Briefs because of the subject lines. If a message from the campus sections was selected to top news, the name of its campus would be added to its title.

Within each treatment, we had two control groups: a good original control group (A1, B1, C1) which used the original subject lines/top news/message order --- as we discussed, these were carefully selected by an experienced editor (the communicator of Brief) according to their criterion on how to design Briefs; a bad random control group (A2, B2, C2) which used random subject lines/top news/message order generated by the system. Figure \ref{fig:per_brief} is a sample personalized Brief for this faculty member if we assigned them to A4 x B5 x C5. 


\subsection{Research Questions and Hypotheses}

We propose hypotheses and questions on these designs' influences on the communication goals. 

\noindent\textbf{A Subject lines.} When subject lines match the employees' preferences, Brief might achieve these strategic goals: 

\noindent  Adding employee-preferred messages on subject lines will increase the newsletter's interest rate (\textbf{H1.1}) / reading time (\textbf{H1.2}) / overall recognition rate (\textbf{H1.3}) / open rate  (\textbf{H1.4}).

On tactical goals, the messages on subject lines should have a greater chance of being seen by the employees compared to the messages not on subject lines:

\noindent  (\textbf{H1.5}) Putting messages on subject lines will increase these messages' recognition rates.

We expect only to see an improvement in the read-in-detail rate for those employee-preferred messages, as the content should be interesting to employees to make them click/read in detail \cite{kim2016click, kessler2019we}:

\noindent  (\textbf{H1.6}) Putting employee-preferred messages on subject lines will increase their read-in-detail rates compared to the messages not on subject lines. 

\noindent\textbf{B Top News.} When top news matched the employees' preferences, Brief might achieve a better interest rate:

\noindent \textbf{H2.1} Putting employee-preferred messages in top news will increase the newsletter's interest rate.

We do not have theories to predict the effect of placing employee-preferred messages in top news on the reading time and overall recognition rate. For example, when putting employee-preferred messages in top news, employees might be motivated to read the rest of the newsletter or only read top news and leave. We propose these questions:

\noindent What is the effect on the newsletter's reading time (\textbf{Q2.2}) and overall recognition rate (\textbf{Q2.3} when we 

\begin{itemize}
    \item put organization-preferred messages in top news
    \item put employee-preferred messages in top news
    \item mix employee-preferred messages and organization-preferred messages in top news
\end{itemize}


We hope that the mixed top news let employees read the organization-preferred messages when reading interesting messages: Mixing organization-preferred messages with the employee-preferred messages in top news will increase the organization-preferred messages' recognition rates (\textbf{H2.4}).

Besides that, we also had the hypotheses similar to those for the subject lines on tactical goals:

\noindent  (\textbf{H2.5}) Putting messages in top news will increase their recognition rates.

\noindent  (\textbf{H2.6}) Putting employee-preferred messages in top news will increase their read-in-detail rates. 

\noindent\textbf{C Message Order.} We were uncertain about the direction and scale of message order's effect. We propose the following questions: what is the effect of sorting messages by employee's preference/organization's preference/zipper order on the overall recognition rate (\textbf{Q3.1}) / reading time (\textbf{Q3.2}) of the newsletter?

\noindent(\textbf{Q3.3}) What is the effect of interleaving messages (sorting messages by the zipper order of employee/organization's preference) on the recognition rates of the organization-preferred messages?

We hypothesize that sorting by employee's preference would make them feel the Brief is more interesting as they will see interesting messages in advance:

\noindent(\textbf{H3.1}) Sorting messages by employee's preference will increase the interest rate of the newsletter.

We summarized our hypotheses and research questions in Table \ref{tab:rq}. There were blanks (marked as to be observed) in this table when we did not find any related work or reasons to make a guess on a significant effect. We just observed what happened in these blanks. Besides the questions above.

\begin{table}
\centering
\caption[Summary of the newsletter's communication goals, hypotheses, and research questions]{Summary of the newsletter's communication goals, hypotheses, and research questions. org-pref messages: organization-preferred messages. emp-pref messages: employee-preferred messages. A1/2, B1/2, C1/2 are control groups. There were blanks (to be observed) in this table when we did not find any related work or reasons to make a guess on a significant effect. NA: not applicable.}
\label{tab:compare}
\resizebox{\textwidth}{!}{%
\begin{tblr}{
  cell{1}{3} = {c=4}{c},
  cell{1}{7} = {c=2}{c},
  cell{3}{1} = {r=4}{},
  cell{7}{1} = {r=5}{},
  cell{12}{1} = {r=5}{},
  vlines,
  hline{1-3,7,12,17} = {-}{},
  hline{4-6,8-11,13-16} = {2-8}{},
}
                       &                                           & \textbf{Strategic Goal }            &                                            &                                                        &                             & \textbf{Tactical Goal }                                                                          &                                         \\
\textbf{Group}         & \textbf{Treatment}                        & {Interest Rate \\of this Brief}     & {Reading time \\of this Brief}             & {Recognition Rate \\of this Brief}                     & {Open Rate \\of this Brief} & {Recognition Rate \\of this message}                                                             & {Read-in-detail Rate \\of this message} \\
{A: \\Subject \\lines} & 1: Original                               & {To be\\observed}                   & {To be\\observed}                          & {To be\\observed}                                      & {To be\\observed}           & {To be\\observed}                                                                                & {To be\\observed}                       \\
                       & {2: Add a \\random\\message}              & {To be\\observed}                   & {To be~\\observed}                         & {To be\\observed}                                      & {To be\\observed}           & {To be\\observed}                                                                                & {To be\\observed}                       \\
                       & {3: Add an \\org-pref \\message}          & {To be \\observed}                  & {To be \\observed}                         & {To be \\observed}                                     & {To be \\observed}          & {H1.5 Increase  \\recognition rate.}                                                             & {To be \\observed}                      \\
                       & {4: Add an \\emp-pref \\message}          & {H1.1 Increase \\ interest \\rate.} & {H1.2 Increase \\reading time.}            & {H1.3 Increase \\overall \\recognition rate.}          & {H1.4 Increase\\open rate.} & {H1.5 Increase  \\recognition rate.}                                                             & {H1.6 Increase  \\read-in-detail rate.} \\
{B: \\Top \\news}      & {1: Original\\top news}                   & {To be\\observed}                   & {To be\\observed}                          & {To be\\observed}                                      & NA                          & {To be\\observed}                                                                                & {To be\\observed}                       \\
                       & {2: Put \\random\\messages}               & {To be\\observed}                   & {To be\\observed}                          & {To be\\observed}                                      & NA                          & {To be\\observed}                                                                                & {To be\\observed}                       \\
                       & {3: Put \\org-pref \\messages}            & {To be \\observed}                  & {Q2.2 How does\\it affect\\reading time?}  & {Q2.3 How does it\\affect overall \\recognition rate?} & NA                          & {H2.5 Increase  \\recognition rate.}                                                             & {To be \\observed}                      \\
                       & {4: Put \\emp-pref \\messages}            & {H2.1 Increase \\ interest \\rate.} & {Q2.2 How does\\it affect\\reading time?}  & {Q2.3 How does\\it affect  overall\\recognition rate?} & NA                          & {H2.5 Increase  \\recognition rate.}                                                             & {H2.6 Increase  \\read-in-detail rate.} \\
                       & {5: Mix \\emp-pref\\/org-pref \\messages} & {To be \\observed}                  & {Q2.2 How does\\it affect\\reading time?}  & {Q2.3 How does it\\affect overall \\recognition rate?} & NA                          & {H2.4 Increase  \\recognition rate \\of org-pref messages. \\H2.5 Increase  \\recognition rate.} & {To be \\observed}                      \\
{C: \\Order}           & {1: Original\\order}                      & {To be\\observed}                   & {To be\\observed}                          & {To be\\observed}                                      & NA                          & {To be\\observed}                                                                                & {To be\\observed}                       \\
                       & {2: Random\\order}                        & {To be\\observed}                   & {To be\\observed}                          & {To be\\observed}                                      & NA                          & {To be\\observed}                                                                                & {To be\\observed}                       \\
                       & {3: Sort \\by org\\-preference}           & {To be \\observed}                  & {Q3.1 How does\\it affect \\reading time?} & {Q3.2 How does it\\affect overall \\recognition rate?} & NA                          & NA                                                                                               & NA                                      \\
                       & {4: Sort \\by emp\\-preference}           & {H3.1 Increase \\ interest \\rate.} & {Q3.1 How does\\it affect \\reading time?} & {Q3.2 How does it\\affect overall\\recognition rate?}  & NA                          & NA                                                                                               & NA                                      \\
                       & {5: Sort by \\mix-order}                  & {To be \\observed}                  & {Q3.1 How does\\it affect \\reading time?} & {Q3.2 How does it\\affect overall\\recognition rate?}  & NA                          & {Q3.3 How does it\\affect  recognition rate\\of org-pref messages?}                              & NA                                      
\end{tblr}}
\end{table}

\section{Study Procedure}
\subsection{Overview}
We collaborated with the editor of Brief to conduct the experiment in the university with 141 employees. The experiment took 8 weeks. The steps were (see Figure \ref{fig:procedure}):

\noindent \textbf{Step 1 (week 1), recruiting and assigning participants}: through Brief and a communication newsletter. \hlc[white]{Each selected participant would be assigned to a treatment combination $A_i\times B_j\times C_k$ through the whole study}.
\begin{figure}[!htbp]
\centering
  \includegraphics[width=0.8\columnwidth]{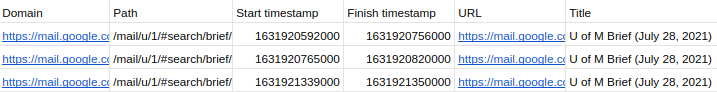}
  \caption[A sample csv file recorded by the plugin]{A sample csv file recorded by the plugin. It records the domain, path (which helps us to discriminate whether it is the experimental Brief), start/end reading timestamp, url, and tab titles.}~\label{fig:plugin}
\end{figure}

\noindent \textbf{Step 2 (week 1), collecting and calculating the employees' preferences}: \hlc[white]{the employees filled in preference surveys and we collected their work-relevance/interest scores for each topic}.

\noindent \textbf{Step 3 (week 2 to 7), collecting and calculating the organization's preferences}:  \hlc[white]{the editor sent the draft of the newsletter to the experiment system every week. The system extracted text and html, then sent the editor a survey to collect each message's topics (up to 4) and work-relevance/importance scores from the organization's perspective}.

\noindent \textbf{Step 4 (week 2 to 7), generating newsletters}: the system sent the original non-personalized newsletter to collect base performance data in week 2 and generated personalized newsletters based on the employees' experimental groups, organization's preference, and employee's preference during weeks 3 to 7. 

\noindent \textbf{Step 5 (week 8), collecting performance metrics and feedback}: the system sent out the end surveys to collect the recognition/read-in-detail/interest data, and the log files of a plugin which tracked open rates and reading time.




\subsection{Recruitment (step 1, week 1)}
The goal of this experiment is to study newsletter reading in a natural context. The scope of this experiment is organizations' newsletters sent to a large number of employees and their regular and occasional readers. We did not study employees who never read Briefs because reading the Brief is not their natural behavior --- being in the experiment could force their open rate and recognition rate to increase from 0 to a substantial level, which could largely interfere with the main outcomes of our interventions. Accordingly, we posted the recruitment message as the first message of a Brief. In the hope of broadening our participants, we also reached out to the communicator network and posted the message at the top of its newsletter.\footnote{We did two things to understand the generalizability in view of the Brief-centric recruitment --- 1) compare the participants from different sources: the 10 other employees recruited from the communication newsletter had a base recognition rate of the week 2's original Brief (29\%) similar to the other 107 employees recruited from Brief (30\%); 2) checking the number of occasional Brief readers: 44 of our participants did not open all the experimental Briefs, and 22 participants opened less than or equal to 60\% of the experimental Briefs (Brief's average open rate).} The newsletters only had their brands without specific messages on their subject lines which could bias the recruitment.\footnote{The messages closest to our recruitment message in that recruitment Brief had topics on Administrative News/Student Stories/Awards and Recognition. According to Table \mbox{\ref{tab:conflict}}, our participants were not more interested in these topics compared to other topics.}

We planned to have at least 100 participants. This number was estimated through 1000 simulations on the generalized linear mixed model power analysis tool SMIR \mbox{\cite{green2016simr}}. The analysis of open rate/interest rate, where we could only collect 5 data points from each employee, has the highest requirement. We target to observe a 15\% change in these metrics with a 20\% standard deviation and an 80\% power. Considering the likely dropout rate, we targeted 140 to 150 participants.

In the signup form, we asked the potential participants whether they were employees of the university and mainly used Gmail and Chrome in reading Briefs and selected those confirmed. We also collected their campuses and job categories. To make our surveys more concise, 20 job families from the university's human resource system were summarized into 7 job categories by two researchers and the Brief editor, according to whether these categories were considered different audience groups of Brief (see Appendix \ref{app6}).

We received 304 responses to our recruitment message and selected participants from this pool. We balanced the number of selected participants from different campuses, job categories, and recruiting sources and contacted 181 employees to set up the study (\hlc[white]{38 did not reply and were not enrolled}). Each employee filled in a preference survey on message's topics and had a 20-minute 1-to-1 zoom meeting with a research team member. In this meeting, we helped the employee: 1) set up a filter rule in their university Gmail, which archived all the original Briefs they received during the experiment into a separate folder. They were told to avoid checking this folder during the study; 2) install a plugin on their Chrome browser. The plugin only recorded the time they spent when they were on a tab with the text ``U of M Brief'' (see figure \ref{fig:plugin}). 141 participants completed the setup process (2 employees could not install the plugin and were not enrolled). The participants were compensated with a \$20 Amazon gift card after setting up the study.

\subsection{Methods}
To define and collect preference data within limited survey questions, we used ``topic'' as a bridge to connect messages
and preferences (we limit each message to 4 topics). 20 topics were summarized by a thematic analysis \cite{braun2012thematic} of 140 messages from 5 previous Briefs (see Appendix \ref{app6}). 5 research team members grouped them, labeled the clusters, and identified the hierarchy. The Brief editor checked the list and suggested two special topics --- ``news from my campus'' and ``news from other campuses''. We summarized the symbols we used in the personalization procedure and their definitions in Table \ref{tab:symbol}.



\subsubsection{Collecting and calculating employees' preferences (step 2, week 1)}
At the setup of this study, we asked the participants to fill out a preference survey (figure \ref{fig:procedure} - week 1). For each topic, we asked the participants to check whether the statements ``I would look up this category for messages interesting to me'' or ``I would look up this category for messages work-relevant to me'' applied to them separately. Let's call $employee_i$'s answers to $topic_j$ as $interest_{i,j}$ and $relevance_{i,j}$, which took the value 0 or 1. We also asked the campus ($campus_i$) and job category ($job_i$) in the survey.
\begin{table}[!htbp]
\centering
\caption[The variables used in the personalization procedure and their definitions]{The variables used in the personalization procedure and their definitions. Except that $org\_importance_{k}$ is the same for all the employees, the employee/organization's preferences were calculated by the inputs from the employees/editor on the left column.}
\label{tab:symbol}
\arrayrulecolor[rgb]{0.753,0.753,0.753}
\scalebox{0.65}{
\begin{tabular}{!{\color{black}\vrule}l|l!{\color{black}\vrule}l|l!{\color{black}\vrule}} 
\arrayrulecolor{black}\hline
\begin{tabular}[c]{@{}l@{}}\textbf{Inputs from }\\\textbf{Employees}\end{tabular}                & \textbf{\textbf{Definition}}                                                                                                          & \begin{tabular}[c]{@{}l@{}}\textbf{Employee's }\\\textbf{Preference}\end{tabular}                                                           & \textbf{Definition}                                                                                                                                                   \\ 
\arrayrulecolor[rgb]{0.753,0.753,0.753}\hline
$campus_i$                                                                                       & $employee_i$'s campus                                                                                                                 & $emp\_interest_{i,k}$                                                                                                                       & \begin{tabular}[c]{@{}l@{}}$employee_i$'s interest score for~$message_k$\\(0 to 1)\end{tabular}                                                                      \\ 
\hline
$job_i$                                                                                          & $employee_i$'s job category   (Appendix B)                                                                                                        & $emp\_relevance_{i,k}$                                                                                                                      & \begin{tabular}[c]{@{}l@{}}$employee_i$'s work-relevance score\\for $message_k$ (0 to 1)\end{tabular}                                                                \\ 
\hline
$interest_{i,j}$                                                                                 & \begin{tabular}[c]{@{}l@{}}whether~$employee_i$~looks for interesting \\messages from $topic_j$ (0 or 1)\end{tabular}                                    & $emp\_pref_{i,k}$                                                                                                                           & \begin{tabular}[c]{@{}l@{}}$employee_i'$s preference on $message_k$ \\($=(emp\_interest_{i,k}+emp\_relevance_{i,k})/2$,\\ 0 to 1)\\\end{tabular}                           \\ 
\hline
$relevance_{i,j}$                                                                                & \begin{tabular}[c]{@{}l@{}}whether~$employee_i$~looks for work-relevant \\messages from~$topic_j$~(0 or 1)\end{tabular}               &                                                                                                                                             &                                                                                                                                                                       \\ 
\arrayrulecolor{black}\hline
\begin{tabular}[c]{@{}l@{}}\textbf{\textbf{Inputs from }}\\\textbf{\textbf{Editor}}\end{tabular} & \textbf{\textbf{Definition}}                                                                                                          & \begin{tabular}[c]{@{}l@{}}\textbf{\textbf{\textbf{\textbf{Organization's }}}}\\\textbf{\textbf{\textbf{\textbf{Preference}}}}\end{tabular} & \textbf{\textbf{\textbf{\textbf{Definition}}}}                                                                                                                        \\ 
\arrayrulecolor[rgb]{0.753,0.753,0.753}\hline
$important_k$                                                                                    & \begin{tabular}[c]{@{}l@{}}$message_k'$s general importance to all the\\employees from the organization's view\\(1 to 4)\end{tabular} & $org\_importance_{k}$                                                                                                                       & \begin{tabular}[c]{@{}l@{}}$message_k$'s general importance score to all\\the employees~ from the organization's view\\($=(important_k-1)/3$, 0 to 1)\end{tabular}  \\ 
\hline
$target\_campus_k$                                                                               & $message_k$'s target campuses (list)                                                                                                  & $org\_relevance_{i,k}$                                                                                                                      & \begin{tabular}[c]{@{}l@{}}organization's work-relevance score for \\$message_k$ given~$employee_i$~(0 to 1)\end{tabular}                                             \\ 
\hline
$target\_job_k$                                                                                  & $message_k$'s target job categories~(list)                                                                                            & $org\_pref_{i,k}$                                                                                                                           & \begin{tabular}[c]{@{}l@{}}organization's preference on~$message_k$ given\\ $ employee_i$\\($=(org\_relevance_{i,k} + org\_importance_{k})/2$,\\ 0 to 1)\end{tabular}                                                        \\ 
\hline
\begin{tabular}[c]{@{}l@{}}$topic\_list_k$\\\end{tabular}                                        & \begin{tabular}[c]{@{}l@{}}$message_k$'s relevant topics (list, up to 4,\\ Appendix A)\end{tabular}                                                                                         &                                                                                                                                             &                                                                                                                                                                       \\
\arrayrulecolor{black}\hline
\end{tabular}
}
\end{table}

Then for $employee_i$ and $message_k$, we calculated the employee's preference on this message ($emp\_pref_{i,k}$) with 3 steps. First, $employee_i$'s interest score for $message_k$ is defined as:
\begin{small}
\begin{equation}
    emp\_interest_{i,k} =
    \begin{dcases}
    0, \text{if } employee_i \ \text{is not interested in other campuses' messages and } campus_i \\ \quad \text{not in} target\_campus_k \\ 
     avg(\{interest_{i,q}|topic_q \in topic\_list_k\}),               \text{otherwise}
\end{dcases}
\end{equation}
\end{small}

Second, we calculated $employee_i$'s work-relevance score $emp\_relevance_{i,k}$ for 

\noindent $message_k$ as:
\begin{small}
\begin{equation}
    emp\_relevance_{i,k} =
    \begin{dcases}
    0, \text{if } employee_i \ \text{does not look for other campuses' relevant messages or  }\\ \quad campus_i \text{not in} target\_campus_k  \\
     avg(\{relevance_{i,q}|topic_q \in topic\_list_k\}),               \text{otherwise}
\end{dcases}
\end{equation}
\end{small}

\hlc[white]{Third,  $employee_i$'s preference on $message_k$ is defined as the average of interest and relevance score, as belo. It is worth noting that max might also work here, but the author also tried using the maximum of interest and relevance score but found the selected messages less interesting and therefore decided to use average here.}
\begin{small}
\begin{equation}
    emp\_pref_{i,k} =  (emp\_interest_{i,k} + emp\_relevance_{i,k})/2
\end{equation}
\end{small}

\subsubsection{\hlc[white]{Collecting and calculating the organization's preferences (step 3, week 2 to 7)}}
During weeks 2 to 7, the Brief editor provided the organization's preference with each message weekly (Figure \ref{fig:procedure}, week 2 to 7 (1) - (5)). The Brief editor sent the draft Brief to the system every week. The system then retrieved the messages (subject lines, titles, content, html, etc.) from the draft Brief, generated the editor survey, and sent it to the editor. The system listened to the editor's responses and loaded the responses to the database. After the responses were successfully loaded, a verification message would be sent to the editor to ensure that the original Briefs could only be sent after the system got the data to generate personalized Briefs. The editor survey contained the following questions for each $message_k$:
\begin{enumerate}
    \item How relevant this message is in building community, pride, common understandings of excellence and mission of the university (from 1: not relevant to 4: very relevant)? ($importance_k$).  
    \item Select the employee categories that might find the message above work-relevant ($target\_job_k$).
    \item Specify the message's relevant topics (select no more than 4 topics) ($topic\_list_k$).
    \item There is an implicit question on target campus ($target\_campus_k$), as the editor suggested that the original Briefs' campus sections had already represented it --- the messages in the top news and U-Wide news are targeted at all campuses.
\end{enumerate}

\hlc[white]{Then we calculate the organization's preference ($org\_pref_{i,k}$) on $message_k$ given $employee_i$ by 3 steps. First, the organization's work-relevance score for $message_k$ given $employee_i$ with job category ($job_i$) and campus ($campus_i$) is}:
\begin{small}
\begin{equation}
    org\_relevance_{i,k} = 1,              if \ campus_i \in target\_campus_k \ \text{and} \ job_i \in target\_job_k; \ 0, otherwise
\end{equation}
\end{small}

\hlc[white]{Second, $message_k$'s general importance score to all the employees is just the standardization of $important_k$. This value is the same for all the employees, as the editor suggested that the important messages should apply to all}:
\begin{small}
\begin{equation}
    org\_importance_{k} = (important_k-1)/3
\end{equation}
\end{small}

\hlc[white]{Third, the organization's preference on $message_k$ given $employee_i$ is defined as}:
\begin{small}
\begin{equation}
    org\_pref_{i,k} =  (org\_relevance_{i,k} + org\_importance_{k})/2
\end{equation}
\end{small}

\subsubsection{Generating newsletters (step 4, week 2 to 7)}
After we calculated these preference data, the system generated personalized Briefs from weeks 3 to 7 (original Brief for week 2) and sent these Briefs to the employees. We give the employees the choice to select to receive the email at 6 AM or 9 AM, given which time is better for them to receive Briefs and read it on their laptop or desktop's Chrome with the plugin we installed (eventually, we did not observe significant differences on the performance metrics between the 6 AM and 9 AM group).

\hlc[white]{With a 4 x 5 x 5 factorial design on the treatments A (subject lines), B (top news), C (message order) below, each participant would be assigned to a treatment combination $A_i \times B_j \times C_k$ through the study randomly. Their Briefs' would be generated according to the criterion in 4.2}, based on the employee preference $emp\_pref$ and organization preference $org\_pref$ we calculated for each message (Figure \ref{fig:procedure}, week 2 to 7 (6)).

\subsection{Collecting Metrics (step 5, week 8)}
At week 8, we sent out the end survey (Figure \ref{fig:procedure}, week 8). Participants were compensated with another \$20 Amazon gift card after submitting the end surveys. We collected the recognition data for the message below: 1) week 2 and week 7's messages in the top news, up to the top 10 messages in the u-wide news, up to the top 2 messages in the participant's campus news; 2) weeks 5 to 7's messages in the top news; 3) weeks 3 to 7's messages on the subject lines.

Specifically, we measured these 6 metrics:

\noindent$\star$ Recognition/read-in-detail rate: the percentage of the investigated messages being self-reported as seen/read-in-detail by the employees in the study's end survey (tactical goal). For example, the percentage of the messages in Top News the employees reported ``seen'' when Top News were all organization-preferred messages. We could not measure the reading time of a single message in Brief because of the technical challenge of tracking specific regions' reading time naturally (our future work). First, many browsers (e.g., Chrome and Gmail) block access to the exact loading time of invisible pixels. Second, there is a lack of low-cost eye-tracking technology (e.g., eye-tracking based on a single computer camera) \cite{ferhat2016low}; and employees might also pay more attention to bulk emails when being recorded by camera.

\noindent$\star$ \hlc[white]{Open rate: the percentage of the investigated Briefs being opened by the employees (strategic goal). For example, the percentage of a Brief being opened when we put the organization-preferred messages on subject lines.}

\noindent$\star$ \hlc[white]{Interest rate: the percentage of the investigated Briefs being rated as ``interesting'' by the employees (strategic goal). }

\noindent$\star$ \hlc[white]{Reading time: the average reading time of the investigated Briefs (strategic goal). }

\noindent$\star$ \hlc[white]{Overall recognition rate: the average of the recognition rates of all the investigated Briefs' messages (strategic goal). }

The recognition data was collected by the question ``Have you seen it in recent Briefs? No/Not Sure/Skimmed/Read fully''. We defined $employee_i$'s \textbf{$recognition$} as 1 if the answer is skimmed or read fully, and \textbf{$read\_in\_detail$} as 1 if the answer is read fully. After that, the survey asked the participants to indicate how interesting each Brief is to them in general from ``1 Not interesting'' to ``4 Very interesting''. The survey also collected plugin data. We would know the reading time of each Brief, and whether the participants opened a Brief or not. We also collected the interest scores (scale 1 to 4) and work-relevant scores (scale 1 to 4) for week 2's messages to study the size of conflicts. The order of these questions is randomized. The participants were asked not to search their inbox while answering these questions. For each experimental group (e.g., the Briefs with random subject lines), we reported and tested:

\noindent\hlc[white]{
 $\star$ \textbf{$recognition\_rate$} and \textbf{$read\_in\_detail\_rate$}: the percentage of the considered messages (e.g, the messages in the top news, the message on subject lines) in the experimental group that got $recognition = 1$ or $read\_in\_detail=1$.}

\noindent\hlc[white]{
 $\star$ \textbf{$overall\_recognition\_rate$}: the percentage of the messages in this experimental group's Briefs with $recognition = 1$.}

\noindent \hlc[white]{
 $\star$ \textbf{$interest\_rate$}: the percentage of Briefs that got an interesting level $\geq$ 3 in that experimental group.}

\noindent \hlc[white]{
$\star$ \textbf{$open\_rate$}: the percentage of Briefs that got $open = 1$ in that experimental group.}

\noindent\hlc[white]{
 $\star$ \textbf{$reading\_time$}: the average of the Briefs' reading time in that experimental group.}

We received 132 responses for the end survey, and 117 of them were complete. There were 15 incomplete responses either because the participants did not complete the surveys, the plugin was deleted or blocked by a Chrome update, or the participants lost access to their devices. This dataset contained the recognition data and read-in-detail data of 4242 messages, and the recognition data, open data, and reading time data of 702 Briefs in total. \hlc[white]{We did received 2 reports of participants forgetting to read in the browser, and their data was excluded.}

\hlc[white]{To avoid spillover effects \mbox{\cite{sinclair2012detecting}}, our participants were scheduled for separate 1-on-1 meetings in the setup, and they were not aware of each other's participation nor their experimental groups. We observed no communication or sharing that would have led to spillover effects. To avoid Hawthorne effects \mbox{\cite{jones1992there}}, we sent out the original Briefs in week 2, and measured the base performance data, which would be later included in our models as a factor. And our participants were in the experiment for 6 weeks, to avoid encouraging participants to pay more attention to read/remember these Briefs, we only analyze the performance data collected in the last week of this experiment.}


\section{Results}
\subsection{Analysis and Overview}
We summarize the performance of each experimental group in Table \ref{tab:result_original} and the results on the hypotheses and questions in Table \ref{tab:result}. We built mixed logistic models to evaluate categorical performance metrics (interest rate, recognition rate, open rate, and read-in-detail rate) and mixed linear models to test numerical performance metrics (reading time) by the afex package, which provided ANOVA table with likelihood-ratio tests for both linear and logistic models \cite{singmann2015package}. We had a random effect based on subjects (from which employee we collected this data point) \cite{barr}, and we selected likelihood-ratio tests because we had many levels on the random effect (number of participants) \cite{barr2013random}. The independent variables include the corresponding experimental groups and the base performance metrics (the average of that performance metric given the corresponding employee's reactions to week 2's original Brief). For the base open rate, we used the number of Briefs they opened in 2021/the number of Briefs they received in 2021 before the experiment. We asked the employees to input queries in their Gmail in the preference survey to retrieve this number. If they have deleted Briefs, they reported their approximate numbers. We excluded the employees who gave the same  interest scores to all the experimental Briefs from the analysis of interest scores and excluded the employees who opened all or did not open any of the experimental Briefs from the analysis of open rates.

\begin{table}[!htbp]
\caption[The performance metrics of each experimental group]{\textbf{The performance metrics of each experimental group}, and the standard deviations of that group's participants' metrics. The blank cells are not applicable or not of interest. non-s: messages not on subject lines. non-t:not in top news.}~\label{tab:result_original}
\centering
\arrayrulecolor{black}
\scalebox{0.5}{
\begin{tabular}{|l|l|l|l|l|l!{\vrule width \heavyrulewidth}l|l|} 
\hline
                                  &                                                                                & \multicolumn{4}{c!{\vrule width \heavyrulewidth}}{\textbf{Strategic Goal}}                                                                                                                                                                                                                                                                                                                                                                                                                                                                              & \multicolumn{2}{c|}{\textbf{Tactical Goal}}                                                                                                                                                                                                                                                                                                \\ 
\hline
\begin{tabular}[c]{@{}l@{}}\textbf{ Group }\end{tabular}                  & \textbf{Treatment}                                                             & \begin{tabular}[c]{@{}l@{}}Interest Rate \\of Brief (\%) \end{tabular}                                                               & \begin{tabular}[c]{@{}l@{}}Reading Time \\of Brief \\(seconds)\end{tabular}                                                          & \begin{tabular}[c]{@{}l@{}}Recognition Rate \\of Brief (\%)\end{tabular}                                                                & \begin{tabular}[c]{@{}l@{}}Open Rate of\\Brief (\%)\end{tabular}                                          & \begin{tabular}[c]{@{}l@{}}Recognition Rate \\of Message (\%)\end{tabular}                                                                                                                   & \begin{tabular}[c]{@{}l@{}}Read-in-detail \\Rate of \\Message (\%)\end{tabular}                                                \\ 
\toprule
{ \begin{tabular}[c]{@{}l@{}}Non-s\\messages\end{tabular}} 
&\begin{tabular}[c]{@{}l@{}}\ \\ \ \\ \ \end{tabular} 
&&&&& \textbf{41\%}, 20\%&\textbf{15\%}, 13\% \\
\hline
\multirow{4}{*}{ \begin{tabular}[c]{@{}l@{}}A:\\Subject\\lines\end{tabular}} 
&\begin{tabular}[c]{@{}l@{}}1: Original \\subject \\line\end{tabular}         
& \begin{tabular}[c]{@{}l@{}}\textbf{68\%}, 22\% \end{tabular}   &  \textbf{143s}, 92s &\textbf{44\%}, 25\% &\begin{tabular}[c]{@{}l@{}}\textbf{65\%}, 21\%\end{tabular} & & \\\cline{2-8}
&\begin{tabular}[c]{@{}l@{}}2: Add a \\random \\message\end{tabular}         
&\begin{tabular}[c]{@{}l@{}}\textbf{69\%}, 27\% \end{tabular} & \textbf{150s}, 105s &\textbf{39\%}, 20\% &\begin{tabular}[c]{@{}l@{}}\textbf{63\%}, 14\%\end{tabular} &\textbf{40\%}, 27\% &\textbf{15\%}, 17\% \\\cline{2-8}
&\begin{tabular}[c]{@{}l@{}}3: Add an \\org-pref \\message\end{tabular}         
& \begin{tabular}[c]{@{}l@{}}\textbf{67\%}, 29\%  \end{tabular}& \textbf{149s}, 96s &\textbf{40\%}, 23\% &\begin{tabular}[c]{@{}l@{}}\textbf{60\%}, 24\%\end{tabular} &\textbf{60\%}, 24\% &\textbf{18\%}, 19\% \\\cline{2-8}
& \begin{tabular}[c]{@{}l@{}}4: Add an\\emp-pref\\message\end{tabular}           
& \begin{tabular}[c]{@{}l@{}}\textbf{68\%}, 31\%  \end{tabular}& \textbf{141s}, 76s &\textbf{36\%}, 20\% &\begin{tabular}[c]{@{}l@{}}\textbf{76\%}, 8\%\end{tabular} &\textbf{56\%}, 27\% &\textbf{24\%}, 19\% \\
\hline
{ \begin{tabular}[c]{@{}l@{}}Non-t\\messages\end{tabular}} 
&\begin{tabular}[c]{@{}l@{}}\ \\ \ \\ \ \end{tabular} 
&&&&& \textbf{37\%}, 25\%&\textbf{13\%}, 14\% \\
\hline
\multirow{5}{*}{ \begin{tabular}[c]{@{}l@{}}B:\\Top\\news\end{tabular}}    
& \begin{tabular}[c]{@{}l@{}}1: Original \\top news\end{tabular}          
& \begin{tabular}[c]{@{}l@{}}\textbf{68\%}, 38\% \end{tabular}&\textbf{146s}, 98s &\textbf{38\%}, 25\% & &\textbf{44\%}, 27\% &\textbf{17\%}, 15\% \\
\cline{2-8}
& \begin{tabular}[c]{@{}l@{}}2: Put\\random \\messages\end{tabular}          
& \begin{tabular}[c]{@{}l@{}}\textbf{58\%}, 29\% \end{tabular} &\textbf{123s}, 68s &\textbf{30\%}, 17\% & &\textbf{32\%}, 16\% &\textbf{9\%}, 12\% \\
\cline{2-8}
& \begin{tabular}[c]{@{}l@{}}3: Put\\org-pref \\messages\end{tabular}          
& \begin{tabular}[c]{@{}l@{}}\textbf{63\%}, 25\% \end{tabular} &\textbf{158s}, 81s &\textbf{46\%}, 23\% & &\textbf{49\%}, 16\% &\textbf{18\%}, 12\% \\
\cline{2-8}
& \begin{tabular}[c]{@{}l@{}}4: Put \\emp-pref \\messages\end{tabular}          
& \begin{tabular}[c]{@{}l@{}}\textbf{76\%}, 21\% \end{tabular} &\textbf{142s}, 82s &\textbf{37\%}, 21\% & &\textbf{49\%}, 21\% &\textbf{22\%}, 18\% \\
\cline{2-8}
& \begin{tabular}[c]{@{}l@{}}5: Mix \\emp-pref\\/org-pref\\messages\end{tabular} 
&\begin{tabular}[c]{@{}l@{}}\textbf{72\%}, 23\% \end{tabular} &\textbf{162s}, 123s &\textbf{49\%}, 19\% & &\begin{tabular}[c]{@{}l@{}}\textbf{53\%}, 17\%\\  \end{tabular}  &\textbf{18\%}, 14\% \\
\hline
\multirow{5}{*}{ \begin{tabular}[c]{@{}l@{}}C:\\Message\\order\end{tabular}}   
& \begin{tabular}[c]{@{}l@{}}1: Original \\message\\order\end{tabular}  
&\begin{tabular}[c]{@{}l@{}}\textbf{66\%}, 24\% \end{tabular} &\textbf{136s}, 66s &\textbf{40\%}, 23\% & & & \\
\cline{2-8}
& \begin{tabular}[c]{@{}l@{}}2: Sort \\messages\\randomly\end{tabular}  
&\begin{tabular}[c]{@{}l@{}}\textbf{69\%}, 28\% \end{tabular} &\textbf{127s}, 77s &\textbf{34\%}, 20\% & & & \\ 
\cline{2-8}
& \begin{tabular}[c]{@{}l@{}}3: Sort \\messages by\\org-pref\end{tabular}  
&\begin{tabular}[c]{@{}l@{}}\textbf{59\%}, 30\% \end{tabular} &\textbf{166s}, 124s &\textbf{46\%}, 25\% & & & \\
\cline{2-8}
& \begin{tabular}[c]{@{}l@{}}4: Sort \\messages by\\emp-pref\end{tabular}  
&\begin{tabular}[c]{@{}l@{}}\textbf{74\%}, 31\% \end{tabular} &\textbf{146s}, 84s &\textbf{41\%}, 20\% & & & \\ 
\cline{2-8}
& \begin{tabular}[c]{@{}l@{}}5: Sort \\messages\\by mix-order\end{tabular}      
&\begin{tabular}[c]{@{}l@{}}\textbf{74\%}, 19\%\end{tabular} &\textbf{157s}, 104s &\textbf{39\%}, 22\% & & & \\
\hline
\end{tabular}
}
\end{table}

\begin{table}[!htbp]
\caption[Results of hypotheses and research questions]{Results of hypotheses and research questions. Format: experimental group mean = control group and its mean + difference between experimental and control groups (p.val). Signif. codes:  `*' 0.05, `+' 0.1, `NS' no significant effect was found. 
Control groups: rnd: random control group; org: original control group; \\
rnd-s: random subject lines' messages;  non-s: messages not on subject lines;\\
rnd-t: random top news' messages; org-t: original top news' messages; non-t: messages not in top news.
P-values were adjusted by Holm-Bonferroni correction \cite{holm1979simple, tian2021online, tian2019addis}.}~\label{tab:result}
\centering
\arrayrulecolor{black}
\scalebox{0.83}{
\resizebox{\textwidth}{!}{%
\begin{tabular}{|l|l|l|l|l|l!{\vrule width \heavyrulewidth}l|l|} 
\hline
                                  &                                                                                & \multicolumn{4}{c!{\vrule width \heavyrulewidth}}{\textbf{Strategic Goal}}                                                                                                                                                                                                                                                                                                                                                                                                                                                                              & \multicolumn{2}{c|}{\textbf{Tactical Goal}}                                                                                                                                                                                                                                                                                                \\ 
\hline
\begin{tabular}[c]{@{}l@{}}\textbf{ Group }\end{tabular}                  & \textbf{Treatment}                                                             & \begin{tabular}[c]{@{}l@{}}Interest Rate \\of Brief (\%) \end{tabular}                                                               & \begin{tabular}[c]{@{}l@{}}Reading Time of \\Brief (seconds)\end{tabular}                                                          & \begin{tabular}[c]{@{}l@{}}Recognition Rate \\of Brief (\%)\end{tabular}                                                                & \begin{tabular}[c]{@{}l@{}}Open Rate of\\Brief (\%)\end{tabular}                                          & \begin{tabular}[c]{@{}l@{}}Recognition Rate \\of Message (\%)\end{tabular}                                                                                                                   & \begin{tabular}[c]{@{}l@{}}Read-in-detail \\Rate of Message (\%)\end{tabular}                                                \\ 
\toprule

\multirow{3}{*}{ \begin{tabular}[c]{@{}l@{}}A:\\Subject\\lines\end{tabular}} 
& Anova P.val & (0.867) &(0.553) &(0.628) & (0.349)&\textbf{(0.001*)} &\textbf{(0.009*)}\\\cline{2-8}
&\begin{tabular}[c]{@{}l@{}}3: Add an \\org-pref \\message\end{tabular}         &   NS                                                                                                                                         &         NS                                                                                                                              &       NS                                                                                                                                       &             NS                                                                                            & \begin{tabular}[c]{@{}l@{}}H1.5 Increase  \\recognition rate.\\\textbf{60\%=non-s(41)+19 }\\\textbf{(0.001*)}\\\textbf{60\%=rnd-s(40)+20 }\\\textbf{(0.007*)}\end{tabular}                                                                     &   NS                                                             \\ 
\cline{2-8}
                                  & \begin{tabular}[c]{@{}l@{}}4: Add an\\emp-pref\\message\end{tabular}           & \begin{tabular}[c]{@{}l@{}}H1.1 Increase \\interest rate.\\ NS\end{tabular}                     & \begin{tabular}[c]{@{}l@{}}H1.2 Increase \\reading time.\\NS \end{tabular}                     & \begin{tabular}[c]{@{}l@{}}H1.3 Increase\\ overall\\recognition rate.\\NS \end{tabular}                        & \begin{tabular}[c]{@{}l@{}}H1.4 Increase \\open rate.\\NS\end{tabular} & \begin{tabular}[c]{@{}l@{}}H1.5 Increase  \\recognition rate.\\\textbf{56\%=non-s(41)+15 }\\\textbf{(0.001*)}\\\textbf{56\%=rnd-s(40)+16 }\\\textbf{(0.026*)}\end{tabular}                                                                    & \begin{tabular}[c]{@{}l@{}}H1.6 Increase  \\read-in-detail rate.\\\textbf{24\%=non-s(15)+9}\\\textbf{(0.008*)}\\rnd-s: NS\end{tabular}      \\ 
\hline
\multirow{4}{*}{ \begin{tabular}[c]{@{}l@{}}B:\\Top\\news\end{tabular}}    &Anova P.val&(0.182) &(0.809) &\textbf{(0.008*)} & &\textbf{(0.001*)} & \textbf{(0.001*)}\\\cline{2-8}  & \begin{tabular}[c]{@{}l@{}}3: Put\\org-pref \\messages\end{tabular}            &       NS                                                                                                                                  &    \begin{tabular}[c]{@{}l@{}}Q2.2 How does\\it affect \\reading time?\\ NS \end{tabular}                                                                                                                                       &        \begin{tabular}[c]{@{}l@{}}Q2.3 How does it\\affect overall\\recognition rate?\\org: NS\\\textbf{46\%=rnd(30)+16 }\\\textbf{(0.000*)}   \end{tabular}                                                                                                                                   &                                                                                                                & \begin{tabular}[c]{@{}l@{}}H2.5 Increase  \\recognition rate.\\\textbf{49\%=non-t(37)+12}\\\textbf{(0.002*)}\\rnd-t: NS\\org-t: NS\end{tabular}                                                                                         &NS                                                                 \\ 
\cline{2-8}
                                  & \begin{tabular}[c]{@{}l@{}}4: Put \\emp-pref \\messages\end{tabular}           & \begin{tabular}[c]{@{}l@{}}H2.1 Increase \\ interest rate.\\org:NS\\\textbf{76\%=rnd(58)+18}\\\textbf{ (0.075+)}\end{tabular} & \begin{tabular}[c]{@{}l@{}}Q2.2 How does\\it affect\\reading time? \\NS \end{tabular}                               & \begin{tabular}[c]{@{}l@{}}Q2.3 How does\\it affect overall\\recognition rate?\\NS \\\end{tabular}                       &                                                                                                                & \begin{tabular}[c]{@{}l@{}}H2.5 Increase  \\recognition rate.\\\textbf{49\%=non-t(37)+12} \\\textbf{(0.001*)}\\\textbf{49\%=rnd-t(32)+17} \\\textbf{(0.008*)}\\org-t: NS\end{tabular}                                                                  & \begin{tabular}[c]{@{}l@{}}H2.6 Increase  \\read-in-detail rate.\\\textbf{22\%=non-t(13)+9} \\\textbf{(0.001*)}\\\textbf{22\%=rnd-t(9)+13} \\\textbf{(0.005*)}\\org-t: NS\end{tabular}  \\ 
\cline{2-8}
                                  & \begin{tabular}[c]{@{}l@{}}5: Mix \\emp-pref\\/org-pref\\messages\end{tabular} &    NS                                                                                                                                          &  \begin{tabular}[c]{@{}l@{}}Q2.2 How does\\it affect \\reading time? \\NS \end{tabular}                                                                                                                                         &          \begin{tabular}[c]{@{}l@{}}Q2.3 How does it\\affect overall\\recognition rate?\\\textbf{49\%=org(38)+11}\\ \textbf{(0.069+)}\\\textbf{49\%=rnd(30)+19}\\\textbf{ (0.002*) } \end{tabular}                                                                                                                                  &                                                                                                                & \begin{tabular}[c]{@{}l@{}}H2.4 Increase  \\recognition rate of \\org-pref messages.\\orn\_B3: NS \\H2.5 Increase  \\recognition rate.\\\textbf{53\%=non-t(37)+16}\\\textbf{(0.001*)}\\\textbf{53\%=rnd-t(32)+21}\\\textbf{(0.050+)}\\org-t: NS\end{tabular} & NS                                                     \\ 
\hline
\multirow{4}{*}{ \begin{tabular}[c]{@{}l@{}}C:\\Message\\order\end{tabular}}    &  Anova P.val& \textbf{(0.088+)}&(0.674) &(0.446) & & &\\\cline{2-8}    & \begin{tabular}[c]{@{}l@{}}3: Sort \\messages by\\org-preference\end{tabular}  &   NS                                                                                                                                           & \begin{tabular}[c]{@{}l@{}}Q3.1 How does\\it affect\\reading time?\\NS\end{tabular}                      & \begin{tabular}[c]{@{}l@{}}Q3.2 Increase/\\decrease overall\\recognition rate?\\NS\end{tabular}            &                                                                                                                &                                                                                                                                                                                                   &                                                                                                                                     \\ 
\cline{2-8}
                                  & \begin{tabular}[c]{@{}l@{}}4: Sort \\messages by\\emp-preference\end{tabular}  & \begin{tabular}[c]{@{}l@{}}H3.1 Increase \\ interest rate.\\NS\end{tabular}                     & \begin{tabular}[c]{@{}l@{}}Q3.1 How does\\it affect \\reading time?\\NS\end{tabular}                     & \begin{tabular}[c]{@{}l@{}}Q3.2 How does it\\affect overall\\recognition rate?\\NS \end{tabular}             &                                                                                                                &                                                                                                                                                                                                   &                                                                                                                                     \\ 
\cline{2-8}
                                  & \begin{tabular}[c]{@{}l@{}}5: Sort \\messages\\by mix-order\end{tabular}       &       NS                                                                                                                                 & \begin{tabular}[c]{@{}l@{}}Q3.1 How does\\it affect \\reading time?\\NS \end{tabular} & \begin{tabular}[c]{@{}l@{}}Q3.2 How does it\\affect overall\\recognition rate?\\NS \end{tabular} &                                                                                                                & \begin{tabular}[c]{@{}l@{}}Q3.3 How does it\\affect recognition rate\\of org-pref messages?\\NS \end{tabular}                                                      &                                                                                                                                     \\
\hline
\end{tabular}
}
}
\end{table}

For the mixed logistic models in this paper, we checked \cite{kassambara2018logistic} 1) whether the numeric independent variables were linearly associated with the dependent variable in logit scale by visually plotting the line of predictor's value - logit of predicted probabilities; 2) multicollinearity (whether GVIF < 2);  3) whether outlier exists (by package dharma \cite{hartig2019dharma}). For the mixed linear model on reading time, we transformed $reading\_time$ and $base\_reading\_time$ by $log10(1+x)$. 3 outliers (the 3 emails were read for more than 30 minutes) were removed. The transformation and outlier removing were needed to satisfy the normality requirement of the model's residuals. We checked the homogeneity of variance by Levene’s Test and checked the normality of residuals by QQPlot \cite{roiger2020just}.

For each model and effect, first, we calculated the average of that performance metric of each experimental group. Second, we checked whether the effect was (marginally) significant by its ANOVA table to see whether there existed significant differences among different experimental groups (treatments) of this effect. If it was significant and we did observe large differences between some experimental groups with the control groups, we conducted pairwise tests between these experimental groups with its control groups (see Table \ref{tab:result} for the effects, treatments, and control groups we examined). The P-values of the pairwise tests were adjusted by the holm-bonferroni method \cite{holm1979simple}. We got the following marginal / significant results on the hypotheses and questions (see Table \ref{tab:result} for the numbers): 


 \noindent   \hlc[white]{\textbf{Interest rate}: 
H2.1 Putting employee-preferred messages in top news marginally increased Brief's interest rate versus putting random messages.}
    
\noindent \hlc[white]{\textbf{Overall recognition rate}:
     Q2.3 Mixing organization/employee-preferred  messages in top news increased Brief's overall recognition rate significantly versus putting random messages, marginally versus putting original messages. }
    
    \noindent \hlc[white]{Putting organization-preferred messages in top news significantly increased Brief's overall recognition rate versus putting random messages.}
    
    \noindent \hlc[white]{\textbf{Recognition rate}:
     H1.5 Putting messages on subject lines significantly increased their recognition rates versus the messages not on subject lines or putting random messages.}
    
    \noindent\hlc[white]{H2.5 Putting messages in top news significantly increased their recognition rates versus the messages not in top news.}

    \noindent \hlc[white]{\textbf{Read-in-detail rate}:
    H1.6 Putting employee-preferred messages on subject lines significantly increased their read-in-detail rates versus the messages not on subject lines.}
    
    \noindent\hlc[white]{H2.6  Putting employee-preferred messages in top news significantly increased their read-in-detail rates versus the messages not in top news or putting random messages.
    }

\subsection{Strategic Goals}
\hlc[white]{\textbf{Interest Rate}. Strategically, we could marginally make employees perceive Brief as more interesting by personalizing top news with their preferred messages. }We plot the average of the employees’ $interest\_rate$ on personalized Briefs of each
experimental group on top news in Figure \ref{fig:top4}. It shows that the average $interest\_rate$ of the employee-preferred top news group (B4) was higher than the random control group of top news (B2) by 18\%. The pairwise tests showed that \textbf{H2.1 was marginally supported versus the random control group}. In the end survey, a participant from B4 commented \textit{``I like the headings or topics at the top''}. A participant from B4 recognized that the top messages were related to their answers in the preference survey and hoped that recipients could update their preferences in the system in the future: \textit{``I liked having focused information. However, I think if you move forward with customized Briefs (which I support), people should get some what regular reminders with the ability to change what items they select to follow.''} A participant from the random group B2 seems to be disappointed: \textit{``I still think there's too much boosterism and fluff and I wish it was more work-related.''} And a participant from group B3 (which prioritized organization-preferred messages) found the content boring: \textit{``Most of it was skimmed. Most of the topics don't apply to me and/or my work. The content overall is generally uninteresting.''} The subject lines and message order's effects on the interest rate are not significant.

\noindent\hlc[white]{\textbf{Overall Recognition Rate}. However, putting employee-preferred messages in top news seemed to be a bad choice for the overall recognition rate} (see Figure \ref{fig:top_recog}, B4). The employees might close the Briefs early if they learned that most of the interesting messages would be at the top positions. As a participant from B4 said \textit{``It was fun to see the things I was interested in at the top. It also let me pay more attention to the beginning of the briefs and then skim the rest. ''}

\hlc[white]{To improve Brief's overall recognition rate, we could mix employee-preferred messages with organization-preferred messages in top news.} With pairwise tests, we found that the overall recognition rates of group B5 (mixing employee/organization preferences) were significantly higher than group B2 (random top news) by 19\% and marginally higher than group B1 (original top news) by 11\% (\textbf{Q2.3}). Seeing interesting content both at the top and in other sections might keep employees reading, though they did feel some of the messages ``irrelevant to them''. A participant from B5 said \textit{``I liked them. Overall I find things interesting; however, they are not really pertinent to my work always.''} It is worth noting that the overall recognition rate of the organization-preferred message group was also significantly higher than the random control group. The employees seemed to keep searching for items of interest if they did not find them in top news. But this searching process might cause disappointment. A participant from B3 commented \textit{``I was a little disappointed because I was expected slightly more tailored content.''} 
\begin{table}[!htbp]
\begin{minipage}{.49\textwidth}
  \centering
  \includegraphics[width=0.8\linewidth]{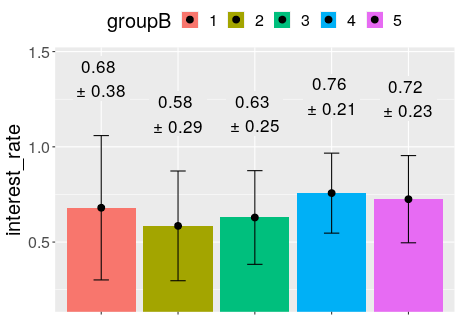}
  \captionof{figure}[The  interest rates on personalized Briefs ]{The  interest rates on personalized Briefs of each experimental group on top news and the standard deviations of the participants'  interest rates in that group.}
  \label{fig:top4}
\end{minipage}%
\hspace{0.05in}
\begin{minipage}{.49\textwidth}
  \centering
  \includegraphics[width=0.85\linewidth]{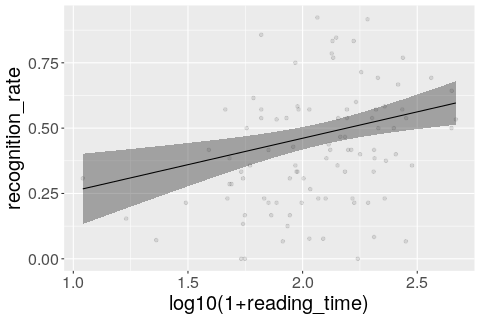}
  \captionof{figure}{Brief's reading time versus its  recognition rate.}
  \label{fig:time_recog}
\end{minipage}%

\end{table}

\begin{table}[!htbp]
\begin{minipage}{.49\textwidth}
  \centering
  \includegraphics[width=0.8\linewidth]{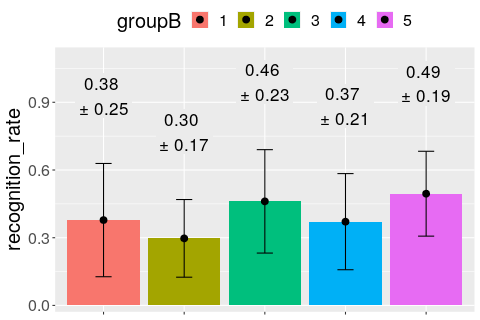}
  \captionof{figure}[The overall recognition rates on personalized Briefs]{The overall recognition rates on personalized Briefs of each experimental group on top news  and the standard deviations of the participants' overall recognition rates in that group.}
  \label{fig:top_recog}
\end{minipage}%
\hspace{0.05in}
\begin{minipage}{.49\textwidth}
  \centering
  \includegraphics[width=0.8\linewidth]{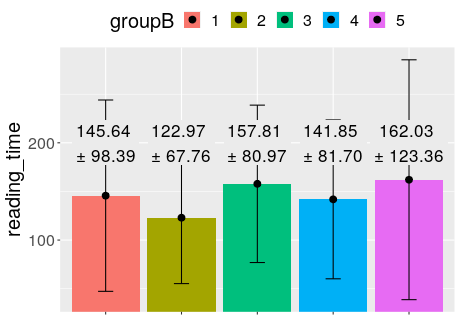}
  \captionof{figure}[The reading time (seconds) on personalized Briefs]{The reading time (seconds) on personalized Briefs of each experimental group on top news  and the standard deviations of the participants' reading time in that group.}
  \label{fig:top_time}
\end{minipage}%
\end{table}

\noindent\textbf{Reading Time}. There were no significant effects of subject lines/top news/message order designs on reading time. On average, the Briefs were read for around 120 to 170 seconds, and the variation of reading time is large. However, we found that the patterns of reading time matched with the patterns of overall recognition rates (see \ref{fig:top_recog} and \ref{fig:top_time}). We plot reading time versus overall recognition rate in Figure \ref{fig:time_recog}. The correlation between reading time (transformed by log10(1+10)) and the overall recognition rate was significant (Chisq=10.46, p.value = 0.001,  coef = 0.20\textpm0.06). This result shows that the gain in awareness is usually accompanied by time costs.

\noindent\textbf{Open Rate}. We did not observe significant differences among subject line groups' open rates. The average $open\_rate$ of group A4 (employee-preferred subject line group) was higher than other subject line groups but the pairwise tests were insignificant. The reason might be that our participants usually would take a quick check of these experimental Briefs during the experiment (though we asked them to treat these Briefs as naturally as possible). This is a limitation of this study --- we only collected 6 weeks' datapoints, because we wanted to collect all the recognition data together by a survey with a reasonable amount of questions at the end. A longer study might find different results on the open rates. However, some participants did indicate that they decided whether to open a Brief or not based on its subject lines --- a participant from A2 said that they left two Briefs unread because their subject lines were ``not at all interesting''.

\subsection{Tactical Goals}
\noindent\hlc[white]{\textbf{Recognition Rate}. Tactically, organizations could make those messages they view as important/relevant be recognized by more employees by putting them on subject lines or top news.} The Anova tests showed that whether a message was on subject lines influences its recognition rate significantly. The pairwise tests showed that either putting organization-preferred messages or employee-preferred messages on subject lines would increase their recognition rates by over 15\% compared to the recognition rates of the messages not on subject lines (see Figure \ref{fig:subonly}). Similarly, either putting organization-preferred messages or employee-preferred messages in top news would increase their recognition rates by over 12\% compared to the recognition rates of the messages not in top news (see Figure \ref{fig:toponly}). It is worth noting that these recognition rates were not significantly higher than the recognition rates of the original top news groups (B1), which indicated an opportunity to learn from the human editor on their design and selection strategies.

\begin{table}[!htbp]
\begin{minipage}{.48\textwidth}
  \centering
  \includegraphics[width=0.75\linewidth]{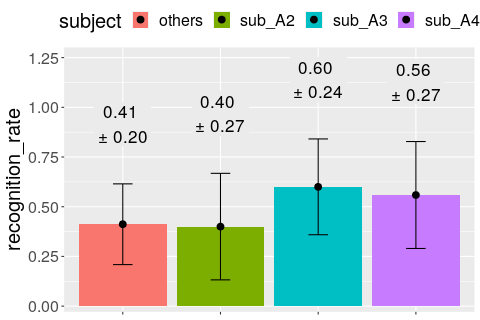}
  \captionof{figure}[The recognition rates of the messages on subject lines]{The recognition rates of the messages on subject lines with respect to group A and non-subject line messages and the standard deviations of the participants'  recognition rates in that group.}
  \label{fig:subonly}
\end{minipage}
\hspace{0.05in}
\begin{minipage}{.48\textwidth}
  \centering
  \includegraphics[width=0.75\linewidth]{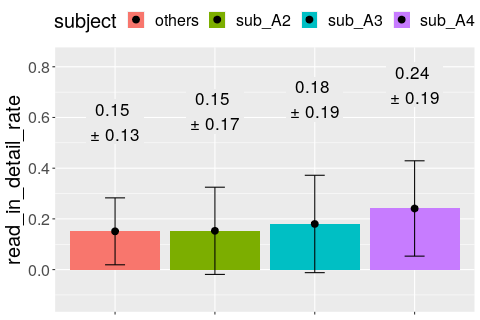}
  \captionof{figure}[The read-in-detail rates of the messages on subject lines]{The read-in-detail rates of the messages on subject lines with respect to group A and non-subject messages and the standard deviations of the participants'  read-in-detail rates in that group.}
  \label{fig:subread}
\end{minipage}
\end{table}

\noindent\hlc[white]{\textbf{Read-in-Detail Rate}. However, organizations could not make employees read the organization-preferred messages in detail.} The read-in-detail rates of those organization-preferred messages on subject lines were not significantly improved (see Figure \ref{fig:subread}). Actually, only the read-in-detail rates of those employee-preferred messages were significantly increased by 9\% when being putting on subject lines or top news (see Figure \ref{fig:subread}, \ref{fig:topread}). The reasons might be that the employees tended to only click the messages they had some interest in. We might need stronger incentives if we would like the employees to read those important-to-organization messages thoroughly.
\begin{table}[!htbp]
\begin{minipage}{.48\textwidth}
  \centering
  \includegraphics[width=0.75\linewidth]{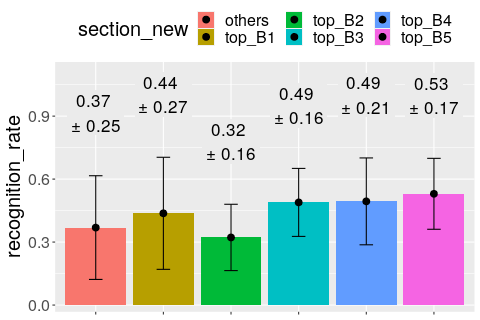}
  \captionof{figure}[The recognition rates of the messages in top news]{The recognition rates of the messages in top news with respect to group B and non-top messages and the standard deviations of the participants'  recognition rates in that group.}
  \label{fig:toponly}
\end{minipage}
\hspace{0.05in}
\begin{minipage}{.48\textwidth}
  \centering
  \includegraphics[width=0.75\linewidth]{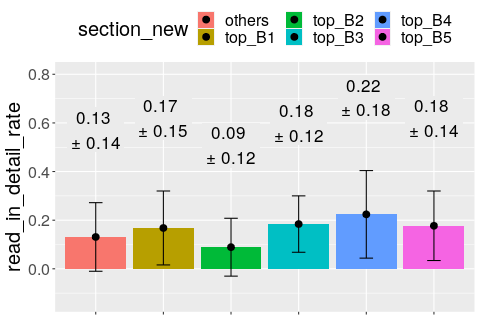}
  \captionof{figure}[The read-in-detail rates of the messages in top news]{The read-in-detail rates of the messages in top news with respect to group B and non-top messages and the standard deviations of the participants'  read-in-detail rates in that group.}
  \label{fig:topread}
\end{minipage}
\end{table}

Mixing organization-preferred messages with employee-preferred messages in top news/other sections did not bring further improvements to their recognition rates. For the messages in top news, though the mixed group B5's organization-preferred messages' average recognition rate was 4\% higher than B3's messages in the corresponding positions (45\% versus 49\%), the difference was not significant. And the difference (3\%) between the recognition rates of group C3 and C5's top 2 organization-preferred messages in the u-wide news sections was also not significant.

\subsection{Organization and Employee's Bulk Message Preferences}
In this section, we discussed where are the preference conflicts on bulk messages between the organization and the employees. Table \ref{tab:conflict} shows a set of messages' topics. For each topic, it shows how the Brief editor labeled the messages in that topic (whether it was important to the organization) in the editor surveys, and the employees' assessments of the work-relevance and interest of a sample message representing that topic (we collected these data in the preference survey at the experiment setup). For example, for the topic fundraising \& development, there were 9 messages during the study period. 5 of them were marked by the editor as important. 45.1\% of the employees felt the corresponding message was relevant to their jobs (18.0\% of the employees felt interesting and 38.5\% of them felt work-relevant).

We arranged a meeting with the Brief editor to discuss Table \ref{tab:conflict}. In that meeting, the editor told us that the frequency of topics (\#messages) is basically a true reflection of the number of topics submitted to them. The editor rejected a small number of the submissions that were too narrowly focused: \textit{``I really only reject maybe 10\% of submissions. We have communicators (in each campus). You know, it's really up to those folks to determine what they feel is important.''}. 


We noticed a number of interesting things from Table \ref{tab:conflict}. First, a large number of messages fell into the topic categories that the editor did not feel were usually important, and the employees generally would find unimportant and/or uninteresting, including award/recognition, student/alumni stories, faculty/staff stories. These were all work-relevant to fewer than 10\% of the employees and interesting to fewer than 20\% of them. 


Second, there are topics that the editor viewed as very important while the employees felt not that interesting or relevant, including university history \& celebrations, policy/admin news/governance, sports \& spirit. Over 60\% of the corresponding messages were viewed as important by the editor, while fewer than 40\% of the employees viewed these categories as work-relevant, and fewer than 30\% felt interesting.

Third, there are topics that the editor viewed as unimportant while the employees felt interesting. Some of them frequently appeared, including climate/eco, program awards/applications, health/covid. The editor told us that some contents were put because the employees might find them interesting: \textit{``I probably select half of those (the messages about events) myself just based on what I think folks will find interesting. I'm thinking both in terms of readership like we'd like them to find something of interest, so they come back and read.''} However, some of them appear only 3 or 4 times, including art \& museums and engineering science research stories. 

We further evaluated the preferences on the original Brief's messages. In week 2's Brief (the original Brief), 58\% of the surveyed messages were tagged as neither interesting nor work-relevant by the employees. The editor identified one message with the title ``U of M Public Engagement Footprint'' very relevant in building community and common understanding ($org\_importance = 4$). This message is from the Provost's office to advocate employees to submit plans for the university’s service, outreach, and community engagement. However, 58\% of the employees found this message neither interesting nor work-relevant. The message ``University and Faculty Senate Meetings'' was tagged as work-relevant to all employees, while 39\% of the employees found this message neither interesting nor work-relevant.

As we look at the results in total, it is clear that employee interest and editor judgment of importance are not perfectly-aligned. This finding reiterates the importance of considering the composition of the newsletter as a whole --- how to have enough relevant and interesting content to encourage reading the important content as well.
\begin{table}[!htbp]
\caption[Summary of the organization and employees' preferences on bulk messages' topics]{Summary of the organization and employees' preferences on bulk messages' topics. The topics are ordered by
\# messages (the number of messages that included a corresponding topic in the experiment). The editor selected $\leq$ 4 topics for each message. \\
\# times\_imp: the number of times that the message got an importance score $\geq$ 3.\\ org\_imp\%: \# times\_imp / \# messages. emp\_rel/int/pref\%: the percentage of employees who tagged the message with the corresponding topic as work-relevant/interesting/either work-relevant or interesting in the preference survey.
}~\label{tab:conflict}
\scalebox{0.6}{
\centering
\begin{tabular}{|p{4.3cm}|p{2cm}|p{2.2cm}|p{1.8cm}|p{1.9cm}|p{1.7cm}|p{1.7cm}|} 
\hline
\textbf{topic }                                                                                     & \textbf{\#messages } & \textbf{\#times\_imp } & \textbf{org\_imp\% } & \textbf{emp\_pref\% } & \textbf{emp\_rel\% } & \textbf{emp\_int\% }  \\ 
\hhline{|=======|}
Talk/ Symposium/ Lectures Announcements                    & 29                  & 1                      & 3.4\%                & 39.3\%                & 4.9\%                & 38.5\%                \\ 
\hline
Student/ Alumni Stories                                    & 27                  & 10                     & 37.0\%               & 18.0\%                & 4.9\%                & 14.8\%                \\ 
\hline
Community~Service/ Social Justice/ Underserved Population  & 21                  & 11                     & 52.4\%               & 78.7\%                & 35.2\%               & 73.0\%                \\ 
\hline
Faculty  Staff Stories                                     & 20                  & 4                      & 20.0\%               & 23.8\%                & 9.0\%                & 17.2\%                \\ 
\hline
Health/ Biology Research Stories                           & 15                  & 8                      & 53.3\%               & 64.8\%                & 9.0\%                & 60.7\%                \\ 
\hline
Climate/ Eco/ Agriculture                                  & 15                  & 6                      & 40.0\%               & 71.3\%                & 18.0\%               & 69.7\%                \\ 
\hline
Health Wellness Resources/ COVID                           & 12                  & 2                      & 16.7\%               & 91.0\%                & 67.2\%               & 73.8\%                \\ 
\hline
Award/ Recognition to University, Faculty, Staff, Students & 11                  & 5                      & 45.5\%               & 23.0\%                & 6.6\%                & 19.7\%                \\ 
\hline
Program  Award Applications/Announcements                  & 10                  & 2                      & 20.0\%               & 85.2\%                & 60.7\%               & 54.9\%                \\ 
\hline
Fundraising  Development                                   & 9                   & 5                      & 55.6\%               & 45.1\%                & 18.0\%               & 38.5\%                \\ 
\hline
History/ Social Science Research Stories                   & 9                   & 2                      & 22.2\%               & 45.9\%                & 15.6\%               & 38.5\%                \\ 
\hline
Policies/ Admin News/ Governance                           & 8                   & 5                      & 62.5\%               & 46.7\%                & 39.3\%               & 14.8\%                \\ 
\hline
Tech Tool Updates/ Workshops                               & 8                   & 0                      & 0.0\%                & 35.2\%                & 26.2\%               & 13.1\%                \\ 
\hline
Sports  Spirit                                             & 6                   & 5                      & 83.3\%               & 27.9\%                & 5.7\%                & 23.8\%                \\ 
\hline
University History/ Celebrations                           & 6                   & 4                      & 66.7\%               & 43.4\%                & 29.5\%               & 22.1\%                \\ 
\hline
Art  Museums                                               & 4                   & 0                      & 0.0\%                & 65.6\%                & 6.6\%                & 63.9\%                \\ 
\hline
University Program Success Stories                         & 4                   & 2                      & 50.0\%               & 39.3\%                & 17.2\%               & 27.9\%                \\ 
\hline
Operations Awareness/ Facility Closures                    & 3                   & 1                      & 33.3\%               & 89.3\%                & 82.0\%               & 49.2\%                \\ 
\hline
Engineering Science Research Stories                       & 3                   & 0                      & 0.0\%                & 54.1\%                & 3.3\%                & 52.5\%                \\ 
\hline
Youth, Children                                            & 0                   & 0                      & 0.0\%                & 36.1\%                & 8.2\%                & 31.1\%                \\
\hline
\end{tabular}}
\end{table}

\hlc[white]{For the engagement with topics versus campuses, only 25\%/32\% of the employees looked for work-relevant/interesting messages from other campuses. When being put in top news, the messages selected from other campuses got a recognition rate (21\%, p-value=0.0001) significantly lower than the messages selected from the employees' own campuses (43\%), and in this case, the employees' preference score $emp\_pref$ does not significantly influence these messages' recognition rate (this calculation has excluded the employees who indicated that they would not look at other campuses' messages). For the messages selected from the employees' own campuses, their recognition rate (43\%) is not significantly different from the messages originally selected from Top News (50\%), and in this case, the employees' preference score is positively correlated with the messages' recognition rate (p-value=0.024, cohen-size=2.259).}

\section{Discussion}
\hlc[white]{We explored 3 kinds of personalization (subject lines, top news, message order) based on 2 stakeholders' preferences (organization, employee), and investigated 2 types of goals (strategic / tactical goals). Overall, the tactical goals are easier to achieve than the strategic goals --- the organization could put whatever they want to promote in the top position and would get a reasonable recognition rate. But for the strategic goals, the organization needs to also consider the employees' information needs, and some strategic goals (reading time, open rate) can't achieve with blanket newsletters.}

\hlc[white]{Our work is different from many work in personalizing working emails \mbox{\cite{nelson2010mail2tag, moody2002reinventing, 10.1145/3290605.3300604}} in that we try to address the challenge of bulk emails in this multi-stakeholder case --- organizations have messages that they want their employees to be aware of while employees make individual judgments on which messages are relevant. Instead of prioritizing the recipients' preferences, we found that the best strategy for organizations is to mix messages they prefer with the messages their employees prefer. To the best of our knowledge, this is the first work focusing on this multi-objective personalization problem in the multi-stakeholder organizational bulk communication environment.}

\subsection{Organizations need to decide which messages to be sent and better communicate why.}
\hlc[white]{Organizations and employees perceive different messages as important/relevant --- this difference might have 2 outcomes. First, organizations might need to know more about their employees. For example, announcing the new Dean for the College of Biological Sciences through Brief is a convenient communication approach for university leaders, because they do not need to spend time figuring out who would be interested in it and how to personalize the contents. However, many of its recipients would perceive this message as irrelevant, which might make them stop reading Brief in the future. In this case, organizations should collect more information to enable better targeting, such as collecting preferences based on message topics (this study), or allowing employees to select interesting message tags} \cite{nelson2010mail2tag}.

\hlc[white]{Second, if organizations decide that some messages are worthwhile for their employees to know about, they need to better communicate to their employees why they need to read those messages. For example, for the messages like ``Board of Regents Meeting Highlights'', the employees often skip reading them (62.5\% of the administrative news was viewed as important by the organization, while only 14.8\% of the investigated employees found this topic interesting). However, the employees might decide to read it if they are aware of, for example, that the board was discussing their salary plans. Potential approaches on this aspect include pricing emails \mbox{\cite{kraut2005pricing, reeves2008marketplace}} (however \mbox{\citeauthor{kraut2005pricing}} found that recipients still did not interpret the prices as emails' importance), indicating expected actions \mbox{\cite{alrashed2019evaluating, 10.1145/2556288.2557182}}, etc.  }

\subsection{There are always tradeoffs --- suggestions.}
\hlc[white]{Within the current framework, we did not find any single optimal solution. Even with the mixed strategy, its interest rate was not as high as when we only put the messages the employees preferred on top news. The most interesting/efficient newsletter for employees would not be the newsletter that could best help organizations convey their messages. However, there are some decisions organizations can make when they know the priority of their communication goals:}

\noindent $\star$ \hlc[white]{Subject lines: for subject lines, organizations could put the messages they perceive as most important/relevant. This approach would bring these messages higher recognition rates. At the same time, at least in our (occasional) reader group, subject lines did not significantly affect the employees' open rate or interest rate.}

\noindent $\star$ \hlc[white]{Top news: to improve the overall recognition rate, a good approach for organizations would be mixing employee-preferred messages and organization-preferred messages in top news.}

\hlc[white]{Besides, there is also a trade-off between bulk communication's cost and performance. Though we did not find any significant effect on reading time, reading time is significantly correlated with the overall recognition rate, which means that organization needs to pay for more employee's time if they want higher performance. Also, longer email reading time is correlated with lower working productivity \mbox{\cite{mark2016email}}. In that sense, organizations should also try to remove unnecessary messages and use personalization to put important/relevant messages upfront.}

\subsection{Limitations and Generalizability}
The limitations of this study included:

\noindent 1) Reordering only: because of the requirement of our collaborator, we did not exclude any message from the studied newsletter. However, newsletters can also be personalized by filtering a subset of relevant messages and this method might work in organizations that allow taking this mechanism.

\noindent2) Measurement of recognition: we trusted our participants that they would select ``Skimmed'' or ``Read fully'' if they have seen a message and select ``No'' or ``Not Sure'' if they did not recognize this message or were uncertain.

    
\noindent3) Selection of participants: our participants were relatively active readers of Briefs. The employees who had stopped reading Briefs might have lower recognition rates, open rates, etc.
    
\noindent4) Technical issues: some plugins were deleted by a Chrome update during the experiment, the participants did not read Briefs with that browser, etc. Our personalization model is based on coarse-grained topics. \footnote{Among the 1404 messages we sent through the original Briefs in week 2 (our predicted employee preference based on topics versus these messages’ employee preference calculated from the interest scores and work-relevance scores collected directly from the participants in the end survey), we achieve a precision of 66\% and a recall of 75\%.}

In short, our study could only be generalizable when: 

\noindent 1) the organization newsletter is sent to a large list of employees;

\noindent \hlc[white]{2) studying newsletters' occasional/regular readers. We keep the newsletter's structure because the editor suggested that their audiences liked its campus structure; however, to how much extent the campus structure influenced the personalization's performance is still left to be studied.}

\subsection{Future Work}
After conducting this study, we see the following future work in improving organizational bulk communication. 

\noindent \hlc[white]{1) Measuring each message's reading time to better understand employees' preferences. It would be useful to run a study with eyetracking devices to collect such reading data, and to develop estimation algorithms based on recipients' interactions with bulk emails' webpages.}

\noindent \hlc[white]{2) Exploring different designing strategies that could help employees understand why they need to read some messages: for example, encourage senders to tag the reasons for sending some messages.}

\noindent \hlc[white]{3) Studying the effectiveness of fine-grained personalization models, enabling employees to update their preferences, exploring tools could better target the recipients (allowing excluding some messages), etc.}

\noindent \hlc[white]{4) Studying how to bring back nonreaders: restore nonreaders' trust on the bulk communication channels.}

\noindent 5) Auto-profiling employees: learning employees' job responsibilities and interests through their job descriptions, therefore we don't need to use surveys in the future.

\section{Conclusion}
This work studied how to use personalization to help the studied organization lead their employees' attention to the bulk messages they perceive as important or relevant for the employees to know (tactical goals) while maintaining the employees' overall positive experiences with these emails (strategic goals). We conducted an 8-week 4x5x5 controlled field experiment with 141 employees of a university and a weekly university-wide newsletter. 

\hlc[white]{We found that tactically, putting organization-preferred messages on subject lines or top news significantly increased their recognition rates, but did not increase their read-in-detail rates significantly. Only the employee-preferred messages' read-in-detail rates were improved. Strategically, mixing the employee-preferred/organization-preferred messages in top news significantly increased the overall recognition rate. Putting employee-preferred messages in top news increased their interest rates marginally. } We further looked into where the preferences on bulk messages' topics conflicted between the employees and the organization. We discussed the limitations and generalizability of this study.

Besides the findings above, this work also provided a basic backend framework for coordinating multiple stakeholders' preferences on organizational bulk emails --- employees could input their preferences through an onboarding survey; communicators and the organization leaders could input their preferences through weekly surveys; the system handles the transformation between text and html, listens to the survey responses, and generates personalized newsletters.

Besides encouraging employees to read important messages, we would also like to help senders consider each message's performance and cost. In the following chapter (6 \& 7), we introduce a system design study on the evaluation platforms of organizational bulk emails. 




\chapter{Message-Level Reading Estimation}
\label{eye_tracking}

\section{Introduction}
From Chapter 3's interviews, we found that to prevent communicators from overwhelming employees, we need to encourage communicators to consider employees' interests and the cost of bulk messages. To enable such consideration of interests and costs, in chapter 7, we would like to measure employees' time spent on bulk messages and show them to the commnicators. Chapter 6 serves as a prior study for Chapter 7. In this chapter, we learn how to estimate whether \textbf{each user} (e.g., an employee who receives a bulk email)  read in detail, skimmed, or skipped each message (\textbf{read level}) by collecting necessary ground truth data and building estimation models. \footnote{Ruoyan Kong, Ruixuan Sun, Charles Chuankai Zhang, Chen Chen, Sneha Patr,
Gayathri Gajjela, and Joseph A Konstan. Getting the most from eye-tracking:
User-interaction based reading region estimation dataset and models. The 2023
ACM Symposium of Eye Tracking Research \& Applications), 2023. URL \url{https://doi.org/10.1145/3588015.3588404}. \cite{eyetrack}} 

\begin{figure}[!htbp]
\centering
  \includegraphics[width=0.40\columnwidth]{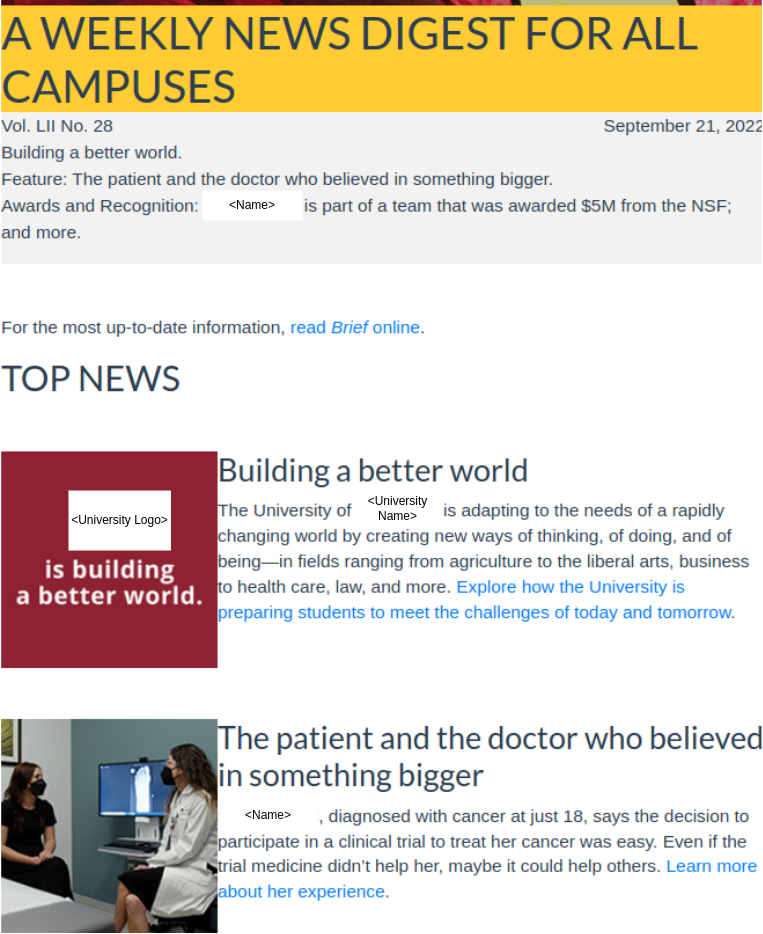}
  \caption[An example e-newsletter page of the study site with 30 messages]{An example e-newsletter page of the study site with 30 messages. By message-level reading estimation, we estimate each message's (like ``Building a better world'')  reading time and read level for each user. }~\label{fig:sample_page}
\end{figure}

Making the estimations above is difficult because digital newsletters (such as the one shown in Fig \ref{fig:sample_page}) often include a variety of different \textbf{messages} within them, and how to estimate employees' time spent on each message is still left to be learned. Eye-tracking \cite{gibaldi2017evaluation, funke2016eye, othman2020eye} would provide a good solution, but users lack the hardware (and willingness) to provide that data. 

Instead, we want to rely on browser instrumentation to track time, scrolling, mouse clicks, etc. However, there is no existing dataset on users web interactions and reading behaviors that we can use to build reading estimation models. In this chapter, we show how collecting a modest set of eye-tracking data and using it to train machine learning models for estimating reading behavior from browser logs can significantly increase the accuracy of classifying \textbf{message-level} reading behaviors.

Our core research question is \textbf{how to estimate message-level reading time and read level based on each user's logged interactions with browsers}. We looked progressively at heuristics, logistic regression, neural network and its variants to identify the best estimation model. For per message in a newsletter, we measured the models' error of estimating reading time and accuracy of classifying read-level category (skip/skim/read-in-detail), see Table \ref{tab: metrics}.

\begin{table}[!htbp]
\centering
\caption[The Metrics' Definitions]{The Metrics' Definitions. Precision/recall is based on the specific skim/detail/read category. We evaluated various models' performance on the users they had not seen in training by cross validation.}~\label{tab: metrics}
\resizebox{0.85\textwidth}{!}{%
\begin{tabular}{|l|l|} 
\hline
\textbf{Metric}                                                                                                        & \textbf{Definition}                                                                                                                                                                                     \\ 
\hline
\begin{tabular}[c]{@{}l@{}}percentage error of \\reading time (per\_error)\end{tabular}                                & \begin{tabular}[c]{@{}l@{}}$per\_error = |true \ reading \ time - estimated \ reading \ time|/true \ reading \ time$\\when $true \ reading \ time \geq 10s$\end{tabular}                                \\ 
\hline
\begin{tabular}[c]{@{}l@{}}absolute error of \\reading time (abs\_error)\end{tabular}                                  & \begin{tabular}[c]{@{}l@{}}$abs\_error = |true \ reading \ time - estimated \ reading \ time|$\\when~$true \ reading \ time < 10s$\end{tabular}                                                      \\ 
\hline
\begin{tabular}[c]{@{}l@{}}accuracy of read-level \\(accuracy)\end{tabular}                                            & \begin{tabular}[c]{@{}l@{}}$accuracy = \sum_{m \in \{messages\}} I(estimated$ $category = true \ category) / |\{messages\}|$~\\$\{messages\}$ are all the messages from all the sessions.\end{tabular}  \\ 
\hline
\begin{tabular}[c]{@{}l@{}}precision of~read-level (skim\_precision,\\detail\_precision, read\_precision)\end{tabular} & e.g., $skim\_precision = \frac{\sum_{m \in \{messages\}} I(true \ category = skim) I(estimated \ category = skim)}{\sum_{m \in \{messages\}} I(estimated \ category = skim)}$                           \\ 
\hline
\begin{tabular}[c]{@{}l@{}}recall of read-level (skim\_recall, \\detail\_recall, read\_recall)\end{tabular}            & e.g., $skim\_recall = \frac{\sum_{m \in \{messages\}} I(true \ category = skim) I(estimated \ category = skim)} {\sum_{m \in \{messages\}} I(true \ category = skim)}$                                  \\
\hline
\end{tabular}
}
\end{table}


The dataset (200k datapoints) is available at \url{https://github.com/ruoyankong/reading_region_prediction_dataset}. The proposed approach enables us to make relatively accurate (27\% error in reading time) message-level reading estimations of bulk emails solely by user web interactions in Chapter 7. 


\section{Related Work}
\subsection{Reading Estimation Based on Eye-tracking Technology}
Eye-tracking technology refers to the applications, devices, and algorithms that use user's eye movement to catch their attention trace \cite{lai2013review}. It has been widely used in information processing tasks, including reading estimation of advertisement \cite{rayner2008eye}, literature \cite{augereau2016estimation}, social media \cite{namara2019you}, recommendations \cite{zhao2016gaze}, etc.

Remote eye-trackers and mobile eye-trackers are often used in reading behavior estimation. The remote eye-trackers use cameras, pupil center, and cornea reflection to track user's gaze position \cite{bohme2006remote}. Example eye-trackers in this category include Tobii-Pro-TX300, which contains separate eye-tracking units, \footnote{https://www.tobii.com/products/discontinued/tobii-pro-tx300}or GazeRecorder, which simply uses Webcam.\footnote{https://gazerecorder.com/}The mobile eye-trackers are often head-mounted devices that users need to wear and use a camera on the visual path to record the view, such as Tobii Pro Glasses.\footnote{https://www.tobii.com/products/eye-trackers/wearables/tobii-pro-glasses-3}

Though eyetrackers provide relatively accurate data on reading estimation, they either ask users to wear/use specific devices or allow access to their webcams. This accesse is hard to get on a large scale for platforms, most of which only have access to browser interaction data. \textbf{Therefore there is a need to build message-level reading estimation models based on available browser data}.

\subsection{Eyetracking Data in Machine Learning}
Beyond reading estimation, eye-tracking data has been used widely in training machine learning models for different goals. \citeauthor{zhao2016gaze} predicted users' gaze on movie recommendations by eye tracking data through hidden markov models and reached an AUC of 0.823 \cite{zhao2016gaze}. \citeauthor{glaholt2011eye} studied participants' decision-making processes with eye-tracking data, and found that participants' preference on art images is positively correlated with their time spent gazing at corresponding types of images \cite{glaholt2011eye}. \citeauthor{buscher2009you} used eyetracking data to generate a model for predicting the visual attention that individual page elements may receive \cite{buscher2009you}. All these studies show the potential of using eyetracking data to generate scalable machine-learning models for predicting user behaviors / preferences.

\subsection{Reading Estimation Based on User Interactions}

Several studies have been done in estimating reading behaviors (e.g., time, focus area) based on user interactions. \citeauthor{seifert2017focus} studied which regions of the web pages users focus on when they read Wikipedia and found using single features (mouse position, paragraph position, mouse activity) cannot reach enough accuracy in estimating reading regions \cite{seifert2017focus}. \citeauthor{huang2012user} and \citeauthor{chen2001can} built linear models based on cursor position to estimate user's gaze coordinates during the web search / browsing \cite{huang2012user, chen2001can}.  \citeauthor{huang2012improving} and \citeauthor{10.1145/2766462.2767721} used cursor position in ranking search results \cite{huang2012improving, 10.1145/2766462.2767721} and found that 60\% of the tested queries were benefited by incorporating cursor data or significantly improve clicking prediction accuracy by 5\%.   \citeauthor{10.1145/2661829.2661909} recorded how users respond to online news and found the distance between mouse cursor and the corresponding news is significantly correlated with user's interestingness \cite{10.1145/2661829.2661909}.

However, most of these models did not consider user patterns. For example, the mouse position feature was treated the same for a user who moves mouse frequently versus a user who moves mouse infrequently \cite{lagun2015inferring}. \citeauthor{hauger2011using}'s work proposes to adjust reading estimations with user behavioral patterns --- but only a linear weight model was evaluated on the ground truth \cite{hauger2011using}. In fact, most of these studies focused on learning statistical correlations or only evaluated the proposed model's performance on groundtruth / heuristics. \textbf{We therefore found a need to evaluate various features and models' performance on reading estimation systematically}.

\section{Methods}
We defined \textbf{``reading estimation''} as 1) estimating reading time: how much time a user spends on reading each message per session (e.g., a user reads a message for 5 seconds); 2) classifying read level: whether a user skipped or skimmed a message, or read it in detail. Below, we introduced how we built features and models.

\subsection{Features}
To consider which interaction features a model solely based on user interactions can use, we identified the data a browser can collect in a reading session (from open to close a page). They can be classified into a 2x4 category: temporary/sessional features x pattern / user / message / baseline features (see Table \ref{tab: eye_features}).

\textbf{Temporary features} are collected per timestamp (each second for our case, e.g., the mouse position at time t). \textbf{Sessional features} are summaries of a reading session (e.g., the total number of seconds a message is visible on window in a reading session). \textbf{User features} represent user's status (e.g., mouse position). \textbf{Message features} represent message's status (e.g., message's position). \textbf{Pattern features} represent user's real-time behavioral patterns (e.g., mouse moving frequency until the latest timestamp). \textbf{Baseline features} represent the estimation given by the baselines --- the heuristics found to be correlated in the previous literature \cite{10.1145/1753846.1753976, seifert2017focus, huang2012user} (e.g., the share of a message on window). See the supplementary material for feature definitions.

\begin{table}[!htbp]
\centering
\caption{The Definitions of Features in Temporary / Sessional Message / User / Pattern / Baseline Category}~\label{tab: eye_features}
\resizebox{\textwidth}{!}{%
\begin{tabular}{|l|l|l|} 
\hline
\textbf{Feature Category}                                                                                                                           & \textbf{Temporary (collected per second)}                                                                                                                                                                                                                                                                                                              & \textbf{Sessional features (collected per reading session)}                                                                                                                                                                                                                                                              \\ 
\hline
\begin{tabular}[c]{@{}l@{}}\textbf{Message (represent }\\\textbf{message's status)}\end{tabular}                                                    & \begin{tabular}[c]{@{}l@{}}the message's position on the window, \\the share of the window at that time-\\-stamp; time gap to the latest time that \\the message is clicked by the user.\end{tabular}                                                                                                                                                  & \begin{tabular}[c]{@{}l@{}}the message's average share of the window during \\the session, average position on window during the \\session; whether the user clicked the message \\during the session; the number of seconds the \\message is visible during the session.\end{tabular}                                   \\ 
\hline
\begin{tabular}[c]{@{}l@{}}\textbf{User (represent }\\\textbf{user's status)}\end{tabular}                                                          & \begin{tabular}[c]{@{}l@{}}the user's mouse position; the time \\gap to the user's latest click.\end{tabular}                                                                                                                                                                                                                                          & not applicable                                                                                                                                                                                                                                                                                                           \\ 
\hline
\begin{tabular}[c]{@{}l@{}}\textbf{Pattern (represent }\\\textbf{user's behavioral }\\\textbf{patterns)}\end{tabular}                               & \begin{tabular}[c]{@{}l@{}}the user's mouse moving frequency in \\the past 2/5/10/infinite (since the \\beginning of the test) seconds in the \\horizontal/vertical direction, mouse \\scrolling frequency in the past 2/5/10\\/infinite seconds; the percentage of \\messages in that newsletter the user\\clicked until that timestamp.\end{tabular} & \begin{tabular}[c]{@{}l@{}}the user's average mouse moving frequency during the \\reading session in the horizontal/vertical direction, the \\average mouse scrolling frequency during the reading \\session; the percentage of messages in that newsletter \\the user clicked during the reading session.\end{tabular}  \\ 
\hline
\begin{tabular}[c]{@{}l@{}}\textbf{Baseline (the }\\\textbf{heuristics found }\\\textbf{correlated in }\\\textbf{previous literature)}\end{tabular} & \begin{tabular}[c]{@{}l@{}}the probability of the message being \\read at that timestamp according to \\baselines (see 3.3.1).\end{tabular}                                                                                                                                                                                                            & \begin{tabular}[c]{@{}l@{}}baselines' estimations on a message's reading time of \\the session.\end{tabular}                                                                                                                                                                                                             \\
\hline
\end{tabular}
}
\end{table}

\subsection{Models}
\begin{figure}[!htbp]
\centering
  \includegraphics[width=1\columnwidth]{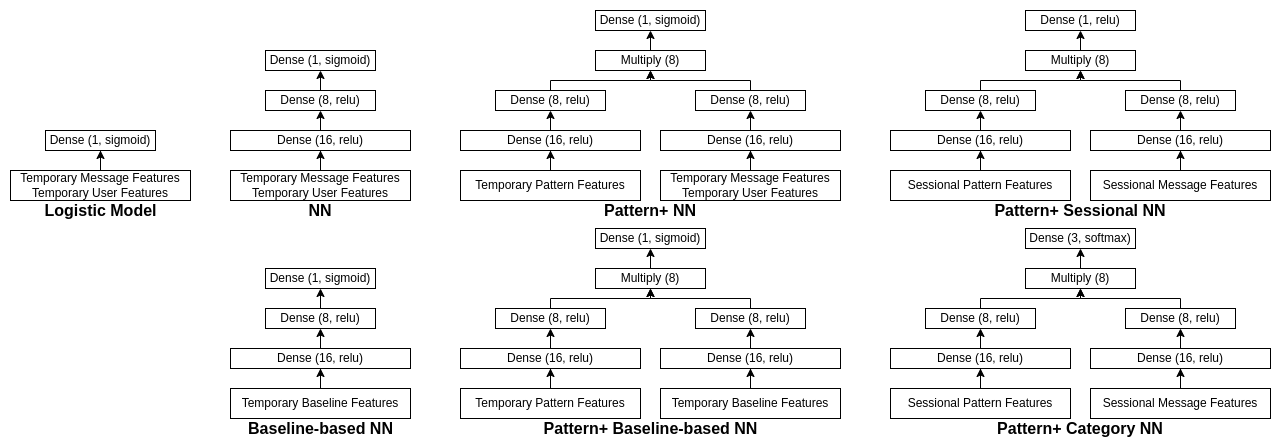}
  \caption[The estimation models and their configurations.]{The estimation models and their configurations. Logistic / NN / Pattern+ NN / Baselin-based NN / Pattern+ Baseline-based NN output $p_{m,t}$ per second to optimize binary loss. Pattern+ Sessional NN outputs $time_m$ per session to optimize absolute error loss. Pattern+ Category NN outputs $read\_level_m$ per session to optimize crossentropy loss.}~\label{fig:model}
\end{figure}

We considered a list of reading estimation models that use the interaction features above (Figure \ref{fig:model}). We explored models from simple heuristics baselines, and logistic regression, to neural networks. We also explored per-timestamp models (make estimations per second and sum the estimations at the end of the sessions) and per-session models (make estimations per session).
\subsubsection{Per-Timestamp Models}
Per-timestamp models estimate the probability $p_{m,t}$ of a message $m$ being read at each timestamp $t$. We summarize per-timestamp models' estimations for each message at the end of each reading session to estimate the reading time and read level. Previous work found that message size, message position, and mouse position are usually found to be correlated with reading behavior \cite{10.1145/1753846.1753976, seifert2017focus, huang2012user}. Thus we started from 3 baselines: 

\noindent \textbf{Baseline 1}: $p_{m,t}$ = message $m$'s current share of window area. 

\noindent \textbf{Baseline 2}: $p_{m,t}$ is weighted by the reverse of $m$'s distance to the window center.

\noindent \textbf{Baseline 3}: $p_{m,t} = 1$ if $m$ is closest to user mouse.

We started with simple logistic regression:

\noindent\textbf{Logistic Regression Model}: it takes temporary message/user features as inputs, $p_{m,t}$ as outputs.

After that, we considered neural networks to learn more complex decision boundaries.

\noindent\textbf{Baseline-based NN}: A fully connected feed-forward neural network that takes temporary baseline features as inputs and $p_{m,t}$ as outputs.

Then we tried using pattern features to learn which neurons' outputs are important.

\noindent\textbf{Pattern+ Baseline-based NN:} A two-tower feed-forward NN that takes baseline features in one tower and pattern features in another, then merges these two towers (multiply), and finally outputs $p_{m,t}$.

We also tried inputting user temporary interaction features directly as below.

\noindent\textbf{Neural Network (NN)}: A fully connected feed-forward NN which takes temporary message / user features.

\noindent\textbf{Pattern+ NN}: A two-tower feed-forward NN that takes temporary message/user features in one tower and pattern features in another, then merges these two towers (multiply), and finally outputs $p_{m,t}$.

\noindent\textbf{Read Level:} The reading time $time_m$ of a message $m$ in a reading session $t_1$ to $t_2$ is defined as $time_m = \sum_{t_1\leq t \leq t_2}p_{m,t}$. Given message $m$'s number of words ($word_m$), its \textbf{$read\_level_m$ (skip/skim/read-in-detail)} for a user is separated by the reading speed 400 words / minute and 200 words / minute. These limits are selected based on the 70\% and 80\% comprehension levels in \cite{nocera2018rapid}. And \textbf{read} means a message is either skimmed or read-in-detail by a user.


\subsubsection{Per-Session Models}

Instead of making one estimation at each timestamp (e.g., predicting whether the user is reading a message per second and summing the predictions of a reading session as the predicted reading time of that message), we can also make one estimation per reading session (e.g., directly estimating a user's reading time on that message).

\noindent\textbf{Pattern+ Sessional NN}:  A two-tower NN that takes sessional message features in one tower and sessional pattern features in another, then merges these two towers (multiply), and finally outputs reading time $time_m$ in a fully-connected ReLu layer.

\noindent\textbf{Pattern+ Category NN}: A two-tower NN that takes sessional message features in one tower and sessional pattern features in another, then merges these two towers (multiply), and finally outputs \textbf{$read\_level_m$} in a fully-connected softmax layer.

We explore with the following questions:

\noindent\textbf{Q1}: Does logistic regression perform better than the heuristic baselines?

\noindent\textbf{Q2:} Do neural network models based on user interactions perform better than the logistic regression model?

\noindent\textbf{Q3:} Does incorporating user patterns improve neural network models' performance on?

\noindent\textbf{Q4:} Are there performance differences between the neural network models that take the user / message features as input with those that take baseline predictions as input?

\noindent\textbf{Q5:} Are there performance differences between the models which make one estimation per timestamp with the models which make one estimation per reading session? \footnote{We defined a reading session as the duration between a user opening a newsletter and closing it or going to other irrelevant tabs (which are not hyperlinks in the newsletter).}

\subsection{Eye-tracking Tests}
We conducted eye-tracking tests to collect interaction data and ground-truth labels. Participants were recruited through 5 mailing lists of the university employees / in-person contacts of the research team \footnote{This study was determined to be non-human subject research by the university's IRB}. We excluded the university's communication professionals, as they might have processed the test messages. The selected participants were invited to the university's usability lab. 

Each participant then received 8 e-newsletters sampled from a diverse set of senders. The e-newsletters in the pool were selected randomly from the
university-wide e-newsletters sent in 2022, containing 3 to 30 messages. During the test, the participants' gaze positions were tracked by Tobii-TX300. They were asked to read as naturally as possible and read at least 1 message of each e-newsletter in detail. They can move mouse, scroll pages, and click on any hyperlinks. All these interactions were caught by browser Javascript and sent to the backend database (Google Firebase) in real-time. They can spend at most 30 minutes reading these e-newsletters and can leave anytime if they finish reading. The research team labeled the message located at the participants' gaze positions per second and calculated the reading time of each message as the ground truth. 


After each eye-tracking test, we retrained our models to check whether we got improvements in model performance. We stopped recruiting when we did not observe further improvements in the models' performance (see section 4). We conducted 9 eye-tracking tests finally, which resulted in a 200k dataset (one data point for each message's reading status per second).

\section{Results}
After collecting and labeling the data, we trained the models and compared their performance. We used the Adam optimizer, a batch size of 64, 50 epochs with early stop. We sampled the split of train/validation/test set for 72 rounds --- for each round, one participant's data were used as test set, and the rest participant's data were split into train set and validation set with a ratio of 7:1 based on reading sessions. The weight of positive samples is set as 20, given that each newsletter contains approximately 20 messages on average, and the user is reading one message at any time.

The average performance of all the proposed models are shown in Table \ref{tab:perf}, and the pair-wise t-tests' results between these models are shown in Table \ref{tab:compare}.

\subsection{What can we do without eyetracking data}
Without models trained by eyetracking data, the heuristic baselines have around 43-46\% percentage error and 2.1 to 2.5 seconds absolute error in reading time estimation. At the same time, these baselines reach 66\% to 69\% precision and 48\% to 59\% recall on read level classification.

The baseline1 (estimated by the share of screen size) has the lowest error among baselines (43\% percentage error and 2.1 seconds absolute error). It can be seen that heuristic baselines are not enough for us to provide an accurate metric on reading time (for Chapter 7). Below we show that with eyetracking data, we can improve the performance of estimations to a reasonable level.

\subsection{What can we do with eyetracking data} 
In summary, \textbf{we can reach approximately 27\% percentage error in reading time estimation with models based on user interactions and eyetracking data}, compared to the ground truth of using an eyetracker, while the heuristic baselines reach around 43-46\% percentage error. Specifically, with Pattern+ NN, which uses user interaction features as input and adjusts their weights with user behavioral patterns, for reading time estimation, we reached 27\% in percentage error and 1.7 seconds in absolute error; for read level classification, we reached 79\% in read category precision and 68\% in read category recall.

Also, we found that per-timestamp neural network models (Baseline-based NN, Pattern+ Baseline-based NN, NN, Pattern+ NN) have relatively higher overall performance compared to other models. Pattern+ Category NN's accuracy is only 47\%, which might indicate that making one estimation on the read level category per session does not utilize information efficiently.

\begin{table}[!htbp]
\centering
\caption[The average performance of all the proposed models]{The average performance of all the proposed models on the testsets (the users in testsets are not in the corresponding train / validation sets). Read: the message is either skimmed or read-in-detail by the user.}~\label{tab:perf}
\scalebox{0.7}{
\begin{tabular}{|c|c|c|c|c|c|c|c|c|c|} 
\hline
\textbf{model}                                                        & \begin{tabular}[c]{@{}c@{}}\textbf{per\_}\\\textbf{error}\\\textbf{(\%)}\end{tabular} & \begin{tabular}[c]{@{}c@{}}\textbf{abs\_}\\\textbf{error}\\\textbf{(s)}\end{tabular} & \begin{tabular}[c]{@{}c@{}}\textbf{accu-}\\\textbf{racy}\\\textbf{(\%)}\end{tabular} & \begin{tabular}[c]{@{}c@{}}\textbf{skim\_}\\\textbf{precision}\\\textbf{(\%)}\end{tabular} & \begin{tabular}[c]{@{}c@{}}\textbf{skim\_}\\\textbf{recall}\\\textbf{(\%)}\end{tabular} & \begin{tabular}[c]{@{}c@{}}\textbf{detail\_}\\\textbf{precision}\\\textbf{(\%)}\end{tabular} & \begin{tabular}[c]{@{}c@{}}\textbf{detail\_}\\\textbf{recall}\\\textbf{(\%)}\end{tabular} & \begin{tabular}[c]{@{}c@{}}\textbf{read\_}\\\textbf{precision}\\\textbf{(\%)}\end{tabular} & \begin{tabular}[c]{@{}c@{}}\textbf{read\_}\\\textbf{recall}\\\textbf{(\%)}\end{tabular}  \\ 
\hline
Baseline1                                                             & 43\%                                                                                  & 2.1s                                                                                 & 83\%                                                                                 & 47\%                                                                                       & 32\%                                                                                    & 39\%                                                                                         & 27\%                                                                                      & 69\%                                                                                       & 50\%                                                                                     \\ 
\hline
Baseline2                                                             & 45\%                                                                                  & 2.5s                                                                                 & 81\%                                                                                 & 36\%                                                                                       & 24\%                                                                                    & 42\%                                                                                         & 31\%                                                                                      & 67\%                                                                                       & 48\%                                                                                     \\ 
\hline
Baseline3                                                             & 46\%                                                                                  & 2.3s                                                                                 & 83\%                                                                                 & 49\%                                                                                       & 43\%                                                                                    & 33\%                                                                                         & 40\%                                                                                      & 66\%                                                                                       & 59\%                                                                                     \\ 
\hline
\begin{tabular}[c]{@{}c@{}}Logistic \\Model\end{tabular}              & 38\%                                                                                  & 2.5s                                                                                 & 82\%                                                                                 & 45\%                                                                                       & 42\%                                                                                    & 64\%                                                                                         & 51\%                                                                                      & 64\%                                                                                       & 58\%                                                                                     \\ 
\hline
\begin{tabular}[c]{@{}c@{}}Baseline-\\based NN\end{tabular}           & 30\%                                                                                  & 1.8s                                                                                 & 86\%                                                                                 & 59\%                                                                                       & 51\%                                                                                    & 67\%                                                                                         & 65\%                                                                                      & 77\%                                                                                       & 68\%                                                                                     \\ 
\hline
\begin{tabular}[c]{@{}c@{}}Pattern+\\Baseline-\\based NN\end{tabular} & 28\%                                                                                  & 1.7s                                                                                 & 87\%                                                                                 & 63\%                                                                                       & 55\%                                                                                    & 69\%                                                                                         & 64\%                                                                                      & 79\%                                                                                       & 70\%                                                                                     \\ 
\hline
NN                                                                    & 27\%                                                                                  & 1.7s                                                                                 & 87\%                                                                                 & 65\%                                                                                       & 54\%                                                                                    & 71\%                                                                                         & 63\%                                                                                      & 81\%                                                                                       & 68\%                                                                                     \\ 
\hline
\begin{tabular}[c]{@{}c@{}}Pattern+ \\NN\end{tabular}                 & 27\%                                                                                  & 1.7s                                                                                 & 87\%                                                                                 & 64\%                                                                                       & 55\%                                                                                    & 71\%                                                                                         & 62\%                                                                                      & 79\%                                                                                       & 68\%                                                                                     \\ 
\hline
\begin{tabular}[c]{@{}c@{}}Pattern+\\Category NN\end{tabular}         & \textbackslash{}                                                                      & \textbackslash{}                                                                     & 47\%                                                                                 & 22\%                                                                                       & 69\%                                                                                    & 34\%                                                                                         & 3\%                                                                                       & 32\%                                                                                       & 72\%                                                                                     \\ 
\hline
\begin{tabular}[c]{@{}c@{}}Pattern+\\Sessional NN\end{tabular}        & 61\%                                                                                  & 3.1s                                                                                 & 78\%                                                                                 & 36\%                                                                                       & 26\%                                                                                    & 30\%                                                                                         & 10\%                                                                                      & 57\%                                                                                       & 34\%                                                                                     \\
\hline
\end{tabular}
}
\end{table}

We then compared the relative pairs of models' performance according to questions Q1 to Q5 (see section 1). The comparison results are shown in Table \ref{tab:compare}. We summarized our answers below.

\begin{table}[!htbp]
\centering
\caption[Model performance comparisons]{Model performance comparisons. Each cell is model1's metric value -\textgreater{} model2's metric value (pvalue). '*': pvalue $\leq$ 0.05; '.': pvalue $\leq$ 0.10. The pvalues were adjusted by the holm-sidak method \cite{cardillo2006holm}. }~\label{tab:compare}
\scalebox{0.6}{
\begin{tabular}{|l|l|l|c|c|c|c|c|c|c|c|c|} 
\hline
\begin{tabular}[c]{@{}l@{}}\textbf{que-}\\\textbf{stion}\end{tabular} & \textbf{model1}                                                 & \textbf{model2}                                                        & \begin{tabular}[c]{@{}c@{}}\textbf{per\_}\\\textbf{error}\\\textbf{(\%)}\end{tabular}  & \begin{tabular}[c]{@{}c@{}}\textbf{abs\_}\\\textbf{error}\\\textbf{(s)}\end{tabular}       & \begin{tabular}[c]{@{}c@{}}\textbf{accu-}\\\textbf{racy}\\\textbf{(\%)}\end{tabular}   & \begin{tabular}[c]{@{}c@{}}\textbf{skim\_}\\\textbf{precision}\\\textbf{(\%)}\end{tabular} & \begin{tabular}[c]{@{}c@{}}\textbf{skim\_}\\\textbf{recall}\\\textbf{(\%)}\end{tabular} & \begin{tabular}[c]{@{}c@{}}\textbf{detail\_}\\\textbf{precision}\\\textbf{(\%)}\end{tabular} & \begin{tabular}[c]{@{}c@{}}\textbf{detail\_}\\\textbf{recall}\\\textbf{(\%)}\end{tabular} & \begin{tabular}[c]{@{}c@{}}\textbf{read\_}\\\textbf{precision}\\\textbf{(\%)}\end{tabular} & \begin{tabular}[c]{@{}c@{}}\textbf{read\_}\\\textbf{recall}\\\textbf{(\%)}\end{tabular}  \\ 
\hhline{|============|}
\multirow{4}{*}{Q1}                                                 & Baseline1                                                       & \begin{tabular}[c]{@{}l@{}}Logistic \\Model\end{tabular}               & \begin{tabular}[c]{@{}c@{}}\textbf{43-\textgreater{}38}\\\textbf{(0.00*)}\end{tabular} & \begin{tabular}[c]{@{}c@{}}\textbf{2.1-\textgreater{}2.5}\\\textbf{(0.00*)}\end{tabular} & \begin{tabular}[c]{@{}c@{}}83-\textgreater{}82\\(0.50)\end{tabular}                    & \begin{tabular}[c]{@{}c@{}}47-\textgreater{}45\\(0.53)\end{tabular}                        & \begin{tabular}[c]{@{}c@{}}\textbf{32-\textgreater{}42}\\\textbf{(0.00*)}\end{tabular}  & \begin{tabular}[c]{@{}c@{}}\textbf{39-\textgreater{}64}\\\textbf{(0.00*)}\end{tabular}       & \begin{tabular}[c]{@{}c@{}}\textbf{27-\textgreater{}51}\\\textbf{(0.00*)}\end{tabular}    & \begin{tabular}[c]{@{}c@{}}69-\textgreater{}64\\(0.32)\end{tabular}                        & \begin{tabular}[c]{@{}c@{}}\textbf{50-\textgreater{}58}\\\textbf{(0.00*)}\end{tabular}   \\ 
\cline{2-12}
                                                                      & Baseline2                                                       & \begin{tabular}[c]{@{}l@{}}Logistic\\Model\end{tabular}                & \begin{tabular}[c]{@{}c@{}}\textbf{45-\textgreater{}38}\\\textbf{(0.00*)}\end{tabular} & \begin{tabular}[c]{@{}c@{}}2.5-\textgreater{}2.5\\(0.62)\end{tabular} & \begin{tabular}[c]{@{}c@{}}81-\textgreater{}82\\(0.28)\end{tabular} & \begin{tabular}[c]{@{}c@{}}\textbf{36-\textgreater{}45}\\\textbf{(0.06.)}\end{tabular}     & \begin{tabular}[c]{@{}c@{}}\textbf{24-\textgreater{}42}\\\textbf{(0.00*)}\end{tabular}  & \begin{tabular}[c]{@{}c@{}}\textbf{42-\textgreater{}64}\\\textbf{(0.00*)}\end{tabular}       & \begin{tabular}[c]{@{}c@{}}\textbf{31-\textgreater{}51}\\\textbf{(0.00*)}\end{tabular}    & \begin{tabular}[c]{@{}c@{}}67-\textgreater{}64\\(0.62)\end{tabular}                        & \begin{tabular}[c]{@{}c@{}}\textbf{48-\textgreater{}58}\\\textbf{(0.00*)}\end{tabular}   \\ 
\cline{2-12}
                                                                      & Baseline3                                                       & \begin{tabular}[c]{@{}l@{}}Logistic\\Model\end{tabular}                & \begin{tabular}[c]{@{}c@{}}\textbf{46-\textgreater{}38}\\\textbf{(0.00*)}\end{tabular} & \begin{tabular}[c]{@{}c@{}}2.3-\textgreater{}2.5\\(0.23)\end{tabular}                    & \begin{tabular}[c]{@{}c@{}}83-\textgreater{}82\\(0.60)\end{tabular}                    & \begin{tabular}[c]{@{}c@{}}49-\textgreater{}45\\(0.53)\end{tabular}                        & \begin{tabular}[c]{@{}c@{}}43-\textgreater{}42\\(0.95)\end{tabular}                     & \begin{tabular}[c]{@{}c@{}}\textbf{33-\textgreater{}64}\\\textbf{(0.00*)}\end{tabular}       & \begin{tabular}[c]{@{}c@{}}40-\textgreater{}51\\(0.40)\end{tabular}                       & \begin{tabular}[c]{@{}c@{}}66-\textgreater{}64\\(0.95)\end{tabular}                        & \begin{tabular}[c]{@{}c@{}}59-\textgreater{}58\\(0.95)\end{tabular}                      \\ 
                                                                      \cline{2-12}
                                                                      & Baseline1                                                       & NN                                                                     & \begin{tabular}[c]{@{}c@{}}\textbf{43-\textgreater{}27}\\\textbf{(0.00*)}\end{tabular} & \begin{tabular}[c]{@{}c@{}}\textbf{2.1-\textgreater{}1.7}\\\textbf{(0.00*)}\end{tabular} & \begin{tabular}[c]{@{}c@{}}\textbf{83-\textgreater{}87}\\\textbf{(0.00*)}\end{tabular} & \begin{tabular}[c]{@{}c@{}}\textbf{47-\textgreater{}65}\\\textbf{(0.00*)}\end{tabular}     & \begin{tabular}[c]{@{}c@{}}\textbf{32-\textgreater{}54}\\\textbf{(0.00*)}\end{tabular}  & \begin{tabular}[c]{@{}c@{}}\textbf{39-\textgreater{}71}\\\textbf{(0.00*)}\end{tabular}       & \begin{tabular}[c]{@{}c@{}}\textbf{27-\textgreater{}63}\\\textbf{(0.00*)}\end{tabular}    & \begin{tabular}[c]{@{}c@{}}\textbf{69-\textgreater{}81}\\\textbf{(0.00*)}\end{tabular}     & \begin{tabular}[c]{@{}c@{}}\textbf{50-\textgreater{}68}\\\textbf{(0.00*)}\end{tabular}   \\ 

\hline
Q2                                                                    & \begin{tabular}[c]{@{}l@{}}Logistic \\Model\end{tabular}        & NN                                                                     & \begin{tabular}[c]{@{}c@{}}\textbf{38-\textgreater{}27}\\\textbf{(0.00*)}\end{tabular} & \begin{tabular}[c]{@{}c@{}}\textbf{2.5-\textgreater{}1.7}\\\textbf{(0.00*)}\end{tabular} & \begin{tabular}[c]{@{}c@{}}\textbf{82-\textgreater{}87}\\\textbf{(0.00*)}\end{tabular} & \begin{tabular}[c]{@{}c@{}}\textbf{45-\textgreater{}65}\\\textbf{(0.00*)}\end{tabular}     & \begin{tabular}[c]{@{}c@{}}\textbf{42-\textgreater{}54}\\\textbf{(0.00*)}\end{tabular}  & \begin{tabular}[c]{@{}c@{}}\textbf{64-\textgreater{}71}\\\textbf{(0.01*)}\end{tabular}                          & \begin{tabular}[c]{@{}c@{}}\textbf{51-\textgreater{}63}\\\textbf{(0.00*)}\end{tabular}    & \begin{tabular}[c]{@{}c@{}}\textbf{64-\textgreater{}81}\\\textbf{(0.00*)}\end{tabular}     & \begin{tabular}[c]{@{}c@{}}\textbf{58-\textgreater{}68}\\\textbf{(0.00*)}\end{tabular}   \\ 
\hline
\multirow{2}{*}{Q3}                                                   & NN                                                              & \begin{tabular}[c]{@{}l@{}}Pattern+ \\NN\end{tabular}                  & \begin{tabular}[c]{@{}c@{}}27-\textgreater{}27\\(0.99)\end{tabular}                    & \begin{tabular}[c]{@{}c@{}}1.7-\textgreater{}1.7\\(0.76)\end{tabular}                    & \begin{tabular}[c]{@{}c@{}}87-\textgreater{}87\\(0.97)\end{tabular}                    & \begin{tabular}[c]{@{}c@{}}65-\textgreater{}64\\(0.98)\end{tabular}                        & \begin{tabular}[c]{@{}c@{}}54-\textgreater{}55\\(0.97)\end{tabular}                     & \begin{tabular}[c]{@{}c@{}}71-\textgreater{}71\\(0.99)\end{tabular}                          & \begin{tabular}[c]{@{}c@{}}63-\textgreater{}62\\(0.99)\end{tabular}                       & \begin{tabular}[c]{@{}c@{}}81-\textgreater{}79\\(0.48)\end{tabular}                        & \begin{tabular}[c]{@{}c@{}}68-\textgreater{}68\\(0.99)\end{tabular}                      \\ 
\cline{2-12}
                                                                      & \begin{tabular}[c]{@{}l@{}}Baseline-\\based NN\end{tabular}     & \begin{tabular}[c]{@{}l@{}}Pattern+ \\Baseline-\\based NN\end{tabular} & \begin{tabular}[c]{@{}c@{}}\textbf{30-\textgreater{}28}\\\textbf{(0.00*)}\end{tabular}                    & \begin{tabular}[c]{@{}c@{}}\textbf{1.8-\textgreater{}1.7}\\\textbf{(0.01*)}\end{tabular} & \begin{tabular}[c]{@{}c@{}}86-\textgreater{}87\\(0.17)\end{tabular}                    & \begin{tabular}[c]{@{}c@{}}59-\textgreater{}63\\(0.12)\end{tabular}     & \begin{tabular}[c]{@{}c@{}}\textbf{51-\textgreater{}55}\\\textbf{(0.10.)}\end{tabular}  & \begin{tabular}[c]{@{}c@{}}67-\textgreater{}69\\(0.24)\end{tabular}                          & \begin{tabular}[c]{@{}c@{}}65-\textgreater{}64\\(0.68)\end{tabular}                       & \begin{tabular}[c]{@{}c@{}}77-\textgreater{}79\\(0.24)\end{tabular}                        & \begin{tabular}[c]{@{}c@{}}68-\textgreater{}70\\(0.24)\end{tabular}   \\ 
\hline
Q4                                                                    & \begin{tabular}[c]{@{}l@{}}Pattern+ \\NN\end{tabular}           & \begin{tabular}[c]{@{}l@{}}Pattern+ \\Baseline-\\based NN\end{tabular} & \begin{tabular}[c]{@{}c@{}}27-\textgreater{}28\\(0.86)\end{tabular}                    & \begin{tabular}[c]{@{}c@{}}1.7-\textgreater{}1.7\\(0.92)\end{tabular}                    & \begin{tabular}[c]{@{}c@{}}87-\textgreater{}87\\(0.73)\end{tabular}                    & \begin{tabular}[c]{@{}c@{}}64-\textgreater{}63\\(0.86)\end{tabular}                        & \begin{tabular}[c]{@{}c@{}}55-\textgreater{}55\\(0.92)\end{tabular}                     & \begin{tabular}[c]{@{}c@{}}71-\textgreater{}69\\(0.86)\end{tabular}                          & \begin{tabular}[c]{@{}c@{}}62-\textgreater{}64\\(0.92)\end{tabular}                       & \begin{tabular}[c]{@{}c@{}}79-\textgreater{}79\\(0.92)\end{tabular}                        & \begin{tabular}[c]{@{}c@{}}68-\textgreater{}70\\(0.66)\end{tabular}   \\ 
\hline
\multirow{2}{*}{Q5}                                                  & \begin{tabular}[c]{@{}l@{}}Pattern+ \\Sessional NN\end{tabular} & \begin{tabular}[c]{@{}l@{}}Pattern+ \\NN\end{tabular}                  & \begin{tabular}[c]{@{}c@{}}\textbf{61-\textgreater{}27}\\\textbf{(0.00*)}\end{tabular} & \begin{tabular}[c]{@{}c@{}}\textbf{3.1-\textgreater{}1.7}\\\textbf{(0.00*)}\end{tabular} & \begin{tabular}[c]{@{}c@{}}\textbf{78-\textgreater{}87}\\\textbf{(0.00*)}\end{tabular} & \begin{tabular}[c]{@{}c@{}}\textbf{36-\textgreater{}65}\\\textbf{(0.00*)}\end{tabular}     & \begin{tabular}[c]{@{}c@{}}\textbf{26-\textgreater{}55}\\\textbf{(0.00*)}\end{tabular}  & \begin{tabular}[c]{@{}c@{}}\textbf{30-\textgreater{}71}\\\textbf{(0.00*)}\end{tabular}       & \begin{tabular}[c]{@{}c@{}}\textbf{10-\textgreater{}62}\\\textbf{(0.00*)}\end{tabular}    & \begin{tabular}[c]{@{}c@{}}\textbf{57-\textgreater{}79}\\\textbf{(0.00*)}\end{tabular}     & \begin{tabular}[c]{@{}c@{}}\textbf{34-\textgreater{}68}\\\textbf{(0.00*)}\end{tabular}   \\ 
\cline{2-12}
                                                                      & \begin{tabular}[c]{@{}l@{}}Pattern+ \\Category NN\end{tabular}  & \begin{tabular}[c]{@{}l@{}}Pattern+ \\NN\end{tabular}                  & \textbackslash{}                                                                       & \textbackslash{}                                                                           & \begin{tabular}[c]{@{}c@{}}\textbf{47-\textgreater{}87}\\\textbf{(0.00*)}\end{tabular} & \begin{tabular}[c]{@{}c@{}}\textbf{22-\textgreater{}65}\\\textbf{(0.00*)}\end{tabular}     & \begin{tabular}[c]{@{}c@{}}\textbf{69-\textgreater{}55}\\\textbf{(0.01*)}\end{tabular}  & \begin{tabular}[c]{@{}c@{}}\textbf{34-\textgreater{}71}\\\textbf{(0.01*)}\end{tabular}       & \begin{tabular}[c]{@{}c@{}}\textbf{3-\textgreater{}62}\\\textbf{(0.00*)}\end{tabular}     & \begin{tabular}[c]{@{}c@{}}\textbf{32-\textgreater{}79}\\\textbf{(0.00*)}\end{tabular}     & \begin{tabular}[c]{@{}c@{}}72-\textgreater{}68\\(0.45)\end{tabular}   \\
\hline
\end{tabular}
}
\end{table}

\noindent\textbf{Q1:} Do machine learning models perform better than the heuristic baselines?

Yes, the logistic model and neural network models perform better than the heuristic baselines on most of the numerical and classification metrics. For example, compared to baseline 1 (heuristic estimation based on window share), NN significantly improved per\_error from 43\% to 27\%, read\_precision from 69\% to 81\%, read\_recall from 50\% to 68\%, etc. Though the logistic model does not perform better than baseline1 on abs\_error, the NN model performs better than the baselines on all the metrics.

\noindent\textbf{Q2:} Do neural network models perform better than the logistic model?

Yes, the neural network model performs better than the logistic model on almost all the numerical and classification metrics (similar performance on detail\_precision). For example, compared to the logistic model, NN significantly improves per\_error from 38\% to 27\%, abs\_error from 2.5s to 1.7s, accuracy from 82\% to 87\%, etc.

\noindent\textbf{Q3:} Do neural network models adjusted by user patterns 
perform better than the single-tower neural network models?

Yes for the NN that takes baselines as input but not for the NN that takes message/user features as input. The baseline neural network adjusted by user pattern features further improves the baseline neural network on the absolute error and skim prediction (marginally) significantly: per\_error (30\% to 28\%), skim\_recall (51 to 55\%). There is no significant performance difference between NN and Pattern+ NN.

\noindent\textbf{Q4:} Are there performance differences between the neural network models that take the user /
message features as input with those that take baseline features as input?

No, the pattern+ neural network that takes baseline features as input performs similarly to the pattern+ NN that takes user / message features as input.

\noindent\textbf{Q5:} Are there performance differences between the models which make one estimation per timestamp with the models which make one estimation per reading session?

Yes, the models that make one estimation for each reading session perform significantly worse than the models that make one estimation each second (see Table \ref{tab:compare}). For example, Pattern+ Sessional NN's per\_error is 61\% compared to Pattern+ NN's 27\%, accuracy is 78\% versus Pattern+ NN's 87\%.

\noindent\textbf{Model Robustness:} Here we report the trends of our models' performance during the data collection process. We recruited participants in 4 rounds. After each round, we retrained the models and evaluated their performance with the training/testing process above. Table \ref{tab:trend} is the performance trend of the Pattern+ Baseline-based NN model. In these 4 rounds, we see improvements mainly in model per\_error (36\% to 28\%), abs\_error (2.1s to 1.7s). We stopped when we did not observe further improvements on per\_error and abs\_error.
\begin{table}[!htbp]
\centering
\caption[The performance trend of the Pattern+ Baseline-based NN model]{The performance trend of the Pattern+ Baseline-based NN model (its average performance in the train/test splits after each round of the data collection process).}~\label{tab:trend}
\scalebox{0.7}{
\begin{tabular}{|c|c|c|c|c|c|c|c|c|c|c|} 
\hline
\multicolumn{1}{|l|}{\textbf{round}} & \multicolumn{1}{l|}{\begin{tabular}[c]{@{}l@{}}\textbf{\#parti-}\\\textbf{cipants}\end{tabular}} & \begin{tabular}[c]{@{}c@{}}\textbf{per\_}\\\textbf{error}\\\textbf{(\%)}\end{tabular} & \begin{tabular}[c]{@{}c@{}}\textbf{abs\_}\\\textbf{error}\\\textbf{(s)}\end{tabular} & \begin{tabular}[c]{@{}c@{}}\textbf{accu-}\\\textbf{racy}\\\textbf{(\%)}\end{tabular} & \begin{tabular}[c]{@{}c@{}}\textbf{skim\_}\\\textbf{precision}\\\textbf{(\%)}\end{tabular} & \begin{tabular}[c]{@{}c@{}}\textbf{skim\_}\\\textbf{recall}\\\textbf{(\%)}\end{tabular} & \begin{tabular}[c]{@{}c@{}}\textbf{detail\_}\\\textbf{precision}\\\textbf{(\%)}\end{tabular} & \begin{tabular}[c]{@{}c@{}}\textbf{detail\_}\\\textbf{recall}\\\textbf{(\%)}\end{tabular} & \begin{tabular}[c]{@{}c@{}}\textbf{read\_}\\\textbf{precision}\\\textbf{(\%)}\end{tabular} & \begin{tabular}[c]{@{}c@{}}\textbf{read\_}\\\textbf{recall}\\\textbf{(\%)}\end{tabular}  \\ 
\hhline{|===========|}
1                                    & 2                                                                                                & 36\%                                                                                  & 2.1s                                                                                 & 87\%                                                                                 & 55\%                                                                                       & 54\%                                                                                    & 65\%                                                                                         & 65\%                                                                                      & 70\%                                                                                       & 67\%                                                                                     \\ 
\hline
2                                    & 5                                                                                                & 30\%                                                                                  & 2.0s                                                                                 & 84\%                                                                                 & 61\%                                                                                       & 52\%                                                                                    & 39\%                                                                                         & 38\%                                                                                      & 74\%                                                                                       & 63\%                                                                                     \\ 
\hline
3                                    & 7                                                                                                & 28\%                                                                                  & 1.7s                                                                                 & 86\%                                                                                 & 67\%                                                                                       & 52\%                                                                                    & 65\%                                                                                         & 67\%                                                                                      & 80\%                                                                                       & 66\%                                                                                     \\ 
\hline
4                                    & 9                                                                                                & 28\%                                                                                  & 1.7s                                                                                 & 87\%                                                                                 & 63\%                                                                                       & 55\%                                                                                    & 69\%                                                                                         & 64\%                                                                                      & 79\%                                                                                       & 70\%                                                                                     \\
\hline
\end{tabular}
}
\end{table}

\section{Discussion}

\subsection{Per-Timestamp and Per-Session Models.}
We found that the models which make one estimation per second (and summarize the estimations at the end of each session) perform significantly better than the models which make one estimation per session. The reason that per-timestamp models outperform per-session models might be that the per-session models fail to utilize a large part of the information that can be collected per timestamp. In the previous work on learning user's reading interests (either through duration data collected by eye-trackers or interaction data), the estimations are often made per session instead of per timestamp \cite{10.1145/2661829.2661909,augereau2016estimation,10.1145/2766462.2767721, li2017towards}. \textbf{Therefore we suggested future work on reading estimation tasks to evaluate both the per-timestamp and per-session models.} Meanwhile, as per-timestamp models need more computing resources compared to per-session models, future studies also need to consider the tradeoffs between model performance and complexity \cite{shen2022classifying, luo2023efficient}, online learning \cite{luo2019spoton, luo2021smarton}, on-device learning \cite{lee2019intermittent, islam2023amalgamated}, etc.

\subsection{Improve accuracy by more features and more data.}We also found that users' behavioral pattern features helped the Baseline-based neural network to further decrease its error on reading time estimation. This result shows that user contexts could be used in improving reading estimation. Besides user behavioral patterns, other potential contexts to be considered included user's intent (e.g., search queries \cite{huang2012improving}), user's preference (e.g., user's genre preference in a movie recommendation system \cite{zhao2016gaze, zhao2016group}, user's friends \cite{10.1145/3543507.3583229}), item's features (e.g., product ratings \cite{shahriar2020online}).

\subsection{Limitations and Future Work}
We only collected eye-tracking data on 9 users therefore the dataset we collected might not catch enough variance on user patterns. Future work should look at the data requirement and where the value of additional users starts to decay significantly \cite{ma2020statistical, luo2022multisource}.  


\section{Conclusion}
We studied how to estimate message-level reading time and
read level of digital newsletters using user interactions extracted by browser JavaScript (e.g., mouse scroll, click, and hover actions). We applied the accurate but hard-to-scale eye-tracking data to learn the association between user interactions and reading time. We conducted 9 eye-tracking tests and collected 200k per-second participants' reading (gaze position) and interaction datapoints. The dataset will be public. 

First, we found that \textbf{having a small set of eye-tracking ground truth data enabled us to build a model with relatively high accuracy on predicting message-level reading time based on user web interactions only.} Specifically, we reached a 73\% accuracy in reading time estimation with a two-tower neural network based on user interactions, while the heuristic baselines have around 54\% accuracy. Second, we found that \textbf{adding users' behavioral patterns into the neural network models further improved their performance on reading estimation}. This study 1) provides examples of generalizing eye-tracking data to build reading estimation approaches that use browser-extracted features only and thus can be applied on a large scale; 2) give insights for future reading estimation studies on designing features / models. We incorporate the reading estimation models in this study into a prototype platform that enables communicators to see the reading time of each bulk message they send in Chapter 7.

\chapter{Exploring Bulk Email Effectiveness Tools for Better Transparency}
\label{study3}

\section{Introduction}
This chapter is a design and evaluation study on how to better support communicators in evaluating organizational bulk emails with a prototype tool called CommTool. For example, whether knowing about the reading time of each message would help communicators decide which content to put, whether knowing about the interest rate of employees from different units would help communicators personalize bulk emails, etc. \footnote{Ruoyan Kong, Irene Ye Yuan, Ruixuan Sun, Charles Chuankai Zhang, and
Joseph A Konstan. Commtool: Supporting organizations to evaluate bulk emails
(under review). 2023 \cite{evaluation}}

We conducted interviews, iterative designs \cite{zimmerman2014research}, and field tests to study whether specific features are useful to communicators. We started with expert interviews (5 communicators of the studied university) to learn about their experiences with evaluating organizational bulk emails and proposed potential features. We found that besides open rate and click rate, communicators would benefit from various metrics to make decisions, such as \textbf{message-level} reading time, read level, interest rate, etc. \footnote{``Message'' here refers to single stories / pieces of information in bulk emails.}
The potential useful features of bulk email evaluation platforms were then summarized from the qualitative analysis of these interviews' transcripts. A prototype of a bulk email evaluation platform (CommTool) was designed iteratively through 7 usability tests. To evaluate the usefulness of the proposed features, we conducted a 2-month field test in the studied university with 5 communicators and 149 recipients (employees of the university). 

Organizations now often send bulk emails via customer relationship management (CRM) platforms like Salesforce and Mailchimp, which provide metrics such as the open rate of emails and click rate of URLs \cite{bernstein2015research, monitor2012campaign}. However, open rate and click rate may not be sufficient tools to support \textbf{communicators} (organizations' employees in charge of designing and sending organizational bulk emails) in making editing / targeting decisions on bulk emails. Open rate can be an overestimation of employees' awareness as employees often open and close bulk emails in a few seconds and skip their contents, or be underestimated if the email has a useful subject or preview. Many organizational bulk emails focus on sharing important information rather than clicking links \cite{kong2021learning}. Within organizations, there is a need to evaluate the time cost associated with bulk emails, as employees spend much time filtering and reading these messages. Organizations also strive to maintain communication channels' reputations.\footnote{In this chapter, we use ``channel'' to refer to a series of bulk communications / newsletters with the same sender identity (the FROM field) and brand (in the title or format, e.g., xxx Brief Dec 1, 2022).}Therefore we learn how communicators currently evaluate organizational bulk emails, identify design opportunities, and build an evaluation platform prototype (CommTool) that aims to support communicators in understanding their audience and designing bulk emails. 

\begin{figure}[!htbp]
\centering
  \includegraphics[width=1\columnwidth]{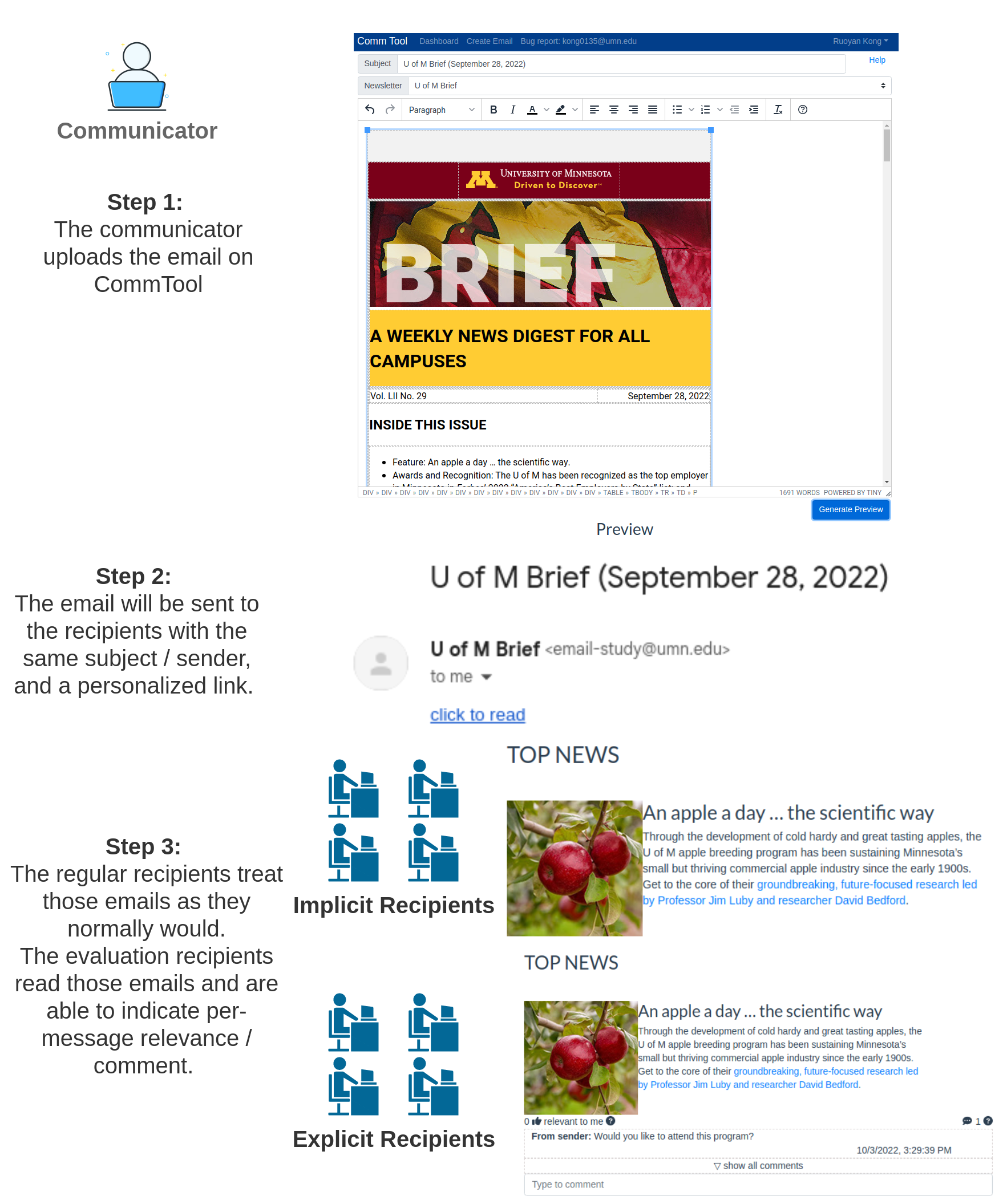}
  \caption{Communicator's workflow of evaluating organizational bulk emails through CommTool. Step 1 to 3.}~\label{fig:workflow1}
\end{figure}

\begin{figure}[!htbp]
\centering
  \includegraphics[width=1\columnwidth]{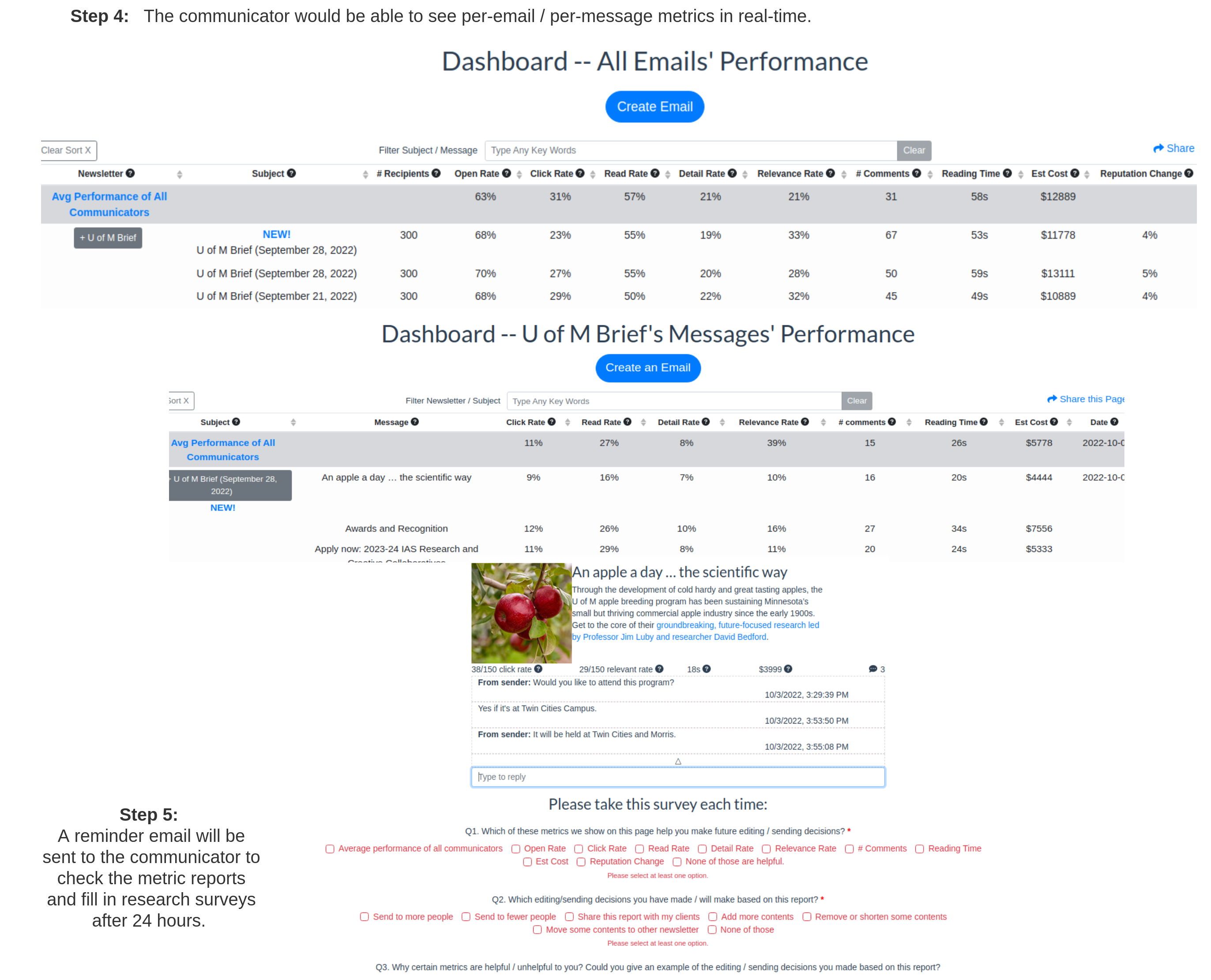}
  \caption{Communicator's workflow of evaluating organizational bulk emails through CommTool. Step 4 to 5.}~\label{fig:workflow2}
\end{figure}

CommTool's workflow is shown in Figure \ref{fig:workflow1} and \ref{fig:workflow2}) (similar to the workflow of the university's current bulk email evaluation platform Salesforce). In step 1, communicators upload emails to CommTool, preview them, and send them to recipients (participants). The recipients receive emails with the same title, sender name, and personalized links in their inboxes (step 2). Then CommTool records recipients' data when they click the links and summarizes various metrics and feedback (step 3). Communicators would be able to check metric reports in real-time (step 4) and also receive a reminder email after 24 hours (step 5).

We posed the following research question:

\noindent\textbf{RQ1:} What are communicators' current practices of evaluating bulk emails and potential features that would help them?

Then we present CommTool, a prototype that implements the feature we proposed to evaluate bulk emails. We carried out a field test to evaluate these features' usefulness and proposed the following questions:

\noindent\textbf{RQ2:} (Feature Evaluation) How do recipients interact with CommTool in the field evaluation?

\noindent\textbf{RQ3:} (Feature Evaluation) What are communicators' experiences with CommTool in the field evaluation? 





\section{Related Work}
As we discussed in Chapter \ref{related_work_chapter}, one challenge of organizational communication is employees' information overload --- employees’ perceptions that they receive more information than they can handle effectively \cite{10.1145/1180875.1180941}.
\citeauthor{davenport2001attention} proposed that employees have an ``attention budget'' --- therefore, the more information they receive from communicators, the more information they will inevitably ignore \cite{davenport2001attention}. \citeauthor{vsliburyte2004internal} proposed that in evaluating the effectiveness of organizational communication, the stakeholders' \textbf{cost} (e.g, time cost) needs to be considered \cite{vsliburyte2004internal}. For example, \citeauthor{jackson2001cost} monitored 15 employees' laptops and found they need 6 seconds to decide whether to open an email and 64 seconds to recover from an email interruption on average \cite{jackson2001cost}. \citeauthor{10.1145/1180875.1180941} conducted a nationwide organization survey and found that email volume is positively correlated with the feeling of email overload at work \cite{10.1145/1180875.1180941}. Given these challenges regarding the effectiveness of organizational bulk emails, many organizations take action (see below) to evaluate their bulk emails' performance.

As reviewed in Chapter \ref{related_work_chapter}, many large business platforms, such as Salesforce, Mailchimp, Revue, Constant Contact, etc., support bulk email senders by evaluating their emails' performance,  \cite{bernstein2015research, monitor2012campaign}. The evaluation features supported by these platforms include the open rates, click rates, number of subscriptions and their trends, A/B Tests, etc.

Besides the business CRM platforms, several marketing studies also shed light on how to evaluate bulk emails (though most of them focused on bulk emails to external customers). \textbf{``Awareness and impact''} of bulk emails are the two concepts most often mentioned. \citeauthor{cruz2008evaluating} interviewed CRM users and learned that reaching and awareness (e.g., knowing the brand) is more important than behavioral metrics (e.g., hits and clicks) in evaluating marketing emails \cite{cruz2008evaluating}. \citeauthor{morgan2012developing} reviewed the metrics of the current CRM platforms and pointed out that they focused on what can be measured quickly (e.g., open rate of marketing emails) instead of what should be measured (e.g., changes in the perceptions of the brand) \cite{morgan2012developing}. \citeauthor{todor2016marketing} strengthened the importance of understanding bulk emails' consequence \cite{todor2016marketing}, like the return of interest, the number of businesses generated, etc. \citeauthor{siano2011exploring} proposed a framework for senders to measure their reputation --- recipient's perception of a specific sender's communication quality \cite{siano2011exploring}. In the engineering field, there is work focusing on the API design and database integration of CRM platforms \cite{stubarev2018development, xu2015infrastructure}.

We noticed a lack of user-centered work on designing platforms for evaluating organizational bulk emails' performance and cost. Their design requirements might be different from the commercial CRM platforms (which focus on external marketing emails). For example, organizations needs to consider their employees' time cost in reading bulk emails, and organizational bulk emails might focus less on conversion rate / return of interest but more on awareness of organizational goals. In this study, we understand organization communicators' experience with evaluating bulk emails, specifying potential useful features, and designing prototypes to better understand which features could support communicators in designing bulk emails.

In the following section, we introduce a formative study for the iterative design process of CommTool (RQ1), then we present the design of CommTool, and at last, we introduce a field test for evaluating CommTool (RQ2, 3).

\section{Formative Study (RQ1)}
We first conducted formative studies to 1) understand communicators' current practice of evaluating bulk emails; 2) identify potential useful features; 3) deploy CommTool iteratively. Our study site is the University of Minnesota, a public university with over 25,000 employees and several campuses. We collaborated with the communicators of its central units (e.g., presidential offices), who are in charge of organizing, designing, and distributing university-wide bulk emails to all staff / faculty. This study is approved by the IRB of the University of Minnesota (STUDY00017138).

\subsection{Expert Interviews}
We conducted expert interviews with 5 university communicators for the purpose of RQ1 --- learn potential features in evaluating bulk emails. We applied an artifact-walkthrough approach  \cite{beyer1999contextual} by which we asked about communicators' practices with specific email cases. During each interview, the communicator selected 5 emails they sent out recently, then answered the following questions:
\begin{itemize}
    \item Communication Goal: Do you come up with the content by yourself, or do your clients provide it? Who are you hoping to reach? How do you hope they react to these messages? What open rate/action rate would you be happy with? How much employee time would be taken according to your estimation?

    \item Communication Practice: How do you decide the ways of writing the content? How do you send it (e.g., in a newsletter, separate email, or distribution channels)? How do you evaluate the effectiveness?
    
    \item Expectations of Evaluation Platforms: Which information do you hope to have when you evaluate this email? Any challenges you have met with the current evaluation platform (Salesforce)?
\end{itemize}

In the end, we introduced several design ideas for the evaluation platform CommTool, and asked the communicators their preferences on these ideas and the reasons.

We interviewed via zoom, and each interview took around 60 minutes. We recruited through the university's communicator forum. The forum director identified a list of potential participants working in the central communication offices. We contacted 6 communicators, and 5 (C1 to C5) agreed to participate in this expert interview.

For data analysis, I conducted the initial open coding of the first interview. The research team (see \cite{evaluation}) met weekly to discuss whether enough repeated themes emerged and whether more interviews were needed. Potential features were summarized using an affinity diagramming approach \cite{holtzblatt1997contextual}. We summarized our findings around RQ1 below.

\noindent\textbf{Observation 1: Communicators design / send / evaluate bulk emails for their clients.} 

Communicators sent those bulk emails for their clients within the university, such as their managers, university leaders, and central offices. They need to communicate with their clients first if they want to make some changes to bulk emails based on the evaluation results. For example, C2 described this editing process:
\begin{quote}
\textit{``So how it usually works is they (the clients) will have an email, and I'll review that and take that information to draft an email. And then they'll approve it, and then I'll send ... Especially when it comes to newsletters where there are multiple parties, it would be helpful to say: you know, based on the research that we've done on this newsletter you know, this kind of content doesn't isn't is appealing to this audience, so we're not going to include it.'' (C2)}
\end{quote}
A bulk email evaluation platform can change the status quo of designing bulk emails only if the platform supports communicators to reach an agreement with their clients on the design. Based on this observation, we proposed the following feature:

\noindent \textbf{Proposed feature 1 -- Share results}: The evaluation platform might enable communicators to share the results and communicate with their clients.

\noindent\textbf{Observation 2: The metrics provided by the current platforms are not sufficiently measuring awareness and relevance.}

Communicators noticed that their open rate was objectively good --- usually above 60\%. This open rate can be deceptive, as employees might not be actively reading those emails. For example, C1 talked about an email sent to over 8100 faculty:

\begin{quote}
    \textit{``There's an open rate of 62\%. It might count as an open rate if they just tap it in their inbox before archiving it. So I'm not sure if that means people actually put their eyes on it, or just like cleared it out.'' (C1)}
\end{quote}

C3 talked about an email asking all employees to check their Human Resources dashboard, but employees seem to pay less attention to it: 

\begin{quote}
    \textit{``So it (the email about checking HR dashboard) has a high open rate, 80\%, but only 10\% of the audience that even got to that dashboard to see that data. It (the email) probably doesn't give them enough information to know if they've needed to pay attention, and we don't have enough information to know if those people are paying attention.'' (C3)}
\end{quote}

In summary, current performance metrics offer little to no insight into whether employees are aware of those bulk emails' contents, nor the  perceived relevance of those emails. Based on this observation, we proposed the following feature:

\noindent \textbf{Proposed feature 2 -- Awareness and relecance}: The evaluation platform might provide metrics around awareness and relevance besides open rate and click rate.

As each bulk email often contains multiple messages, the realization of these features implies that the granularity of those metrics should not only be limited to per email but also per message:

\noindent \textbf{Proposed feature 3 -- Message-level metrics}: The evaluation platform might provide both email-level and message-level metrics.

\noindent\textbf{Observation 3: Communicators sometimes need custom feedback.}

Besides metrics in fixed formats, communicators also mentioned various formats of feedback they want to have. The examples mentioned by communicators include whether the recipients already knew the contents, whether the recipients were busy when receiving those emails, the job codes of the recipients who found those emails relevant, the recipients' devices, etc.

\begin{quote}
    \textit{``We've (the communication unit) had discussions about like what time of day people want to receive emails, is it first thing in the morning, or are they overwhelmed by first thing in the morning.'' (C4)}
\end{quote}

\begin{quote}
    \textit{``The challenge we have for employee benefits, for example, and some other communications is really about which job codes are relevant, for example, some of the vaccine requirement information for specific employee groups.'' (C3)}
\end{quote}

Based on this observation, we proposed the following feature:

\noindent \textbf{Proposed feature 4 -- Custom feedback}: The evaluation platform might enable communicators to collect custom feedback.

\noindent\textbf{Observation 4: Communicators need employee's interest to better target messages.}

Communicators mentioned that, especially for communications sent to a large group, their audience is so mixed that communicators do not know which employees might be interested in which messages. With access to such interest information for employee groups, communicators may be able to better target their messages. For example, C2 talked about this challenge when they are sending communications widely:

\begin{quote}
    \textit{``It's hard to know when you're kind of trying to send to a large group of people, a bunch of different stuff --- you don't always know what it is they want to see, or what they're really interested in. If there is some way to fine-tune it (the newsletter) to opportunities, information, news, and stories that different people are really interested in, that would be helpful.''}(C2)
\end{quote}

Therefore we proposed the following feature:

\noindent \textbf{Proposed feature 5 -- Employee group's interests}: The evaluation platform might enable communicators to understand different employee groups' (job categories, units, etc.) interests.

\noindent\textbf{Observation 5: Communicators realize that bulk email has impacts but cannot measure them.}

Communicators understood that bulk emails may have costs, both monetarily - in paying for employees' time - or impacts to the channel's reputation. Communicators also thought that sharing cost information may help them persuade clients to remove some content, while also expressing concern about the extent to which cost data would influence clients. Currently, communicators do not have access to such cost measures in the current bulk email evaluation platform. C2 talked about how they perceived the potential use of cost data:

\begin{quote}
    \textit{``We don't know, how much time people spend on email and how much money the university loses --- if some people (clients) are very driven by data, and that can help. But when you're working with somebody who just really wants, who really thinks whatever they have to say are important to their to the audience, I don't know if that is going to help.'' (C2)}
\end{quote}

C5 mentioned that their audience stopped reading emails from some channels, and they used other channels instead.
\begin{quote}
    \textit{``We found that if it's like a director of undergraduate studies that's sending out a survey, having it in that professor's name will increase opens tremendously rather than just from the department which people stop reading.'' (C5)}
\end{quote}

With these observations in mind, we proposed the following feature:

\noindent \textbf{Proposed feature 6 -- Communication's impact}: The evaluation platform might provide communicators with the impact of bulk communication (e.g., employee's time cost, reputation cost).

As the communicators mentioned, we are uncertain about the extent to which clients would give this cost data, and we plan on observeing that during the field test.

\noindent\textbf{Observation 6: The evaluation platform should be light-weight.}

Communicators mentioned that a challenge of getting deep evaluation results is that the process is complicated, and they are already overwhelmed by daily work.

\begin{quote}
    \textit{``The current way of doing (deep evaluation) is focus group (usability lab). It is expensive, ineffective, and it's a challenge for us that we have no such internal service on feedback sessions.'' (C3)}
\end{quote}

Therefore, a bulk email evaluation platform should match the following feature:

\noindent \textbf{Proposed feature 7 -- Lightweight use experience}: The platform is lightweight --- it should take no significant effort for communicators to evaluate bulk emails deeply beyond getting their main job done.

\subsection{Design of CommTool.}
Based on the features above, we designed a prototype platform to explore these features' usefulness in supporting communicators in evaluating bulk emails.  An initial version of CommTool was iteratively deployed and tested by the research team.

\noindent\textbf{1) Workflow:} CommTool's workflow is shown in Figure \ref{fig:workflow1}, \ref{fig:workflow2}) (similar to the workflow of the university's current bulk email evaluation platform Salesforce). In step 1, communicators upload emails to CommTool, preview them, and send them to recipients (our study participants). As a prototype aimed at exploring diverse metrics' usefulness, CommTool does not support present emails. All the emails will be sent out immediately when the communicators hit the send button. The recipients receive emails with the same title, sender name, and personalized links in their inboxes (step 2). The personalized links enable recipients to open these links on any device without login into CommTool, so we can record their data (step 3). Then CommTool collects data from recipients and summarizes metric reports. Communicators would be able to check metric reports in real-time (step 4). And also, after 24 hours (to give recipients enough time to react), communicator gets a reminder email to check reports and fill in research surveys (step 5).

\noindent\textbf{2) Regular and Evaluation Recipient Group:}An important step in calculating these metrics is that CommTool splits the recipients into a regular recipient group and an evaluation recipient group. The recipients stayed in one group throughout the whole study. The regular recipients were asked to react to the emails they received as normally as they could --- CommTool uses their log data to calculate the open rate, reading time, read rate, read-in-detail rate, click rate, and estimated cost. The evaluation recipients were asked to read those emails and indicate whether each message was relevant to them and leave comments under the messages if applicable. We split the recipients into these two groups because ``indicating relevance'' and ``commenting'' are time-consuming and require more attention from the recipients. We do not want these actions to interfere with the calculation of awareness metrics. The definitions of the metrics we used are summarized in table \ref{tab: metric_def}. 

\begin{table}[!htbp]
\centering
\caption[Metric Definitions]{Metric Definitions.  Email-level metrics are shown in the email dashboard. Message-level metrics are shown in the message dashboards and report dashboards.}~\label{tab: metric_def}
\arrayrulecolor[rgb]{0.753,0.753,0.753}
\scalebox{0.6}{
\begin{tabular}{!{\color{black}\vrule}l|l!{\color{black}\vrule}l|l!{\color{black}\vrule}} 
\arrayrulecolor{black}\hline
\multicolumn{2}{!{\color{black}\vrule}c!{\color{black}\vrule}}{\textbf{Email-Level Metrics }}                                                                                                                                                                                                                                                   & \multicolumn{2}{c!{\color{black}\vrule}}{\textbf{Message-Level Metrics.}}                                                                                                                                                                                          \\ 
\arrayrulecolor[rgb]{0.753,0.753,0.753}\hline
\begin{tabular}[c]{@{}l@{}}Click\\Rate\end{tabular}        & \begin{tabular}[c]{@{}l@{}}The percentage of the regular recipients who \\at least clicked one link in this email.\end{tabular}                                                                                                                                                 & \begin{tabular}[c]{@{}l@{}}Click\\Rate\end{tabular}         & \begin{tabular}[c]{@{}l@{}}The percentage of the regular recipients\\who at least clicked one link in this message.\end{tabular}                                                                  \\ 
\hline
\begin{tabular}[c]{@{}l@{}}Read\\Rate\end{tabular}         & \begin{tabular}[c]{@{}l@{}}The percentage of the regular recipients who \\at least skimmed one message of this email or \\read it in detail.\end{tabular}                                                                                                                       & \begin{tabular}[c]{@{}l@{}}Read\\Rate\end{tabular}          & \begin{tabular}[c]{@{}l@{}}The percentage of regular recipients who\\skimmed this message or read it in detail.\end{tabular}                                                                      \\ 
\hline
\begin{tabular}[c]{@{}l@{}}Detail\\Rate\end{tabular}       & \begin{tabular}[c]{@{}l@{}}The percentage of the regular recipients who \\at least read one message of this email in detail.\end{tabular}                                                                                                                                       & \begin{tabular}[c]{@{}l@{}}Detail\\Rate\end{tabular}        & \begin{tabular}[c]{@{}l@{}}The percentage of the regular recipients\\who read this message in detail.\end{tabular}                                                                                \\ 
\hline
\begin{tabular}[c]{@{}l@{}}Reading\\Time\end{tabular}      & \begin{tabular}[c]{@{}l@{}}The average number of seconds the regular \\recipients who opened this email spent \\reading this email.\end{tabular}                                                                                                                                & \begin{tabular}[c]{@{}l@{}}Reading\\Time\end{tabular}       & \begin{tabular}[c]{@{}l@{}}The average number of seconds the regular\\recipients who opened this email spent\\reading this message.\end{tabular}                                                  \\ 
\hline
\begin{tabular}[c]{@{}l@{}}Estimated\\Cost\end{tabular}    & \begin{tabular}[c]{@{}l@{}}The estimated money cost of this email: the \\average reading time * open rate * \$40 hour rate\\* the actual number of recipients of this newsletter. \\6 seconds are added for each recipient's time on\\making read / unread decisions.\end{tabular} & \begin{tabular}[c]{@{}l@{}}Estimated\\Cost\end{tabular}     & \begin{tabular}[c]{@{}l@{}}The estimated money cost of this message:\\the average reading time * open rate * \$40\\hour rate * the actual number of recipients.\end{tabular}                         \\ 
\hline
\begin{tabular}[c]{@{}l@{}}Relevance\\Rate\end{tabular}    & \begin{tabular}[c]{@{}l@{}}The percentage of the evaluation recipients who at\\least indicated one message of this email that is\\relevant to them.\end{tabular}                                                                                                                   & \begin{tabular}[c]{@{}l@{}}Relevance\\Rate\end{tabular}     & \begin{tabular}[c]{@{}l@{}}The percentage of the evaluation recipients\\who indicated that this message is relevant.\end{tabular}                                                                    \\ 
\hline
\# Comments                                                & \begin{tabular}[c]{@{}l@{}}The number of comments the evaluation recipients \\left on this email.\end{tabular}                                                                                                                                                                     & \# Comments                                                 & \begin{tabular}[c]{@{}l@{}}The average number of seconds the regular\\recipients who opened this email spent\\reading this message.\end{tabular}                                                  \\ 
\hline
\begin{tabular}[c]{@{}l@{}}Open\\Rate\end{tabular}         & \begin{tabular}[c]{@{}l@{}}The percentage of the regular recipients who \\opened this email.\end{tabular}                                                                                                                                                                       & \begin{tabular}[c]{@{}l@{}}Who are\\Interested\end{tabular} & \begin{tabular}[c]{@{}l@{}}The percentage of different categories\\(departments, job categories) of recipients\\who are interested (clicked/read/indicated\\relevant) in this message.\end{tabular}  \\ 
\hline
\begin{tabular}[c]{@{}l@{}}Reputation\\Change\end{tabular} & \begin{tabular}[c]{@{}l@{}}This email's influence on this newsletter's reputation:\\this newsletter's predicted future open rate - this \\email's open rate.\end{tabular}                                                                                                          & \multicolumn{1}{l}{}                                        &                                                                                                                                                                                                      \\
\arrayrulecolor{black}\hline
\end{tabular}
}
\end{table}

In the following, we introduce the design details of CommTool. The match between those designs with the proposed features is summarized in Table \ref{tab:commtool_feature}.

\begin{table}[!htbp]
\centering
\caption{Match between designs and proposed features}~\label{tab:commtool_feature}
\begin{tabular}{|l|l|} 
\hline
\textbf{Design}                                                                                      & \textbf{Proposed Feature}                                                                        \\ 
\hline
Split Email into Messages                                                                            & 3) Message-level metrics                                                                         \\ 
\hline
\begin{tabular}[c]{@{}l@{}}Measure Message's Performance \\with Various Metrics\end{tabular}         & \begin{tabular}[c]{@{}l@{}}2) Awareness and relevance; \\6) Communication's impact\end{tabular}  \\ 
\hline
\begin{tabular}[c]{@{}l@{}}Measure Message's Performance \\on Different Employee Groups\end{tabular} & 5) Employee group's interests                                                                    \\ 
\hline
Collect Comments                                                                                     & 4) Custom feedback                                                                               \\ 
\hline
Share Reports                                                                                        & 1) Share results                                                                                 \\ 
\hline
Automation of Evaluation Process                                                                     & 7) Lightweight use experience                                                                    \\
\hline
\end{tabular}
\end{table}

\noindent\textbf{3) Split Email into Messages:} To provide message-level metrics, in CommTool, when communicators upload an html email, the email would be automatically split into sections in preview (by recognizing html tags like h2, h3, etc.), as displayed in Figure \ref{fig:message_level}. Given the html length of each message, CommTool recognizes whether each section is just a title (like the ``TOP NEWS'' in Figure \ref{fig:message_level}) or is a message with contents (like the ``An apple a day ... '' in Figure \ref{fig:message_level}). CommTool adds surveys under each of those messages --- ``survey'' here means we would collect metrics and feedback for these messages. CommTool also enables communicators to remove / add / merge sections, and add / remove surveys in case the split is incorrect.

\begin{table}
\begin{minipage}{.48\textwidth}

\centering
  \includegraphics[width=1\columnwidth]{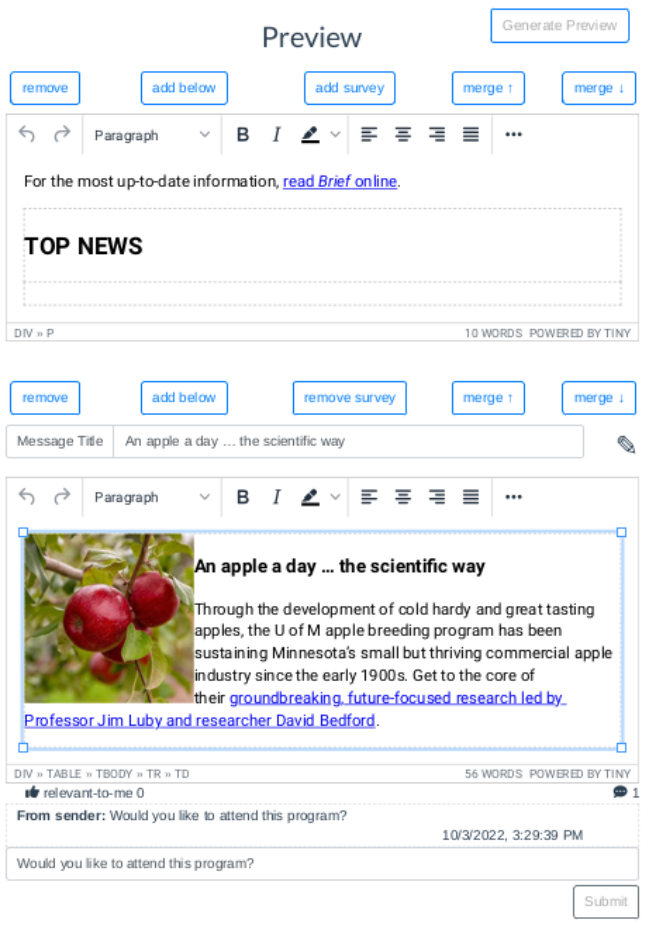}
  \captionof{figure}[CommTool can automatically split an html email]{CommTool can automatically split an html email into single messages during editing.}~\label{fig:message_level}

\end{minipage}%
\hspace{0.05in}
\begin{minipage}{.46\textwidth}
\centering
  \includegraphics[width=1\columnwidth]{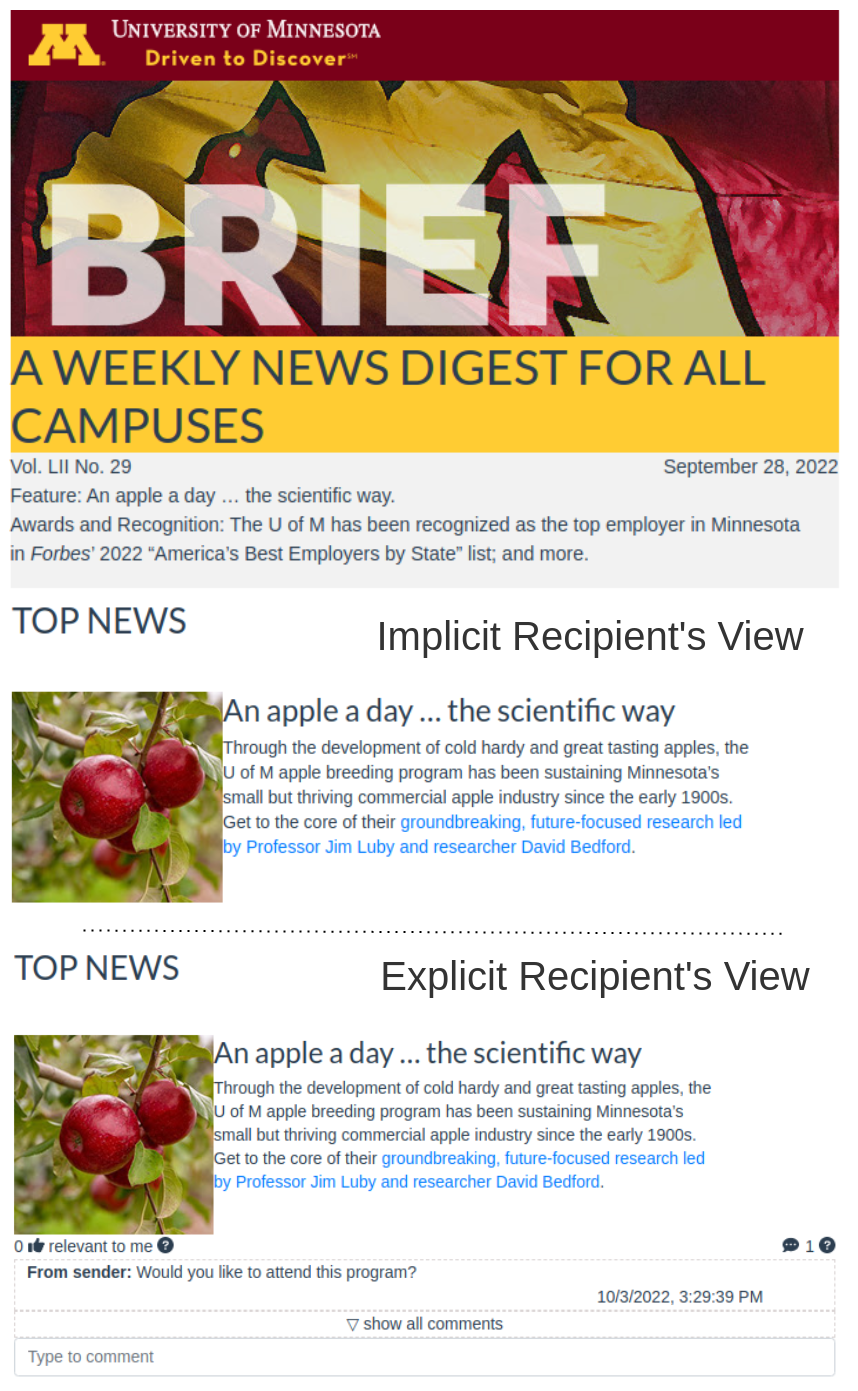}
  \captionof{figure}[The email received by regular recipients is just like the original email]{The email received by regular recipients is just like the original email. The email received by evaluation recipients would include ``relevant-to-me'' buttons and comment areas.}~\label{fig:evaluation_regular}
\end{minipage}%
\end{table}

\begin{sidewaysfigure}[!htbp]
\centering
  \includegraphics[width=1\columnwidth]{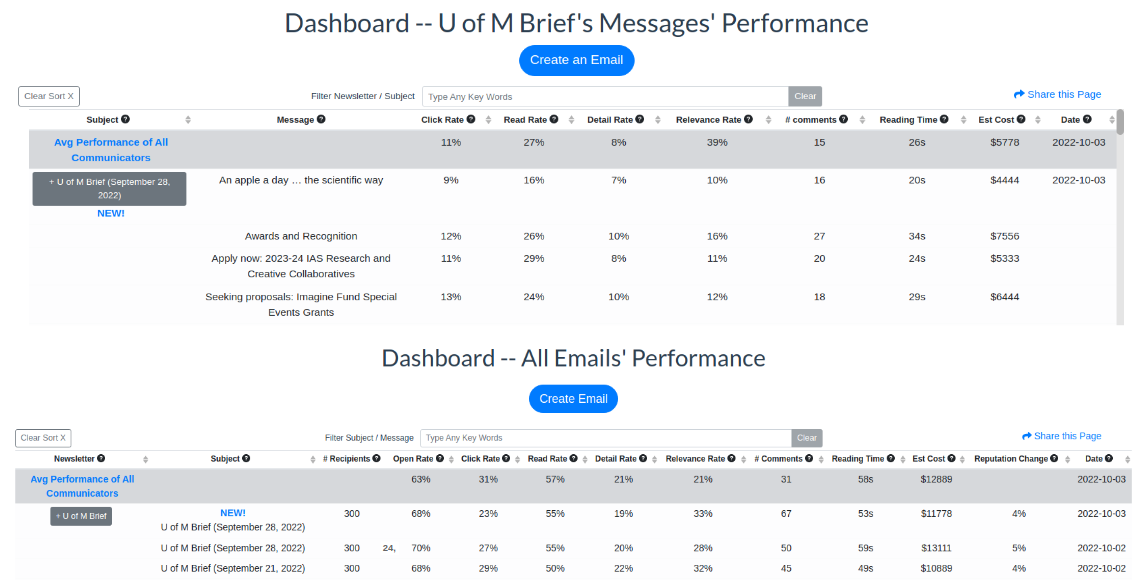}
  \caption[The email/message dashboards of CommTool]{The email/message dashboards of CommTool. The message dashboard (top) shows message-level metrics. The email dashboard (bottom) shows email-level metrics. }~\label{fig:dashboard}
\end{sidewaysfigure}

\noindent\textbf{4) Measure Message's Performance with Various Metrics:} CommTool supports communicators in collecting the awareness / impact / relevance metrics on both message level and email level. For awareness, we calculate the recipients' open rate for each email, average reading time, read (skim or read-in-detail) rate, and read-in-detail rate of each email / message. For impact, we calculate the estimated cost of each email/message and each email's estimated impact on its channel's reputation. For relevance, we calculate the percentage of recipients who viewed an email/message as relevant (either work-relevant or personally interesting) to them, or clicked them. The definitions of metrics are summarized in Table \ref{tab: metric_def}. In the dashboard of email's performance and dashboard of message's performance (see Figure \ref{fig:dashboard}), communicators would be able to see the metrics above in real-time (and they will receive a reminder email to check these metrics after 24 hours of sending the email).

Specifically, for email-level reading time, CommTool uses the average time the regular recipients are active on the email page.\footnote{``Active'' here means that the window is visible, and if there are no interactions (scroll / click / mouse hover) in the past minute a window will pop up to check whether recipients would like to stay on the page.}6 seconds are added for each recipient's time on making read / unread decisions, based on \cite{jackson2001cost}'s finding that 70\% of emails were opened within 6 seconds in the workplace. For message-level reading time, CommTool estimated a recipient's reading time for each message based on their interactions with the emails' pages (see Chapter 6 \cite{eyetrack}).

\begin{figure}[!htbp]
\centering
  \includegraphics[width=0.8\columnwidth]{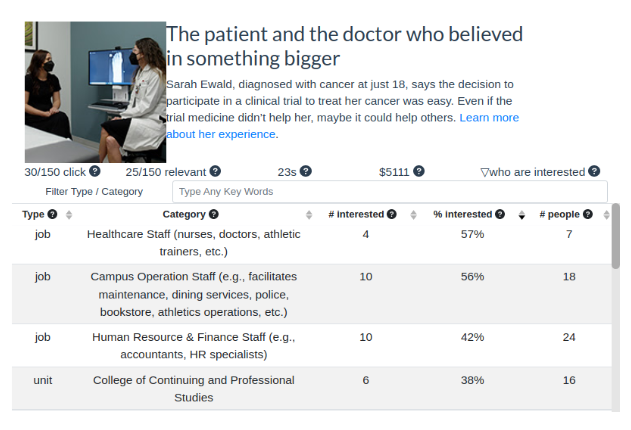}
  \caption[The report dashboard of a sample email]{The report dashboard of a sample email. Communicators would be able to see the number of regular recipients who clicked the message's hyperlinks / the number of regular recipients, the number of evaluation recipients who viewed it as relevant / the number of evaluation recipients, average reading time, estimated cost, and the interest rate split by unit / job categories.}~\label{fig:group_interest}
\end{figure}

\noindent\textbf{5) Measuring Message's Performance on Different Employee Groups:} On the dashboard page which shows the sample email (Figure \ref{fig:group_interest}), when communicators click ``Who are interested?'', a detailed view of different unit/job's interest rate will be shown. ``Interested'' here is defined as either a regular recipient clicked / read a message or an evaluation recipient indicating that the message is relevant.

\begin{figure}[!htbp]
\centering
  \includegraphics[width=0.9\columnwidth]{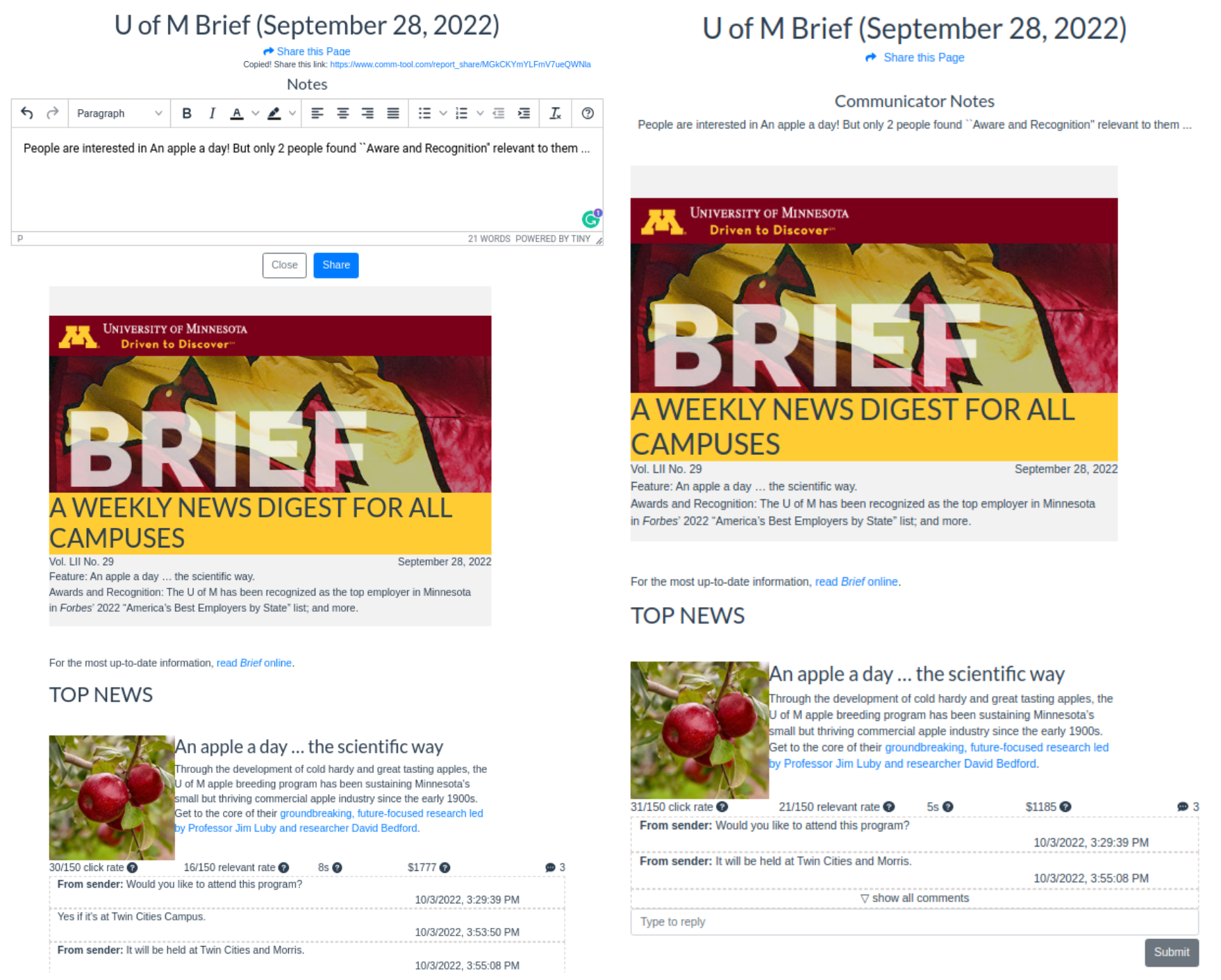}
  \caption[Share report]{Share Report. Communicators can summarize the report in notes and share that link with their clients. Their clients can comment as senders with shareable links.}~\label{fig:share}
\end{figure}

\noindent\textbf{6) Collect Comments:} CommTool also enables communicators to collect custom feedback. Communicators can add their questions in the comment area (e.g., ``Would you like to attend this program'' in Figure \ref{fig:message_level}). Communicators' questions would be pinned in the comment area and highlighted by ``from sender'' (anonymously). evaluation recipients would see these pinned questions and answer them anonymously if they want to (see Figure \ref{fig:share}).

\noindent\textbf{7) Share Reports:} CommTool enables communicators to share the results with their clients. On each dashboard (message performance, email performance, recipient view), there is a ``share'' button; communicators can click this button, add high-level summaries to that dashboard, and get the shareable links (see Figure \ref{fig:share}). Their clients do not need to register in CommTool to view these reports but can also post comments as the senders of that email with the shareable links.

\noindent\textbf{8) Automation of Evaluation Process:} The whole workflow of evaluation (see Figure \ref{fig:workflow1} and \ref{fig:workflow2}) only needs communicators to copy and paste their email to CommTool, click send, then they will receive results --- in the usability tests, this process took less than few minutes.\\

 After the research team reached an agreement on the usefulness of the platform, we invited communicators to conduct a 30-min usability test. The usability test aims to understand CommTool's workflow's intuitiveness and its metrics' clarity. During the usability tests, the communicators were asked to explore the platform and complete the following tasks: 1) create a channel and send an email from that channel; 2) find the performance reports of those emails and tell us how they interpret those metrics; 3) share the reports with their clients. The value of the metrics was simulated during the usability tests. We observed whether they met any difficulties in completing the tasks, and whether they understood the metric reports. We then revised the platform according to their feedback. We stopped inviting communicators when the newly invited communicator used the platform smoothly without guidance. We conducted 7 usability tests in the end. We made the following changes to CommTool during the usability tests:

\noindent\textbf{Visibility of Comment Area. }Communicators wanted the platform to be more communicator-focused. One area of concern was that the recipients who commented might naturally be more disgruntled, or have more negative things to say. Based on that, the comment area was changed to only be fully accessible to the corresponding communicator, and recipients were not able to see each others' comments (except comments from senders).

\noindent\textbf{Definition of Reading Time / Read Rate / Detail Rate. } Communicators indicated that they would like the reading time calculation to exclude recipients who did not open emails, because that information is already included in the open rate. For the calculation of the read rate and the read-in-detail rate, however, they prefer to consider all the recipients; then they can intuitively compare the open rate, read rate, and read-in-detail rate and use those rates together  to measure recipients' awareness.

\noindent\textbf{Provide Definition Tips of Metrics. }As CommTool provides various metrics, communicators could not recall definitions of each metric, and asked for explanations during the usability tests. Therefore, we provided tip buttons next to each metric in the reports to give detailed definitions.

\section{Field Test (RQ 2,3)}
To evaluate the proposed features' usefulness, we conducted a 2-month field test on CommTool with 5 communicators and 149 recipients (employees of the university).
\subsection{Methods}
The participating communicators were recruited from the communicators we contacted in the expert interviews and usability tests from the central communication offices. Before the field test, we asked each communicator to select a university-wide bulk email channel (sent at least monthly) to participate in the test. Before the field test, we asked the communicators in the past month, on average, how many times they communicated with their clients or shortened content during editing that channel.

The recipients were recruited through 1) the  communicators putting the recruitment message on the selected channels; 2) other mailing lists of the university; 3) the research team distributing the study posters to 38 buildings' poster boards and offices; 4) the university's slack channels. We recruited 149 recipients in total. We split the recipients into evaluation/regular groups randomly (the recipients would stay in the assigned group throughout the whole study) and gave corresponding instructions (see 3.3). Then we gave the communicators the list of recipients to be excluded from the channels during the field test (to avoid these recipients receiving duplicate emails).

During the field test, each time when the communicators sent an email under the selected channel, they would go to CommTool to send the same one to the list of recipients. We logged the recipients' interactions with each email. At the first 9 a.m. after 24 hours after they sent out each email, CommTool will send the communicators a reminder email, asking them to check the reports of that email and answer the research surveys.\footnote{The communicators asked for a regular time of receiving reports.}The research surveys contained 3 questions: 1) Which of these metrics we show on this page help you make future editing / sending decisions? 2) Which editing/sending decisions have you made / will make based on this report? 3) Why are certain metrics helpful/unhelpful to you? Could you give an example of the editing / sending decisions you made based on this report? We also logged whether the communicators read each report, how much time they spent reading, and whether they shared the reports with their clients. 

After the field test, we conducted 30-min interviews with the communicators, with the following questions: 1) What’s your general experience with CommTool? What's its pros and cons compared to your current evaluation platforms? 2) Which features of CommTool do you find helpful / useless? Why? 3) Which decisions do you usually make based on the reports (could you give a specific example)? How do you perceive the cost and performance tradeoff of these emails? 4) Do you have any other features you would like to see in such a bulk email evaluation tool? Would you like to keep using CommTool in the future? Why? The interviews were analyzed following the same practice of expert interviews.

\subsection{Recipients' Interaction with CommTool (RQ2)}

\begin{figure}[!htbp]
\centering
  \includegraphics[width=0.8\columnwidth]{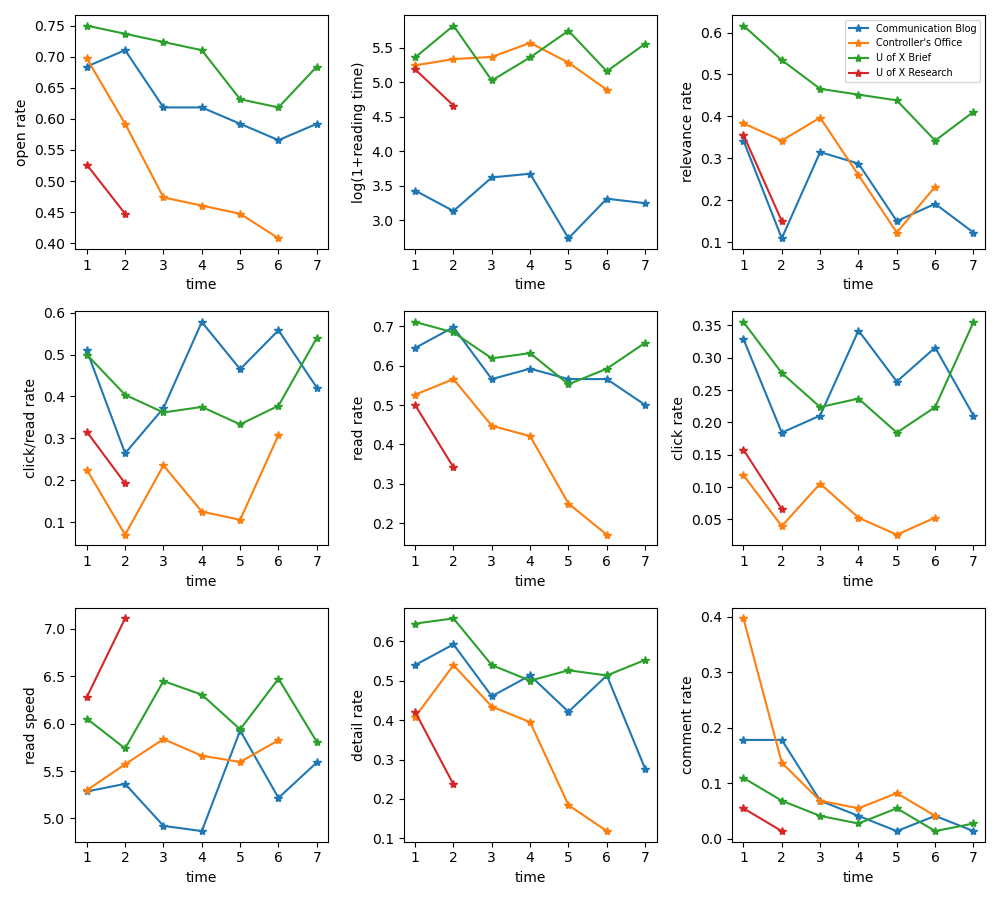}
  \caption[Summary statistics of recipients' reactions with each email in different channel]{Summary statistics of recipients' reactions with each email in different channel. Time refers to the n-th email sent in that channel. Read speed is defined as n\_word / log(1+reading time).}~\label{fig:summary}
\end{figure}

Figure \ref{fig:summary} shows the statistics of recipients' interactions with per email. The level of open rate of each channel is close to their original level (around 70\% for Brief, 60 \% for Communication Blog, 50 \% for Controller's Office, and Research is a brand new channel that is never sent before). 

Most emails were skimmed. Recipients who opened these emails spent around 60 seconds reading long channels like (Brief, Research, and Controller's Office) and around 10 seconds reading the short channel (Communication Blog). The email read rates (of at least reading one message in that email) were below 70\%, and the detail rates were below 60\%. Besides Brief, the relevance rates were below 40\%, and the comment rates were mostly below 10\%. The read rate of these channels varied from 20\% to 70\%, while the detail rate of these channels varied from 10\% to 60\%. The click rates are below 35\%. The relevance rates varied between 10\% to 60\%. The comment rates are generally below 20\%. These results show that the channels' reputation needs to be protected.

Recipients' comments could be categorized into 3 topics: the reasons that they feel the corresponding messages are interesting / uninteresting personally; how the messages are work-relevant to them; feedback on the content / designs.

\noindent\textbf{Why a message is interesting / uninteresting to me personally: } employees told the senders about why they feel a message is interesting / uninteresting to themselves. The first reason is that the employees have some / no similar experience with the message topic. For example, an employee found a message about Alzheimer research interesting because \textit{``My father-in-law has Alzheimer's so this is very relevant to my personal life.''} An employee found a staff story from another campus uninteresting because \textit{``Don't have a connection to that campus.''}


The second reason for a message being interesting is that the employees like to see that the university is paying attention to the corresponding topics. For example, an employee is glad to see a message about first-generation student celebration because \textit{``It's helpful to have visibility into central university focuses on first generation students.''}

\noindent\textbf{Why a message is work-relevant to me: } employees tell the sender when they feel a message work-relevant to them. The work-relevance when mentioned in 2 cases: 1) the message is directly related to the employee's job: \textit{``I find virtual events here that I attend for education hours as required by my Dept's Flexible Work Agreement''}; 2) the employee know someone in their units would be relevant: \textit{``This is professionally relevant to me because the unit I work in has polar explorers.''}  (the message discusses climate-related studies)

\noindent\textbf{Feedback on the content / design: } Recipients give suggestions on the specific content they want to know / how they feel the content could be designed better. For example, an employee suggested \textit{``It would be good to provide more information regarding HOW to set suppliers up with electronic payments - even if it is just "have the supplier contact us directly via phone (\#) or email (address).''} under a message about special check handling. Another employee commented on a long message \textit{``A lot of text with less bold/bullet/other ways to highlight. I would not have made it through the whole thing.''}

\subsection{Communicators' Experience with CommTool (RQ3)}
In this section, we report communicators' experience with CommTool. In the below, Communication Blog's communicator is P1; Brief's communicator is P2; Controller's Office's communicator is P3; Research's 2 communicators are P4 and P5.

\subsubsection{Overall Experience}~\\
\noindent\textbf{Number of times of reading reports}: Communicators used CommTool frequently. They sent 2 to 7 emails throughout the study and, on average, visited CommTool 5 times for each email. The message dashboards were visited most frequently (53 times) and shared with clients 7 times. The email dashboards were visited 40 times and shared with clients 4 times. The detailed report dashboards were visited 29 times and shared 2 times (see Table \ref{tab:summary}).

\begin{table}[!htbp]
\centering
\caption[Summary Metrics of Communicators' Feedback]{Summary Metrics of Communicators' Feedback. The percentages show the rates that the metrics were selected as useful by the communicators in the surveys (averaged by all the channels' statistics).}~\label{tab:summary}
\arrayrulecolor{black}
\scalebox{0.65}{
\begin{tabular}{|ll|ll|ll|} 
\hline
\textbf{Email Dashboard}                                                             &                                                                                  & \textbf{Message Dashboard}                                                             &                                                                                  & \textbf{Report Dashboard}                                                            &                                                                                   \\
\begin{tabular}[c]{@{}l@{}}Total time of email\\dashboards being read\end{tabular}   & 40                                                                               & \begin{tabular}[c]{@{}l@{}}Total time of message\\dashboards being read\end{tabular}   & 53                                                                               & \begin{tabular}[c]{@{}l@{}}Total time of report\\dashboards being read\end{tabular}  & 29                                                                                \\
\begin{tabular}[c]{@{}l@{}}Total time of email \\dashboards being shared\end{tabular}  & 4                                                                                & \begin{tabular}[c]{@{}l@{}}Total time of message \\dashboards being shared\end{tabular}  & 7                                                                                & \begin{tabular}[c]{@{}l@{}}Total time of report \\dashboards being shared\end{tabular} & 2                                                                                 \\ 
\arrayrulecolor[rgb]{0.502,0.502,0.502}\hline
\begin{tabular}[c]{@{}l@{}}\textbf{Email Dashboard}\\\textbf{Metrics}\end{tabular}   & \begin{tabular}[c]{@{}l@{}}\textbf{\% selected}\\\textbf{as useful}\end{tabular} & \begin{tabular}[c]{@{}l@{}}\textbf{Message Dashboard}\\\textbf{Metrics}\end{tabular}   & \begin{tabular}[c]{@{}l@{}}\textbf{\% selected}\\\textbf{as useful}\end{tabular} & \begin{tabular}[c]{@{}l@{}}\textbf{Report Dashboard}\\\textbf{Metrics}\end{tabular}  & \begin{tabular}[c]{@{}l@{}}\textbf{\% selected}\\\textbf{as useful}\end{tabular}  \\
Click Rate                                                                           & 48\%                                                                             & Click Rate                                                                             & 57\%                                                                             & Comments                                                                             & 35\%                                                                              \\
Relevance Rate                                                                       & 44\%                                                                             & Read Rate                                                                              & 51\%                                                                             & Relevance Rate                                                                       & 24\%                                                                              \\
Open Rate                                                                            & 33\%                                                                             & \begin{tabular}[c]{@{}l@{}}Average performance \\of all communicators\end{tabular}     & 46\%                                                                             & Who are interested?                                                                  & 15\%                                                                              \\
\begin{tabular}[c]{@{}l@{}}Average performance \\of all communicators\end{tabular}   & 30\%                                                                             & Relevance Rate                                                                         & 38\%                                                                             & Reading Time                                                                         & 8\%                                                                               \\
Read Rate                                                                            & 23\%                                                                             & Reading Time                                                                           & 35\%                                                                             & Click Rate                                                                           & 4\%                                                                               \\
Detail Rate                                                                          & 18\%                                                                             & Detail Rate                                                                            & 15\%                                                                             & Est Cost                                                                             & 0\%                                                                               \\
Reading Time                                                                         & 15\%                                                                             & \# Comments                                                                            & 12\%                                                                             & None of those are helpful.                                                           & 0\%                                                                               \\
\# Comments                                                                          & 12\%                                                                             & Est Cost                                                                               & 11\%                                                                             &                                                                                      &                                                                                   \\
Reputation Change                                                                    & 11\%                                                                             & None of those are helpful.                                                             & 0\%                                                                              &                                                                                      &                                                                                   \\
Est Cost                                                                             & 4\%                                                                              &                                                                                        &                                                                                  &                                                                                      &                                                                                   \\
None of those are helpful.                                                           & 0\%                                                                              &                                                                                        &                                                                                  &                                                                                      &                                                                                   \\ 
\hline
\begin{tabular}[c]{@{}l@{}}\textbf{Email Dashboard}\\\textbf{Decisions}\end{tabular} & \begin{tabular}[c]{@{}l@{}}\textbf{\% taken}\\\textbf{this action}\end{tabular}  & \begin{tabular}[c]{@{}l@{}}\textbf{Message Dashboard}\\\textbf{Decisions}\end{tabular} & \begin{tabular}[c]{@{}l@{}}\textbf{\% taken}\\\textbf{this action}\end{tabular}  & \begin{tabular}[c]{@{}l@{}}\textbf{Report Dashboard}\\\textbf{Actions}\end{tabular}  & \begin{tabular}[c]{@{}l@{}}\textbf{\% taken}\\\textbf{this action}\end{tabular}   \\
Add more contents                                                                    & 27\%                                                                             & Add more contents                                                                      & 36\%                                                                             & Add more contents                                                                    & 24\%                                                                              \\
None of those                                                                        & 21\%                                                                             & \begin{tabular}[c]{@{}l@{}}Share this report \\with my clients\end{tabular}            & 29\%                                                                             & \begin{tabular}[c]{@{}l@{}}Remove or shorten\\some contents\end{tabular}             & 17\%                                                                              \\
\begin{tabular}[c]{@{}l@{}}Remove or shorten \\some contents\end{tabular}            & 17\%                                                                             & \begin{tabular}[c]{@{}l@{}}Remove or shorten\\some contents\end{tabular}               & 25\%                                                                             & Send to more people                                                                  & 7\%                                                                               \\
\begin{tabular}[c]{@{}l@{}}Share this report \\with my clients\end{tabular}          & 15\%                                                                             & None of those                                                                          & 25\%                                                                             & \begin{tabular}[c]{@{}l@{}}Share this report\\with my clients\end{tabular}           & 7\%                                                                               \\
Send to more people                                                                  & 12\%                                                                             & Send to more people                                                                    & 16\%                                                                             & None of those                                                                        & 7\%                                                                               \\
Send to fewer people                                                                 & 12\%                                                                             & \begin{tabular}[c]{@{}l@{}}Move some contents\\to other newsletter\end{tabular}        & 13\%                                                                             & Send to fewer people                                                                 & 4\%                                                                               \\
\begin{tabular}[c]{@{}l@{}}Move some contents\\to other newsletter\end{tabular}      & 8\%                                                                              & Send to fewer people                                                                   & 8\%                                                                              & \begin{tabular}[c]{@{}l@{}}Move some contents\\to other newsletter\end{tabular}      & 0\%                                                                               \\
\arrayrulecolor{black}\hline
\end{tabular}
}
\end{table}

\noindent\textbf{Ease of use}: In interviews, communicators reflect that the platform is lightweight as they only spent 5 to 10 minutes initiating an evaluation on CommTool: 

\textit{``Everything about the tool, the basics, way things flow, the way you do. Things all seem logical and very useful. (P3)''} 

Communicators are generally satisfied with the readability of dashboards: \textit{``This is a nice overview for quick scanning. (P5)''} 

P1 felt that the designs of the 3 dashboards were too similar, but the reminder email sent to them after 24 hours provided a quick channel to check and differentiate the dashboards: 

\textit{``I'm glad for the email the next morning that links to all 3 of those dashboards, because I don't know if I'd be able to figure out, because the first in the second dashboard looks so similar.'' (P1)}

\noindent\textbf{Comparison with existing platform Salesforce}: Communicators feel that Salesforce provides more help in editing email templates, but CommTool provides more information in evaluating emails: 

\textit{``Salesforce has templates which everybody loves. It  has a glamorous interface. CommTool is simpler. It had metrics that people should care about and should be able to track, and at some point, some senior leader is going to ask them for them.  I think it is helpful, especially for people who didn't have a lot of experience with their audience or who were working with completely unknown audiences. (P3)''} 

P1 mentioned that CommTool provided a channel to retrieve various data legally from employees: 

\textit{``Apple messed things up last year. They pull out the tracking pixel before it gets to your inbox --- so the open rates were wrong. In CommTool, people are those who agreed to share this information.'' (P1)}

\noindent\textbf{Sharing reports:} Communicators shared the message dashboards 7 times and shared the email dashboards 4 times. These dashboards provided communicators with the reason for removing/extending specific messages. The per-message reading time specifically gets clients' attention: 

\textit{``Because they are interested in how long people took to read it. They've always had the number of eyes on the open rates. They didn't know that that existed before I started working with them, so they were like, we can tell how many people read the message. Yes? you can. so now they're all comparing numbers with each other, our message did this. They want to go further with metrics like that, how long people are engaged with the content, what their feedback is, and so forth.'' (P3)}

The detailed report page was shared less because clients care more about numbers: \textit{``When I'm talking with the people, they've all seen the message. They've helped write the message in most instances. So they're passed that part, and they want to see the numbers. Plus everyone I work with is a person they're all accountants. So when they see words and pictures, it freaks out a little bit, but when they see numbers they're like. Oh, I love it.'' (P3)}~\\

\subsubsection{Experience with Specific Features}~\\
\noindent\textbf{Message - level versus email - level metrics}  (feature 3). 
Message metric dashboards were visited more times (53 visits) compared to email metric dashboards (40 visits), see Table \ref{tab:summary}. P2 mentioned that the message dashboard \textit{``is useful if you wanted some more detailed analysis of your readership.''} The report dashboards were visited less (29 visits) --- P2 reflected that communicators who run large newsletters are too busy to look at the details on the report dashboard \textit{``I would not have time to go through all those comments, especially on a list that's 20,000 strong. ''}

\noindent\textbf{Awareness vs relevance vs cost vs reputation metrics (feature 2, 6)}.
The relevance metric was selected as useful most times, and the awareness metric followed. The cost and reputation metrics were less likely to be used.

Specifically, the email click rate is most usually to be selected as useful (48\% of times) and followed by the relevance rate (44\%). P2 told us that after seeing the email relevance rate, they took more time to  \textit{``consider relevance for audience members when developing content.''}.

Besides, the message-level reading time and read rate are also often found to be helpful. In interviews, P3 said that: 

\textit{``I pay attention to how much time they spent on it when that's available. That is the part of this tool that most interested me. So I could see, Did they engage with the content? And did they spend some time there?'' (P3)} 

Communicators care less about detail rate because most messages only ask people to have some sense of the content: 

\textit{``In-depth reading doesn't really matter much just because it's (the email content) a summary. They (the recipients) can take action based on the email, click in or read the full article.'' (P1)} 

The finance newsletter cares more about read-in-detail rate because they are sending important information that every employee should know: 

\textit{``I'm more interested in the detail than the I mean. All the tools will give you how many eyes looked at it? Right? I want to know how long they spent on it. You know, the University's budget is between 3 and 4 billion dollars a year. It's a lot of money, and most of it is spent on human costs, you know salaries and etc. Every employee should want to care, and the fact that person felt that it wasn't relevant to them meant that they didn't connect with the institution as an entity that has to watch its money, and watch how we behave with it.'' (P3)}

Communicators do trust CommTool's estimation on per-message reading time: 

\textit{``From what I could tell, and I had my own team members look at the email and sort of track. How long it took them to read the sections carefully, and I compared the 2 and I. It seemed very accurate, based on their experience.'' (P2)} 

\textit{``The top article in the top section has about a minute the one spent on it, and then the number gets kind of small as you go down, which makes sense to me because some of these things are meant to scan and other things. It's like you only have this much time to get those attention. So if they spend the most of their attention on the top thing, and then scroll the rest. That seems reasonable.'' (P4)}

Communicators did review estimated cost but found it less helpful because of the organization's structure: 

\textit{``(We) discussed estimated cost, and while interesting, it didn’t seem to influence leaders' decision to send.'' (P5)} 

Also as we just discussed above, communicators felt some cost does make sense given the value of the information:

\textit{``It (the cost) was frightening. So now, every time I send out an email, I'm like, how much are we actually costing by making people read this? Then I remember we don't send out fluff, or like party invitations, or come to our sale we're sending out. You have work to do. This is related to your work. So we would incur that cost.'' (P3)}

Communicators did not look at the reputation metrics directly but checked out the trend of metrics themselves and had their own judgments:

\textit{``I think it would be most effective in looking at trends over time.'' (P1)} 

\textit{``I'm going to have to see the metrics over time before I make any decisions about changing up the newsletter.'' (P2)} 

\textit{``I did look at the like week to week or message to message on my own like improvement, and if that went up a little bit, I was happy. If it went down, figure out what was in my content that made it go down. I like the predicted reputation change, although I'd want to know how it's actually calculated. But in really simple terms.'' (P3)}

Communicators looked at the average performance of other communicators but did not use it to make decisions --- they preferred to make decisions based on their own data: 

\textit{``I knew there was a total average. I knew my messages were probably different than other messages, so I didn't pay attention to comparing them very much.'' (P1)}

\noindent\textbf{Custom feedback} (feature 4). Comments gave communicators direct help with writing and formatting (see 4.2 for employees' comments). P3 indicates that based on the comments they received, they want to make the content clear: \textit{``We've removed a layer of formality from our messaging. We try to be brief, be bright. Taxes shouldn't be scary.'' (P3)}

We also notice that bi-directional communication gets more feedback than single-direction communication. In our study, an email that the communicator has back-and-forth conversations with recipients gets 107 comments while the others are below 40. Some communicators did not use this feature because they did not notice this feature and said that they would reply:

\textit{``If they're (recipients) asking a direct comment like a direct question. (P1)''} 

P3, who got much feedback, gave us suggestions on how to get more replies: 

\textit{``If you ask them unexpected questions, or you ask them in a less formal way, they're more likely to respond. So I think that's why I got a lot of comments.'' (P3)}

\noindent\textbf{Group-level interest} (feature 5). P4 and P5 from the research newsletter particularly look at ``who are interested'' to ``Consider relevance/interest for audience members and which audiences respond to the types of messages when developing content''. But they felt that the ``who are interested'' feature could be more useful if they could setup groups themselves: 

\textit{``Whenever you have that level of detail, it's interesting. I'm seeing all these different colleges. And that's great. But it's not maybe necessarily useful for us to know that. For us, it would be interesting to see the breakdown between staff and students and faculty.'' (P4)}.

These findings suggest that 1) current bulk email evaluation platforms, besides open / click rate, should take more active actions on collecting the relevance data (like by adding a small relevance to me button under each email), skim-level awareness, and trend of metrics; 2) the evaluation platform should allow message-level custom feedback, and especially encourage the bi-direction communication between senders and recipients; 3) if we want to control bulk emails' time cost, we need to build the mechanism that can influence organization leaders.

\subsection{Email's Performance versus Reputation}
In this section, we reported the correlation between email's performance and its channel's reputation. We defined reputation's change as $reputation_t = open\_rate_{t+1} - open\_rate_t$ (the difference between an email sent at time $t$'s open rate to its next open rate). We run ordinary least square models (with constant) between $reputation_t$ and each $metric_t$. The coefficients' values, p-values (for the hypothesis $coefficient \neq 0$), and confidence intervals are reported in Table \ref{tab: coef}.

We found click rate positively correlated with reputation change (coef=0.221, p-val=0.049). Though none of the reading time, read rate, or detail rate correlated with reputation change, the ratio between click rate to read rate is positively correlated with reputation change (coef=0.169, p-val=0.026). This result matches the intuition --- the emails that interest the recipients who read them can build up their reputation. Interestingly, the comment rate was negatively correlated with reputation change (coef=-0.260, p-val=0.041*). According to the contents of the comments we received, a potential explanation is that most recipients only commented when they found an email irrelevant or confusing (about contents, designs, intentions, etc.); therefore, these recipients are less likely to open the next email.

\begin{table}[!htbp]
\centering
\caption[Coefficients between email's performance with its reputation's change]{Coefficients between email's performance with its reputation's change. CI: the [0.025,0.975] confidence interval of the corresponding coefficients.}~\label{tab: coef}
\arrayrulecolor[rgb]{0.8,0.8,0.8}
\scalebox{0.65}{
\begin{tabular}{!{\color{black}\vrule}c|c|c!{\color{black}\vrule}c|c|c!{\color{black}\vrule}c|c|c!{\color{black}\vrule}} 
\arrayrulecolor{black}\hline
\textbf{metric} & \textbf{coef} & \textbf{p-val}                                                     & \textbf{metric}     & \textbf{coef} & \textbf{p-val} & \textbf{metric} & \textbf{coef} & \textbf{p-val}                                                        \\ 
\arrayrulecolor[rgb]{0.8,0.8,0.8}\hline
open rate       & -0.038        & 0.768                                                              & log(1+reading time) & -0.007        & 0.542          & relevance rate  & 0.026         & 0.775                                                                 \\ 
\hline
click/read rate & 0.169         & \begin{tabular}[c]{@{}c@{}}0.026*\\CI{[}0.023, 0.315]\end{tabular} & read rate           & 0.027         & 0.811          & click rate      & 0.221         & \begin{tabular}[c]{@{}c@{}}0.049*\\CI{[}0.001, 0.441]\end{tabular}    \\ 
\hline
read speed      & 0.007         & 0.786                                                              & detail rate         & 0.055         & 0.629          & comment rate    & -0.260        & \begin{tabular}[c]{@{}c@{}}0.041*\\CI{[}-0.508, -0.012]\end{tabular}  \\
\arrayrulecolor{black}\hline
\end{tabular}
}
\end{table}

\section{Discussion}
In this paper, we studied how to support communicators in evaluating organizational bulk emails. We conducted expert interviews to understand potential useful features. Then with an iterative design approach, we deployed the system CommTool. At last, we evaluated these proposed features in a 2-month field test with 149 employees and 5 communicators.

We found that communicators wanted more details like message-level metrics to help them better evaluate and target bulk messages (RQ1). Employees were not paying enough attention to the tested channels (RQ2). Communicators used CommTool to better understand their audience and also suggested improvements on features like cost estimation, group-level interest, etc (RQ3) In the below, we discussed the observations we found interesting.

\subsection{Influence of Organization Structure on Evaluating Bulk Emails}
We observed that the organization's structure influences communicators' use of bulk email evaluation platforms. In the field test, though communicators had access to the evaluation reports and felt that certain email designs could be changed, they did not always have the final say on whether a message should be sent out and they needed to get back to their clients, the organization leaders. The separation of information access and decision-making brought extra costs for CommTool's users in communicating back and forth with their clients if they want to make changes. CommTool tried to reduce this cost by enabling communicators to share reports with their clients. However, this approach does not significantly influence leaders' decisions. Communicators suggested that leaders would need the reports to be more appealing: 

\textit{``If that (the report) was going to be used to show your leadership, it might provide something easier to prepare, more appealing, like a presentation'' (P5)}. 

Another high-level solution to this challenge is to transfer part of the decision power to communicators. As \citeauthor{jensen2009specific} proposed in their paper about organization's knowledge control and structure \cite{jensen2009specific}, \textit{``If the knowledge valuable to a particular decision is to be used in making that decision, there must be a system for partitioning out decision rights to individuals who already have the relevant knowledge and abilities or who can acquire or produce them at the lowest cost''}. For example, the organization leaders could allow communicators to remove some messages themselves if the history reports show that those types of messages are not interesting to the audience.

\subsection{Making positive suggestions}
One interesting observation is that communicators found the message-level reading time useful but cost less helpful, while cost is approximately a rephrase of reading time in CommTool. Communicators reflected that the cost displayed is too ``frightening'' to lead them to make any decisions based on that. Intuitively, reading time and cost can be viewed as the positive side and negative side of the same concept. The fact that users are more likely to take actions based on the positive side match with the previous findings on psychology --- for example, \citeauthor{hayashi2012pedagogical} and  \citeauthor{kim2007pedagogical} both found that students/participants were more motivated to learn when they received positive feedback from conversational agents \cite{hayashi2012pedagogical}. This observation suggests that bulk email evaluation platforms could benefit from making the reports' presentation more positive to encourage communicators to take action in designing / targeting.



\section{Conclusion}
In this chapter, we studied how to support communicators evaluating organizational bulk emails to improve organizational communication's effectiveness. We started with expert interviews with 5 communicators. We found that communicators wanted more detailed information like message-level reading time, relevance rates, comments, etc to help them understand their audience. Then with an iterative design approach, we deployed an organizational bulk email evaluation platform CommTool. CommTool gives communicators convenient real-time access to not only email-level but also message-level reading time, read rate, read-in-detail rate, relevance rate, comments, group-level interests, etc. We then conducted a 2-month field test with 5 communicators and 149 employees to evaluate the proposed features. We collected employees' interactions through the field test and communicators' feedback through surveys during the field tests, as well as interviews after the test. We found that communicators liked the message-level reading time, read rate, and relevance rate provided by CommTool, which helped them to better understand their audience's interests in each message. Comments helped communicators in designing messages though communicators might be too busy to check all of the comments. Communicators trust the reputation and cost algorithm, but felt that they preferred to observe the trends themselves and use that as reputation instead of a number, and their clients felt that some messages should be sent out at such costs. 

The limitations of this study include that 1) though the tested channels are selected from the newsletters that target the whole organization, their mailing list is not the same as our participant list and the measured metrics would not be the same as these channels' performance on their real mailing list; 2) we conducted our study in a large organization with a top-down communication system; organizations with different structures (such as a flattened management structure) might have different findings (for example, their communicators might have more say in deciding which messages to be sent out).

With these limitations, this study contributes to improving organizational communication by understanding communicators' expectations, and deploying and testing a bulk email evaluation platform with diverse message-level and email-level metrics on awareness, relevance, cost, and reputation. Based on the test, this study suggests bulk email platforms add certain message-level metrics such as reading time and read rate, and gives specific recommendations on how to present these metrics to make them helpful for communicators.




\chapter{Discussion and Conclusion}
\label{conclusion_chapter}

In this chapter, I summarize the contributions of the dissertation, the implications of our findings, and identify future work. By conducting surveys, interviews, controlled field experiments, and iterative design studies, this work sheds light on how organizational bulk email system works, and at which control points we can insert interventions to make this system more effective.

\section{Contribution Summary and Future Work}
We started this thesis from our personal experience with the study site --- why, as an employee, we receive hundreds of pieces of non-targeted and non-personalized information from the university through bulk emails every week. In a follow-up meeting with 9 communicators of the study site, we learned that the study site's organizational bulk email system might be ineffective and not what the university wanted it to be. Then we looked back to previous work, and found much knowledge on how organizational communication's effectiveness is impacted by various stakeholders' actions, but less knowledge on organizational bulk emails, especially from a multi-stakeholder perspective. Therefore we identified a need to understand this system's effectiveness and its stakeholders, and to dig into the technology opportunities for making this system better.

In Chapter 3, we first conducted a mixed-methods study to understand organizational bulk email system and its stakeholders. Through a survey of 162 employees of our study site, we learned that the current organizational bulk email system is ineffective as only 22\% of the information communicated was retained by those employees. Then through the artifact walkthroughs with 6 communicators, 9 recipients, and 2 managers, we learned that the failure of the study site's organizational bulk email system was systemic — the organizational bulk email system had many stakeholders, but none of them necessarily had a global view of the system or the impacts of their own actions. The senders would like their messages to be distributed broadly. The communicators lacked tools for personalizing, targeting, and evaluating bulk emails, and as a result, often distributed messages too widely. The employees, overwhelmed by large quantities of information, only want to read messages they perceive as relevant to themselves. And nobody was watching the millions of dollars of people's time consuming in this process. The limitation of this work is that this is a qualitative case study of one study site. The observations may not be generalizable across organizations. It is possible that stakeholders in other organizations have different practices/perspectives on bulk email systems with respect to their organizations' cultures/structures. Therefore a question to be studied in the future is to collaborate with more organizations to study whether our findings in this case study are common. 

In Chapter 4, based on the findings of the empirical study, we further proposed an economic model to describe the value, cost, and actions of this system's stakeholders and how their diverse perspectives cause ineffectiveness. We summarized two promising interventions. On the recipient side, we need mechanisms for encouraging employees to read high-level information. On the sender side, we need to let senders consider the value and cost of their bulk messages, and design tools for communicators to support them in making targeting and designing decisions. The limitation is that we only propose a model without justification. However, the proposed model enables us to clarify the relationships between this system's stakeholders and identify a bunch of potential interventions, enable to do the next two studies. There is future work to be explored in the calibration and implementation of this model, including 1) calibrating and measuring the value and costs proposed in the model; 2) designing tools to make the change of reputation observable to communicators; 3) predicting employees' preferences on each bulk message, such as to learn from employees' job descriptions; 4) learning how to weigh different stakeholders' opinions.

In Chapter 5, we studied the recipient-side interventions --- the personalization designs of organizational bulk emails that can lead employees' attention to the messages important to the organization, while maintaining employees' positive experiences with these bulk emails, then they continue to read these emails in the future. In an 8-week field experiment with a university newsletter, we implemented a 4x5x5 factorial design on personalizing subject lines, top news, and message order based on both the employees' and the organization's preferences. We found that mixing important-to-organization messages with employee-preferred messages in top news could improve the whole newsletter's recognition rate. This work also provided a basic backend framework for communicators in personalizing organizational bulk emails. The limitation of this work is that the personalization is reordering only --- because of the requirement of our collaborator, we did not exclude any message from the studied newsletter. The future work includes 1) personalizing newsletters by filtering a subset of relevant messages in organizations that allow taking this mechanism; 2) exploring different designing strategies that could help employees understand why they need to read some messages: for example, encouraging senders to tag the reasons for sending some messages; 3) enabling employees to update their preferences; 4) studying how to restore nonreaders' trust on the bulk communication channels

In Chapters 6 \& 7, we studied the sender-side interventions --- a communication prototype tool that supports communicators in making editing and targeting decisions on organizational bulk emails. To enable such evaluation, we first developed a novel neural network technique to estimate how much time each message is being read using recipients' interactions with browsers only, which improved the estimation accuracy from 54\% (heuristics) to 73\%, based on 200k ground truth data points we collected through eye-tracking tests. For limitation, we only collected eye-tracking data on 9 users therefore the dataset we collected might not catch enough variance on user patterns. Future work should look at where the value of additional users starts to decay significantly. 

Then we iteratively designed and deployed a prototype of an organizational bulk email evaluation platform (CommTool), which enables communicators to learn the performance and cost of each bulk message. We evaluated the usefulness of different features in CommTool through a 2-month field test with 5 communicators and 149 organization employees.
  We found that 1) the message-level metrics such as reading time and read rate helped communicators in designing; 2) email's performance is correlated with its channels' future reputation; 3) CommTool does not influence the organization leaders' decisions. We summarized with suggestions on designing organizational bulk email evaluation platforms that consider leadership and provide positive feedback. The limitations of this study include that 1) though the tested channels are selected from the newsletters that target the whole organization, their mailing list is not the same as our participant list; 2) organizations with different structures (such as a flattened management structure) might have different findings. As we found that showing the cost to bulk email senders does not influence their decisions, there is future work left on reducing overwhelming bulk communication in organizations, such as allocating communication budget according to communication channels' performance, enabling better filtering mechanisms on the recipient side, etc.



In short, organizational bulk email system is a complex multistakeholder system where different participants have limited capabilities to see the impact of their actions on the organization as a whole. We present a set of studies aimed at improving this system toward an integrated system that considers diverse stakeholders' actions and opinions. This thesis provided: 1) in-depth knowledge of how an organizational bulk email system works and its stakeholders' perspectives, including its stakeholders' roles, practices, and perspectives; 2) an empirical evaluation of several mechanisms for encouraging organizational bulk email recipients to pay more attention to messages deemed as important by the organization; 3)
the design and empirical evaluation of a tool that incorporates various feedback features on bulk email's relevance to employees and the attention they receive.

\section{Discussion}
Based on our studies, we highlighted that organizational bulk email systems should be designed differently from the existing commercial bulk email systems based on their nature as a system in the workplace and for organizations.

\subsection{Nuances, Fluidity, and Exceptions as a Workplace System}
As a social technology system in workplaces, organizational bulk email system is full of nuances, fluidity, and exceptions \cite{ackerman2000intellectual}.

``Nuances'' refers to the depth of details needed for understanding the contexts to make predictions and decisions \cite{ackerman2000intellectual}. All the stakeholders in organizational bulk email system have very nuanced perceptions of the context. For employees, whether they perceive a message as work-relevant depends on their job responsibilities, which can be complicated and obscure. For example, in Chapter 3, we discussed a bulk email reminding Webex users to transfer their documents. Strictly, this information is only relevant to those employees who have not transferred their files and would need to use those files in the future. This granularity of information is hard to collect (especially implicitly without asking employees), and is case-by-case --- emails with different contents would definitely need different context. It is obvious that an effective organizational bulk email system, ultimately, would build on certain knowledge management systems that can identify the required contexts smartly and collect them automatically \cite{maier2011knowledge}. For example, a knowledge management system that stores employees' job responsibilities and preferences to decide which bulk messages are relevant to them.

``Fluidity'' refers to the changes in stakeholders' perspectives and practices  \cite{ackerman2000intellectual}. For example, employees' actions toward bulk emails might depend on their schedules. In Table \ref{tab:disagree}, employee R6 skipped a bulk email because they were too busy that month and found the email relevant when we asked them to read it. Employees' job range is also fluid. Using the ``voicemail down'' message in Chapter 5 as an example, this kind of message might be relevant to an employee if they are planning for a voicemail message on Sunday but might be irrelevant if they have no such plans next time. Also, for the senders, there are less common criteria on how to send a bulk message (see section \ref{study1}). One communication channel can be used to send important and good-to-know messages at the same time, which would cause employees' confusion about the importance of the messages from that communication channel. Given the fluidity of organizational bulk email system, it is important for the knowledge management system to collect contexts in real time. For example, updating employees' time schedules and controlling the quantities of information they receive according to that.

``Exceptions'' --- the revision of general guidelines in practice --- are normal in workplace \cite{suchman1984procedures}. Indeed, the stakeholders in organizational bulk email systems do not completely follow rational standards. Whether and how a bulk message would be sent not only depends on its objective value, but also the perceptions and level of its senders. For example, in Chapter 5, the message \textit{``UMD Chancellor Search''}, though mainly relevant to a single campus, was sent to the employees across all the campuses as a single email, because the sender perceived the value of their message being distributed widely. Therefore organizational bulk email systems should have back-channels that allow stakeholders to negotiate when an exception can be made and evaluate the impacts of exceptions (e.g., the impact on channel reputation of sending a good-to-know message to all the employees).

\subsection{Key Aspects for Designing Organizational Bulk Email Systems \cite{intelligent}}
Given the nuances, fluidity, and exceptions in organizational bulk email system, the agents designed to support this system should be socially-embedded,  which means that it will 1) analyze the social contexts of this system; 2) give the right incentive to each stakeholder that can steer their actions toward the whole organization’s benefit. It should consider the employees' time cost of reading bulk emails, the communicators' time cost of designing and targeting bulk emails, and the leaders' value of getting important information out. 
Specifically, the agent should consider these aspects: organizational structure, economics, data sources, and decision-making process.

For organizational structure, the agent need be aware of different units and their responsibilities (e.g., the University Relation department is in charge of our study site's internal / external communications), the management style (top-down, bottom-up, distributed, etc.), the key stakeholder groups, and the IT infrastructure and its limitations (see Chapter 3).

Then the agent needs to understand the general economics of the bulk email system --- all the stakeholders’ value and costs with the corresponding bulk messages, such as employees' value of receiving job-relevant or interesting messages, employees' time costs, communicators' time costs, information producers' value of letting employees know about important messages, etc (see Chapter \ref{economic_model}).

Given all the data collected above, a socially-embedded agent would look at each of these costs and value in providing feedback and suggestions on how to achieve the disparate goals of the organization --- maximizing the desired value while minimizing both the cost and the loss of reputation (and long-term value) if employees become less likely to read future messages. For example, a bulk email agent can select communication channels for a message according to the channels' effectiveness and the messages' potential impact on the channels' reputation. The agent can also personalize bulk emails based on all the involved stakeholders’ value (Chapter \ref{study2}). We mainly explored personalizing message order in this thesis. Besides that, the agent can also personalize distribution mechanisms by sending message to different defined communication groups via different channels. For example, \citeauthor{beringer2012collaborative} designed a system with a user interface that enabled the organization senders to send a message of a certain message type within one of the collaborative conversation channels \cite{beringer2012collaborative}. The agent can also personalize the sending time by predicting how busy or interrupted the corresponding employee is when they receive the message \citeauthor{danninger2007can}.  

Besides personalization, the agent can influence the university leaders' and central offices’ communication behaviors by tracking performance and controlling the budget (Chapter \ref{study3}). We explored visualizing costs in this thesis. The other way is to formalize this through budget allocations -- units could have communications budgets that force them to make decisions about what's worth communicating broadly (and what's not).

In short, with this thesis, we now understand organizational bulk email systems' stakeholders' value, and how personalization and visibility tools could be used to influence employees' and senders' behavior toward the whole organization's benefits. And give the nuances, fluidity, exceptions, and conflicts in organizational bulk email system, there is still work to be done to manage this system's knowledge, quantify its metrics, and make decisions on designing / targeting organizational bulk emails.





\appendix
\chapter{Appendix for Chapter 3}
\label{app4}

\subsection{Communicator Interview Protocol}

See Table \ref{tab:gatekeeper_protocol}.
\begin{table}[!htbp]
\small
\centering
\arrayrulecolor[rgb]{0.8,0.8,0.8}
\resizebox{\textwidth}{!}{%
\begin{tabular}{|l|r|p{13cm}|} 
\hline
\textbf{Part}  & \multicolumn{1}{l|}{} & \textbf{Questions}                                                                                                                                                  \\ 
\arrayrulecolor{black}\hline
1.Practice     & 1                     & How do you send bulk emails for your unit and why do you send it in that way      ?                                                          \\ 
\arrayrulecolor[rgb]{0.8,0.8,0.8}\hline
               & \multicolumn{1}{l|}{} & Who are your client? What are their requests? Do they send emails themselves?                                                                                          \\ 
\hline
               & \multicolumn{1}{l|}{} & Who offer the mailing list and content?                                                                                                                             \\ 
\hline
               & 2                     & When do you send bulk emails to all employees versus subgroups?                                                                                                                 \\ 
\hline
               & 3                     & When do you decide to put a message in a bulletin of newsletter versus an individual bulk email?                                                                            \\ 
\hline
               & 4                     & Besides newsletters and individual bulk emails, what are the other mechanism you use to communicate? \\ 
\hline
               & 5                     & What happened with the bulk emails you sent? Are you aware of how often they read or do they understand it carefully?                                                  \\ 
\hline
2. Email Cases & 1                     & The title, goal, and recipient of the email                                                                                                                          \\ 
\hline
               & 2                     & Which channel did you use to send it?                                                                                     \\ 
\hline
               & 3                     & Who do you imagine should read this? Who do you imagine did read this? Are they the same?                                                                           \\ 
\hline
               & 4                     & Is who should read this the same as who the email was sent to?                                                                                                 \\ 
\hline
               & 5                     & Who should read in detail and who can just scan it?                                                                                                                 \\ 
\hline
               & 6                     & Who do you imagine should take actions (click links, take surveys, reply)? Did they take the actions?                                                                \\ 
\hline
               & 7                     & Do you have experience that some recipients asked about/forgot messages that you've already sent in an email? When will you send it for multiple times?                  \\ 

\hline
3. Assessment   & 1                     & What's your sense of how email communication between university and employees work (well or poorly)?                                                                \\
\hline
\end{tabular}}
\arrayrulecolor{black}
\caption{Interview Protocol of Communicators.}
\label{tab:gatekeeper_protocol}
\end{table}

\subsection{Recipients Interview Protocol}
See Table \ref{tab:command} for search commands for different types of emails. The following are the questions we asked the recipient for each email:
\begin{itemize}
    \item Did you label/trash this email? Did you open this email or leave it unread? Why?
    \item Did you recognize the sender?
    \item If you opened it, did you scan or read it in detail? Why?
    \item If you didn't open it, open it now, do you find anything important to you?
    \item Did this email require actions? Did you take actions and how soon did you take? Why?
    \item Rate the importance/urgency/relevance of this email from 1 (lowest) - 5 (highest).
\end{itemize}

\chapter{Appendix for Chapter 5}
\label{app6}
\section{The summary of bulk messages' topics}
\begin{table}[!htbp]
\caption[The summary of bulk messages' topics]{The summary of bulk messages' topics of Brief and employees' job categories.}~\label{tab: c5_1}
\centering
\arrayrulecolor[rgb]{0.8,0.8,0.8}
\scalebox{0.6}{
\begin{tabular}{!{\color{black}\vrule}l|l|l!{\color{black}\vrule}l!{\color{black}\vrule}} 
\arrayrulecolor{black}\hline
\multicolumn{3}{!{\color{black}\vrule}c!{\color{black}\vrule}}{\textbf{Appendix A: Message Topics }}                                                                                                                                                                       & \multicolumn{1}{c!{\color{black}\vrule}}{\textbf{Appendix B: Job Categories}}                                                                                           \\ 
\arrayrulecolor[rgb]{0.8,0.8,0.8}\hline
\begin{tabular}[c]{@{}l@{}}Talk/Symposium/\\Lectures Announcements\end{tabular}  & \begin{tabular}[c]{@{}l@{}}Community Service/ Social \\Justice/Underserved Population\end{tabular}  & Sports  Spirit                                                           & \begin{tabular}[c]{@{}l@{}}Administration \& Advancement \\\& Communication Staff\end{tabular}                                                                 \\ 
\hline
\begin{tabular}[c]{@{}l@{}}Operations Awareness\\/Facility Closures\end{tabular} & \begin{tabular}[c]{@{}l@{}}Award/Recognition to University, \\Faculty, Staff, Students\end{tabular} & Youth, Children                                                          & \begin{tabular}[c]{@{}l@{}}Campus Operation Staff (e.g., facilitates \\maintenance, dining services, police,\\~bookstore, athletics operations.)\end{tabular}  \\ 
\hline
\begin{tabular}[c]{@{}l@{}}Health Wellness \\Resources/COVID\end{tabular}        & \begin{tabular}[c]{@{}l@{}}Program \& Award Applications\\/Announcements\end{tabular}               & Art  Museums                                                             & \begin{tabular}[c]{@{}l@{}}Faculty, Teaching \& Research Staff, \\Librarians, Museum Curators \\and Directors, etc.\end{tabular}                               \\ 
\hline
\begin{tabular}[c]{@{}l@{}}Fundraising \& \\Development\end{tabular}             & \begin{tabular}[c]{@{}l@{}}University Program \\Success Stories\end{tabular}                        & \begin{tabular}[c]{@{}l@{}}Policies/Admin News\\/Governance\end{tabular} & \begin{tabular}[c]{@{}l@{}}Healthcare Staff (nurses, doctors, \\athletic trainers, etc.)\end{tabular}                                                          \\ 
\hline
\begin{tabular}[c]{@{}l@{}}Climate/Eco/\\Agriculture\end{tabular}                & \begin{tabular}[c]{@{}l@{}}Engineering Science \\Research Stories\end{tabular}                      & Faculty  Staff Stories                                                   & \begin{tabular}[c]{@{}l@{}}Human Resource \& Finance Staff \\(e.g., accountants, HR specialists)\end{tabular}                                                  \\ 
\hline
\begin{tabular}[c]{@{}l@{}}University History\\/Celebrations\end{tabular}        & \begin{tabular}[c]{@{}l@{}}History/Social Science \\Research Stories\end{tabular}                   & Student/Alumni Stories                                                   & Information Technology Staff                                                                                                                                   \\ 
\hline
\begin{tabular}[c]{@{}l@{}}Tech Tool Updates\\/Workshops\end{tabular}            & \begin{tabular}[c]{@{}l@{}}Health/Biology Research \\Stories\end{tabular}                           &                                                                          & \begin{tabular}[c]{@{}l@{}}Student Services Staff (advisors, \\student union staff, financial aid staff, etc.)\end{tabular}                                    \\
\arrayrulecolor{black}\hline
\end{tabular}
}
\end{table}


  \section{A sample personalized Brief}~\label{tab: c5_2}
  

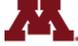

**U of M Brief (November 3, 2021) 15 things to do in November at the Landscape Arboretum**
1 message

brief-study University of Minnesota <brief-study@umn.edu>

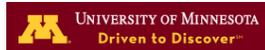

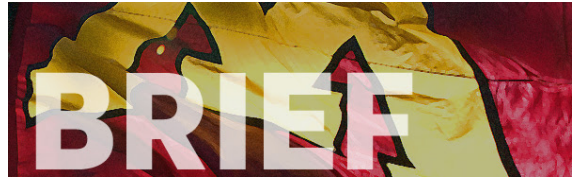

A WEEKLY NEWS DIGEST FOR ALL CAMPUSES

| | |
|---|---|
| Vol. LI No. 33 | November 3, 2021 |

INSIDE THIS ISSUE

- Nov. 12 - Learn about lactation resources at the U.
- U of M establishes nation-leading tuition assistance program for Native American students.
- Nov. 11 - Native American Heritage Month: Virtual conversation with President Joan Gabel and Karen Diver.
- Nov. 3 - Torch & Shield Award Banquet (Crookston).
- 15 things to do in November at the Landscape Arboretum.

Read *Brief* online and see *Brief Alerts* for timely resources.

TOP NEWS

**Nov. 12 - Learn about lactation resources at the U**

Wondering how to support the lactation needs of your employees, colleagues, or students? Returning to work or school after having a baby? Join in this webinar to learn about lactation resources on campus. Hosted by the Lactation Advocacy Committee, the Student Parent Help Center, and the Women's Center. Noon-12:45 p.m.

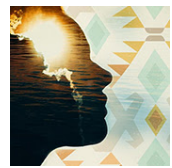

**U of M establishes nation-leading tuition assistance program for Native American students**

The University of Minnesota announced a significant expansion of Native American student tuition support, a new initiative that will be among the nation's most comprehensive free and reduced tuition programs for Native American students. Starting in fall 2022, the University will provide free or reduced tuition on any of its campuses to first-year undergraduate students and Tribal college transfer students who are also enrolled citizens in one of the state's 11 federally recognized Tribal Nations.

**Nov. 11 - Native American Heritage Month: Virtual conversation with President Joan Gabel and Karen Diver**

Celebrate Native American Heritage Month and join in a conversation between President Joan Gabel and Karen Diver, the U of M's first senior advisor to the president for Native American affairs. Find out what Diver hopes to accomplish in her new role—one of only a few such positions in higher education across the country—and learn

how the University and Native American communities and Tribal Nations are working together to build meaningful partnerships. Noon-1 p.m.

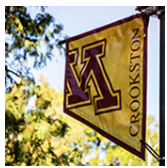

**Nov. 3 - Torch & Shield Award Banquet (Crookston)**

The Torch & Shield Award recognizes individuals who have provided leadership and who have aided in the development of the University of Minnesota Crookston, Northwest Research and Outreach Center, and Extension. It is the highest award an individual can receive from the campus. The banquet begins at 5:30 p.m., Kiehle Auditorium. Learn more and see this year's recipients.

### 15 things to do in November at the Landscape Arboretum

Watch the changing colors of the seasons, explore the Winter Lights display or an art exhibit, pick up new cooking or crafting skills, or just go for a hike at the Arboretum. See 15 things to do in November at the Arb.

## U-WIDE NEWS

### Resilience ahead of the holiday season

According to a recent study, 73 percent of people experience stress that impacts their mental health. As the holiday season approaches, Bakken Center for Spirituality and Healing instructor Mariann Johnson recommends taking a step back to appreciate our resilience and grace during the next few months and beyond.

### Awards and Recognition

The U of M Medical School and the College of Biological Sciences have received $8.5M to expand on their work studying non-dividing cells that increase with age; Rachel Hardeman has been named to the Advisory Committee to the Director of the Centers for Disease Control and Prevention; *U in the News* features highlights of U faculty and staff cited in the media. Awards and Recognition >

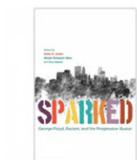

**U of M featured virtual events**

Nov. 3 - Silent and deadly, yet preventable: Stroke prevention, disparities, and links to COVID-19
Nov. 5 - Sparked: George Floyd, Racism, and the Progressive Illusion
Nov. 8 - Building Interdisciplinary Collaborations for Population Dynamics COVID Research
Nov. 9 - The stress of ambiguous loss in a time of pandemic and change
Nov. 9 - What does it mean to be a global campus?
Nov. 9 - TRUTH Project: decolonization and community-engaged scholarship with Tadd Johnson
Nov. 15 - Quie & Peterson Global Health Lecture: While Our Attention Was on COVID: Spillover Effects from a Pandemic

See the full Events Calendar >

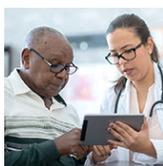

**Getting to the root of health disparities in kidney disease**

Across the United States, 15 percent of adults are estimated to have chronic kidney disease. This widespread disease doesn't take an even toll, however. Kidney failure is up to 3.4 times greater in various racial and ethnic minorities. Through the Advancing Kidney Health through Optimal Medication Management initiative, researchers are developing modules that can both train pharmacists and provide tools to educate patients to increase the safe and effective use of medications to treat the disease.

### Native Voices and Visions

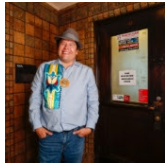

A "Native Voices and Visions" feature highlights the incredible accomplishments of six individuals who are a part of our Native American community. Learn more about Mary Owen, a physician dedicated to increasing the number of Native American and Indigenous healthcare professionals; Korina Barry, an alum who is building Indigenous power through the all-Indigenous organization NDN Collective; and Šišókaduta Joe Bendickson, an instructor helping to preserve the Dakota language for future generations.

**Meet your students where they are**

Half of all University of Minnesota students use the Canvas Student App to access their courses from mobile devices. Instructors can learn about which course materials and activities are suited for mobile. Register for Meet your students where they are: Design your Canvas course to be mobile-friendly, an exploratory session that will provide instructors with the student perspective of course materials and activities.

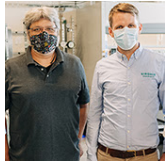

**Startup scales eco-friendly ingredients for cleaning, personal care**

In 2015, researchers in Paul Dauenhauer's laboratory were exploring new technologies for using biomass to make the same chemical building blocks needed to manufacture plastics, rubber, detergents, and more that are normally derived from fossil fuels. In the process, they discovered something exciting. One of the plant-based compounds did more than replace its petroleum-based counterpart; it actually worked better. This ultimately led to the startup company Sironix Renewables.

## Crookston

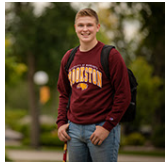

**Crookston student Olander named recipient of FFA award**

Sophomore Ben Olander, Staples, MN, has been named a recipient of a 2021 Agricultural Proficiency Award during the 94th National FFA Convention & Expo. Agricultural Proficiency Awards honor students who have developed specialized skills that they can apply toward their future careers. Olander received the award in the category of Agricultural Mechanics Design and Fabrication—Entrepreneurship/Placement. Learn more about Ben Olander.

## Duluth

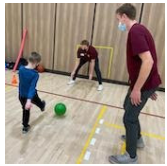

**Collaboration between UMD and area schools provides needed teacher training**

UMD students in the Developmental Adapted Physical Education (DAPE) minor will fill a growing need for physical education teachers who are trained to teach children with special needs. Under Professor Daehyung Lee, DAPE minors have been working with area students as part of a practicum class. "This partnership is making the DAPE program really special—and I believe also very sustainable," says Lee.

**Video: Treyvon Cahalan, UMD Football**

As the center for UMD Football, Treyvon Cahalan is in the trenches making line calls. In the classroom, he's pursuing a degree in marketing while looking ahead to an MBA. Although he describes himself as "a quieter, shyer guy," he's not afraid to communicate on the field. Watch this video about Treyvon Cahalan.

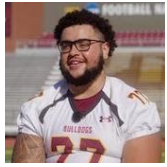

### Student turned diplomat

[Erin Cain completed an internship with the U.S. State Department in summer 2021.](#) Cain, majoring in political science and philosophy with a minor in German studies, worked virtually with the U.S. consulate in Frankfurt, Germany. "I think the state department made me more interested in how problems are expressed globally," she says.

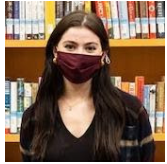

## Morris

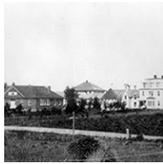

### Nov. 6 - Teach-In: Impacts and History of Native American Boarding Schools

Before there was a University of Minnesota presence in Morris, the site housed an American Indian boarding school first run by the Sisters of Mercy community of the Catholic Church and later operated directly by the U.S. government. The campus was transferred to the State of Minnesota in 1909, with the stipulation that at successor educational institutions, American Indian students be admitted on terms of equality with other students and attend "free of charge for tuition," a policy still honored. [Learn more about this history at this Teach-In.](#) Presentations will be available at 109 Imholte Hall and via Zoom.

### Nov. 4-6 - Theatre Discipline presents *The Wolves*

The [University of Minnesota Morris Theatre Discipline will present the critically acclaimed play The Wolves.](#) A slice-of-life play written by Sarah DeLappe, which was a finalist for the 2017 Pulitzer Prize for Drama, *The Wolves* explores the undeniable power of friendship, community, power, politics, individuality, and loss among nine teenage girls on a soccer team.

## Rochester

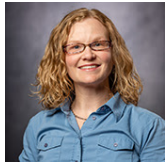

### In pursuit of inclusive pedagogies and equitable participation in STEM classrooms

UMR's interdisciplinary [Center for Learning Innovation](#) drives educational innovation with a faculty research focus on student learning and development. UMR associate professor Kelsey Metzger's recently completed "Warming up the cold call: In pursuit of inclusive pedagogies and equitable participation in STEM classrooms" is soon to be published in *American Biology Teacher*.

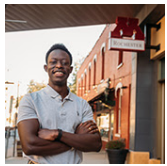

### Personal experiences push UMR student to pursue a career in medicine

Pal Koak, a third-year student at the University of Minnesota Rochester, is a testament to the values of patience and determination as well as the resilience of the human spirit. Koak is now using his energy and passion to pursue a health sciences degree with the goal of one day becoming a physician-scientist. [Read more at *Med City Beat*.](#)

Twin Cities

### Nov. 16 - Kusske Lecture & Dialogue Series: Frank Gehry

Internationally acclaimed architect Frank Gehry will be the inaugural speaker in the College of Design's new Kusske Lecture & Dialogue Series. Gehry is widely regarded as one of the most influential architects of our time. No charge and open to the public. 7 p.m...

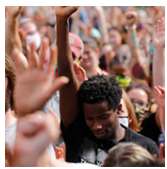

### Nov. 17 exhibit opening - 'Documenting a Reckoning: The Murder of George Floyd'

The death of George Floyd at the knee of a police officer brought demonstrations and protests that swept the Twin Cities and the world. It inspired photojournalists and other photographers to create images with lasting impact. This exhibit brings 54 of those images together to help create a broader sense of the influence of these events on our lives, as documented by professional, community, and student photographers representing diverse viewpoints. 5-7:30 p.m., Elmer Andersen Library.

### New guidance will help designers create safe, effective bike facilities

Many cities across the U.S. are rapidly adding separated bike lanes (SBLs) to their transportation networks. The popularity of these specialized pathways has increased because of their many benefits: reducing crash risk, increasing safety and comfort, and encouraging more people to use bicycles as transportation. Despite the growing demand for SBLs, current design guidance was incomplete. In a recent research project, U of M researchers filled gaps in SBL design guidance.

### Nov. 17 - Panel: Using Rubrics to Drive Writing Instruction

Well-designed rubrics make performance expectations explicit and demonstrate what instructors value when it comes to student writing. This panel brings together faculty who have used rubrics and grading sheets to inform their instruction and to develop activities to help students meet the learning goals of their courses. Noon-1:30 p.m., 530A Bruininks Hall.

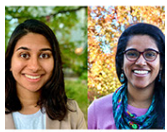

### Students further community connection education

Aditi Kulkarni and Prasanna Vankina, second-year medical students, created an orientation module centered around the history, culture, and geography of the Twin Cities for first-year medical students at the University.

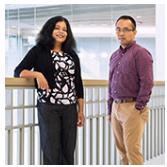

### Understanding the developing brain

School of Public Health biostatistics faculty members Saonli Basu and Mark Fiecas are co-leading the data analytics core at the newly opened Masonic Institute for the Developing Brain. In this Q&A, Basu and Fiecas talk about their combined expertise in the statistical and computational tools that handle big data gism them involved in the institute and their own quest to head off mental health struggles in adolescents.

### Mapping out the life and ideas of a champion of the humanities

Edward Said (pronounced "Sah-EED") played many roles during his eventful life. To Timothy Brennan, professor of cultural studies and comparative literature in the College of Liberal Arts, the late intellectual's greatest accomplishment was in promoting the humanities and demonstrating the value of academic experts to societal discussions. Earlier this year, Brennan published the first comprehensive biography of Said: *Places of Mind: A Life of Edward Said*.

### Rooted in research

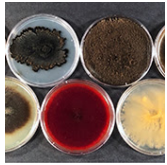

It turns out, the way emerald ash borer (EAB) manages to kill trees is just as complicated as the labyrinthine markings the beetle leaves behind. Nickolas Rajtar—a PhD student in the lab of Robert Blanchette, a professor in the Department of Plant Pathology—is unsnarling the twisty-turny mess of EAB by taking a closer look at the fungal communities that trail behind the pest.


This employee (from the Crookston campus) is interested in art while the organization wants them to know about resources and social justice updates. The employee is in A4 x B3 x C5 category (prioritizing employee-preferred messages in the subject line, and organization-preferred messages in the top news, with a zip order of organization-preferred and employee-preferred messages in the rest sections. 
\chapter{Glossary and Acronyms}
\label{app_glossary}


\section{Glossary}
\label{jargonapp}
Here we define a list of terms we used in this thesis and show the first positions they appeared in this thesis.

\begin{itemize}
    \item Bulk / Mass Email: email that is sent to a large group of recipients (1.1).
    \item Organizational (Internal) Email: emails whose senders and recipients are within the same organization (2.2).
    \item Organizational Bulk Email: bulk emails whose senders and recipients are within the same organization (1.1).
    \item External Bulk Email: the bulk emails sent to the
recipients outside organizations (2.3).
    \item Bulk Email Platform: the business platforms supporting bulk email senders in designing and targeting bulk emails, such as Salesforce, Mailchimp (2.3.3).
    \item Organizational Bulk Email System: the systems that generate, design,
distribute, process, and evaluate organizational bulk emails (1.1).
    \item Communicators (information gatekeeper): the staff in charge of designing and distributing organizational bulk emails (1.1).
    \item Communicators’ clients (information producer, original sender): the original sources of organizational bulk emails, who request the communicators to distribute the information, e.g., leaders (1.1).
    \item Employees (information recipient): the organization’s staff who receive organizational bulk emails from the communicators (1.1).
    \item Management: the direct managers of the employees (1.1).
    \item Senders: communicators and their clients (1.1).
    \item Newsletter: the periodic organizational bulk emails (4.4).
    \item Bulk Message: the single story / piece of information in bulk emails (1.1).
    \item (Communication) Channel: a bulk email's sender and brand. They could be newsletters’ names or university leaders’ (offices’) names. They are characteristics visible to recipients before opening (5.3).
    \item Reputation of channel: the recipients’ perception of the credibility, quality, and relevance of the emails sent through this channel (5.2).
    \item Email Tree: the process of distributing a message through
the management structure from higher-level managers down to lower-level management and eventually to individual employees (5.3.3).
    \item Bulk Message's Value: different stakeholders' understandings on the value of a bulk message being work-relevant / important / interesting (5.4.4)
    \item Bulk Message's Cost: communicators' time on editing bulk messages, employees' time on reading bulk messages, organization's money cost (5.4).
    \item Email-level metrics: the performance / cost metrics of a bulk email (8.1).
    \item Message-level metrics: the performance / cost metrics of a bulk message (8.1).
    \item Message-level reading time: recipient's time spent on reading a message (7.1).

\end{itemize}













\chapter{Other Publications}
\label{other_publication}



\section{Other Conference Papers and Book Chapters}
\begin{itemize}
    \item Guy Aridor, Duarte Gon calves, Daniel Kluver, Ruoyan Kong, and Joseph Konstan. The economics of recommender systems: Evidence from a field experiment on movielens. In ACM Transactions on Economics and Computation (EC2023).,
2023 \cite{aridor2022economics}
    \item Yunzhong He, Cong Zhang, Ruoyan Kong, Chaitanya Kulkarni, Qing Liu, Ashish
Gandhe, Amit Nithianandan, and Arul Prakash. Hiercat: Hierarchical query
categorization from weakly supervised data at facebook marketplace. In Companion Proceedings of the ACM Web Conference 2023, WWW ’23 Companion, page 331–335, New York, NY, USA, 2023. Association for Computing Machinery. ISBN 9781450394192. doi: 10.1145/3543873.3584622. \cite{10.1145/3543873.3584622}
    \item Charles Chuankai Zhang, Mo Houtti, C. Estelle Smith, Ruoyan Kong, and Loren
Terveen. Working for the invisible machines or pumping information into an empty
void? an exploration of wikidata contributors’ motivations. Proc. ACM Hum.-
Comput. Interact., 6(CSCW1), apr 2022. doi: 10.1145/3512982 \cite{10.1145/3512982}
    \item Hongke Zhao, Qi Liu, Yong Ge, Ruoyan Kong, and Enhong Chen. Group preference aggregation: A nash equilibrium approach. In 2016 IEEE 16th International
Conference on Data Mining (ICDM), pages 679–688. IEEE, 2016 \cite{zhao2016group}
    \item Ruoyan Kong and Joseph A Konstan. Socially-embedded agents in organizational
contexts – bulk email as an example. In Intelligent Systems in the Workplace:
Design, Applications, and User Experience. Springer Publishing, 2023 \cite{intelligent}
\end{itemize}

\section{Other Workshop Papers}
\begin{itemize}
    \item Ruixuan Sun, Ruoyan Kong, Qiao Jin, and Joseph A. Konstan. Less can be more:
Exploring population rating dispositions with partitioned models in recommender
systems. In Adjunct Proceedings of the 31th ACM Conference on User Modeling,
Adaptation and Personalization, UMAP ’23 Adjunct, New York, NY, USA, 2023.
Association for Computing Machinery \cite{less}
    \item Ruoyan Kong, Zhanlong Qiu, Yang Liu, and Qi Zhao. Nimblelearn: A scalable
and fast batch-mode active learning approach. In 2021 International Conference
on Data Mining Workshops (ICDMW), pages 350–359. IEEE, 2021 \cite{kong2021nimblelearn}
    \item Ruoyan Kong, Ruobing Wang, and Zitao Shen. Virtual reality system for invasive
therapy. In 2021 IEEE Conference on Virtual Reality and 3D User Interfaces
Abstracts and Workshops (VRW), pages 689–690. IEEE, 2021 \cite{kong2021virtual}
\end{itemize}













\bibliography{references.bib}
\end{document}